\documentclass[twocolumn,amsmath,amssymb]{revtex4}
\usepackage{hyperref}

\usepackage[dvips]{graphicx}

\newcommand{\eh}{\ensuremath{158 A}~GeV }

\newcommand{\Pb}{Pb+Pb~}

\begin{document}

\title{Energy Dependence of Multiplicity Fluctuations\\
 in Heavy Ion Collisions at the CERN SPS}

\author{
C.~Alt$^{9}$, T.~Anticic$^{23}$, B.~Baatar$^{8}$,D.~Barna$^{4}$,
J.~Bartke$^{6}$, L.~Betev$^{10}$, H.~Bia{\l}\-kowska$^{20}$,
C.~Blume$^{9}$,  B.~Boimska$^{20}$, M.~Botje$^{1}$
% ...
\\
\vspace{0.5cm}
\noindent
$^{1}$NIKHEF, Amsterdam, Netherlands. \\
% ...
$^{4}$KFKI Research Institute for Particle and Nuclear Physics, Budapest, Hungary.\\
% ...
$^{6}$Henryk Niewodniczanski Institute of Nuclear Physics, Polish Academy of Sciences, Cracow, Poland.\\
% ...
$^{8}$Joint Institute for Nuclear Research, Dubna, Russia.\\
$^{9}$Fachbereich Physik der Universit\"{a}t, Frankfurt, Germany.\\
$^{10}$CERN, Geneva, Switzerland.\\
% ...
$^{20}$Institute for Nuclear Studies, Warsaw, Poland.\\
% ...
$^{23}$Rudjer Boskovic Institute, Zagreb, Croatia.\\
% ...
}

\begin{abstract}

Multiplicity fluctuations of positively, negatively and all charged hadrons
in the forward hemisphere were studied in central Pb+Pb collisions 
at $20A$, $30A$, $40A$, $80A$ and $158A$ GeV. The multiplicity distributions 
and their scaled variances $\omega$ are presented in dependence of 
collision energy as well as of rapidity and transverse momentum.
The distributions have bell-like shape and  their scaled variances
are in the range from 0.8 to 1.2 without any significant structure in their
energy dependence.  No indication of the critical point in fluctuations are 
observed.
The string-hadronic model UrQMD significantly overpredicts the mean,
but approximately reproduces the scaled variance of the multiplicity 
distributions. The predictions of the statistical hadron-resonance gas 
model obtained within the grand-canonical and canonical ensembles 
disagree with the measured scaled variances.
The narrower than Poissonian multiplicity fluctuations measured in 
numerous cases may be explained by the impact of conservation laws on fluctuations in relativistic {systems}.

\end{abstract}

\maketitle

\section{Introduction}

{
In matter of high energy densities ($\approx$ 1~GeV/fm$^3$) a
phase transition is expected between hadrons and a state of
quasi-free quarks and gluons, the quark gluon plasma (QGP)~\cite{Collins:1974ky,Shuryak:1980tp}.
Measurements indicate that this critical energy density
is exceeded at top SPS~\cite{Margetis:1994tt,Heinz:2000bk} and RHIC~\cite{Adams:2005dq,Adcox:2004mh,Back:2004je,Arsene:2004fa} energies during the
early stage of heavy ion collisions. Moreover, the energy
dependence of various observables shows anomalies at low
SPS energies which suggest the onset of deconfinement
around 30$A$ GeV beam energy in central Pb+Pb collisions~\cite{Afanasiev:2002mx, Alt:2007fe,Gazdzicki:1998vd}. }

It was predicted \cite{Gazdzicki:2003bb} that the onset of deconfinement {can} lead 
to a non--monotonic behaviour of multiplicity fluctuations.
Lattice QCD calculations suggest furthermore the existence of a critical point in the phase diagram of strongly interacting matter which separates the 
line of first order phase transition at high baryo-chemical potentials and low temperature from a crossover at low baryo-chemical
potential and high temperature. 
An increase of multiplicity fluctuations near the critical point of strongly interacting matter is expected~\cite{Stephanov:1999zu}.

In statistical models the widths of the multiplicity distributions depend on the conservation laws which
the system obeys.
Even though for different statistical ensembles the mean multiplicity is the same for sufficiently large volumes 
this is not necessarily so for
higher moments of the multiplicity distribution hence multiplicity fluctuations~\cite{Begun:2004gs}.
Fluctuations are largest in the grand-canonical ensemble, 
where all conservation laws are fulfilled only on
average and not on an event-by-event basis.
The multiplicity fluctuations are much smaller in the canonical ensemble,
where the electric and baryonic charges as well as strangeness are globally conserved.
The smallest fluctuations are obtained within the micro-canonical ensemble, 
for which the charges as well as total energy and momentum are conserved.
It should be underlined that in
non-relativistic gases the situation is very different, namely
particle number is conserved in the micro-canonical and canonical ensembles and
consequently  the {total} multiplicity in these ensembles does not fluctuate.

These theoretical considerations motivated vigorous 
theoretical~\cite{Begun:2004gs, Begun:2004pk, Begun:2006uu, Lungwitz:2007uc,Konchakovski:2007ah} 
and experimental 
studies of multiplicity fluctuations in high energy nuclear collisions.

Results on the centrality dependence of multiplicity fluctuations in \Pb collisions
obtained by the NA49~\cite{Alt:2006jr} and WA98~\cite{Aggarwal:2001aa}
collaborations at top SPS energy show an increase
of multiplicity fluctuations with decreasing centrality of the collision in the forward hemisphere.
A similar increase of multiplicity fluctuations is observed at midrapidity by the 
PHENIX~\cite{Mitchell:2005at,Homma:2007qh} collaboration at RHIC energies.

Transverse momentum fluctuations~\cite{Anticic:2003fd} also show a non-monotonic 
dependence on system size. They increase from p+p to Si+Si and peripheral
Pb+Pb collisions and decrease from peripheral to central Pb+Pb collisions.
Possible relations to multiplicity fluctuations are discussed
in~\cite{Mrowczynski:2004cg,Cunqueiro:2005hx}. Preliminary results of NA49 on the
energy dependence of transverse momentum fluctuations~\cite{Grebieszkow:2007xz} in central Pb+Pb collisions indicate a
constant behaviour. 


This paper presents the dependence 
of multiplicity fluctuations 
on energy as well as on rapidity and transverse momentum for
the most central \Pb collisions at $20A$, $30A$, $40A$, $80A$ and $158A$ GeV
as measured by the NA49 experiment at the CERN SPS.

The paper is organized as follows. 
In chapter~\ref{c_measure} the notation and definitions are presented.
In chapter~\ref{c_NA49} the NA49 experiment and the experimental procedure of selecting 
events and tracks used for this analysis
is described.
In chapter~\ref{mult_res} the experimental results on multiplicity 
fluctuations are shown as a function of energy, rapidity and transverse momentum~\cite{Lungwitz:2007cy}.
These results are compared to the predictions of 
the hadron-resonance gas model~\cite{Begun:2006uu} and the string-hadronic
model UrQMD~\cite{Bass:1998ca}   in chapter~\ref{c_modcomp}. 
Furthermore the measurements are also discussed with respect to the search for 
the onset of deconfinement and the critical point.
The paper ends with a summary in chapter~\ref{c_summary}.

\section{Measure of Multiplicity Fluctuations}\label{c_measure}

Let $P(n)$ denote the probability
to observe a particle multiplicity $n$ ($\sum_n P(n) = 1$)
in a high energy nuclear collision. 

The scaled variance $\omega$ used in this paper as a measure 
of multiplicity fluctuations {is commonly used in elementary and heavy 
ion collisions, both for theoretical (see e.g. Refs.~\cite{Mishustin:2006ka,Konchakovski:2007ss,Lungwitz:2007uc,Begun:2006uu}) 
and experimental (see e.g. Refs.~\cite{Heiselberg:2000fk,Aggarwal:2001aa,Alt:2006jr,Mitchell:2005at}) studies.
It} is defined as
\begin{equation}\label{omega}
\omega=\frac{Var(n)}{\left<n\right>}=\frac{\left<n^2\right>-\left<n\right>^2}{\left<n\right>},
\end{equation}
where $Var(n)=\sum_n (n-\left<n\right>)^2 P(n)$ and $\left<n\right>=\sum_n n P(n)$ 
are variance and mean of the multiplicity distribution, respectively.

In a superposition model $\omega$ is the same in $A+A$ collisions as in nucleon-nucleon interactions at the same energy per nucleon provided 
the number of particle producing sources does not fluctuate from event to event.
{String-hadronic models predict similar values of $\omega$ for p+p and Pb+Pb 
collisions~\cite{Konchakovski:2007ss,Lungwitz:2007uc}.
In a hadron-gas model~\cite{Begun:2006uu} the scaled variance converges quickly to a constant value with 
increasing volume of the system. 
In the special case of 
a} hadron-gas model in the grand-canonical formulation~\cite{Begun:2006uu}, neglecting quantum effects and resonance 
decays, the multiplicity distribution is a Poisson one, namely
\begin{equation}\label{poisson}
P(n)=\frac{\left<n\right>^n}{n!} e^{-\left<n\right>}.
\end{equation}
The variance of a Poisson distribution is equal to its mean, 
and thus the scaled variance is $\omega=1$, independent of mean multiplicity.

If there are no particle correlations in momentum space
and the single particle distribution is independent of particle multiplicity
the scaled variance {of an arbitrary multiplicity distribution} observed in a limited acceptance is 
related to the scaled variance 
in the full phase-space ("4$\pi$") as (see appendix~\ref{a_acc_dep} and Refs.~\cite{Begun:2004gs,Begun:2006uu} for derivation):  
\begin{equation}\label{wscale}
\omega_{acc}=(\omega_{4\pi} -1) p +1,
\end{equation}
where $p$ denotes the fraction of particles measured in the corresponding acceptance.
Note that the dependence described by Eq.~\ref{wscale} is violated if 
effects like resonance decays, 
quantum statistics and energy- momentum conservation introduce correlations in momentum space~\cite{Hauer:2007im}. 

In the following the scaled variances of the multiplicity distributions of positively, negatively 
and all charged hadrons are denoted as $\omega(h^+)$, $\omega(h^-)$ and $\omega(h^\pm)$,
respectively.

\section{The NA49 Experiment}\label{c_NA49}

\begin{figure*}
\includegraphics[width=15cm]{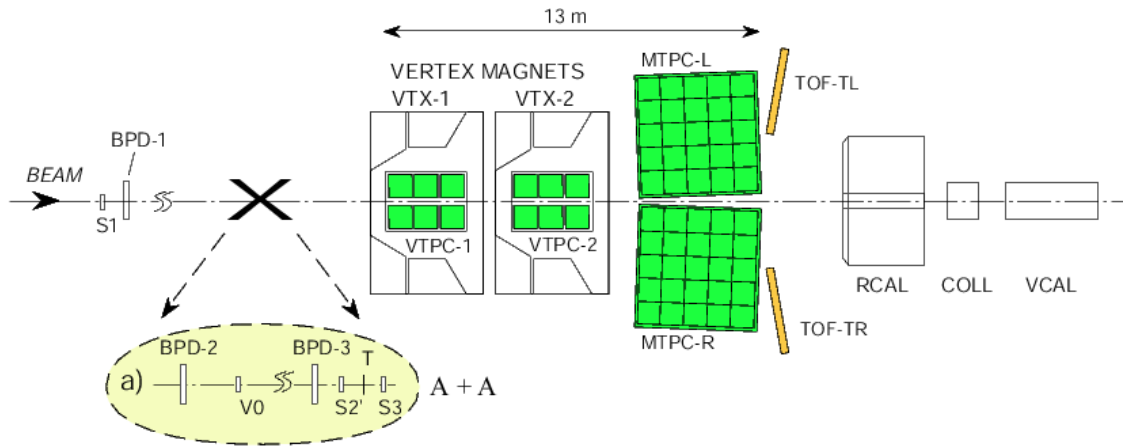}
\caption{\label{na49_setup}(Color online) Setup of the NA49 experiment for Pb+Pb collisions, see text for more details.}
\end{figure*}
The NA49 detector~\cite{Afanasev:1999iu} (see Fig.~\ref{na49_setup})
is a large acceptance fixed target hadron spectrometer. Its main devices are four large 
volume time projection chambers (TPCs).
Two of them, called vertex-TPCs (VTPC-1 and 2), are located in two superconducting dipole magnets (VTX-1 and 2) with a total bending power up to $7.8$ Tm. 
The magnetic field used at \eh (B(VTX-1)$\approx 1.5$ T) and B(VTX-2)$\approx 1.1$ T) was scaled down in proportion to the beam energy for lower energies.
The other two TPCs (MTPC-L and MTPC-R), called main-TPCs,
are installed behind the magnets on the left and the
right side of the beam line allowing precise particle tracking.
The measurement of the energy loss $dE/dx$ in the detector gas provides
particle identification in a large momentum range. It is complemented 
by time of flight (TOF) detectors measuring particles at mid-rapidity.
In this analysis $dE/dx$ information is used only to reject electrons.

The target is located $80$ cm upstream of the first vertex TPC. 
{The target thickness is $0.2$ mm ($0.224$ $\rm{g/cm^2}$) for $20A$ -- $80A$ GeV and $0.3$ mm 
($0.336$ $\rm{g/cm^2}$) for $158A$ GeV.
Using $7.15$ barn as the inelastic cross-section for Pb+Pb collisions this yields an interaction
probability of $0.46\%$ and $0.7\%$, respectively. The interaction length of the strong interaction
for Pb ions in a Pb target is $4.26$ cm.}

Three beam-position-detectors (BPDs) allow a precise determination of the 
point where the beam hits the target foil. 
The centrality of a collision is determined by measuring 
the energy of projectile spectators
in the downstream veto calorimeter (VCAL, see section~\ref{centsel}). 
The acceptance of the veto calorimeter is adjusted at 
each energy by a proper setup of the collimator
(COLL).

\subsection{Data Sets and Event Selection}
\label{ev_sel}
In this publication the results for central Pb+Pb collisions 
at $20A$, $30A$, $40A$, $80A$ and $158A$ GeV
are presented.
The numbers of events used from these data sets are given in Table~\ref{datasets}. 

\begin{table}
\begin{tabular}{|c|c|}
\hline
energy (GeV)& number of events\\
\hline
\hline
$20A$&	6602\\
\hline
$30A$&	8219\\
\hline
$40A$&	21995\\
\hline
$80A$&	2307\\
\hline
$158A$&	5493\\
\hline
\end{tabular}
\caption{\label{datasets}Statistics for the $1\%$ most central collisions used for this analysis at different beam energies.}
\end{table}

In order to get a "clean" sample of events excluding for instance 
collisions outside the target or event pileup,
the following event selection criteria are applied to data:
\begin{itemize}
\item The fit of the interaction point, based on the reconstructed tracks, was successful.
\item The position of the fitted interaction point is close to the position obtained from the beam position detectors.
\item At least $10\%$ of all tracks are used for the reconstruction of the interaction point.
The reconstruction of the interaction point was optimized for precision by selecting long and well
measured tracks in an iterative procedure.
\end{itemize}

The event cuts have a small influence on $\omega$, 
the results differ by less than $1\%$ when only the cut requirement of a successful
fit of the main vertex is used.

Beam lead ions which do not interact strongly in the target produce delta electrons both in
the target foil and the detector gas. These electrons may curl up in the TPCs, increase their occupancy and might therefore
reduce the reconstruction efficiency. In order to avoid this effect only those 
events are selected for the analysis in which there are no beam ions passing through the detector within the read-out time of the event.

\subsection{Centrality Selection}\label{centsel}

Fluctuations in the number of participants lead to an increase of multiplicity fluctuations.
In a superposition model the total multiplicity $n$ is the sum of the number of particles 
produced by $k$ particle production sources:
\begin{equation}\label{mult}
n=\sum_i{n_{i}^{so}},
\end{equation}
{where the summation index $i$ runs over the sources.
Under the assumption of statistically identical sources 
the scaled variance $\omega$ of the multiplicity distribution has two contributions.}
The first is due to the fluctuations of the number of particles emitted by a single source $\omega_{so}$, 
the second is due to the fluctuations in the number of sources $\omega_k$ (see appendix~\ref{a_partfluct} for
derivation):
\begin{equation}\label{omega_supp}
\omega=\omega^{so}+\left<n^{so}\right>\cdot \omega_k,
\end{equation}
where $\left<n^{so}\right>$ is the mean multiplicity of hadrons from a single source.
The fluctuations in the number of sources $\omega_{k}$ can be attributed to fluctuations in the number of projectile and target participants.
In order to minimize the fluctuations of the number of participants
the centrality variation in the ensemble of events should be as small as possible, for which very central collisions are best suited.

\begin{figure}
\centerline{\includegraphics[width=9cm]{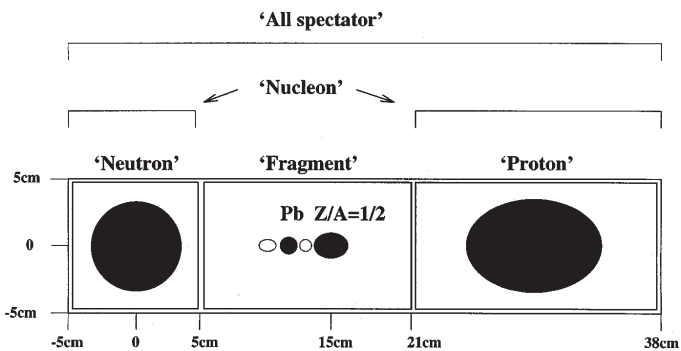}}
\caption{\label{collimator}
A sketch of the horizontal deflection for charged
particles at the front face of the iron collimator for the \eh magnetic field setting. The broadened
distribution of each species is due to the Fermi motion of
nucleons or fragments; additionally, the oval shapes are due to
the deflection of charged particles in the magnetic field. The
sizes of the distributions correspond to one standard deviation.
The open circles in the fragment acceptance represent particles
of Z/A other than one half~\cite{Appelshauser:1998tt}.}
\end{figure}

\begin{table}
\begin{tabular}{|c|c|c|c|c|}
\hline
energy&	\multicolumn{3}{c|}{collimator}&	ring calorimeter\\
&	\multicolumn{2}{c|}{x (cm)}&	y (cm)&	x (cm)\\
\hline
\hline
20&	&	&	&		10\\
\hline
30&	&	&	&		10\\
\hline
40&	$-13$&	$+47$&	$\pm 12$&	17\\
\hline
80&	$-13$&	$+47$&	$\pm 12$&	17\\
\hline
158&	$-5$&	$+38$&	$\pm 5$&	17\\
\hline
\end{tabular}
\caption{\label{coll_settings}Settings of the collimator and the ring calorimeter defining the acceptance of the veto calorimeter for different energies 
with respect to the position of neutrons with zero transverse momentum. See the text for more details.}
\end{table}

In order to fix the number of projectile participants
the NA49 experiment uses the energy in the projectile spectator domain as a measure of centrality, called projectile centrality below.
The downstream veto calorimeter~\cite{DeMarzo:1983gd} of NA49, originally designed for NA5, measures the energy 
carried by the particles in the 
projectile spectator phase space region~\cite{Appelshauser:1998tt}.
A collimator in front of the calorimeter is located $25$~m downstream from the target and is adjusted for each energy in such a way that all projectile 
spectator protons, neutrons and fragments can reach the veto calorimeter. 
For $158A$ GeV the hole in the collimator extends $\pm 5$~cm in vertical direction
and $-5$~cm and $+38$~cm in horizontal direction taking into account the deflection of charged spectators by the magnetic field 
(Fig.~\ref{collimator}, Table~\ref{coll_settings}).
Due to a larger spread of spectators, the hole of the collimator is larger for $40A$ and $80A$ GeV. For $20A$ and $30A$ the collimator is removed
and the ring calorimeter (RCAL in Fig.~\ref{na49_setup}) positioned $18$~m downstream from the target serves as a collimator.

The settings of the hole in the collimator and the position of the ring calorimeter for the different 
energies is shown in Table \ref{coll_settings}.
The zero point is the point where neutrons with no transverse momentum would pass the collimator.
The collimator is not symmetric around the
zero point because the nuclear fragments and spectator protons carry positive charge and are deflected by the 
magnetic field in 
positive x direction. The last column in the table is the position of the center of the ring calorimeter. 
Its hole has a radius of 28 cm.

\begin{figure}
\includegraphics[width=9cm]{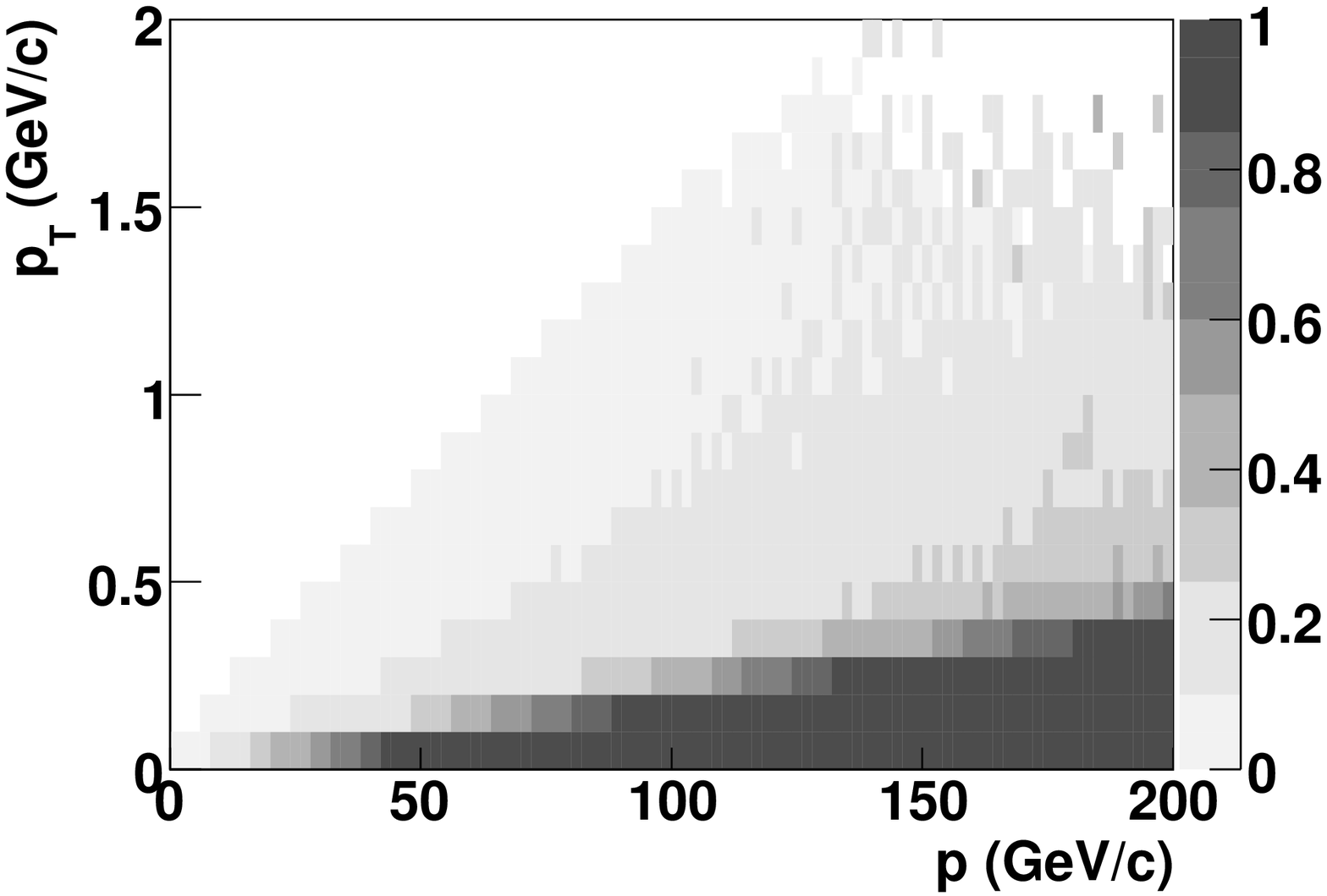}
\includegraphics[width=9cm]{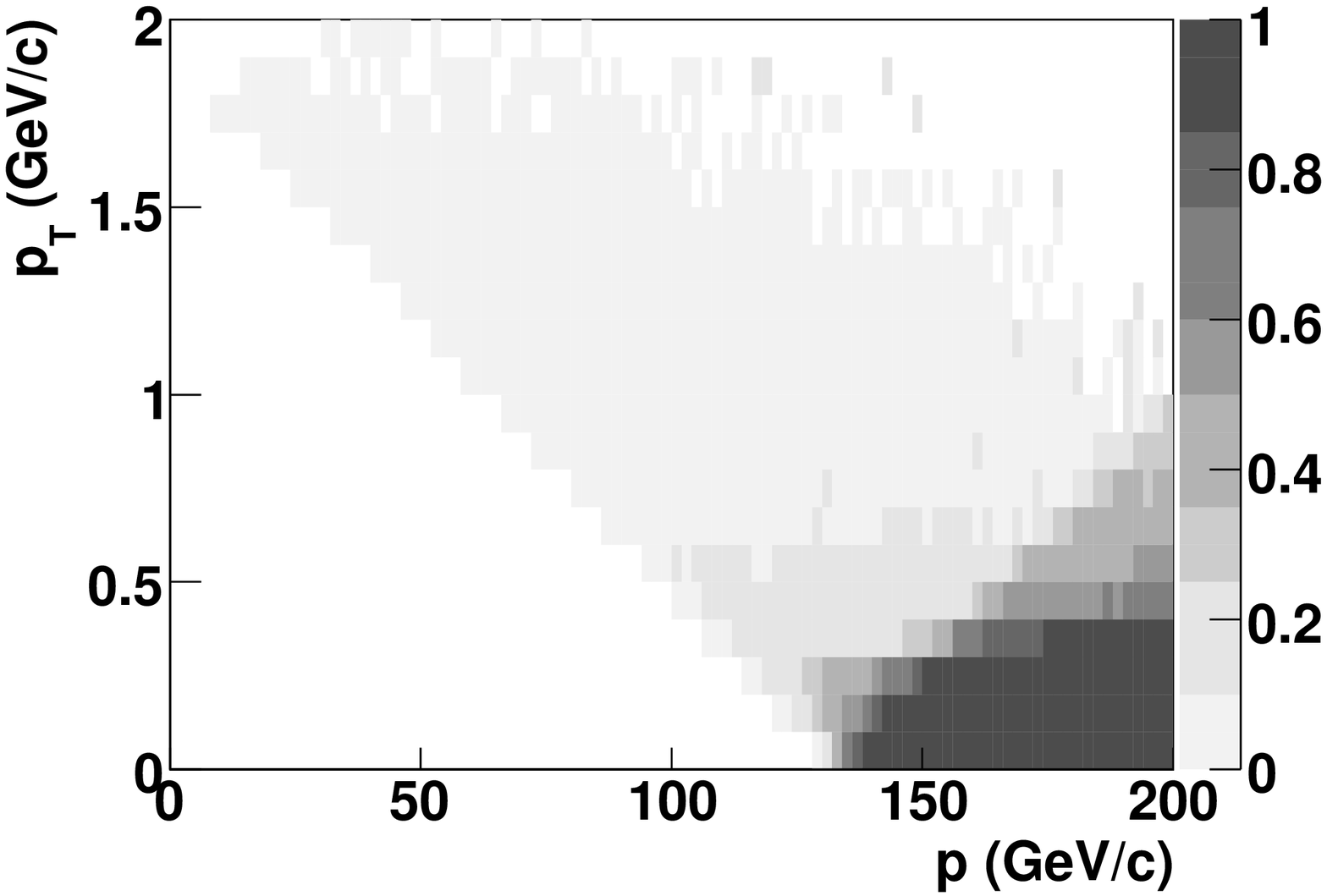}
\caption{\label{veto_acc}Acceptance of the veto calorimeter for neutral (top) and positively charged (bottom) main vertex particles at $158A$~GeV as a function of total 
momentum $p$ and transverse momentum $p_T$.}
\end{figure}

The acceptance of the veto calorimeter for neutral and positive particles for \eh is shown in Fig.~\ref{veto_acc}. Acceptance tables in $p$, $p_T$ and
$\phi$ can be obtained at~\cite{accTables}.

Due to the geometry of the collimator and the magnetic field, a small number of positive and neutral non-spectator particles can hit the veto calorimeter. 
For positively charged particles, the acceptance of the TPCs and the veto calorimeter overlap partly. The maximum amount of a possible auto-correlation is estimated by
a comparison of $\omega(h^+)$ for UrQMD events selected by their veto energy to UrQMD events with a zero impact parameter
in the forward region (Fig.~\ref{w_hp}) and found to be smaller than $3\%$.

The acceptance of the veto calorimeter for negatively charged particles is very small because they are bent by the magnetic field into the direction 
opposite to the one of
the positively charged particles, and the collimator is adjusted to detect positively charged and neutral projectile spectators.

The projectile centrality $C_{Proj}$ of an event with a veto energy $E_{Veto}$ is defined as the percentage of all inelastic events which are as central
or more central than the given event according to the energy deposited in the veto calorimeter by the projectile spectator nucleons.
Smaller $C_{Proj}$ correspond to more central events.
Using the fraction of inelastic cross section 
$C_{trig}=\frac{\sigma_{trig}}{\sigma_{inel}}$ accepted by the trigger
($\sigma_{trig}$ is derived from the target thickness and the interaction rate, $\sigma_{inel}$ is assumed to
be 7.15 barn) and the veto energy distribution
$C_{Proj}$ is given by:
\begin{equation}
C_{Proj}=C_{trig} \cdot \frac{\int_{0}^{E_{Veto}}{dN/dE_{Veto,trig} dE_{Veto}} }{\int_{0}^{\infty} {dN/dE_{Veto,trig} dE_{Veto}}},\label{ctrig_eq}
\end{equation}
where $dN/dE_{Veto,trig}$ is the veto calorimeter energy distribution for a given trigger.

\begin{figure}
\includegraphics[width=9cm]{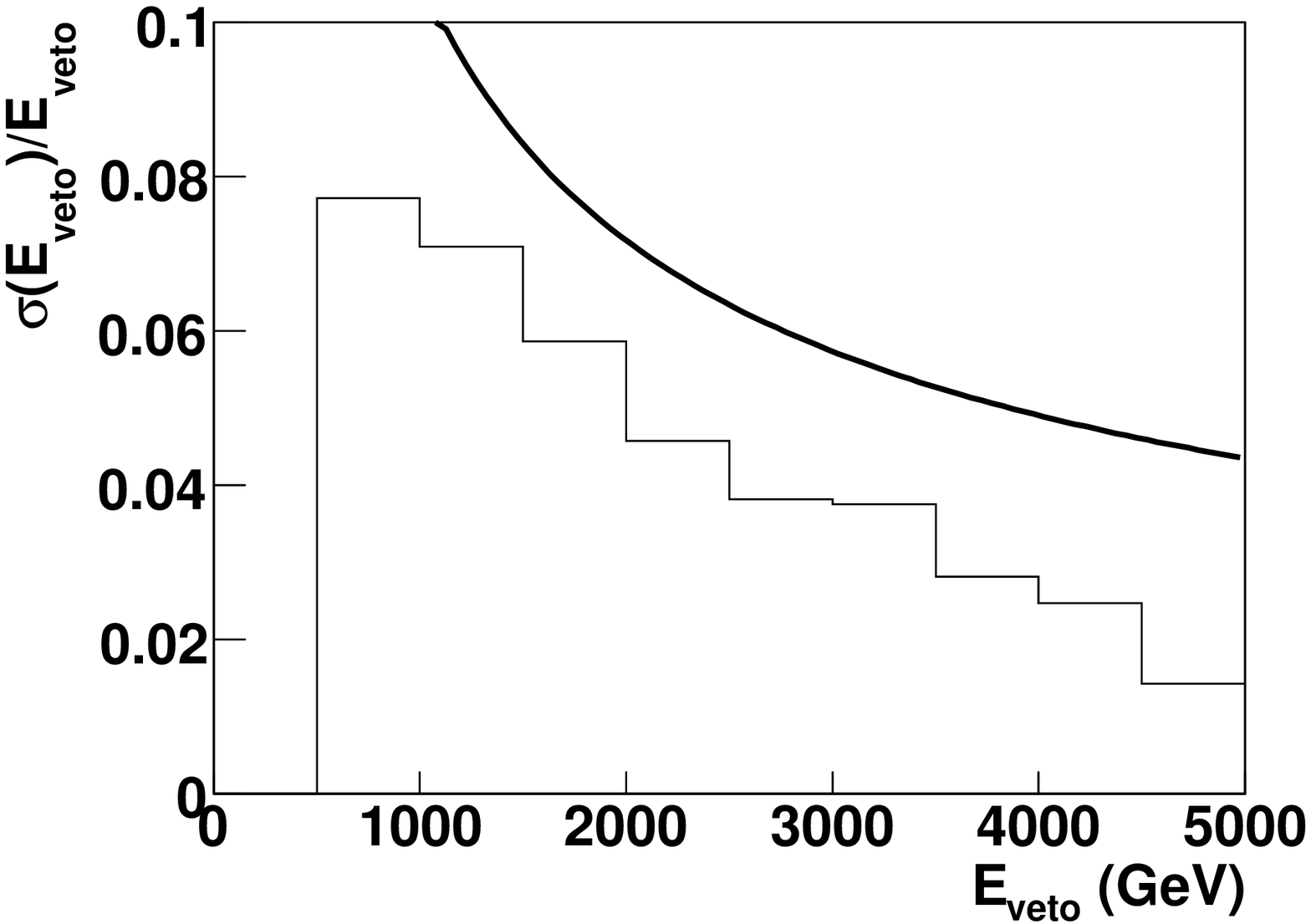}
\includegraphics[width=9cm]{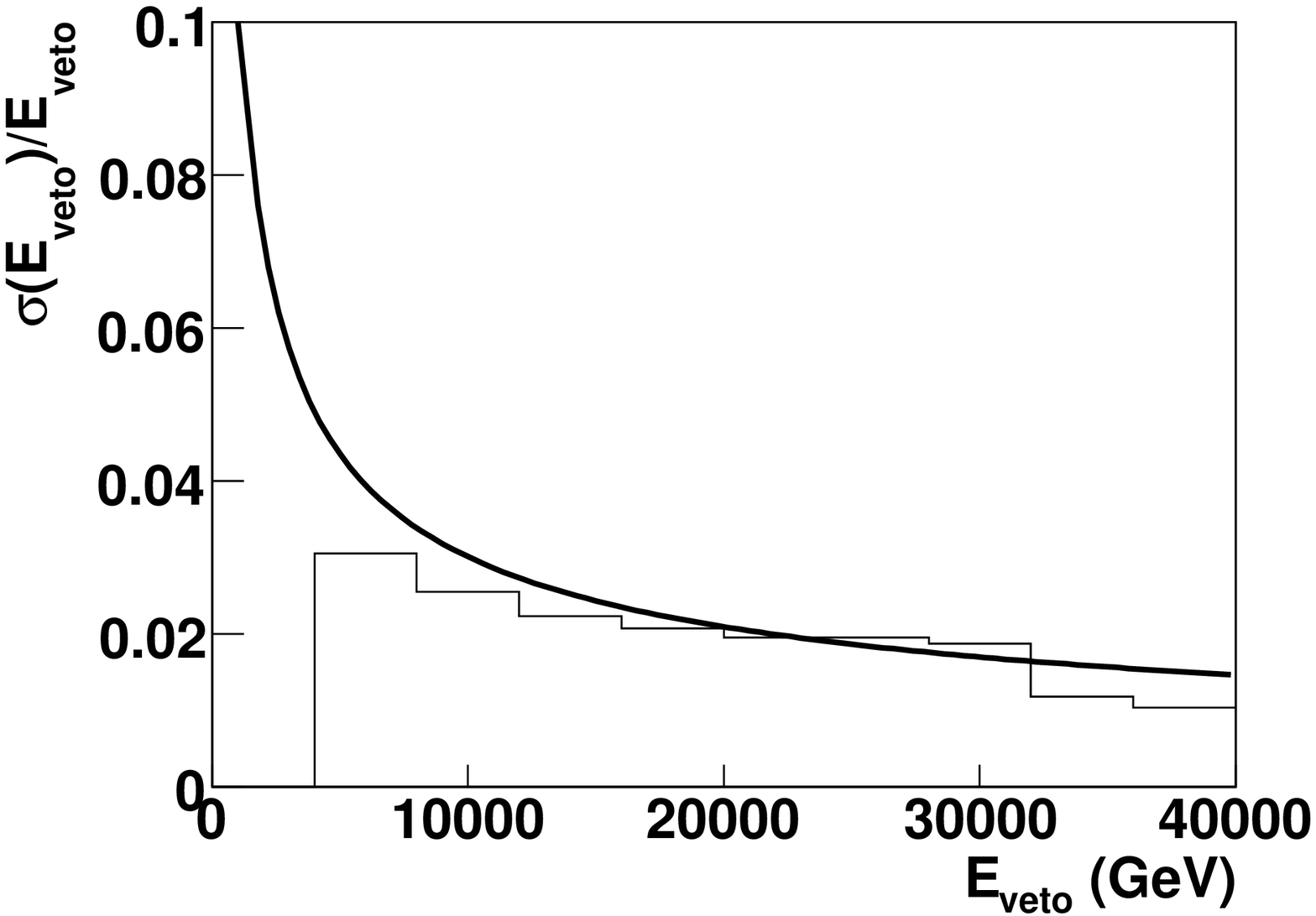}
\caption{\label{vetores_shield}Resolution of the veto calorimeter estimated by a SHIELD simulation (histogram) compared to the parametrization Eq.~\ref{vetores_m}
(solid line) for $20A$ (top) and $158A$~GeV (bottom).}
\end{figure}

The finite resolution of the veto calorimeter causes additional fluctuations in the number of participants.
Based on the analysis of the NA49 Pb+Pb data the resolution of the veto calorimeter was estimated in~\cite{Alt:2006jr} to be:
\begin{equation}
\frac{\sigma(E_{Veto})}{E_{Veto}} \approx \frac{2.85}{\sqrt{E_{Veto}}}+ \frac{16}{E_{Veto}}.\label{vetores_m},
\end{equation}
where $E_{Veto}$ is in units of GeV.
In order to check this parametrization, the distribution of the spectators was simulated by the SHIELD model~\cite{Dementev:1997ca}. 
The SHIELD model delivers both spectator nucleons and nuclear fragments, in contrast to most string hadronic models, which only produce spectator nucleons. 
A simulation performed at $20A$ and $158A$ GeV 
including the geometry of the NA49 detector and the non-uniformity of the veto calorimeter confirms the parametrization given by Eq.~\ref{vetores_m} as an upper 
limit (see Fig.~\ref{vetores_shield}).

The veto calorimeter response can in principle change with time (aging effects, etc.). Therefore a time dependent calibration of the veto energy was applied.
The contribution of this correction to $\omega$ turned out to be very small ($<1\%$, see Table~\ref{syserr_t}).

\begin{figure}
\centerline{\includegraphics[height=8cm]{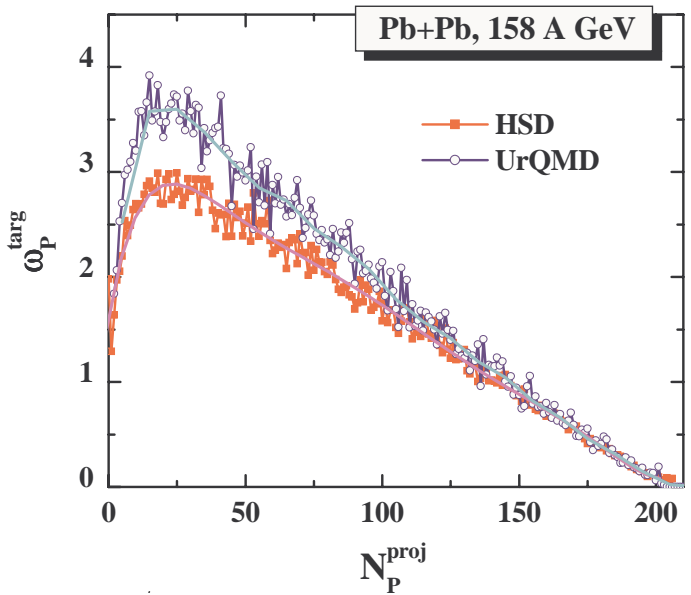}}
\caption{\label{nptarg_fluct}(Color online) Scaled variance of the number of target participants for a fixed number of projectile participants in the UrQMD 
and HSD models. The plot is taken from~\cite{Konchakovski:2005hq}.}
\end{figure}

When fixing the projectile centrality $C_{Proj}$ (Eq.~\ref{ctrig_eq}), thereby fixing the number of projectile participants $N_P^{Proj}$, 
the number of target participants
$N_P^{Targ}$ can still fluctuate.
Thus the total number of participants is not rigorously constant and could contribute to fluctuations.
The fluctuations of the number of target participants obtained by UrQMD and HSD simulations~\cite{Konchakovski:2005hq},
expressed as their scaled variance $\omega_P^{Targ}=Var(N_P^{Targ})/\left<N_P^{Targ}\right>$, are shown in Fig.~\ref{nptarg_fluct}.
For non-central collisions the number of target participants strongly fluctuates, even for a fixed number of 
projectile participants.
{
This is consistent with the increase of $\omega$ with decreasing centrality observed in the forward
hemisphere~\cite{Alt:2006jr,Gazdzicki:2005rr}. However, alternative
explanations also exist~\cite{Rybczynski:2004zi,Cunqueiro:2005hx}.
}

For further analysis the $1\%$ most central collisions (according to their veto energy) are selected in order to minimize the flucutations in the number of participants. 
For these very central collisions, the fluctuation in the number of target participants is expected to be smallest and its scaled variance
$\omega_P^{Targ}$ is expected to be about $0.1$ (see Fig.~\ref{nptarg_fluct}) for an estimated number of target participants of $N_P^{Targ} \approx 192$.


In order to estimate the effect on $\omega$ of target participant fluctuations and
non-spectator particles in the veto calorimeter, the energy dependence of the scaled variance of the multiplicity distribution is 
calculated in the UrQMD 1.3 model both for collisions with zero
impact parameter and for collisions selected according to their veto energy. The resulting difference of $\omega$ in the forward acceptance (see section~\ref{trsel})
is smaller than $2\%$ for 
negatively, smaller than $3\%$ for positively and smaller than $4\%$ for all charged hadrons.
In the midrapidity region the influence of the fluctuations of target participants on $\omega$ is expected to be much larger. Indeed, the 
differences of $\omega$ increase to 
up to $6\%$ for negative, up to $9\%$ for positive and up to $13\%$ for all charged hadrons.

In order to check the influence of the centrality selection, $\omega$ was also determined for the $0.5\%$ most central collisions. The change compared
to the values obtained for the $1\%$ most central collisions is smaller than $3\%$ for positive, $2\%$ for negative and $5\%$ for all charged hadrons.

\subsection{Track Selection}\label{trsel}

Since detector effects like track reconstruction efficiency might have a significant influence on multiplicity 
fluctuations, it is important to select a 
sample of well defined tracks for the analysis. 
The following track selection criteria are used for this analysis and are 
explained in this section:
\begin{itemize}
\item Number of potential points (the number of points a track can have according to its geometry) in the TPCs: $> 30$.
\item The ratio of the number of reconstructed points to the number of potential points: $> 0.5$.
\item Sum of the number of reconstructed points in VTPC-1 and 2: $> 5$.
\item Sum of the number of reconstructed points in VTPC-2 and MTPCs: $> 5$.
\item The track is extrapolated to the plane of the target foil. This point must be closer than 4 cm in x- and 2 cm in y- direction to the interaction point of the
collision.
\item In order to exclude electrons from the analysis, a cut on the energy loss ($dE/dx$) in the detector gas was applied. All tracks with an energy
loss more than 0.2 minimum ionising units higher than the pion $dE/dx$ (in the region of the relativistic rise of the Bethe-Bloch formula) are rejected.
\end{itemize}
{
The reconstruction efficiency is calculated using the embedding method.
Events containing a few tracks were generated and processed by the simulation software. The resulting raw data were embedded into real
events. The combined raw data were reconstructed and the input tracks were matched
with the reconstructed ones.
Embedding simulations show a significant decrease of reconstruction efficiency with increasing event multiplicity in the midrapidity
region at $158A$~GeV using the track selection criteria described above. Therefore for this energy an additional cut was used, 
namely that tracks should have at least $5$ reconstructed points both in VTPC-2 and in the MTPCs. For these tracks no significant dependence
of reconstruction efficiency on track multiplicity is observed.}

Reconstruction inefficiencies mostly occur for tracks with a very low number of points in the TPCs
or for tracks which only have points in the VTPC-1 or in the main TPC. These tracks are not used for this analysis.

{In the following the longitudinal motion of particles is characterized by the rapidity in the center of mass system assuming pion mass of the 
particle. This measure is called pion rapidity and is denoted as $y(\pi)$.}

\begin{figure}
\includegraphics[height=4.5cm]{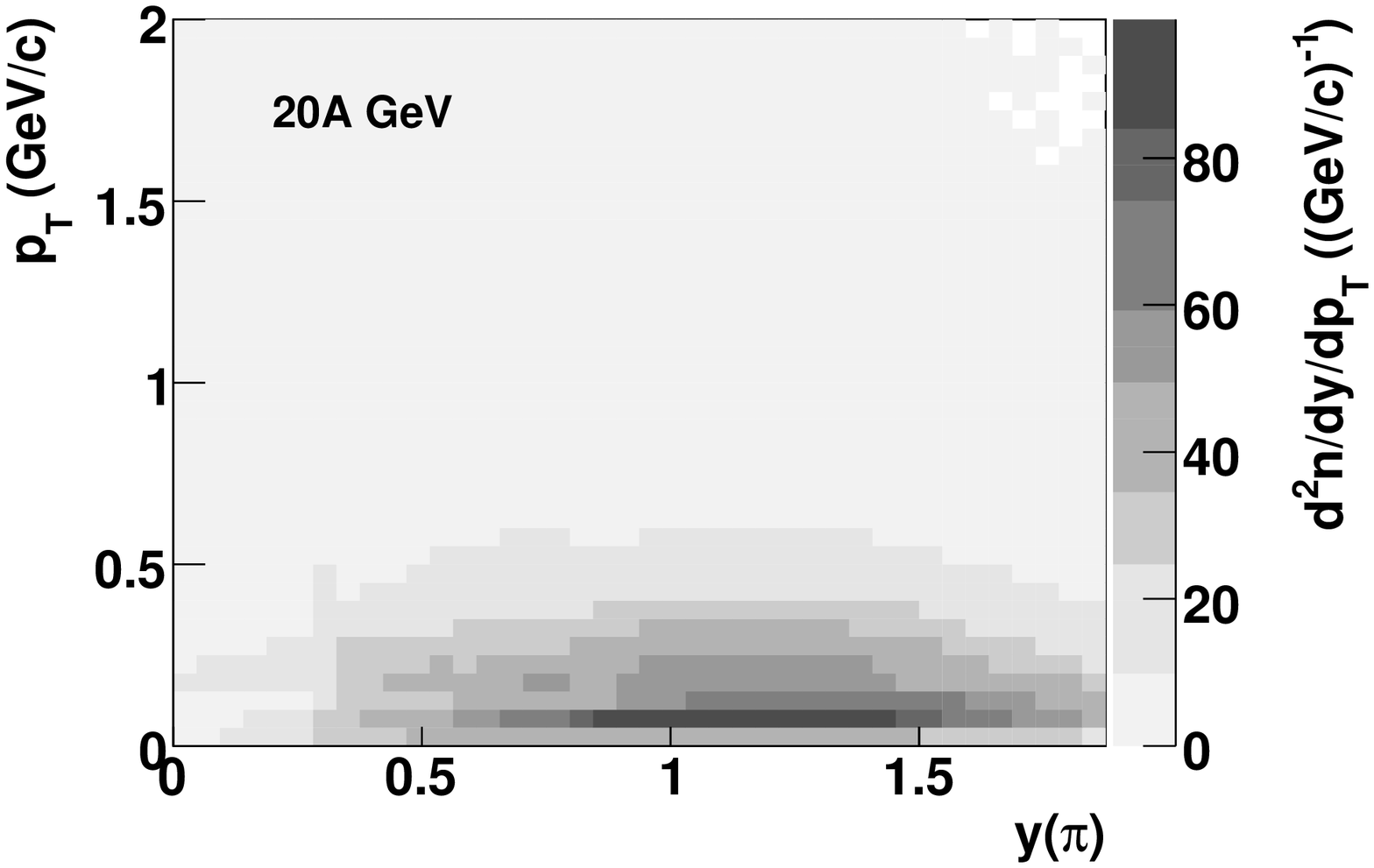}
\includegraphics[height=4.5cm]{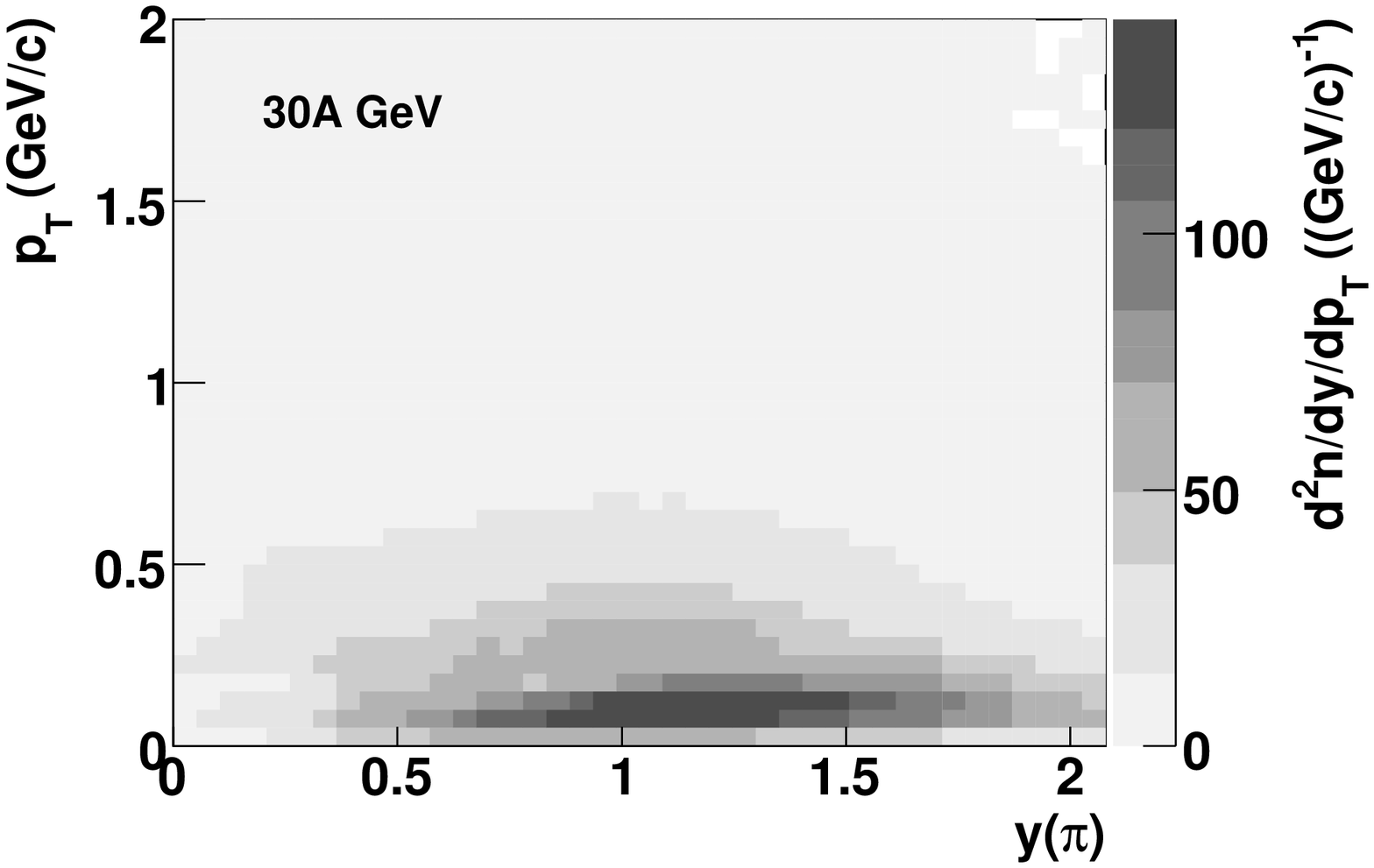}\\
\includegraphics[height=4.5cm]{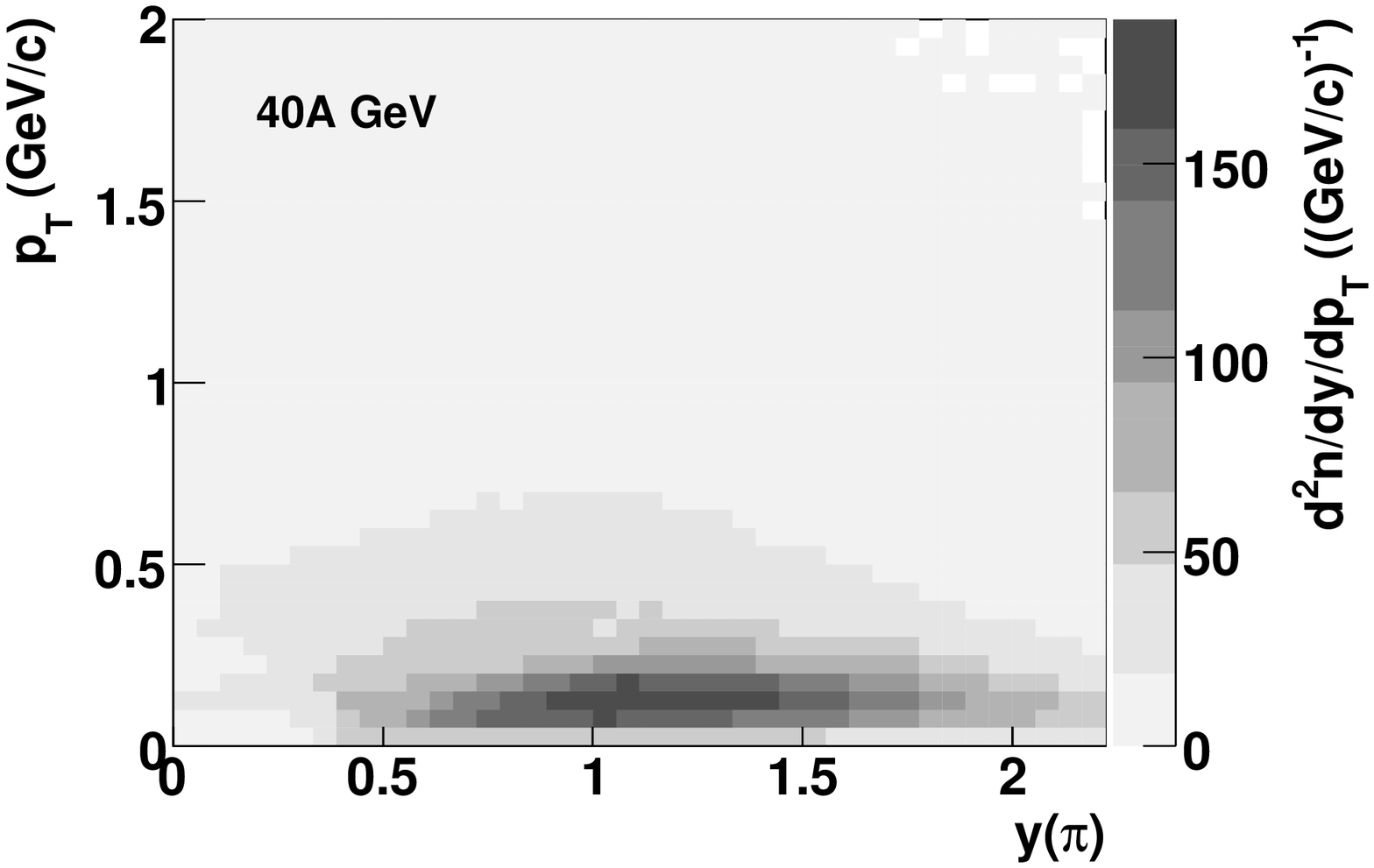}
\includegraphics[height=4.5cm]{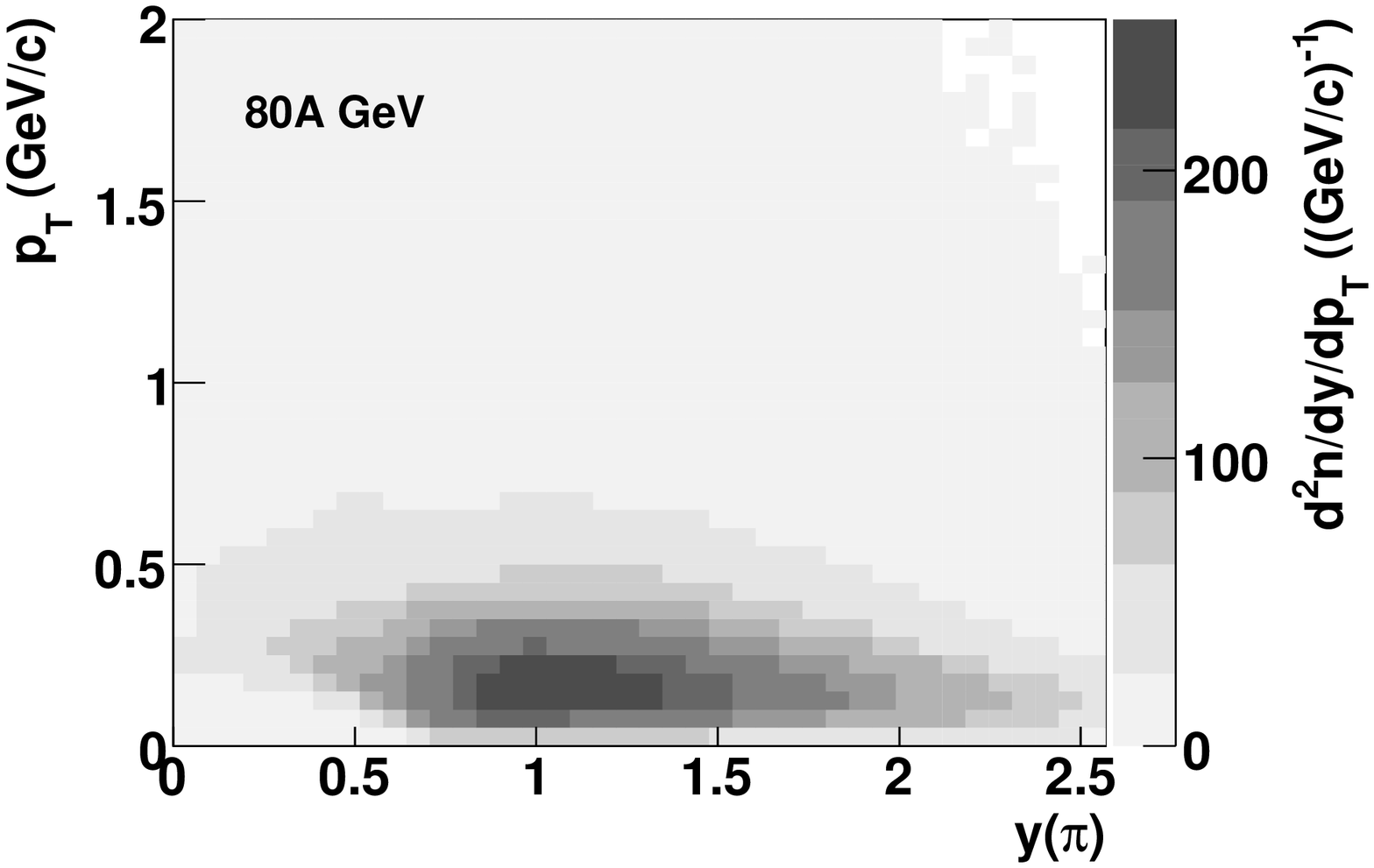}\\
\includegraphics[height=4.5cm]{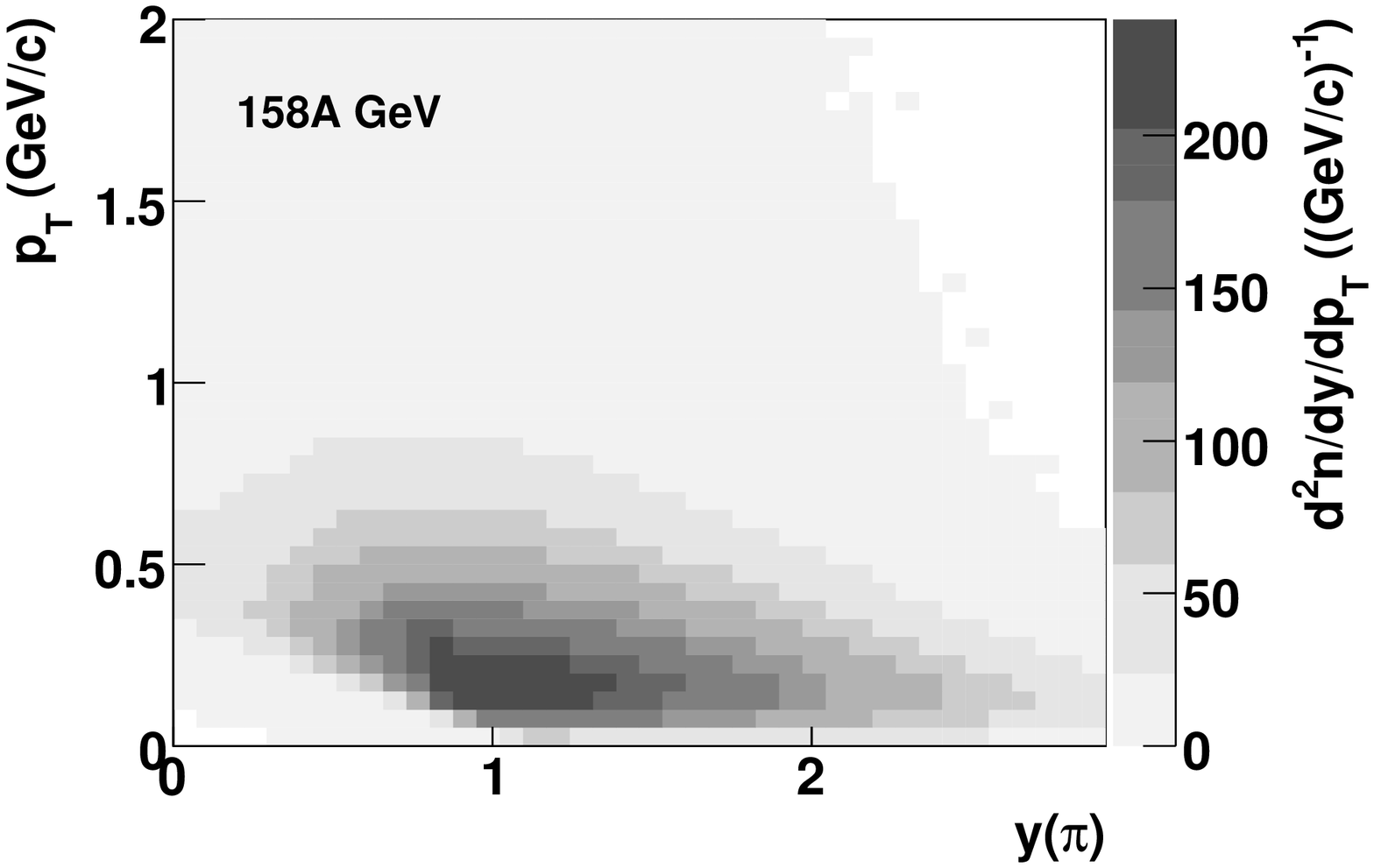}\\
\caption{\label{accept_ypt}Distribution of detected negatively charged particles which fulfill the track selection criteria as a function of 
$y(\pi)$ and $p_T$ for $20A$ (top), $30A$, $40A$, $80A$ and $158A$ GeV (bottom).}
\end{figure}

The distributions of the registered tracks after applying the track selection criteria are shown in Fig.~\ref{accept_ypt} as a function of 
pion rapidity $y(\pi)$ and transverse momentum $p_T$. 
Acceptance tables in $y(\pi)$, $p_T$ and $\phi$ can be obtained from~\cite{accTables}. 
Only tracks in the rapidity interval starting at midrapidity and ending at beam rapidity are used. 

\begin{figure}
\includegraphics[width=9cm]{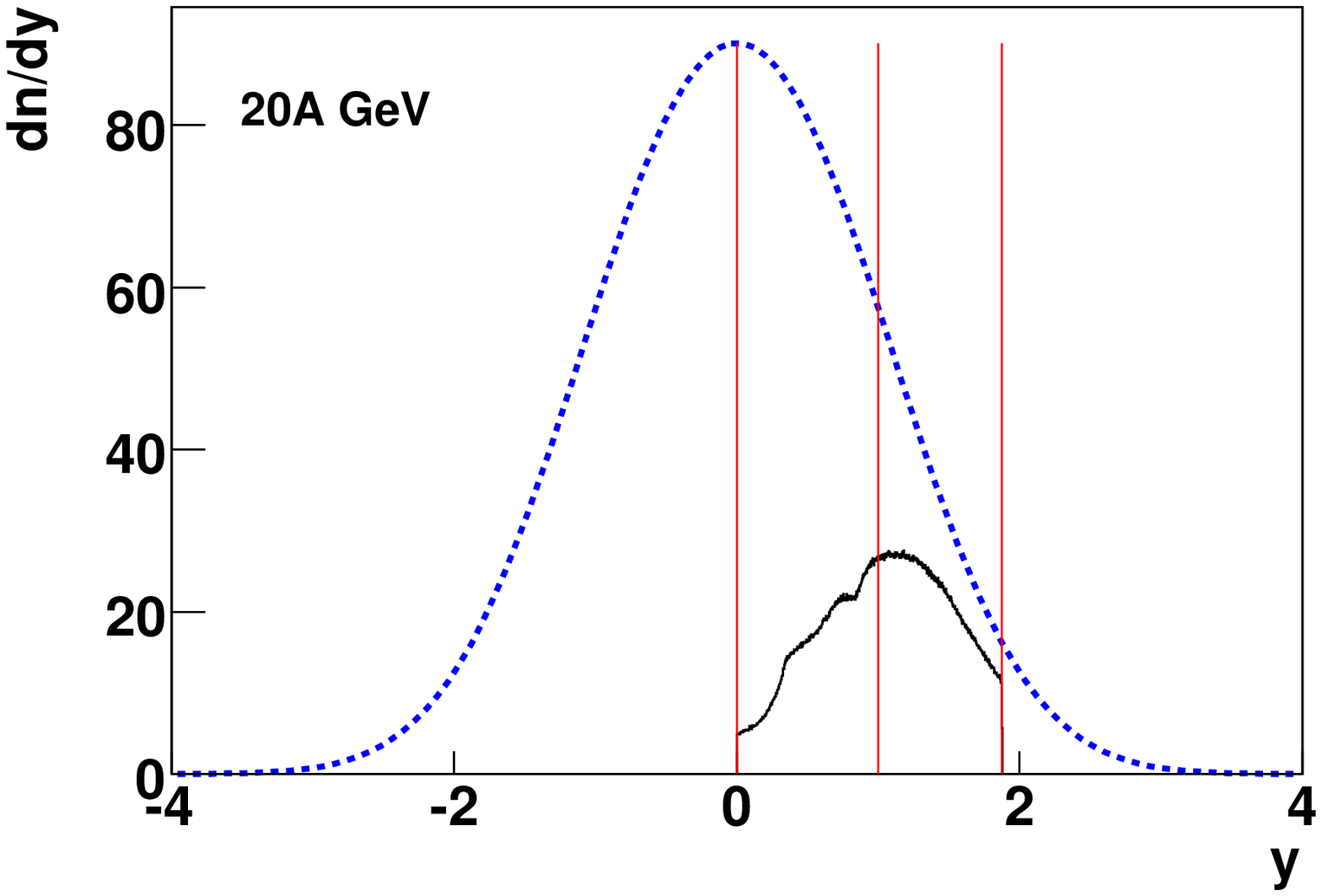}
\includegraphics[width=9cm]{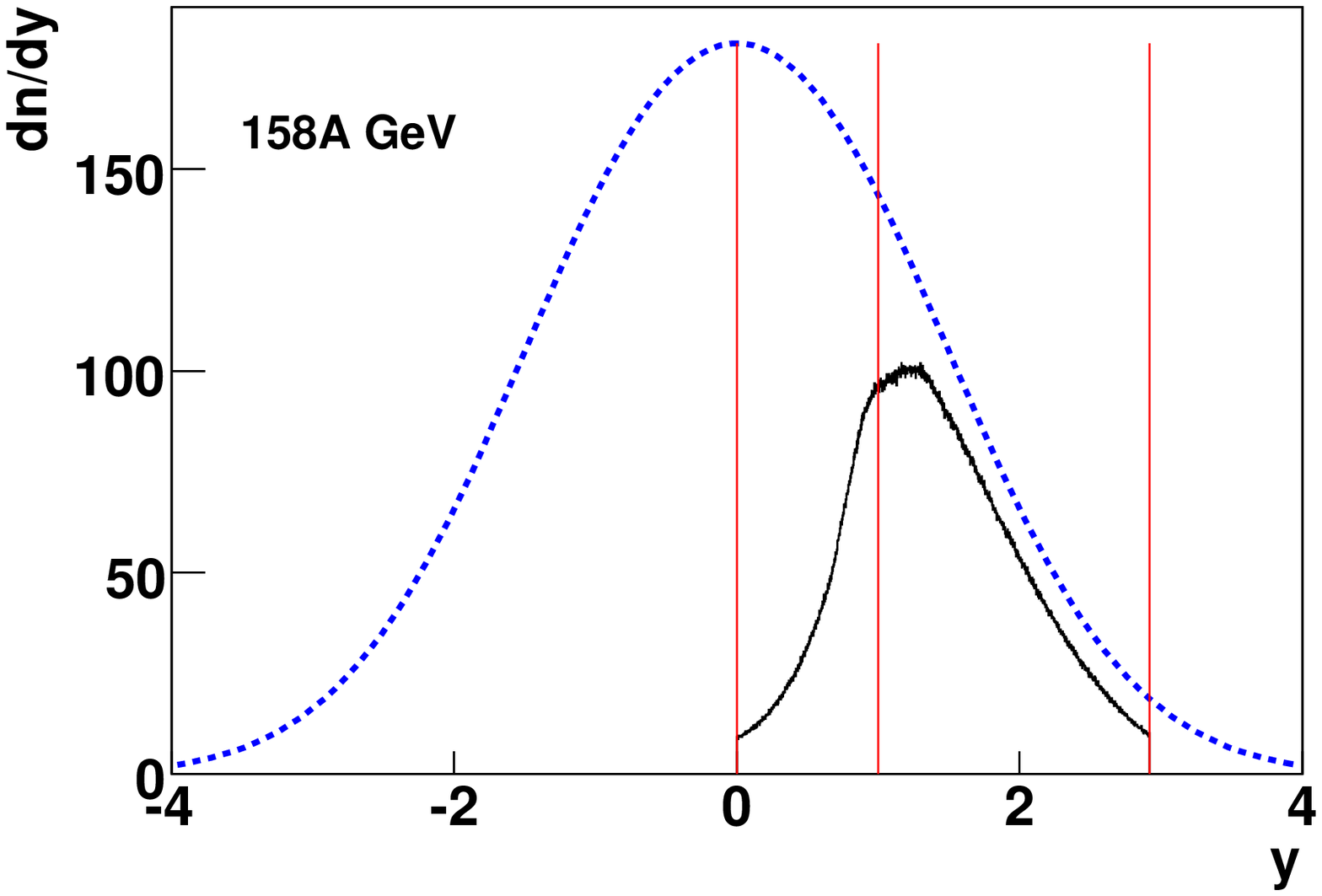}
\caption{\label{ydist}(Color online) Dashed line: Double-Gauss parametrization of the rapidity distribution of negatively charged
 pions and kaons in Pb+Pb collisions
at $20A$ (top)~\cite{Alt:2007fe} and $158A$ GeV (bottom)~\cite{Afanasiev:2002mx}. The solid line is the measured $y(\pi)$ distribution
with the track selection criteria described in section~\ref{trsel}. The vertical lines indicate the limits of the rapidity intervals
$y(\pi)=0$, $y(\pi)=1$ and $y(\pi)=y_{beam}$ used for this analysis.}
\end{figure}

\begin{figure}[htb]
\includegraphics[width=10cm]{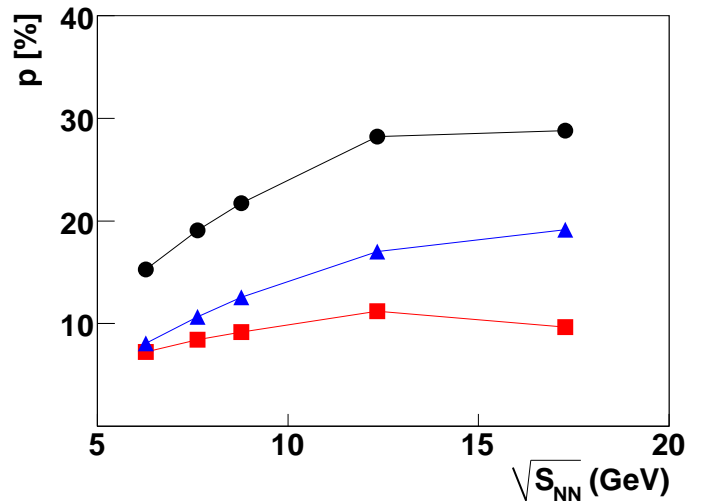}
\caption{\label{accept_ed}(Color online) Fraction $p$ of total negatively charged main vertex pion, kaon and anti-proton multiplicity which is accepted and 
reconstructed
as a function of collision energy. 
Circles: $0<y(\pi)<y_{beam}$, boxes: $0<y(\pi)<1$, triangles: $1<y(\pi)<y_{beam}$.}
\end{figure}

In order to study the multiplicity fluctuations differentially, the pion rapidity interval $0<y(\pi)<y_{beam}$ is divided into two parts,
the "midrapidity" ($0<y(\pi)<1$) and the "forward rapidity" ($1<y(\pi)<y_{beam}$) region (see Fig.~\ref{ydist}). 
The fractions of total charged particle multiplicity falling into the different
rapidity intervals are given in Table~\ref{acc_table} and Fig.~\ref{accept_ed}. 
{The values are calculated using the VENUS event generator~\cite{Werner:1993uh} as input for a GEANT simulation.
The tracks produced by GEANT are converted into detector signals
and reconstructed by the NA49 reconstruction chain. For the determination of the acceptance the negatively charged main vertex pions, 
kaons and anti-protons are used.} 
In both regions a similar number of particles is detected by NA49. In the forward acceptance the
particles are mostly passing through both the vertex- and the main- TPCs and are therefore
efficiently reconstructed for all collision energies. 
According to the UrQMD model the fluctuations in the number of target participants contribute mostly to
the particle number fluctuations in the target hemisphere and the midrapidity region.
Their influence on $\omega$ in the forward
region ($y(\pi)>1$) {can be estimated by the difference in scaled variance between $b=0$ and veto selected collisions
(see section~\ref{centsel}) and is about $1-2\%$.}

Note that the acceptance used for this analysis is larger than the one used for the 
preliminary data shown in \cite{Lungwitz:2006cx,Lungwitz:2006cy}.

\begin{table}
\begin{tabular}{|c||c|c|c|c|c|}
\hline
energy&	$0<y(\pi)$&	$0<y(\pi)$&	$1<y(\pi)$&	$\sigma(y)(\pi^-)$\cr
&	$<y_{beam}$&	$<1$&	$<y_{beam}$&\cr
\hline
\hline
20&	$15.3 \%$&	$7.2 \%$&	$8.1 \%$&	$1.01$\\
\hline
30&	$19.1 \%$&	$8.4 \%$&	$10.7 \%$&	$1.08$\\
\hline
40&	$21.7 \%$&	$9.2 \%$&	$12.6 \%$&	$1.1$\\
\hline
80&	$28.2 \%$&	$11.2 \%$&	$17 \%$&	$1.23$\\
\hline
158&	$28.8 \%$&	$9.6 \%$&	$19.2 \%$&	$1.38$\\
\hline
\end{tabular}
\caption{\label{acc_table}Fraction (in percent) of negatively charged main vertex pions, kaons and anti-protons in different rapidity intervals
for different collision energies which are accepted and reconstructed.
In addition, the width of the rapidity distribution of negatively charged pions is given~\cite{Alt:2007fe,Afanasiev:2002mx}.}
\end{table}

\subsection{Systematic Errors}\label{syserr}

\begin{table}
\begin{tabular}{|c||c|c|c|c|}
\hline
&	$\Delta\omega^+(\%)$&	$\Delta\omega^-(\%)$&	$\Delta\omega^\pm(\%)$\\
\hline
\hline
event selection&	1.5&	1&	1.5\\
\hline
calorimeter resolution&	1&	0.5&	1.5\\
\hline
calorimeter calibration&	0.5&	1&	1\\
\hline
track selection&	1.5&	1&	3\\
\hline
total systematic error&	2.4&	1.8&	3.8\\
\hline
\hline
$0.5\%$ vs. $1\%$ most central&	3&	3&	5\\
\hline
\end{tabular}
\caption{\label{syserr_t}
Maximum change $\Delta\omega$ of the scaled variance $\omega$ of the multiplicity distribution for positively,
negatively and all charged hadrons when applying a correction or neglecting a cut. The systematic errors are calculated by adding
the error contributions in quadrature.
The last row shows the change of $\omega$ resulting from a change in the centrality selection from $1\%$ to $0.5\%$.}
\end{table}

\begin{figure}[h]
\includegraphics[width=9cm]{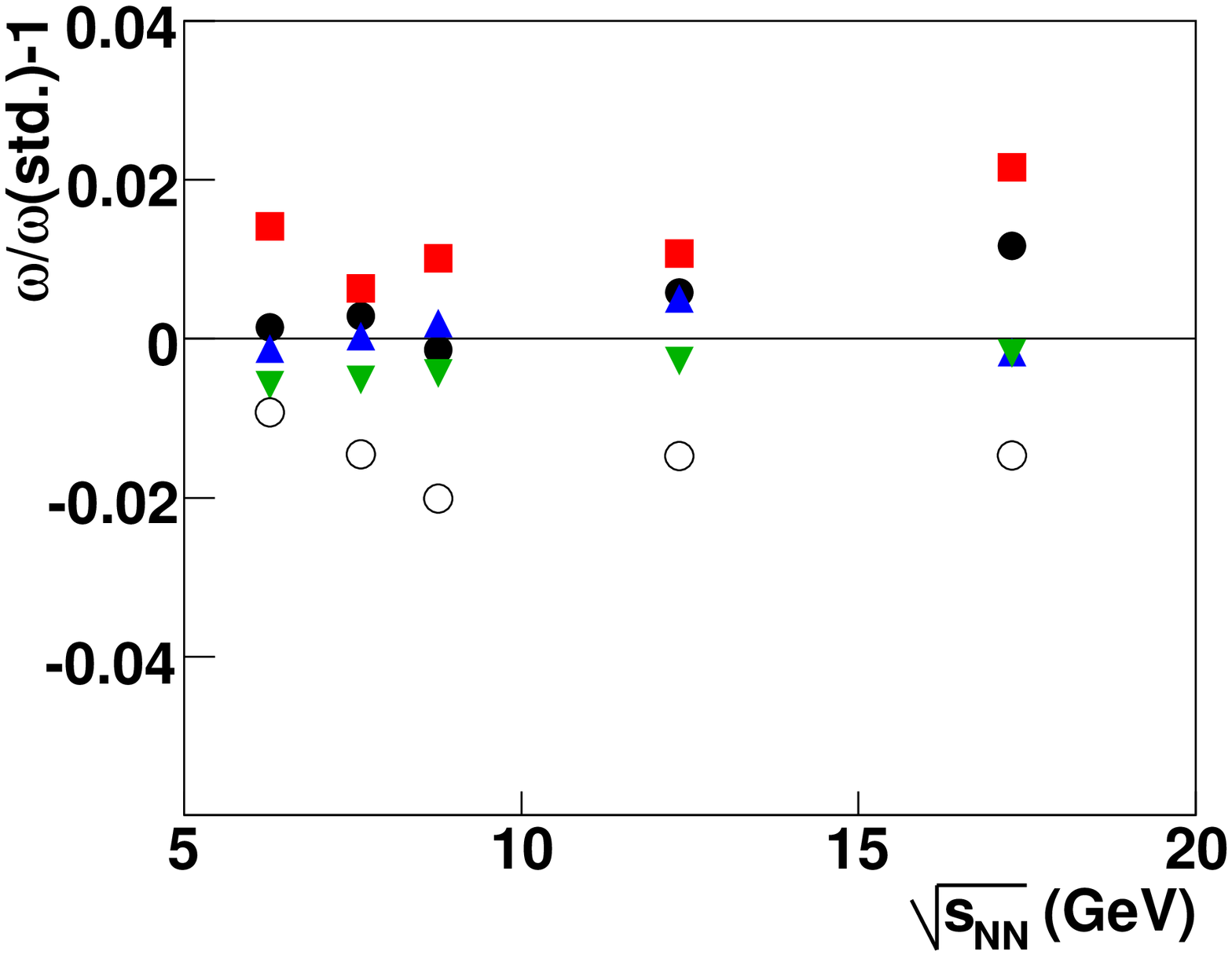}
\includegraphics[width=9cm]{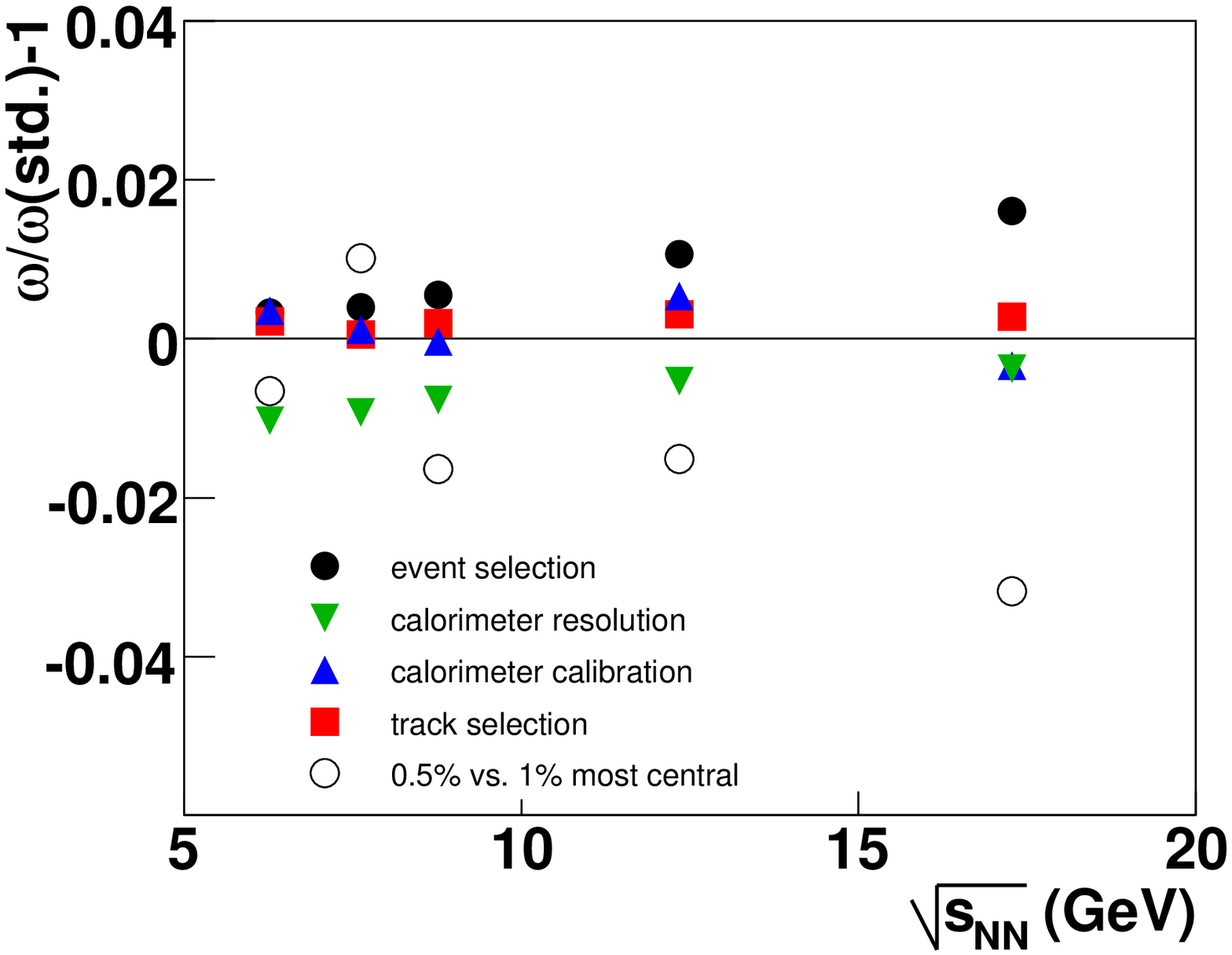}
\caption{\label{syserr_hp}(Color online) Systematic errors and difference between the $0.5\%$ and the $1\%$ most central
collisions of the scaled variance $\omega$ of the multiplicity distribution for positively charged hadrons at midrapidity ($0<y(\pi)<1$, top) 
and forward acceptance ($1<y(\pi)<y_{beam}$, bottom) as a function of collision energy. $\omega(std.)$ corresponds to the value obtained when using the 
standard event and track selection criteria and no correction for the veto calorimeter resolution.}
\end{figure}

\begin{figure}[h]
\includegraphics[width=9cm]{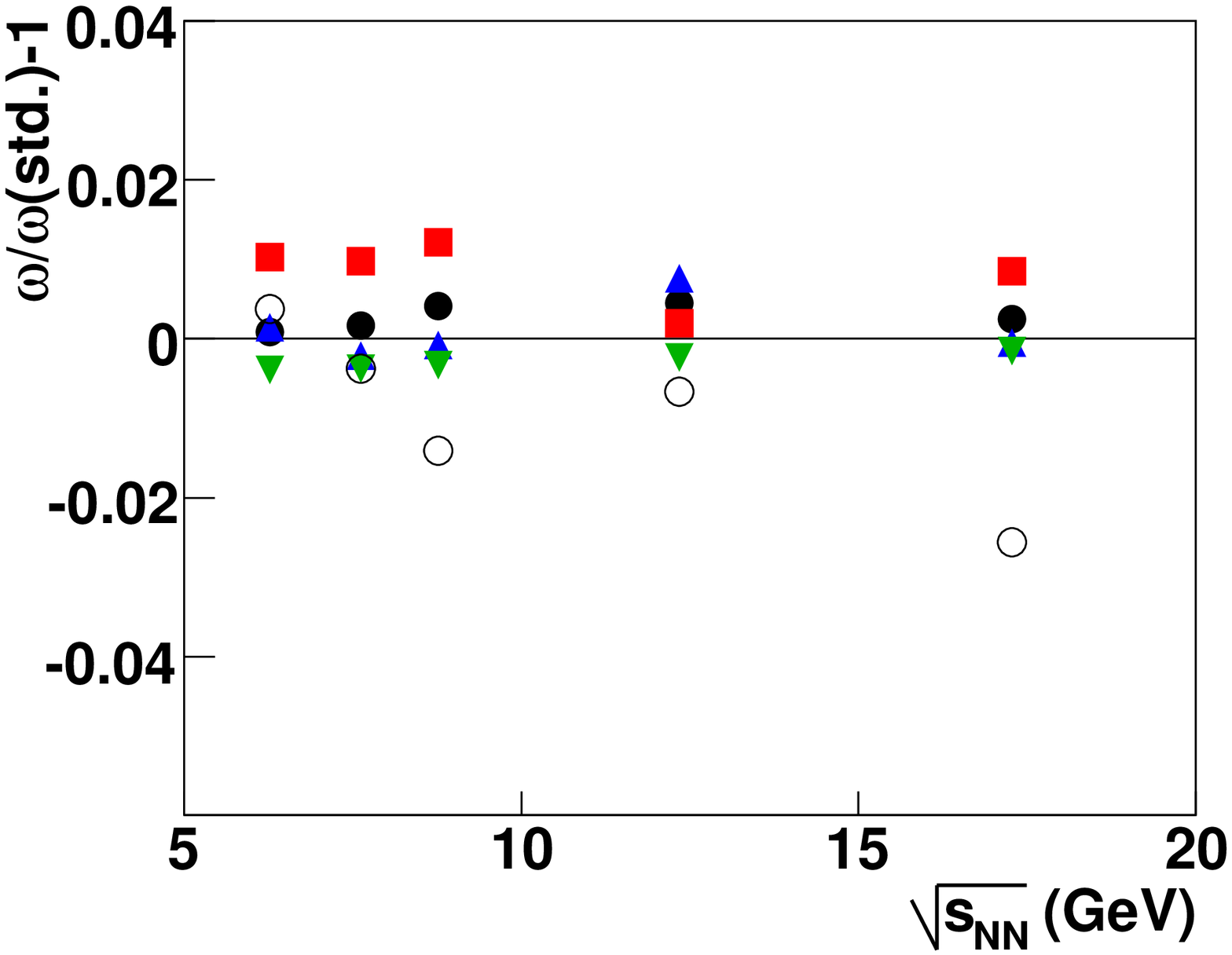}
\includegraphics[width=9cm]{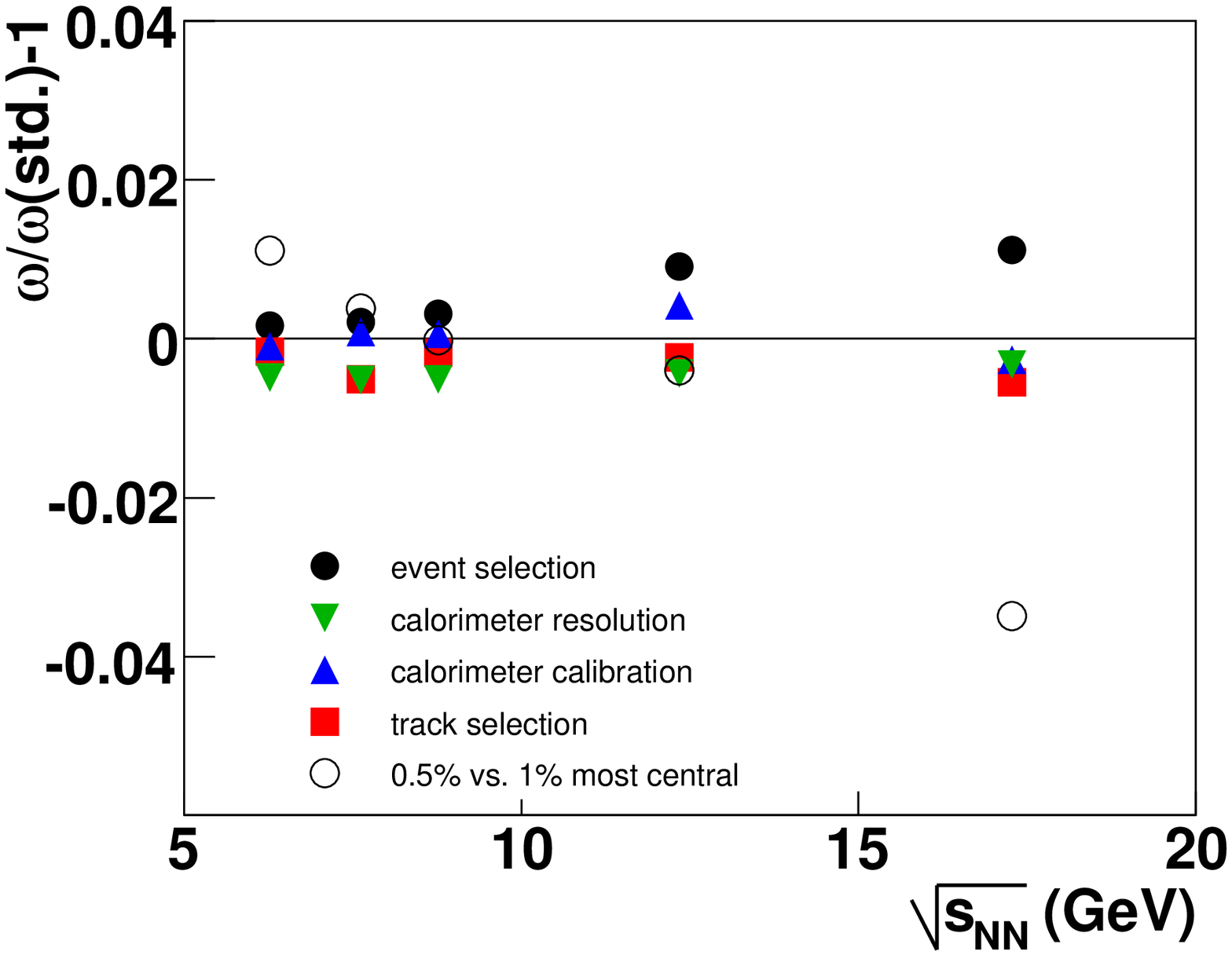}
\caption{\label{syserr_hm}(Color online) Systematic errors and difference between the $0.5\%$ and the $1\%$ most central
collisions of the scaled variance $\omega$ of the multiplicity distribution 
for negatively charged hadrons at midrapidity ($0<y(\pi)<1$, top)
and forward acceptance ($1<y(\pi)<y_{beam}$, bottom) as a function of collision energy. $\omega(std.)$ corresponds to the value obtained when using the 
standard event and track selection criteria and no correction for the veto calorimeter resolution.}
\end{figure}

\begin{figure}[h]
\includegraphics[width=9cm]{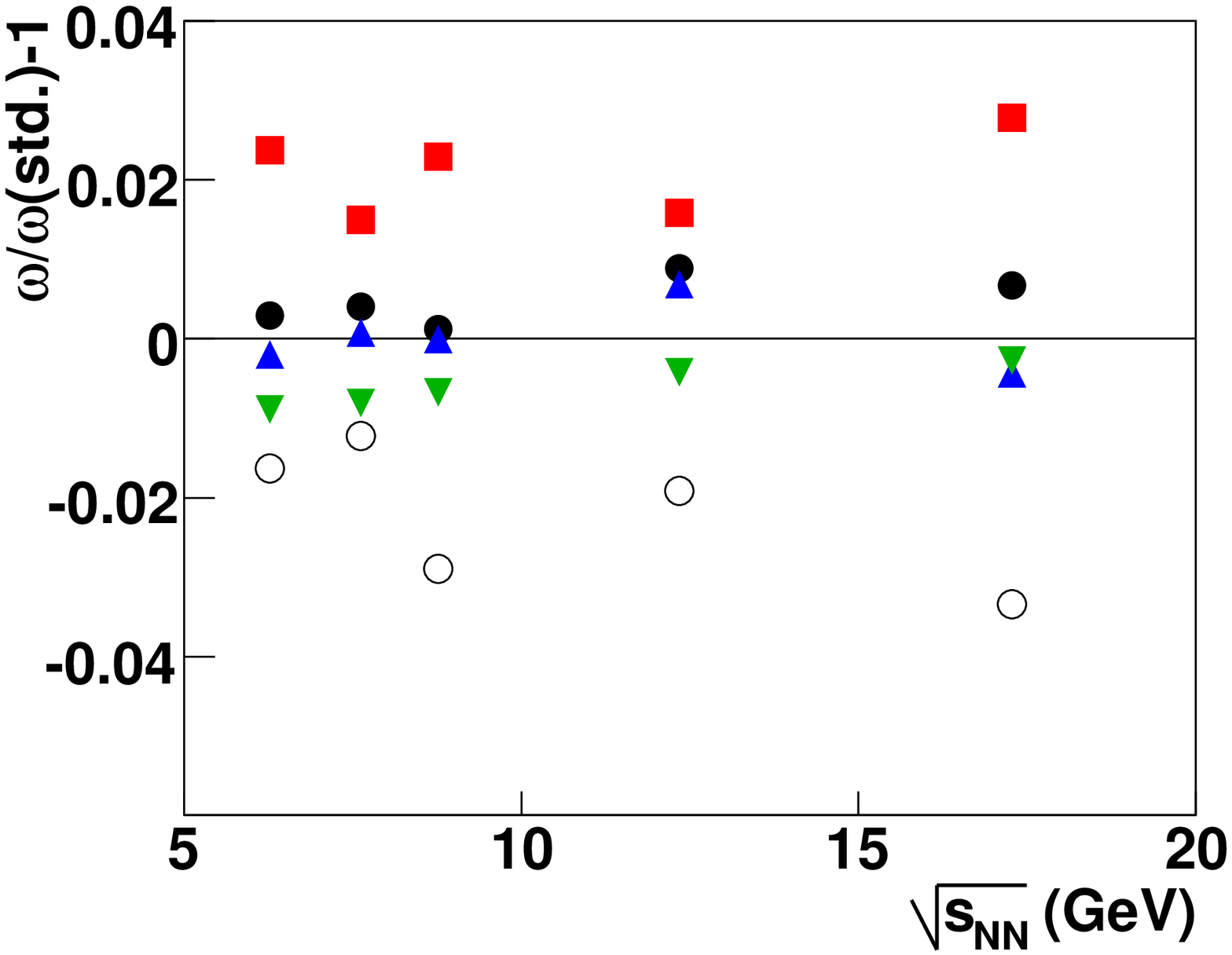}
\includegraphics[width=9cm]{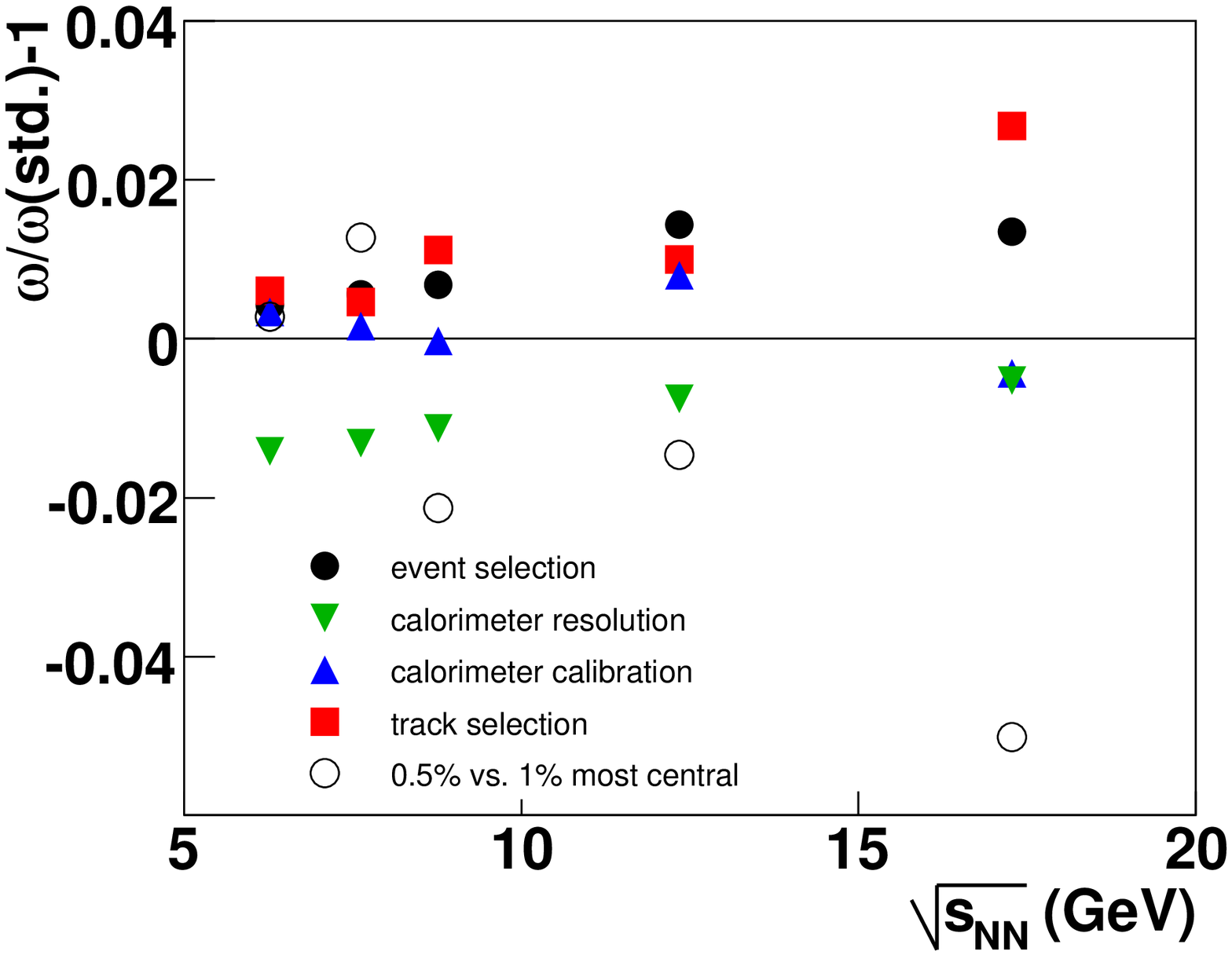}
\caption{\label{syserr_hpm}(Color online) Systematic errors and difference between the $0.5\%$ and the $1\%$ most central
collisions of the scaled variance $\omega$  of the multiplicity distribution
for all charged hadrons at midrapidity ($0<y(\pi)<1$, top)
and forward acceptance ($1<y(\pi)<y_{beam}$, bottom) as a function of collision energy. $\omega(std.)$ corresponds to the value obtained when using the 
standard event and track selection criteria and no correction for the veto calorimeter resolution.d}
\end{figure}

The influence of the selection criteria described above on the scaled variance $\omega$ of the multiplicity distribution has been studied
and the results are presented in Table~\ref{syserr_t} and Figs.~\ref{syserr_hp}-\ref{syserr_hpm}.
The event selection criteria described in section~\ref{ev_sel} change $\omega$ by up to $2\%$ 
compared to the value obtained
when not applying these cuts.
The finite resolution of the veto calorimeter causes additional fluctuations in the number of projectile participants and therefore increases
the measured $\omega$.
In a superposition model the effect of the veto calorimeter resolution is estimated to be \cite{Alt:2006jr}:
\begin{equation}\label{delta} 
\delta=\frac{\left<N\right> \cdot Var(E_{Veto})}{(E_{beam} \cdot N_P^{Proj})^2},
\end{equation}
where $E_{beam}$ is the total energy per projectile nucleon.
The parametrization Eq.~(\ref{vetores_m}), which serves as an upper limit of the resolution of the calorimeter, 
was used to determine the potential influence of
the resolution on $\omega$.
For the very central collisions selected for this analysis the measured $\omega$ is found to increase due to the finite calorimeter resolution
by less than $1.5\%$. Therefore a correction for this effect is not applied.
In order to take possible aging effects of the calorimeter (see section~\ref{centsel})
into account, a time dependent calibration is applied to the measured veto energy. However, the effect of this
correction is very small, $\omega$ changes by less than $1\%$.
Track selection criteria are applied to remove electrons and tracks not originating from the main interaction point.
The value of $\omega$ is changed by less than $1.5\%$ for positively and negatively and less than $3\%$ for all charged hadrons when removing these cuts.

{Embedding simulations demonstrated that the reconstruction efficiency shows no significant decrease with increasing particle multiplicity. Therefore
no systematic error due to reconstruction efficiency was attributed. The overall reconstruction efficiency is about $95\%$ and is included in the 
calculation of the acceptances (Fig.~\ref{accept_ed}, Table~\ref{acc_table}).}

{The total systematic error is calculated by adding the contributions of the different error sources in quadrature.
It is $2.4\%$, $1.8\%$ and $3.8\%$ for positively, negatively and all charged hadrons, respectively.}

In order to estimate the effect of centrality selection, also the $0.5\%$ most central collisions are studied. 
The result for $\omega$ for this stricter selection
is up to $5\%$ different from that obtained for the $1\%$ most central collisions. 
As the centrality selection is a well defined procedure
and can be repeated in model calculations, the difference of $\omega$ for the $0.5\%$ and $1\%$ most central collisions is not
considered as part of the systematic error.

\section{Results on Multiplicity Fluctuations}\label{mult_res}

In this chapter results on multiplicity fluctuations for negatively, positively and all charged hadrons
are presented for \Pb collisions at $20A$, $40A$, $80A$ and $158A$~GeV.
In order to minimize the fluctuations in the number of participants, the $1\%$ most central collisions according to the energy of projectile spectators 
measured in the veto calorimeter are selected (see section~\ref{centsel}).
The rapidity interval $0<y(\pi)<y_{beam}$ used for this analysis is divided into two subintervals,
$0<y(\pi)<1$ ("midrapidity") and $1<y(\pi)<y_{beam}$ ("forward rapidity", see section~\ref{trsel}).

In the following the errors indicated by vertical lines with attached horizontal bars correspond to the statistical errors only, the thick horizontal bars are 
the statistical and systematic errors added in quadrature.

\subsection{Multiplicity Distributions}

The multiplicity distributions for the different energies, charges and rapidity intervals 
as well as the ratios of the measured multiplicity distributions to a Poisson distribution with the same mean multiplicity
are shown in Figs.~\ref{mult_dist_facc_hp}-\ref{mult_dist_hpm}. 
For the ratio to the Poisson distributions only points
with statistical errors smaller than $20\%$ are shown.
All multiplicity distributions have a bell-like shape, and no significant tails or events with a very high or very low multiplicity are observed.
The ratios of measured multiplicity distributions to the corresponding Poisson distributions are symmetric around their 
mean value.

The measured multiplicity distributions are narrower than the Poisson ones 
in the forward acceptance for positively and negatively charged hadrons at all energies. In the midrapidity acceptance the measured distributions are
wider or similar to the Poisson ones. 
The distributions for all charged hadrons are broader than the ones for positively and negatively charged particles separately. 

\subsection{Energy Dependence of $\omega$}

\begin{figure}
\includegraphics[height=6cm]{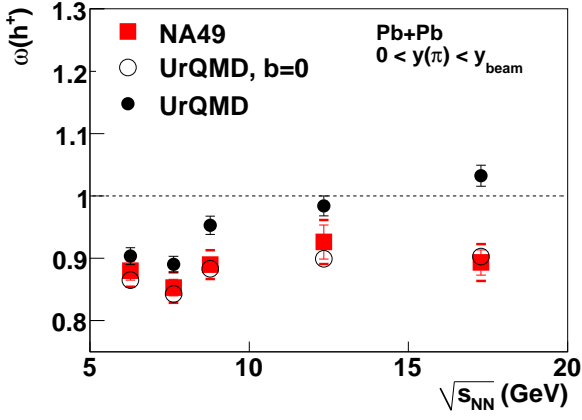}
\includegraphics[height=6cm]{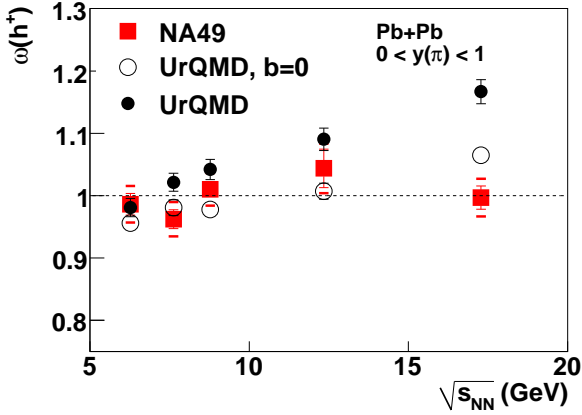}\\
\includegraphics[height=6cm]{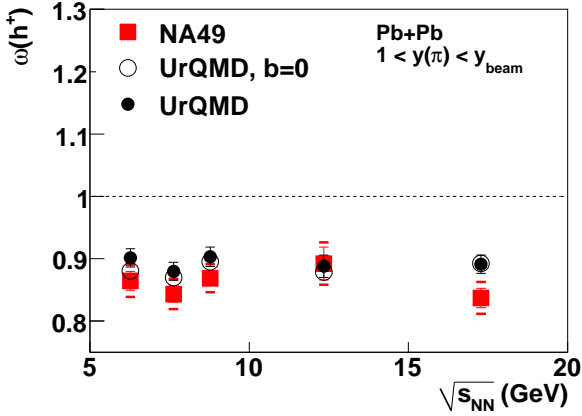}\\
\caption{\label{w_hp}(Color online) Scaled variance $\omega$ of the multiplicity distribution
of positively charged hadrons produced in central \Pb collisions as a function of collision energy. 
Top: full experimental acceptance, middle: midrapidity, bottom: forward rapidity.}
\end{figure}

\begin{figure}
\includegraphics[height=6cm]{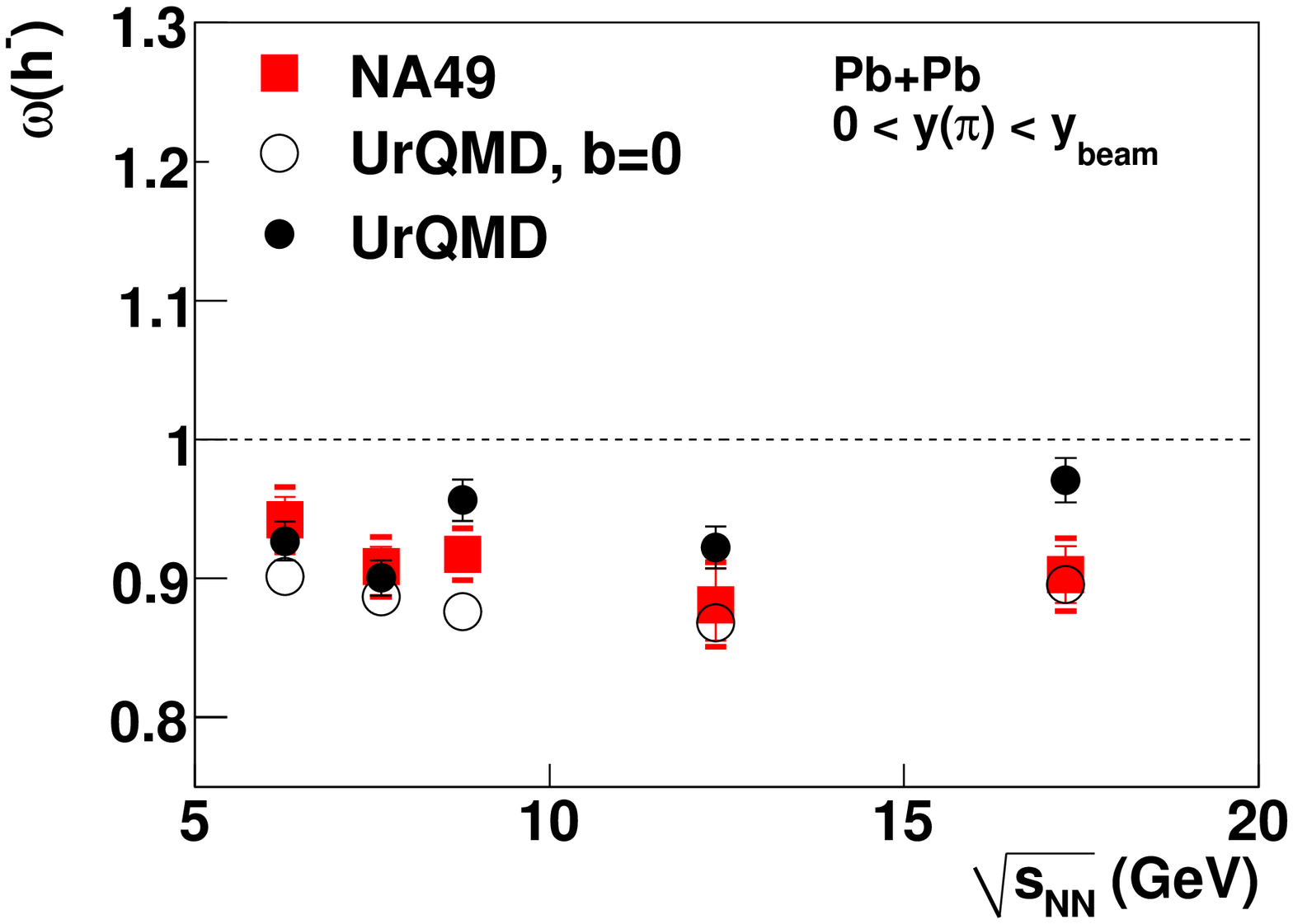}
\includegraphics[height=6cm]{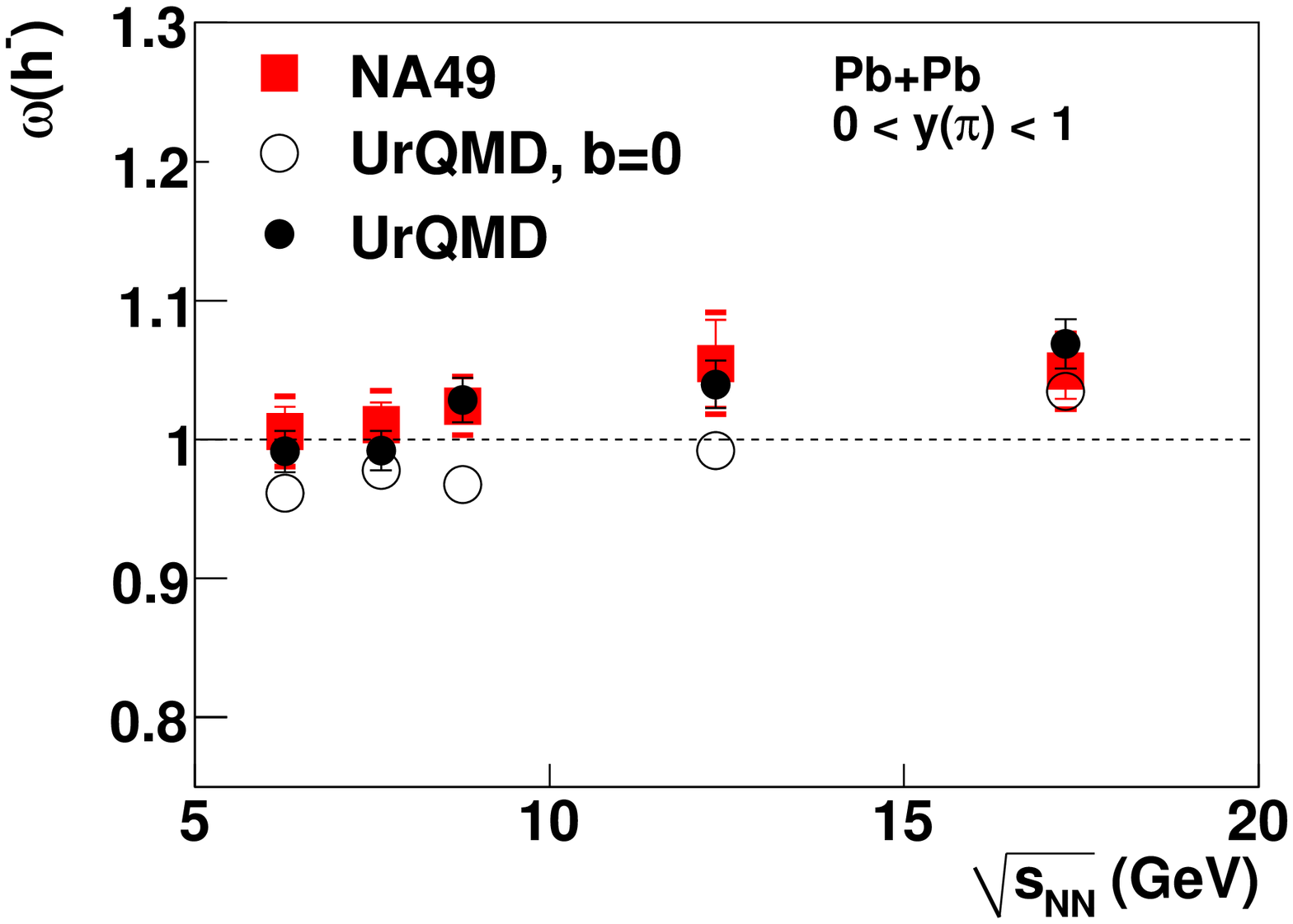}\\
\includegraphics[height=6cm]{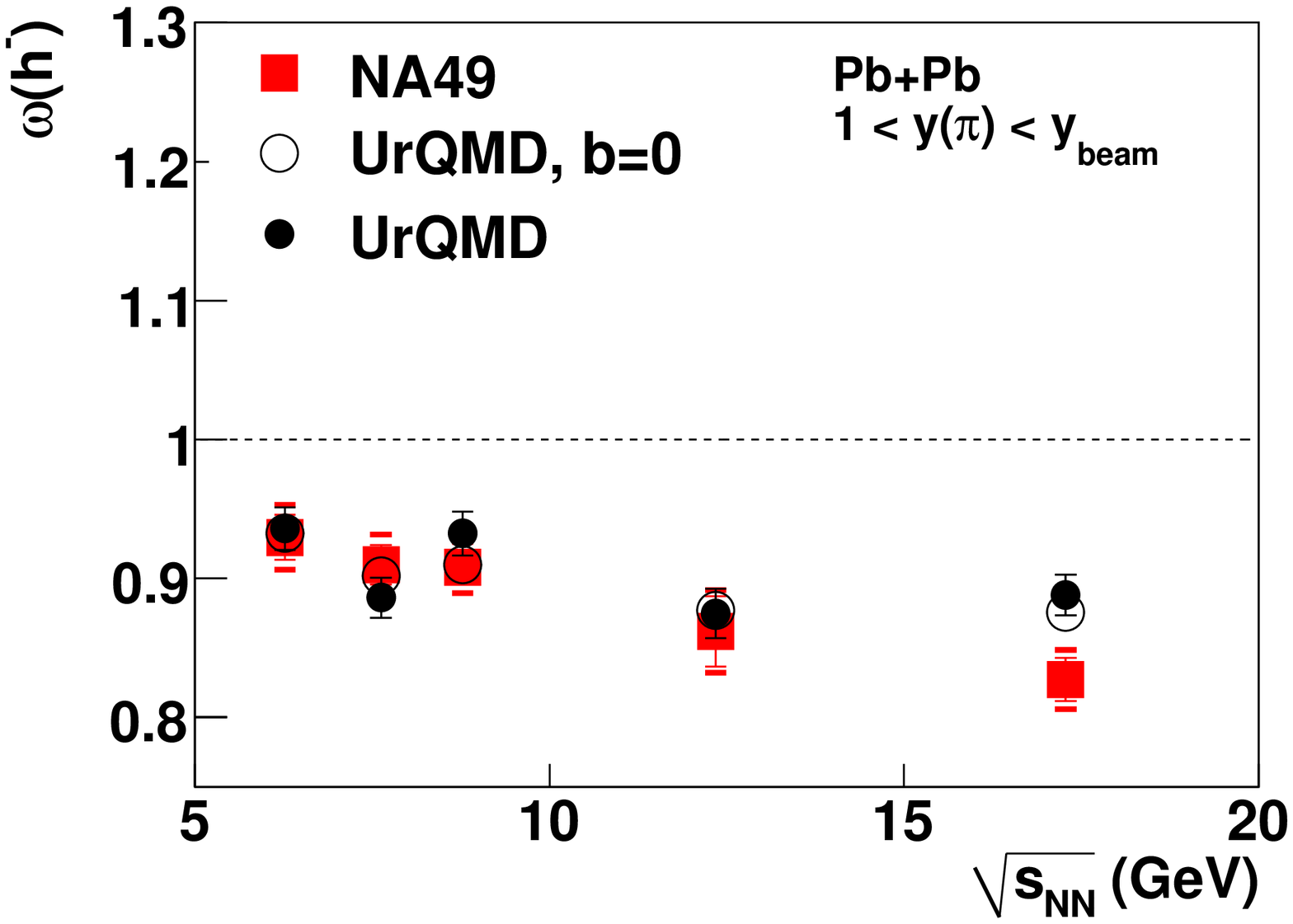}\\
\caption{\label{w_hm}(Color online) Scaled variance $\omega$ of the multiplicity distribution
of negatively charged hadrons produced in central \Pb collisions as a function of collision energy. 
Top: full experimental acceptance, middle: midrapidity, bottom: forward rapidity.}
\end{figure}

\begin{figure}
\includegraphics[height=6cm]{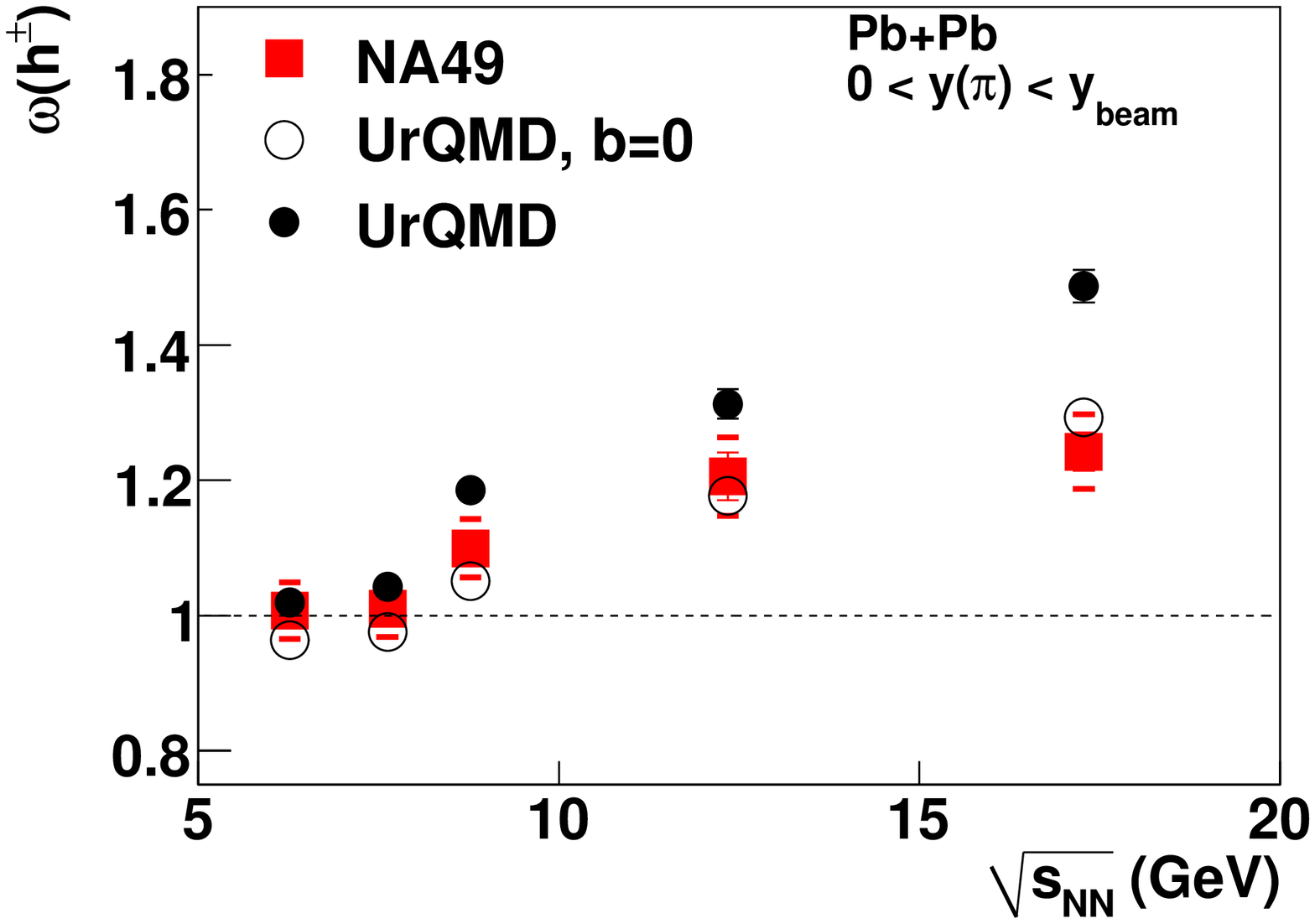}
\includegraphics[height=6cm]{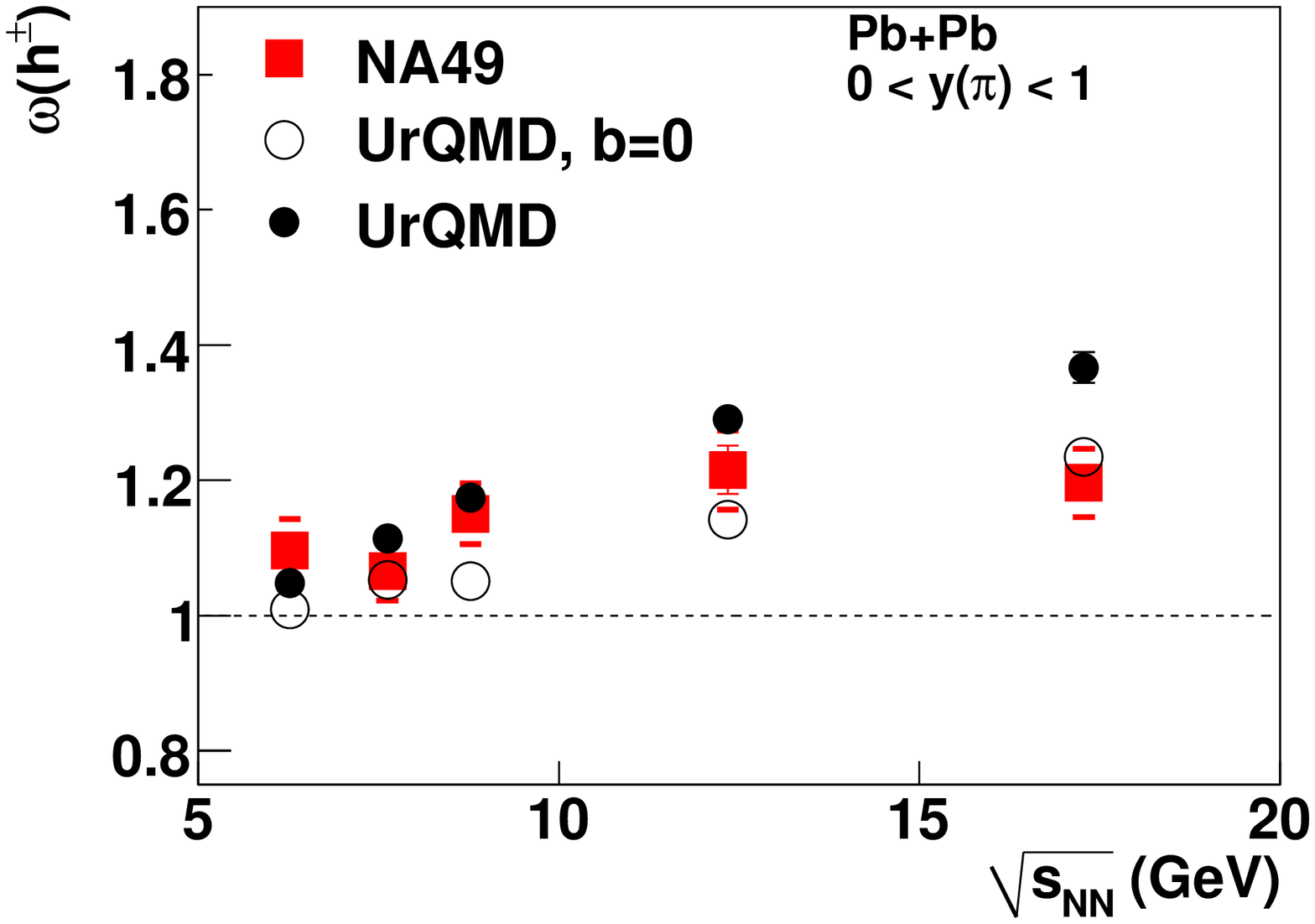}\\
\includegraphics[height=6cm]{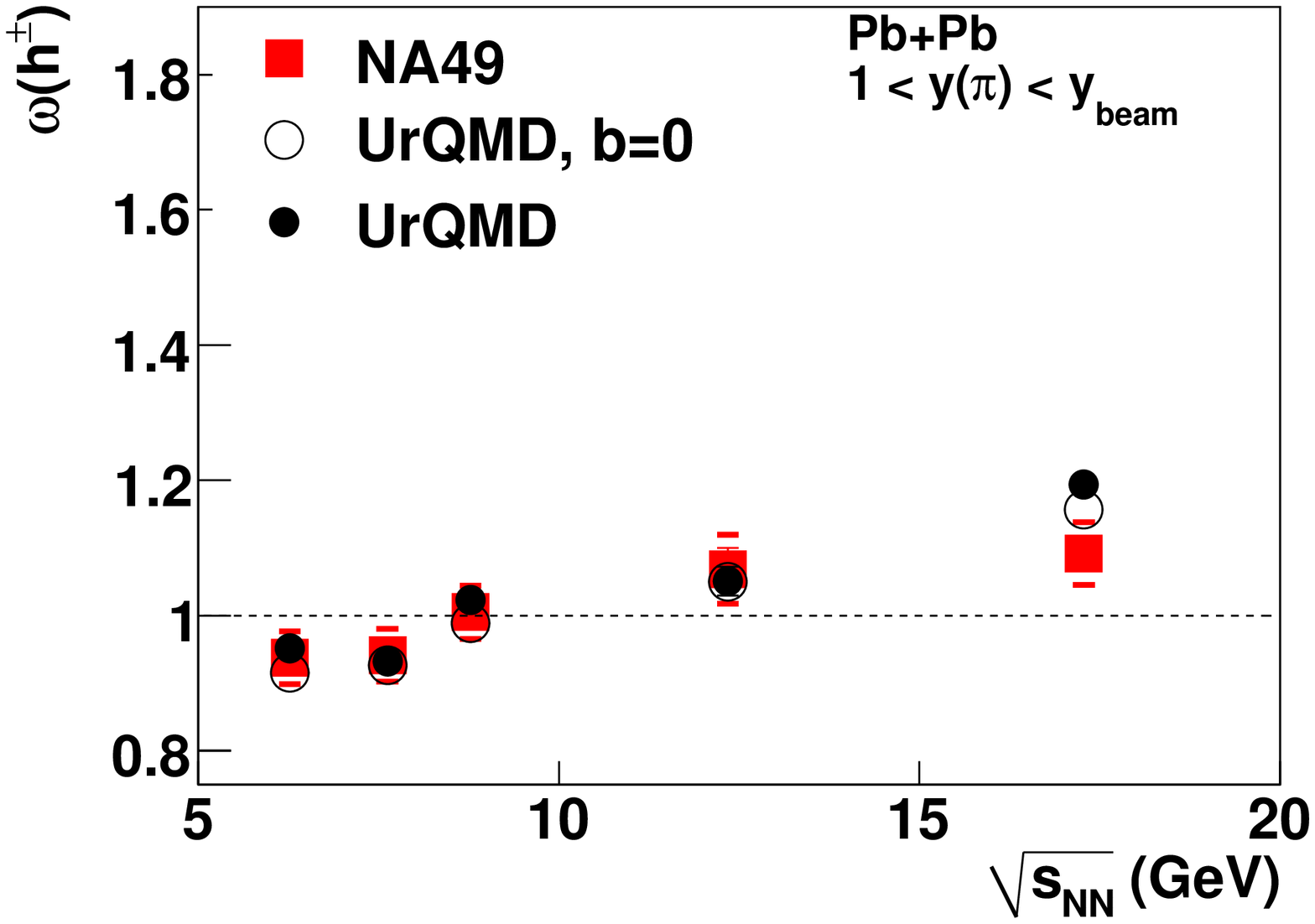}\\
\caption{\label{w_hpm}(Color online) Scaled variance $\omega$ of the multiplicity distribution
of all charged hadrons produced in central \Pb collisions as a function of collision energy. 
Top: full experimental acceptance, middle: midrapidity, bottom: forward rapidity.}
\end{figure}

The energy dependence of the scaled variance $\omega$ of the multiplicity distributions
for negatively, positively and all charged particles for three rapidity intervals is
shown in Figs.~\ref{w_hp}-\ref{w_hpm}, the numerical values are given in Table~\ref{w_num}.

For positively and negatively charged hadrons the values of $\omega$ are similar and
smaller than 1 in the very forward region ($1<y(\pi)<y_{beam}$) at all energies. At midrapidity they are larger than 1. 
For all charged particles $\omega$ is larger than for each charge separately.

No significant structure or non-monotonic behaviour is observed in the energy dependence of $\omega$.


\begin{figure}
\includegraphics[height=6cm]{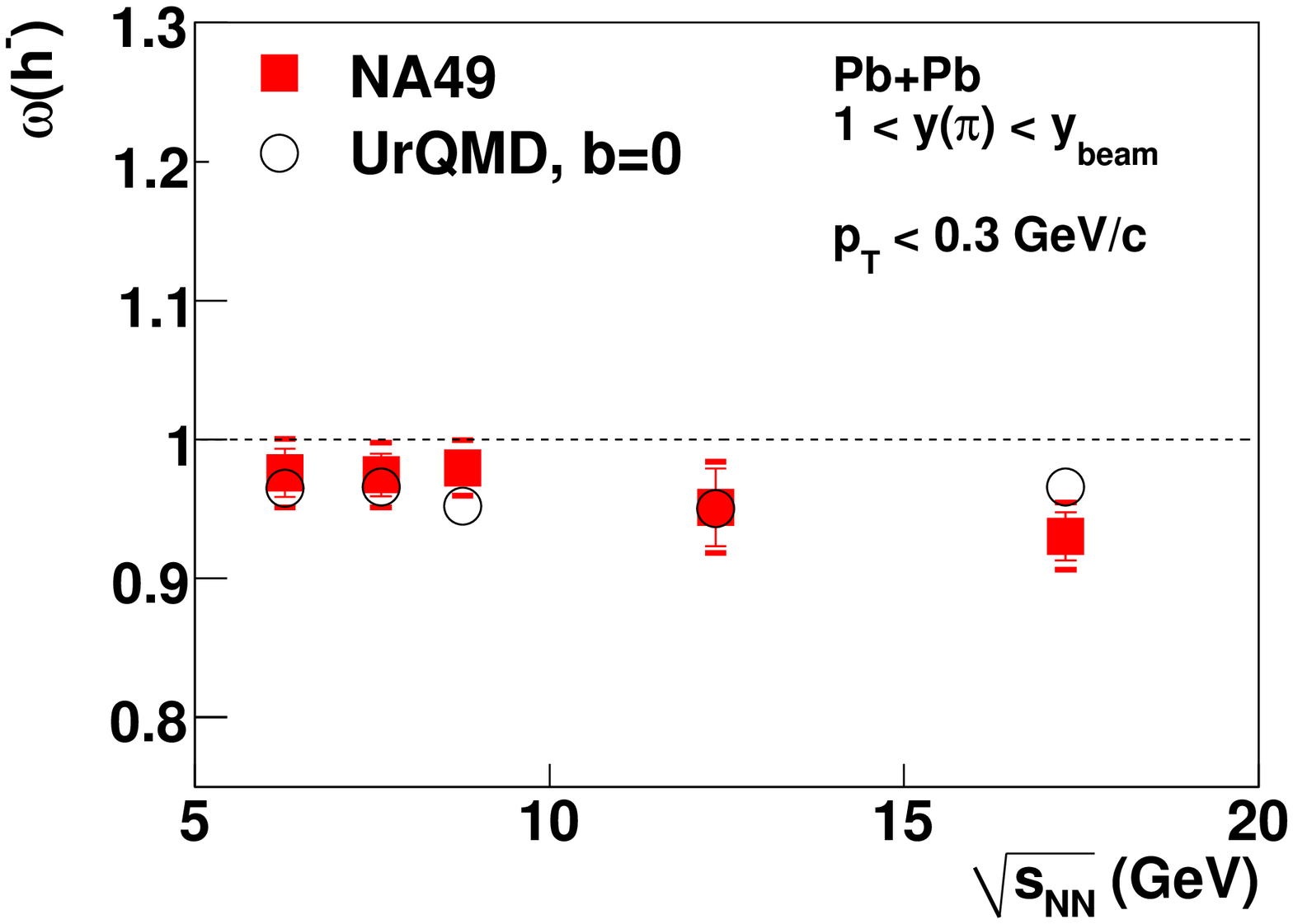}
\includegraphics[height=6cm]{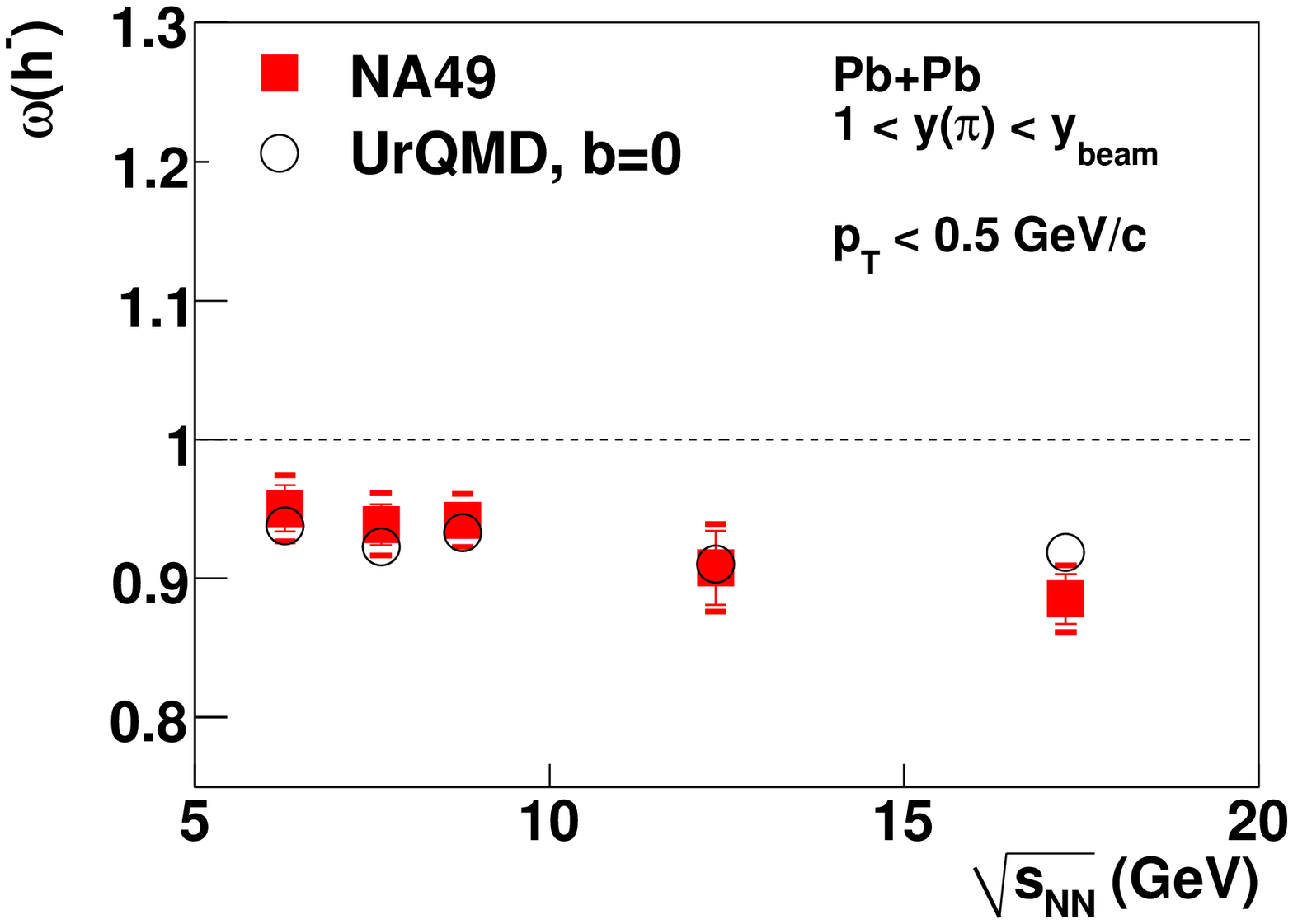}\\
\caption{\label{w_lowpt}(Color online) Scaled variance $\omega$ of the multiplicity distribution
of negatively charged hadrons with low transverse momentum at forward rapidities
produced in central \Pb collisions as a function of collision energy. 
Top: $p_T<0.3$ GeV/c, bottom: $p_T<0.5$ GeV/c.}
\end{figure}

Signatures of the critical point are expected to occur mostly 
at low transverse momenta~\cite{Stephanov:1999zu}. The energy dependence of multiplicity 
fluctuations for low
transverse momentum particles is shown in Fig.~\ref{w_lowpt}. No non-monotonic behaviour is observed.

\subsection{Rapidity Dependence of $\omega$}

\begin{figure}
\includegraphics[height=4.3cm]{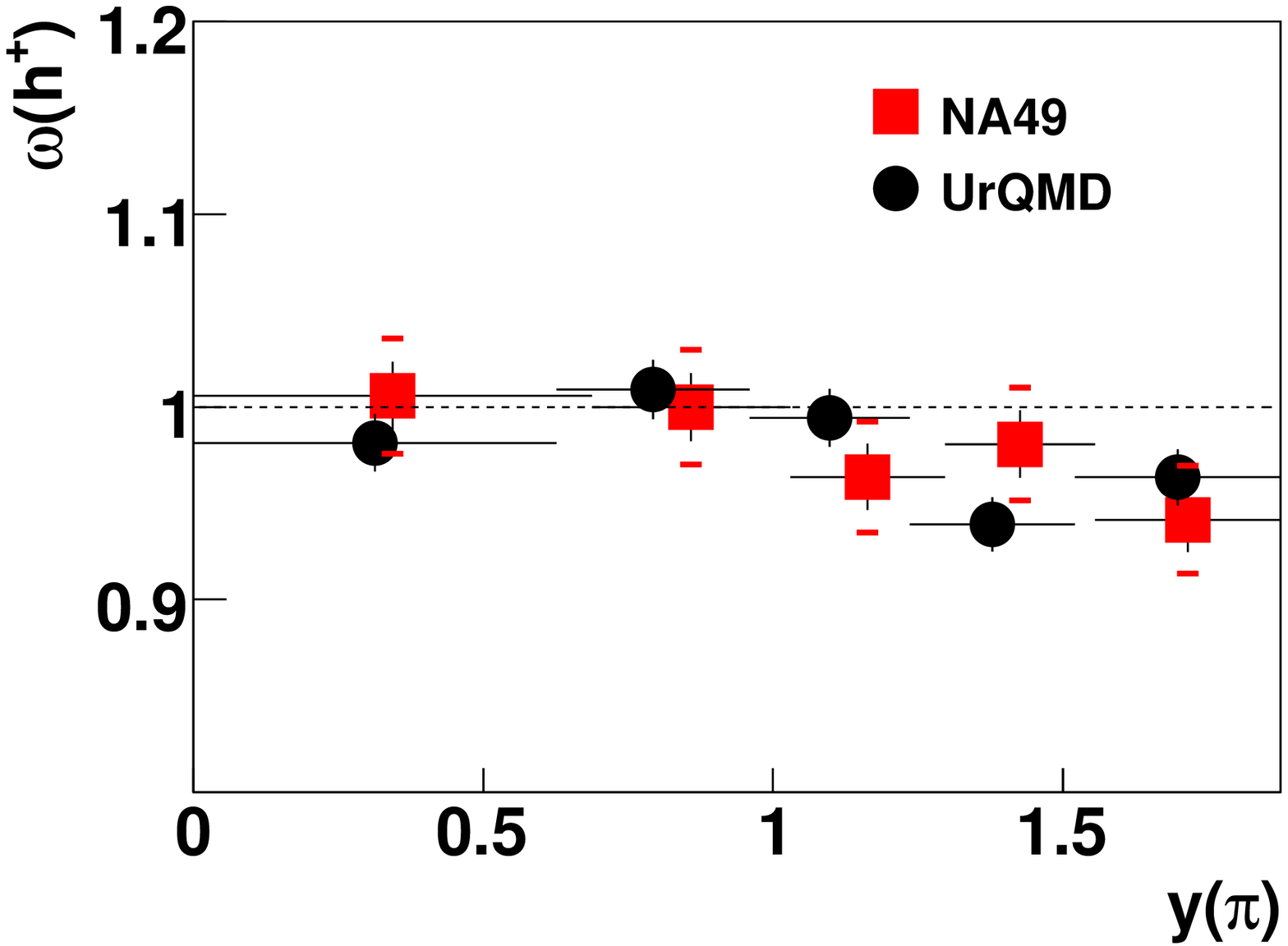}\\
\includegraphics[height=4.3cm]{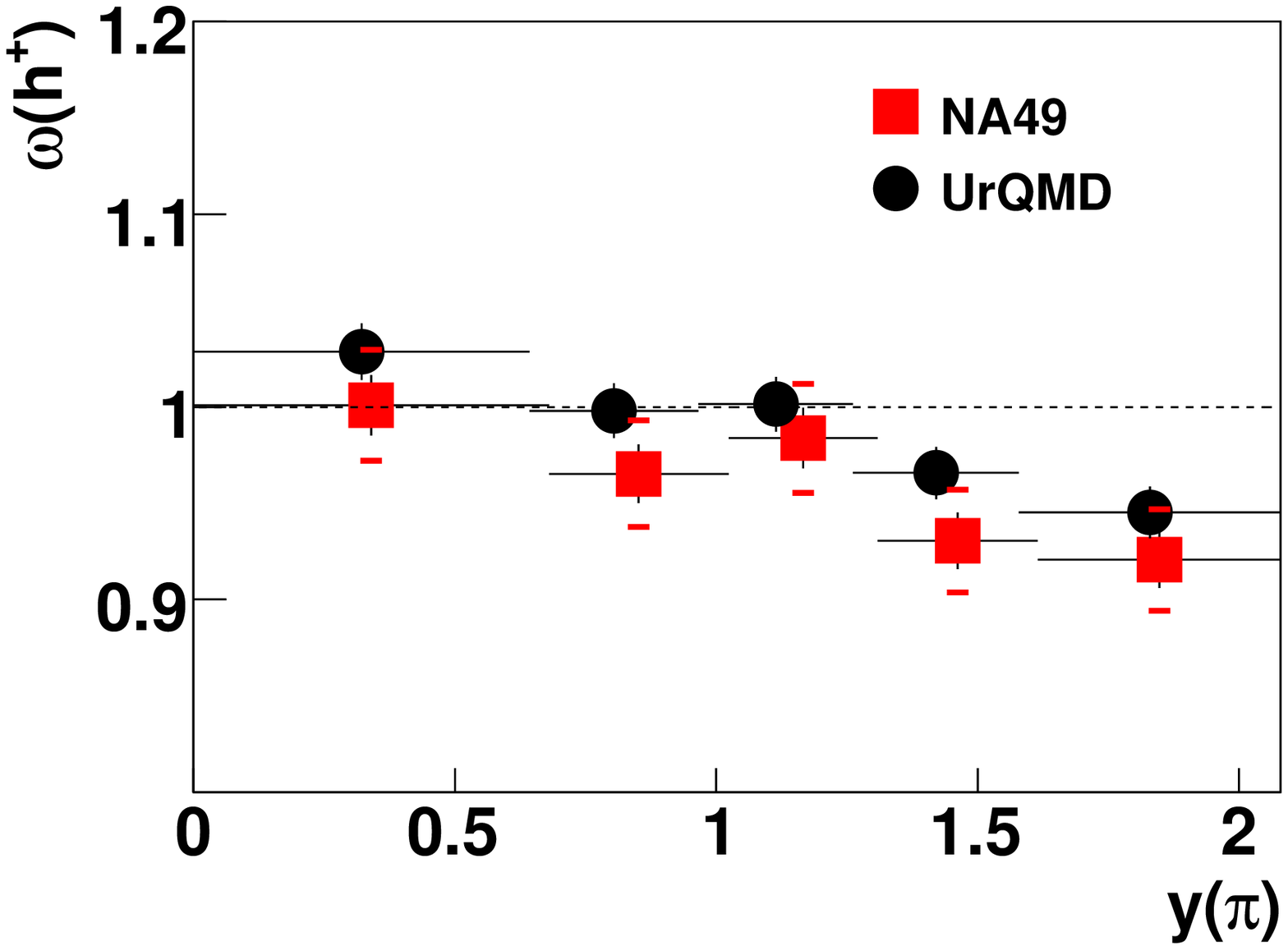}\\
\includegraphics[height=4.3cm]{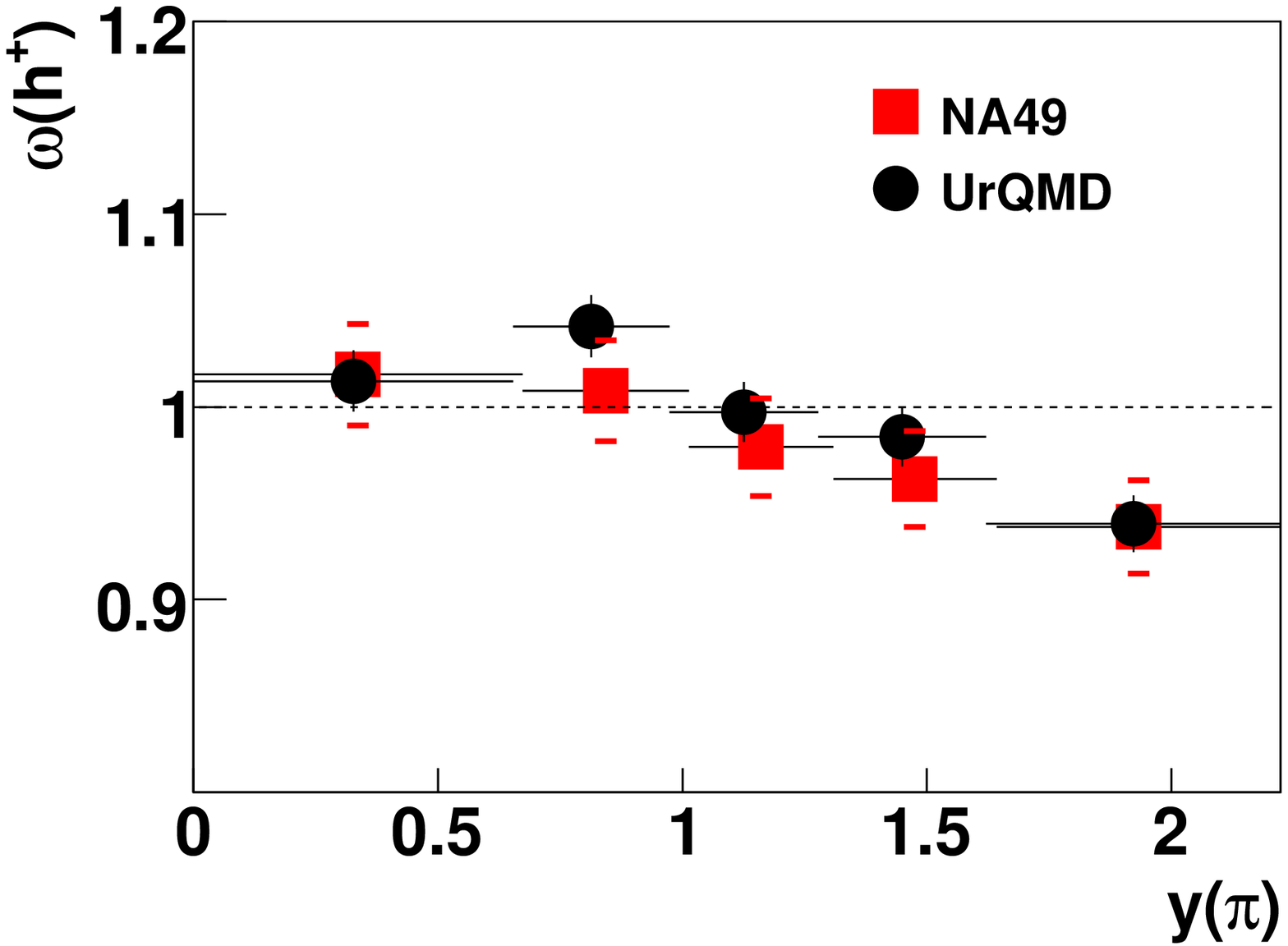}\\
\includegraphics[height=4.3cm]{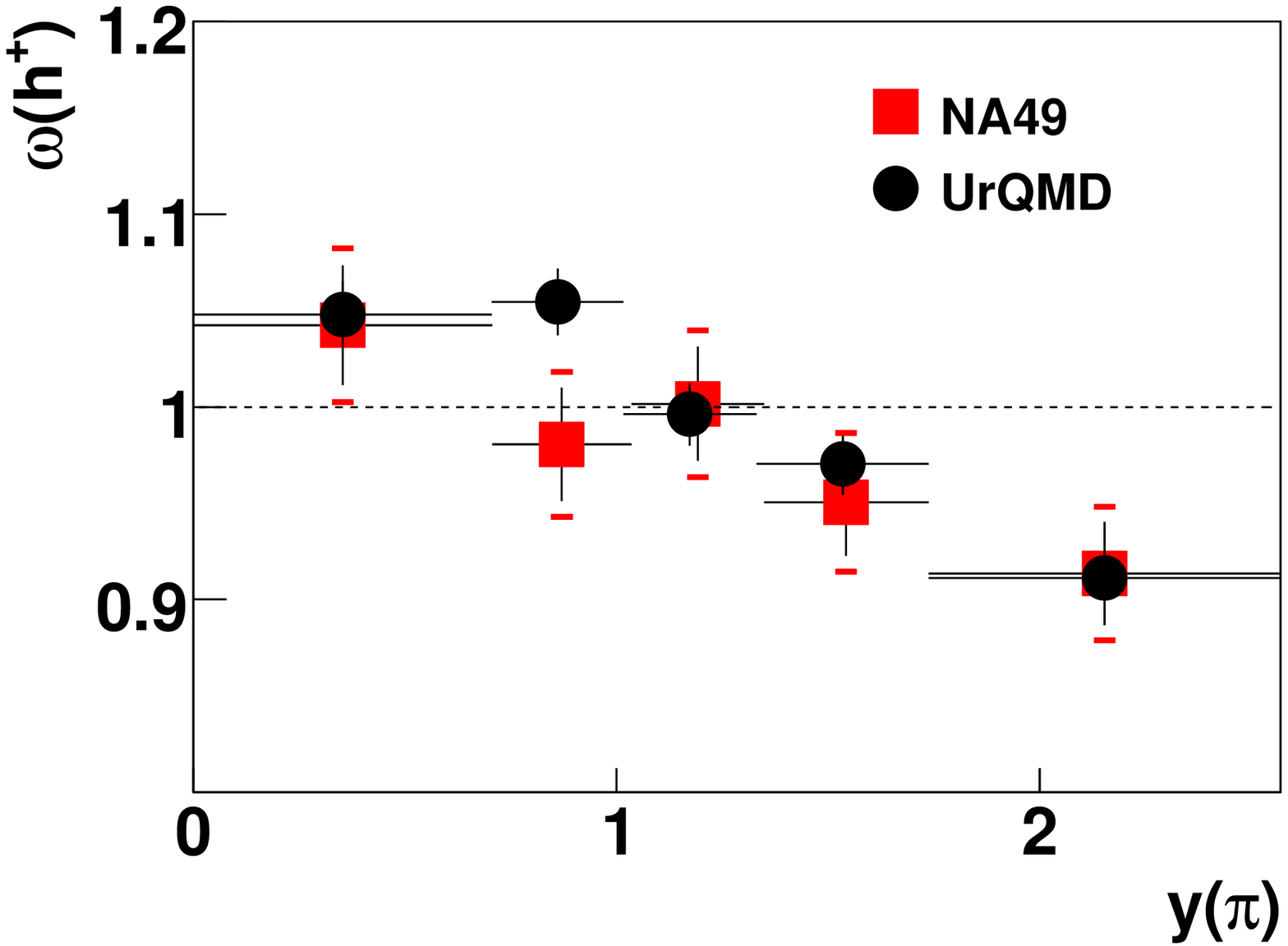}\\
\includegraphics[height=4.3cm]{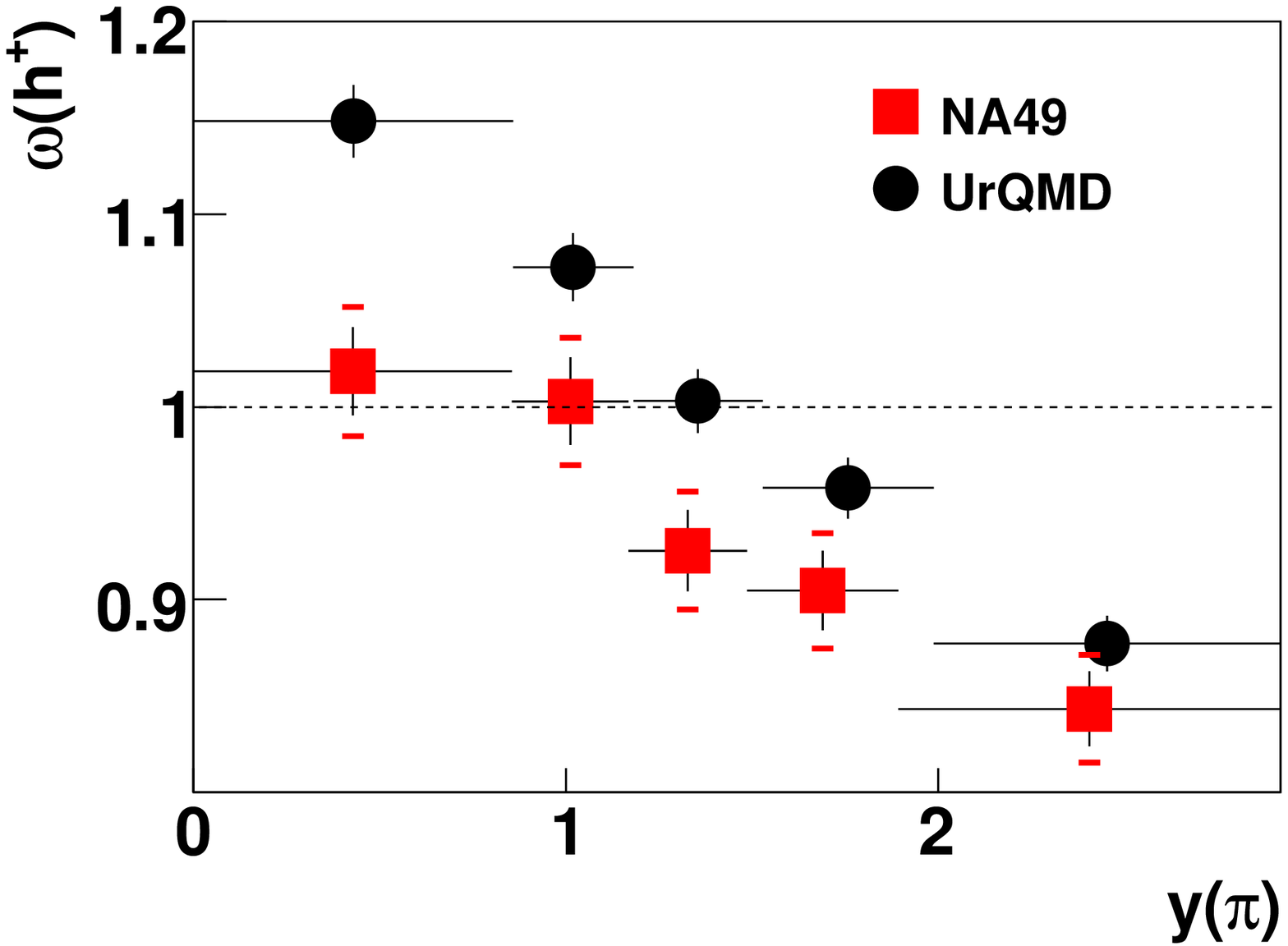} 
\caption{\label{ydep_hp}(Color online) Rapidity dependence of the scaled variance $\omega$ of the multiplicity distribution
of positively charged hadrons in central \Pb collisions 
at $20A$ (top), $30A$, $40A$, $80A$ and $158A$ GeV (bottom) compared to UrQMD predictions with a centrality selection similar 
to the one for the experimental data. The rapidity bins are constructed in such 
a way that the mean multiplicity in each bin is the same.}
\end{figure}

\begin{figure}
\includegraphics[height=4.3cm]{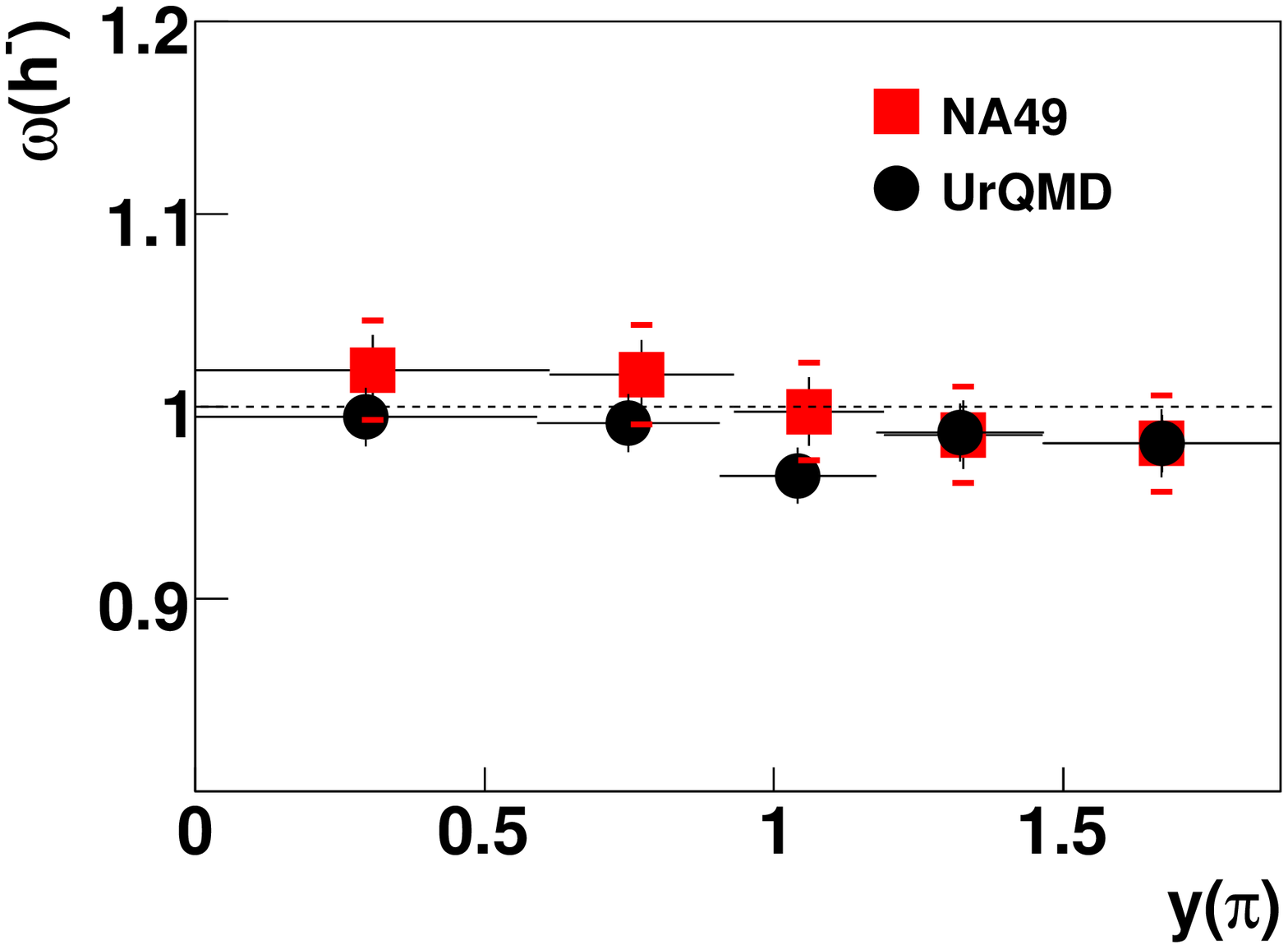}\\
\includegraphics[height=4.3cm]{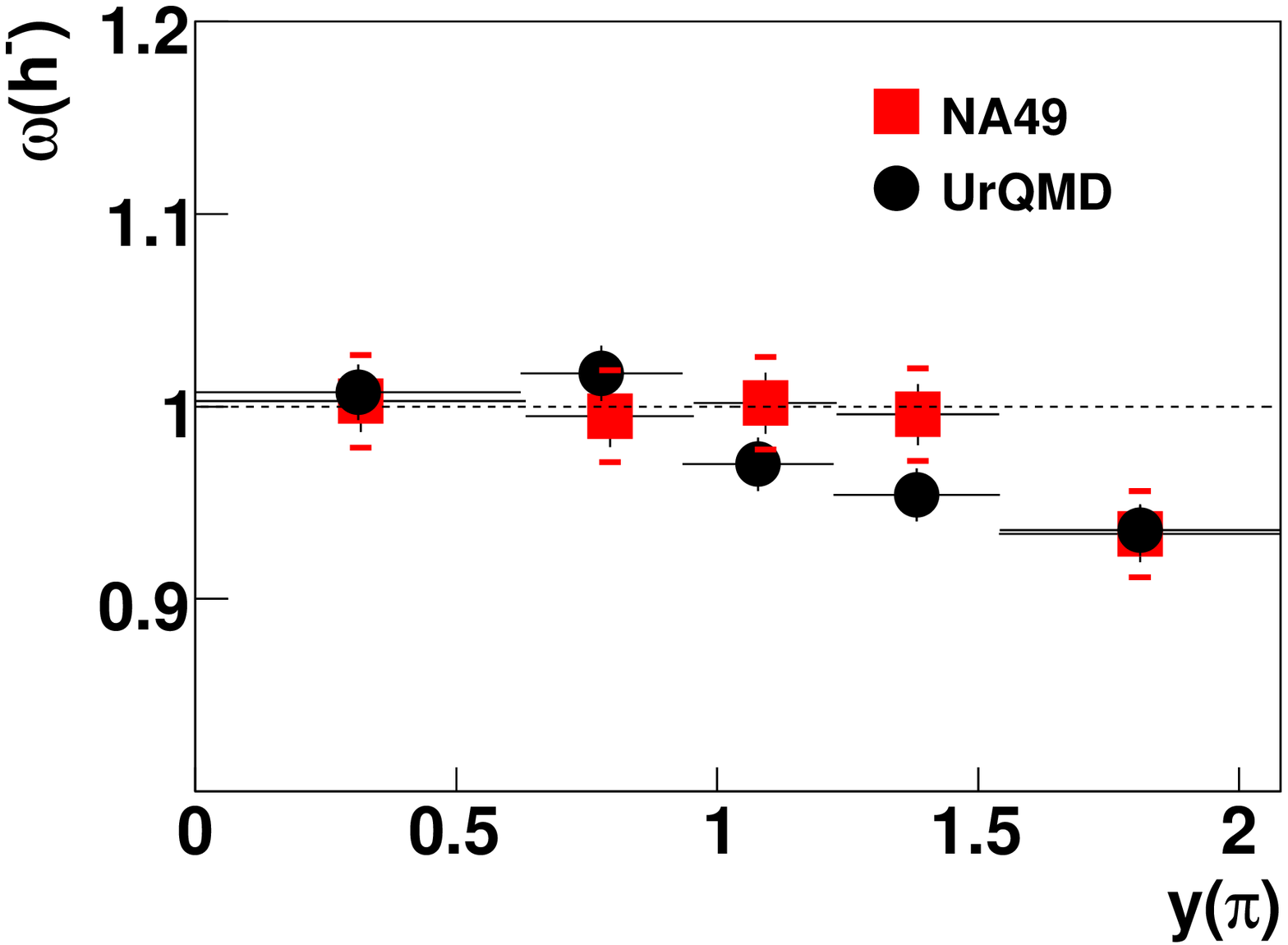}\\
\includegraphics[height=4.3cm]{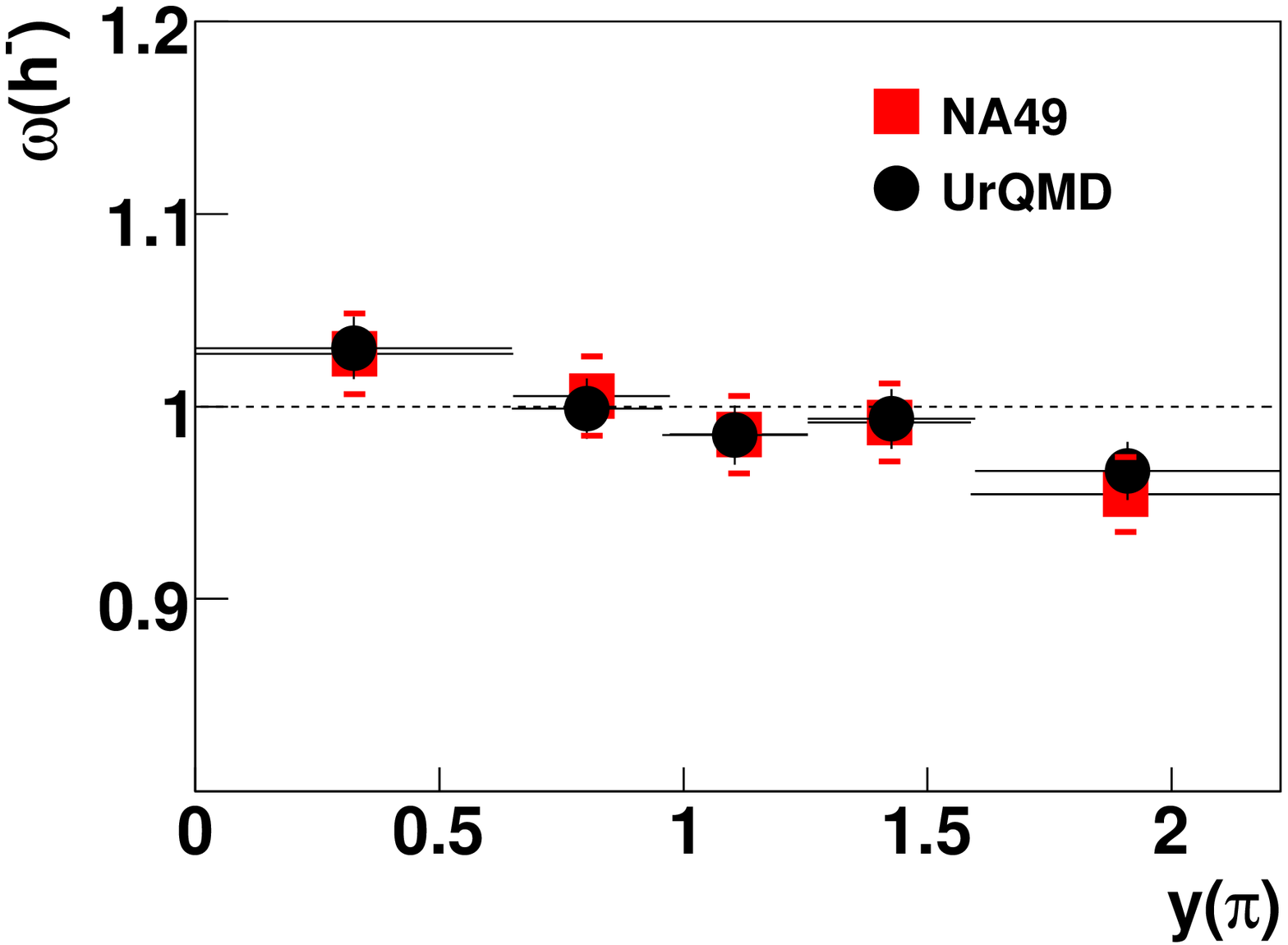}\\
\includegraphics[height=4.3cm]{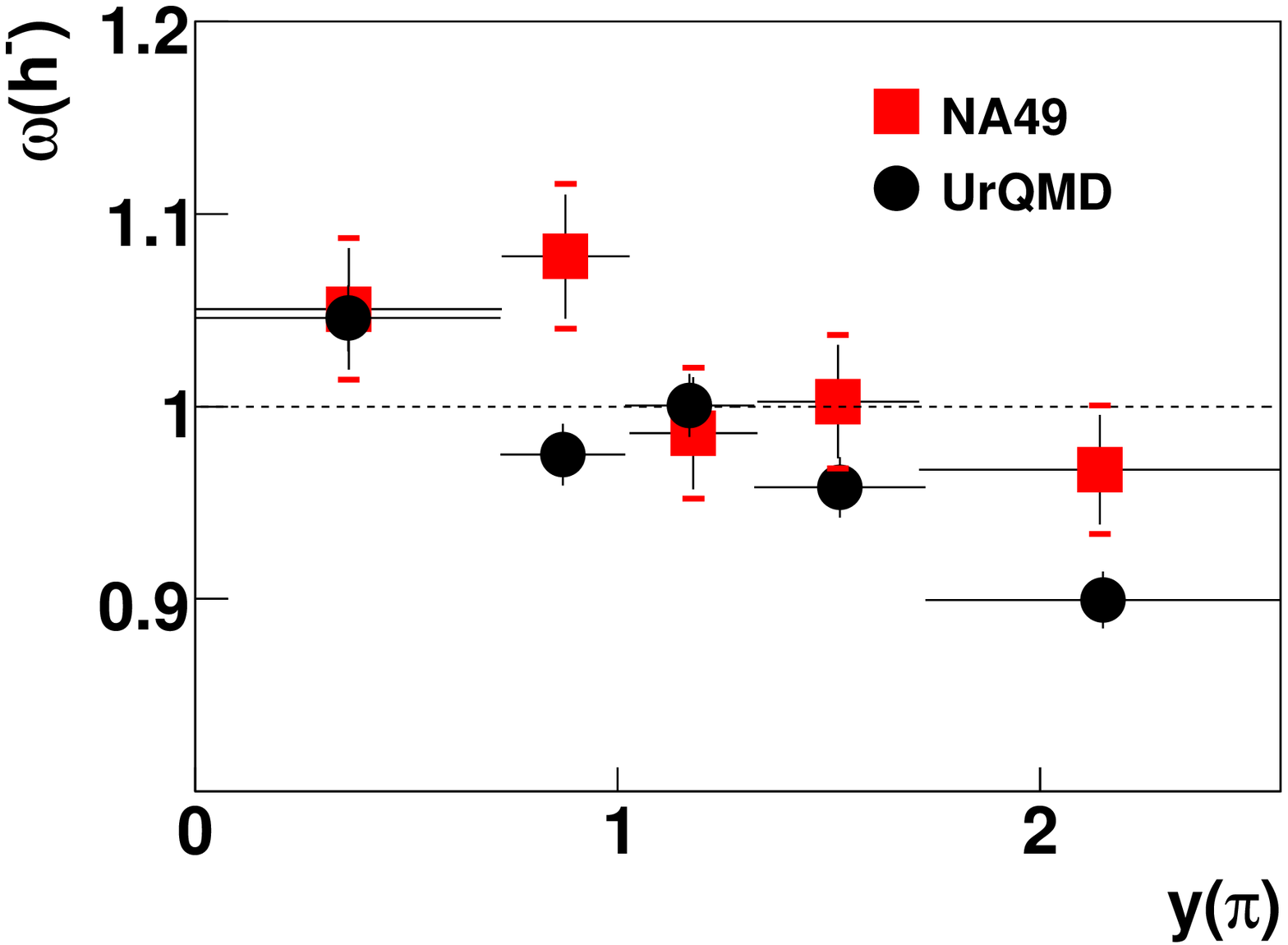}\\
\includegraphics[height=4.3cm]{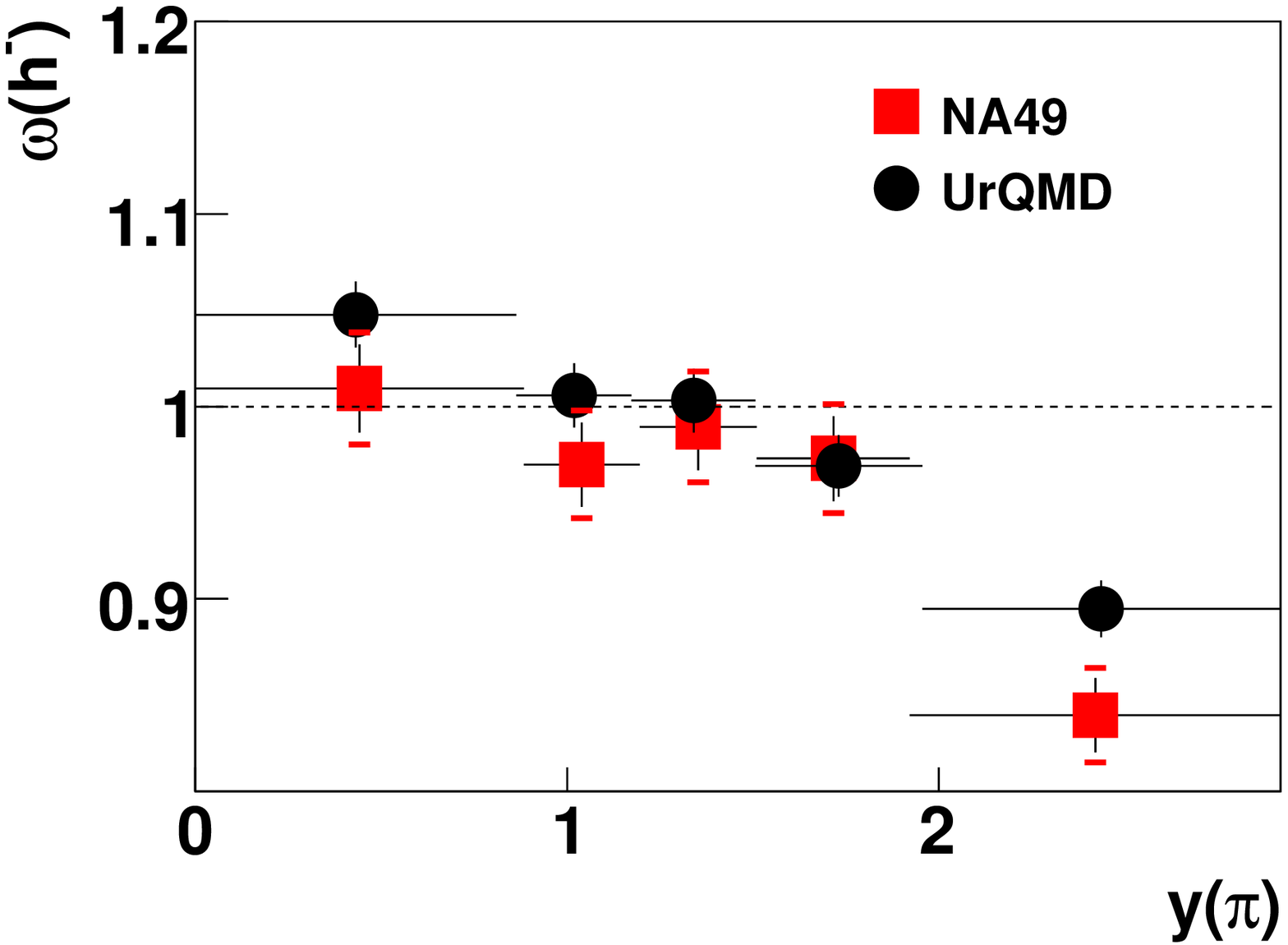}
\caption{\label{ydep_hm}(Color online) Rapidity dependence of the scaled variance $\omega$ of the multiplicity distribution
of negatively charged hadrons in central \Pb collisions 
at $20A$ (top), $30A$, $40A$, $80A$ and $158A$ GeV (bottom) compared to UrQMD predictions with a centrality selection similar 
to the one for the experimental data. The rapidity bins are constructed in such 
a way that the mean multiplicity in each bin is the same.}
\end{figure}

\begin{figure}
\includegraphics[height=4.3cm]{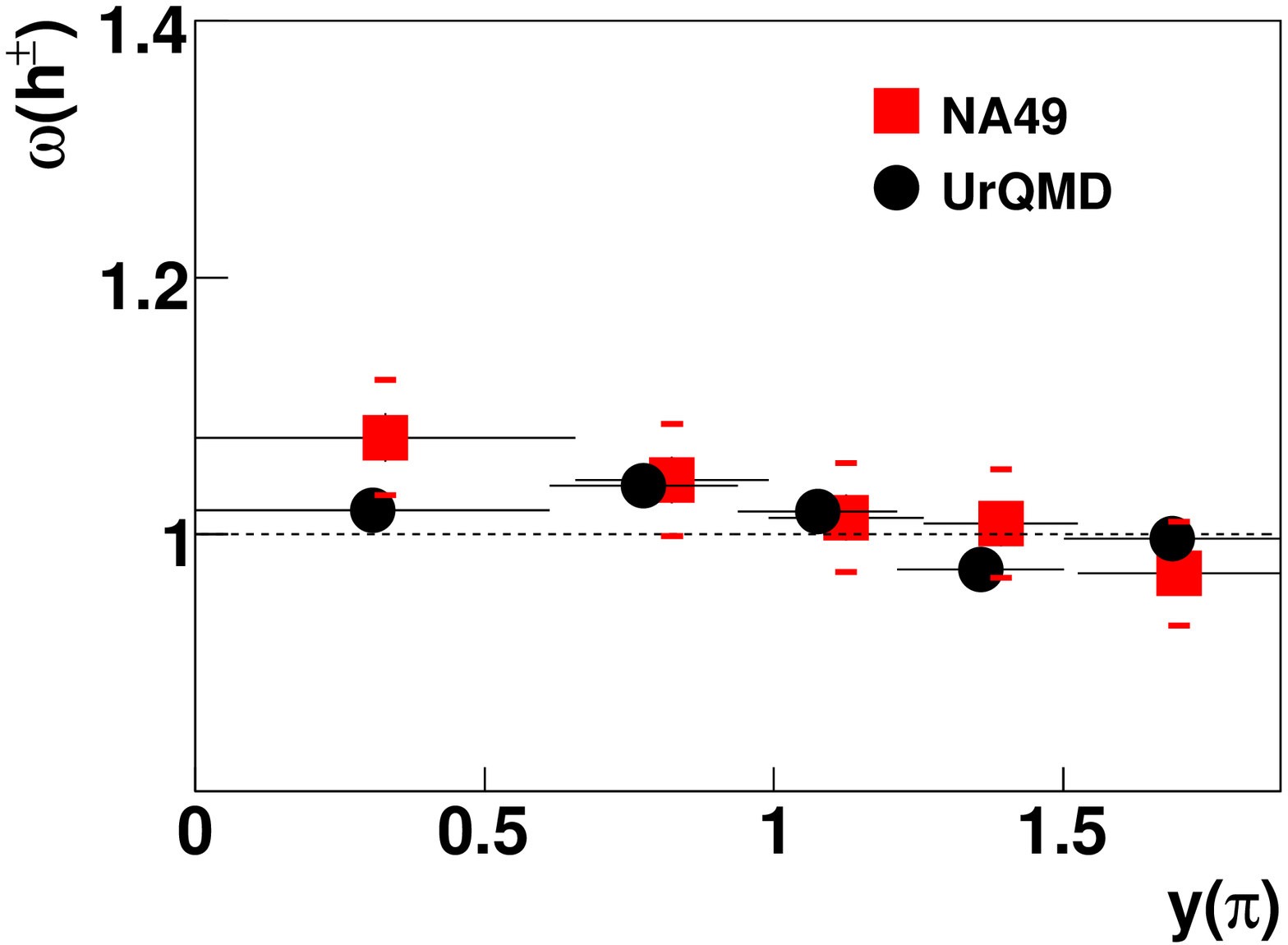}\\
\includegraphics[height=4.3cm]{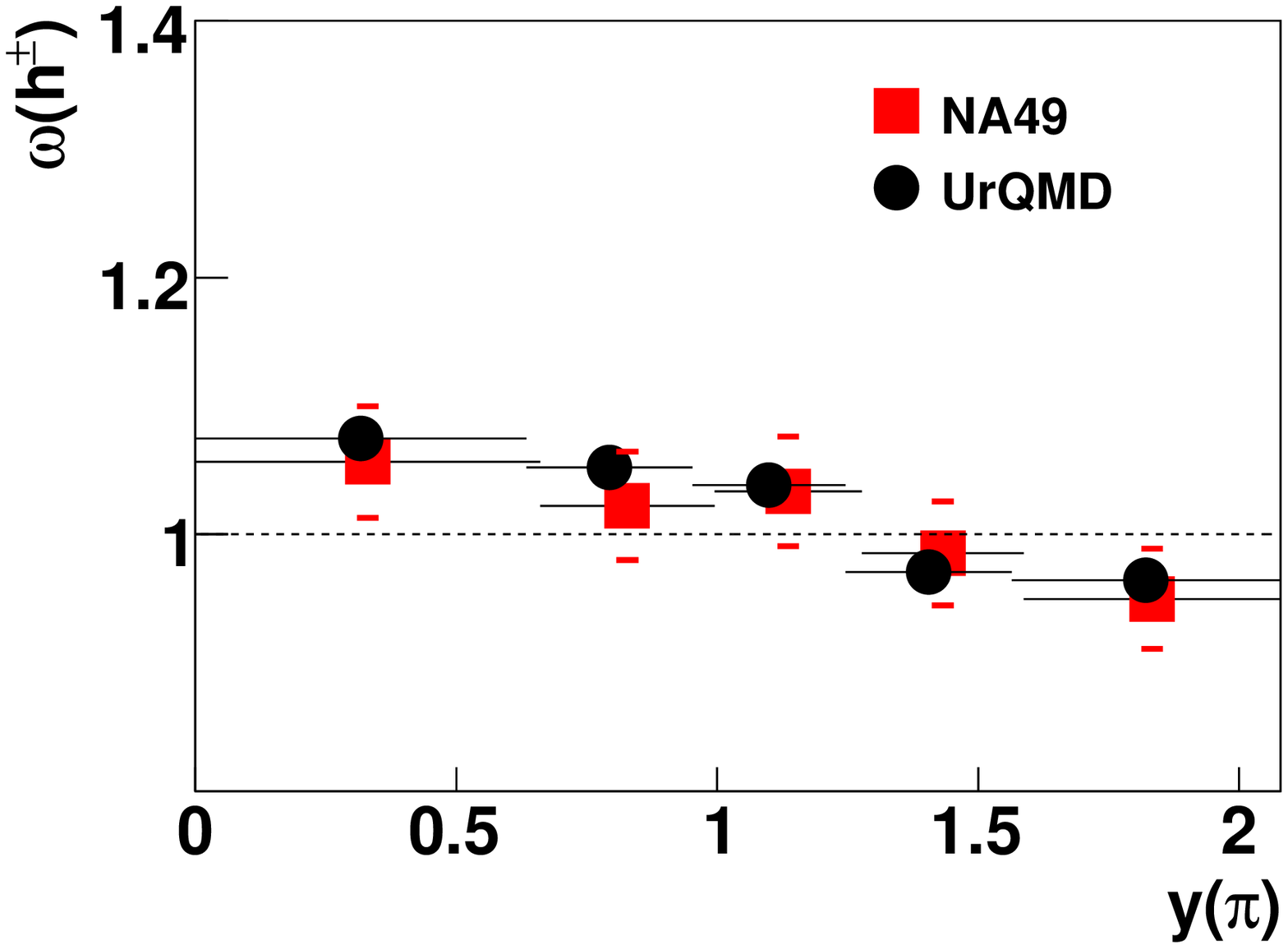}\\
\includegraphics[height=4.3cm]{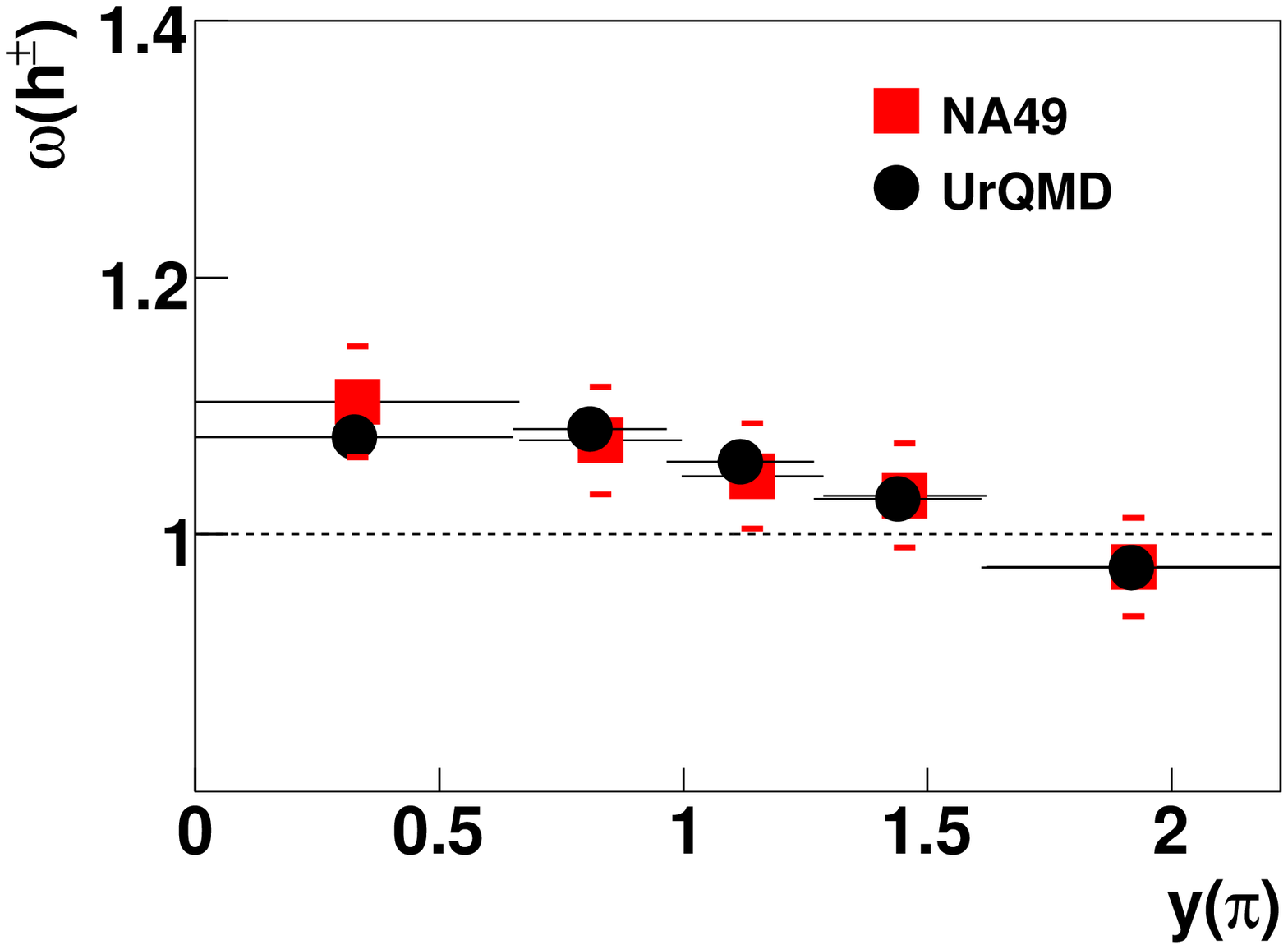}\\
\includegraphics[height=4.3cm]{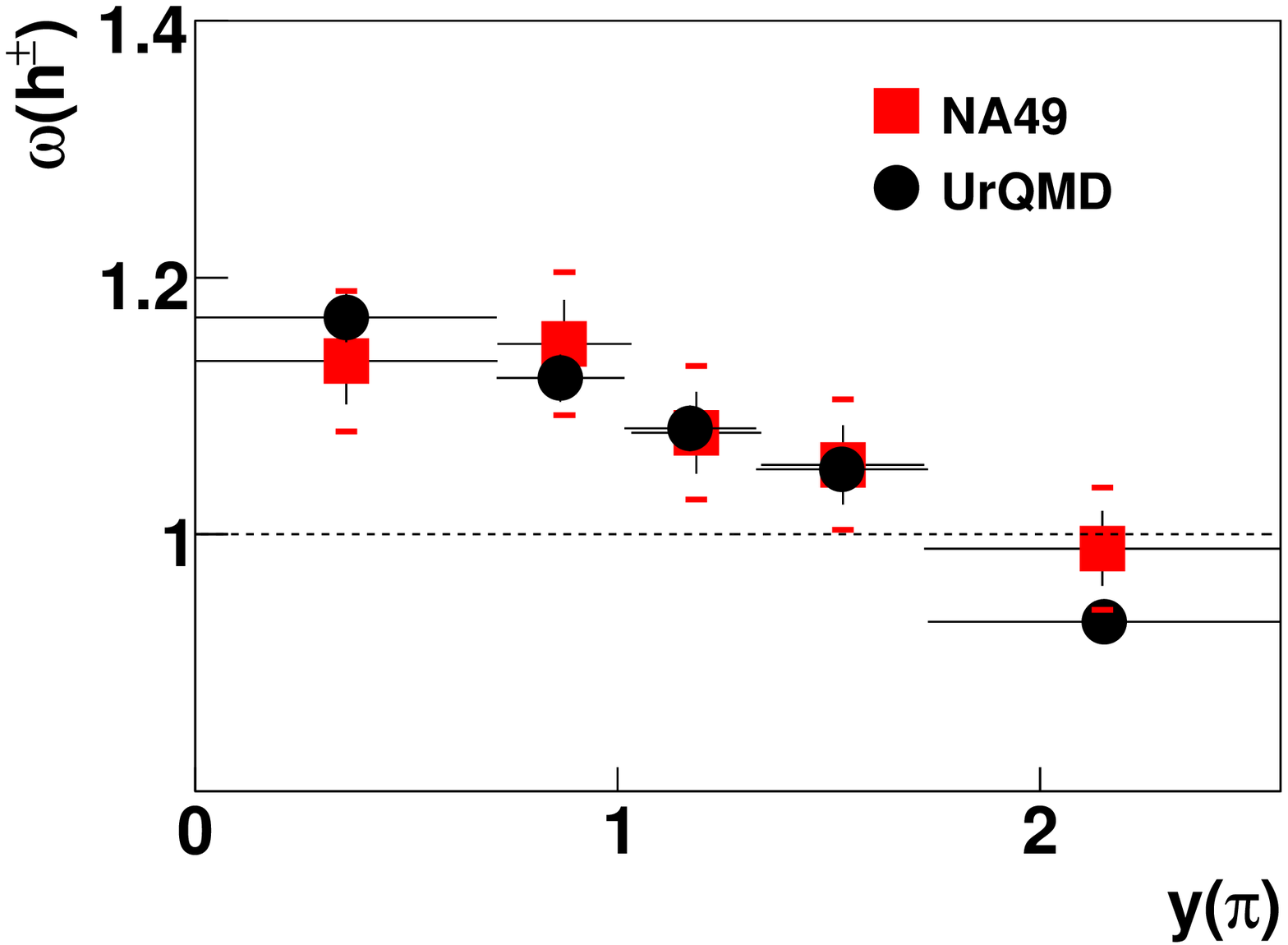}\\
\includegraphics[height=4.3cm]{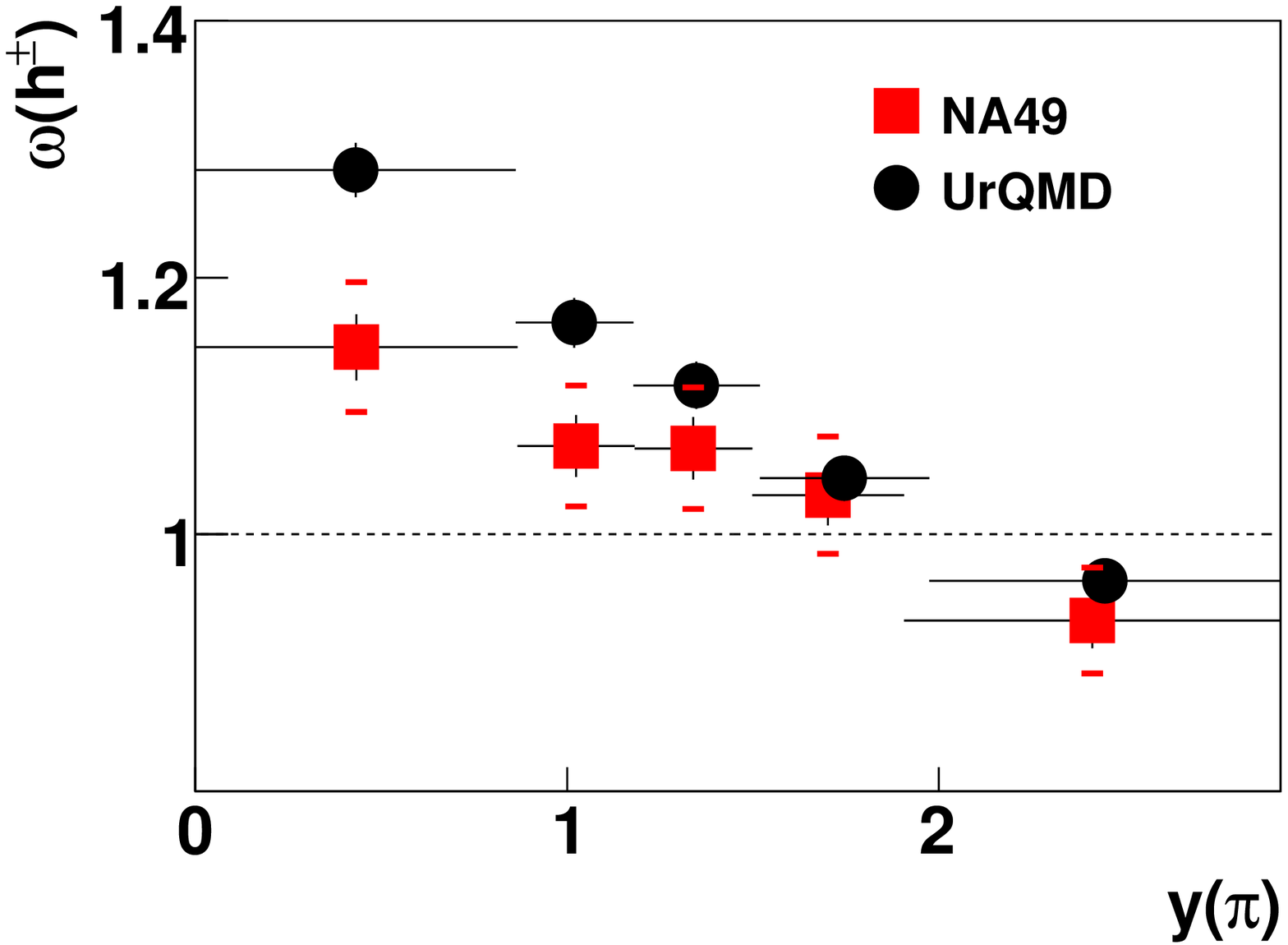}
\caption{\label{ydep_hpm}(Color online) Rapidity dependence of the scaled variance $\omega$ of the multiplicity distribution
of all charged hadrons in central \Pb collisions 
at $20A$ (top), $30A$, $40A$, $80A$ and $158A$ GeV (bottom) compared to UrQMD predictions with a centrality selection similar 
to the one for the experimental data. The rapidity bins are constructed in such 
a way that the mean multiplicity in each bin is the same.}
\end{figure}

The rapidity dependence of the scaled variance $\omega$ of the multiplicity distributions
for $20A$, $30A$, $40A$, $80A$ and $158A$ GeV central Pb+Pb collisions is shown in
Figs.~\ref{ydep_hp}-\ref{ydep_hpm}. 
In order to remove the "trivial" dependence of $\omega$ on 
the fraction of accepted tracks (see Eq.~\ref{wscale})
the rapidity bins $y_c-\Delta y < y < y_c + \Delta y$ are constructed in such a way that 
the mean multiplicity in each bin is the same. 

If there were no correlations in momentum space and the single particle spectra are independent of particle multiplicity, 
the resulting values of $\omega$ shown in Figs.~\ref{ydep_hp}-\ref{ydep_hpm} 
would be independent of rapidity. This is not the case,
the experimental data show an increase of $\omega$ towards midrapidity for all charges and energies. 

\subsection{Transverse Momentum Dependence of $\omega$}

\begin{figure}
\includegraphics[height=6cm]{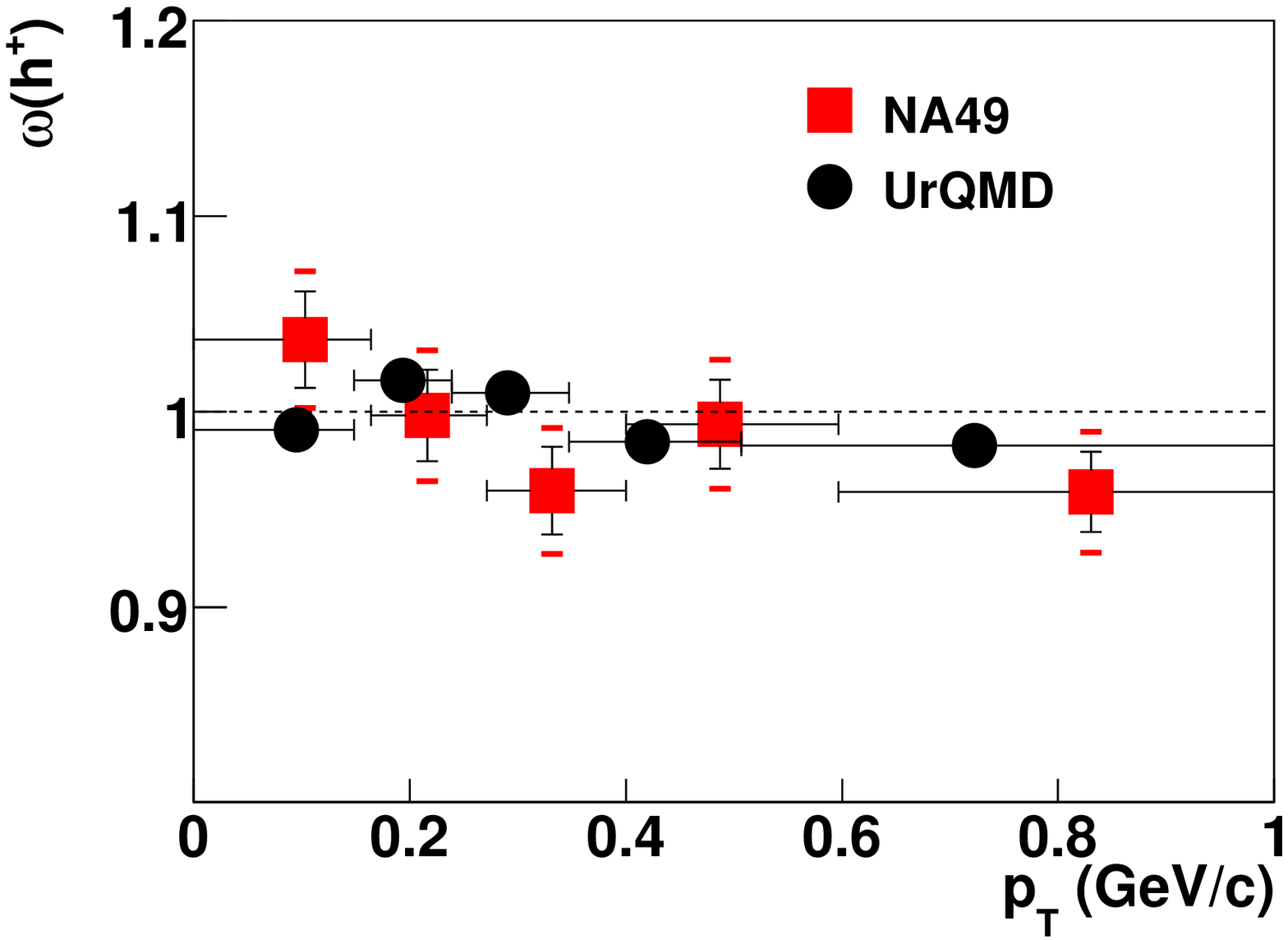}\\
\includegraphics[height=6cm]{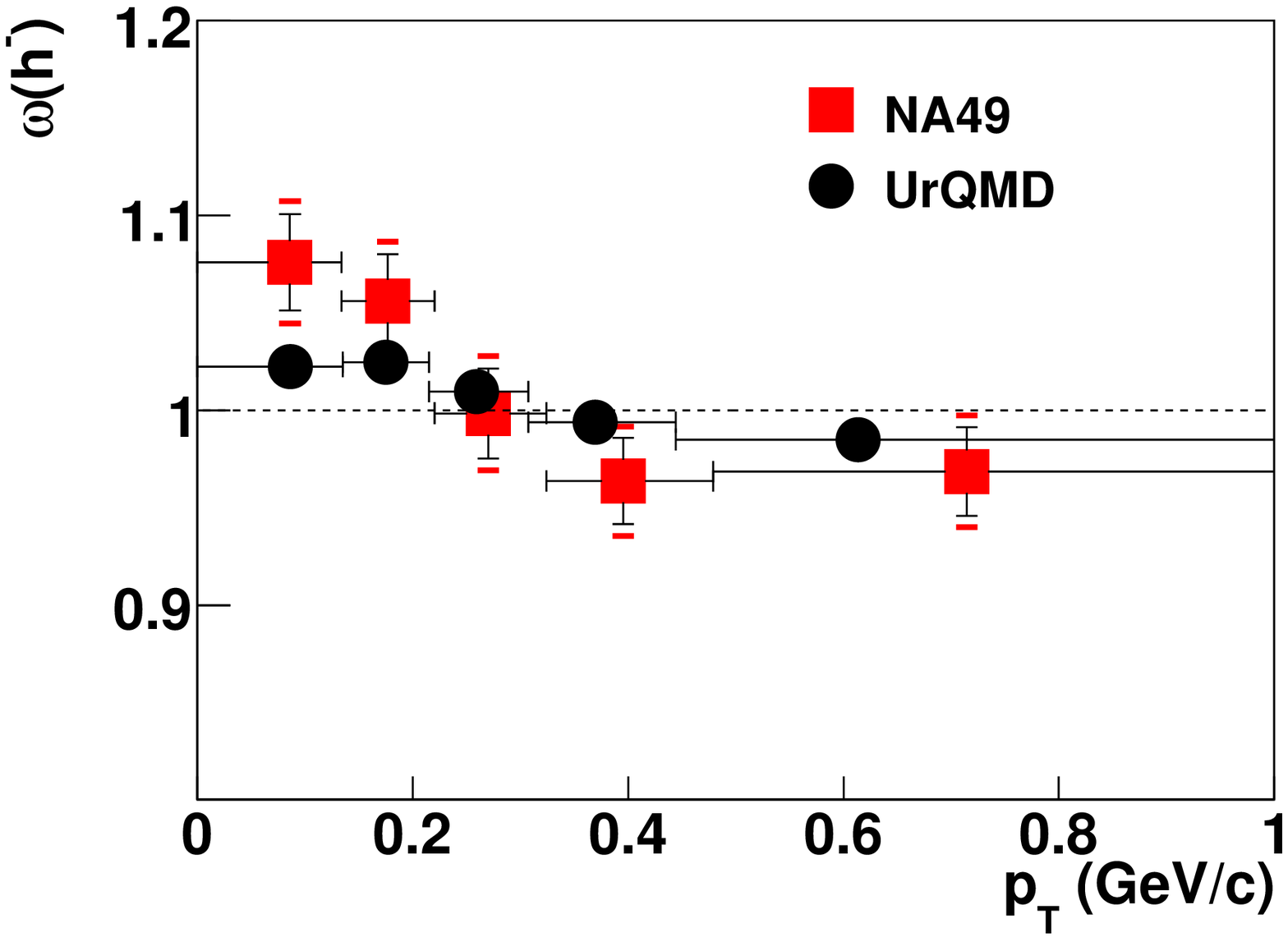}\\
\includegraphics[height=6cm]{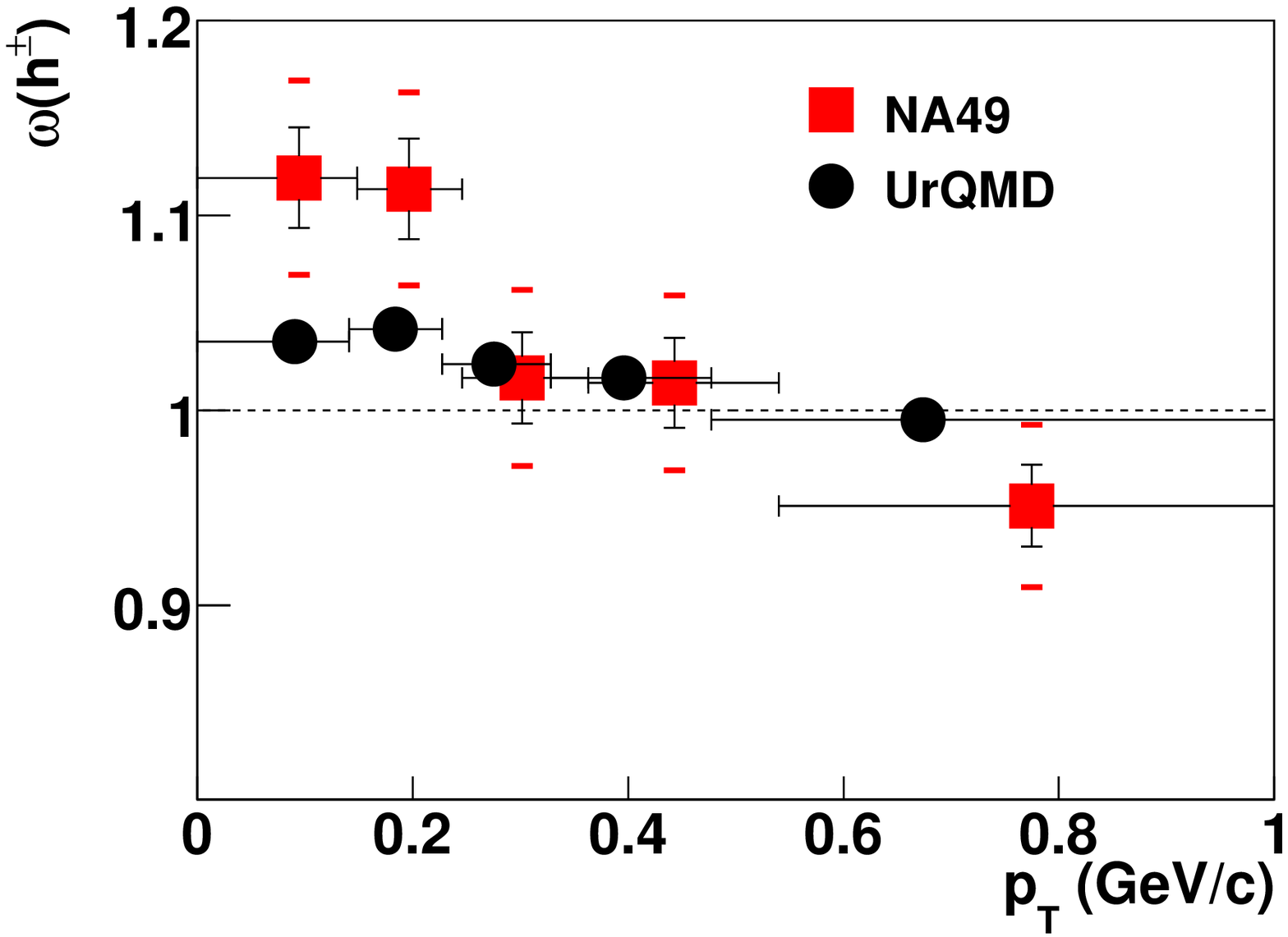}
\caption{\label{ptdep}(Color online) Transverse momentum dependence of the scaled variance of the multiplicity distribution
of positive (top), negative and all charged (bottom) hadrons
in the rapidity interval $1.25<y(\pi)<1.75$ in central \Pb collisions at \eh. 
}
\end{figure}

The transverse momentum dependence of $\omega$ at top SPS energy is shown in Fig.~\ref{ptdep}. The transverse momentum range of
$0-1.5$ $\rm{GeV/c}$ is divided into five bins in such a way that the mean multiplicity in each bin is the same. The horizontal position of 
the points in Fig.~\ref{ptdep} correspond to the center of gravity of the transverse momentum distribution in the transverse momentum range 
of the corresponding bin.
Only a small rapidity interval in the forward acceptance
($1.25<y(\pi)<1.75$) is used for this study. A larger rapidity interval might cause a bias because the acceptance in rapidity is different for different 
transverse momenta. 

An increase of $\omega$ with decreasing transverse momentum, which is more pronounced for $\omega(h^-)$ than for $\omega(h^+)$,
is found. Only the top SPS energy is shown because 
at lower energies the azimuthal acceptance of
the NA49 detector is much smaller and therefore $\omega$
would approach one due to the small multiplicity.

\section{Model Comparison}\label{c_modcomp}

\subsection{Hadron-resonance gas model}\label{c_statmod}

In a hadron-resonance gas model an equilibrium state of hadrons and hadronic resonances is assumed.
Three different statistical ensembles are considered, namely the grand-canonical, the canonical and the micro-canonical ensemble,
which differ by the conservation laws which are taken into account.
In the grand-canonical ensemble conservation laws are not obeyed on an event-by-event basis,
whereas in the canonical ensemble the total baryon number, strangeness and electrical charge have to 
be conserved in each event.
In the micro-canonical ensemble the total energy and momentum are conserved in addition.

\begin{figure}
\includegraphics[width=9.5cm]{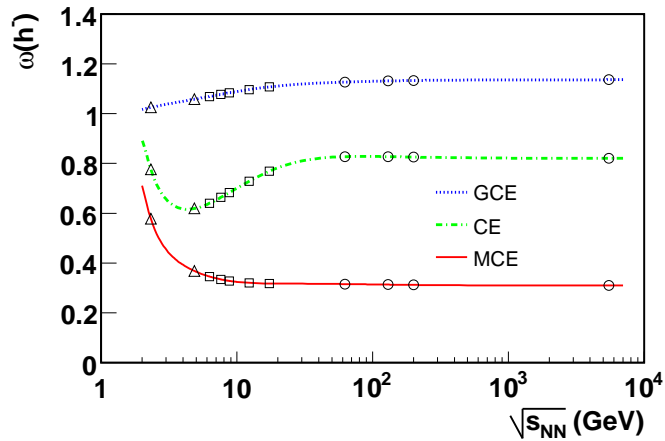}
\caption{\label{w_statmod_4pi}(Color online) Predictions of a hadron-resonance 
gas model for the scaled variance $\omega$ of the multiplicity distribution
in full phase space for negatively charged hadrons.
The parameters of the ensemble, ($T$, $\mu_B$) are the values of the chemical freeze-out
obtained by a hadron-gas model fit to produced particle ratios at different energies.  
Results are shown for the grand-canonical (GCE), canonical (CE) and micro-canonical ensemble (MCE).
The plot is taken from Ref.~\cite{Begun:2006uu}.}
\end{figure}

In \cite{Begun:2006uu} the fluctuations of particle multiplicity in full phase-space 
were calculated for
these three different statistical ensembles in the infinite volume limit. The energy dependence of multiplicity
fluctuations is introduced via the chemical freeze-out parameters $T$ (temperature) and 
$\mu_B$ (baryo-chemical potential), which have been determined by hadron-resonance gas model
fits at all energies to the mean particle multiplicities. 
Quantum statistics and resonance decays are included in the model calculations.
The scaled variance $\omega$ of the multiplicity distribution 
of negatively charged hadrons is shown in Fig.~\ref{w_statmod_4pi} as a function of collision energy.

The results for $\omega$ in the micro-canonical, canonical 
and grand canonical ensemble are very different
at high collision energies. The well known equivalence of statistical ensembles 
in the large volume limit only holds for mean values, 
not for multiplicity fluctuations.

The value of $\omega$ is the largest in the grand-canonical ensemble. In the micro-canonical ensemble 
it is the smallest, the canonical ensemble lies in between. 
In the canonical and micro-canonical ensemble for positively and negatively charged particles separately
narrower than Poisson ($\omega < 1$) multiplicity fluctuations are expected.
The difference between the grand-canonical, canonical and micro-canonical ensemble show the importance of a proper treatment of 
conservation laws for modelling multiplicity fluctuations.

In order to compare the hadron resonance gas model predictions with experimental data, 
$\omega$ calculated in full phase space
is extrapolated to the experimental acceptance
using Eq.~\ref{wscale}.
Although quantum effects and resonance decays introduce correlations in momentum space, Eq.~\ref{wscale}
is the only presently known way to compare the predictions of the grand-canonical and canonical ensemble
to the experimental data.
For the micro-canonical ensemble the energy and momentum 
conservation introduces stronger correlations in momentum space~\cite{Hauer:2007im}. Therefore 
Eq.~\ref{wscale} cannot serve as a reasonable approximation.
Resonance decays introduce only a weak correlation in momentum space 
for positively and negatively charged hadrons, because only a small number
of resonances decay into two particles with the same charge. 
In contrast a large number of resonances decay into two oppositely charged hadrons,
therefore Eq.~\ref{wscale} is not valid for all charged hadrons.

\begin{figure}
\includegraphics[height=6cm]{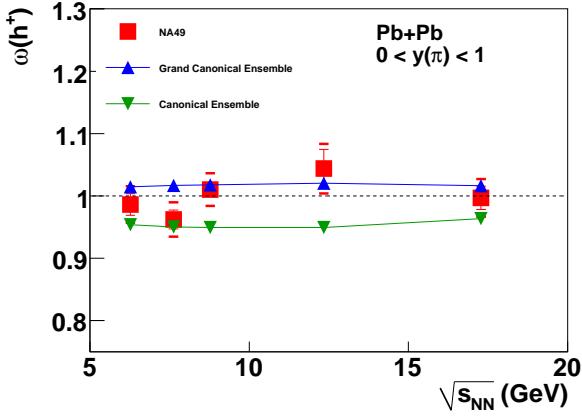}
\includegraphics[height=6cm]{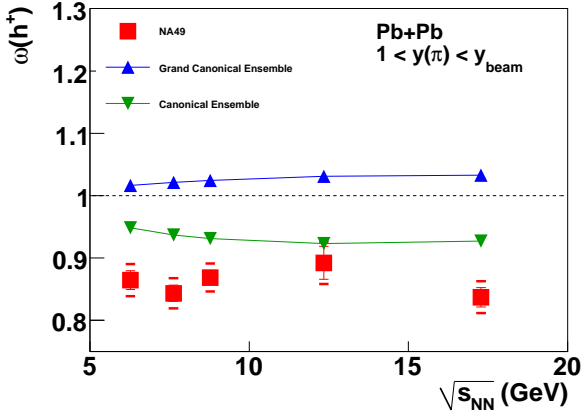}\\
\caption{\label{w_statmod1}(Color online) Scaled variance $\omega$ of the multiplicity distribution of positively  
charged hadrons produced in central \Pb collisions as a function of collision 
energy
in midrapidity (top) and forward (bottom) acceptance 
compared to predictions of a grand canonical and canonical ensemble~\cite{Begun:2006uu}.}
\end{figure}

\begin{figure}
\includegraphics[height=6cm]{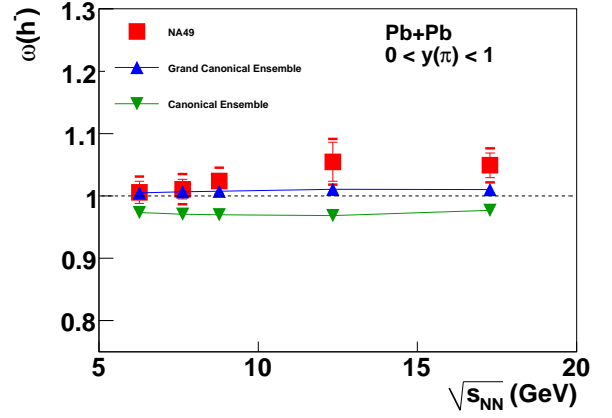}
\includegraphics[height=6cm]{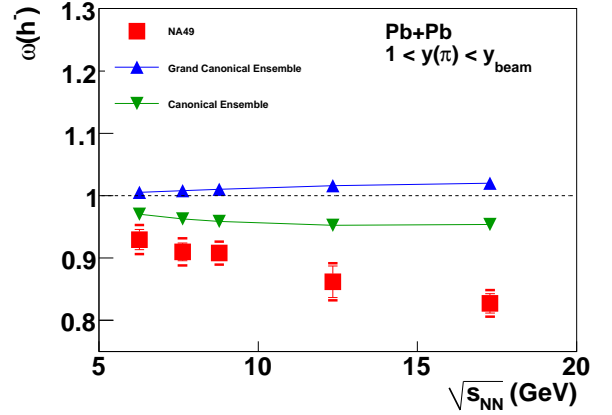}\\
\caption{\label{w_statmod2}(Color online) Scaled variance $\omega$ of the multiplicity distribution of negatively charged hadrons produced in central 
\Pb collisions as a function of collision 
energy
in midrapidity (top) and forward (bottom) 
acceptance compared to predictions of 
a grand canonical and canonical ensemble~\cite{Begun:2006uu}.}
\end{figure}

At forward rapidity ($1<y(\pi)<y_{beam}$; Figs. \ref{w_statmod1} and \ref{w_statmod2}, bottom), the fluctuations are overpredicted by both 
the canonical and the grand canonical models. 
However, the canonical model is closer to data.
A micro-canonical ensemble predicts smaller fluctuations 
than the canonical model, but a quantitative comparison with data is not possible yet, because
correlations in momentum space do not allow to extrapolate 
to the experimental acceptance using Eq.~\ref{wscale}.

At midrapidity $\omega$ of the data (squares in Figs.~\ref{w_statmod1} and \ref{w_statmod2}, top) is higher than in the forward region. 
{In contrast to the experimental data the fluctuations in the number of target participants are not included in the hadron-gas model.
From comparison of UrQMD simulations for $b=0$ collisions and collisions selected according to their veto energy it can be
estimated that
the target participant fluctuations increase $\omega$ by up to $9\%$ in the midrapidity region.}

\begin{figure}
\includegraphics[height=5.9cm]{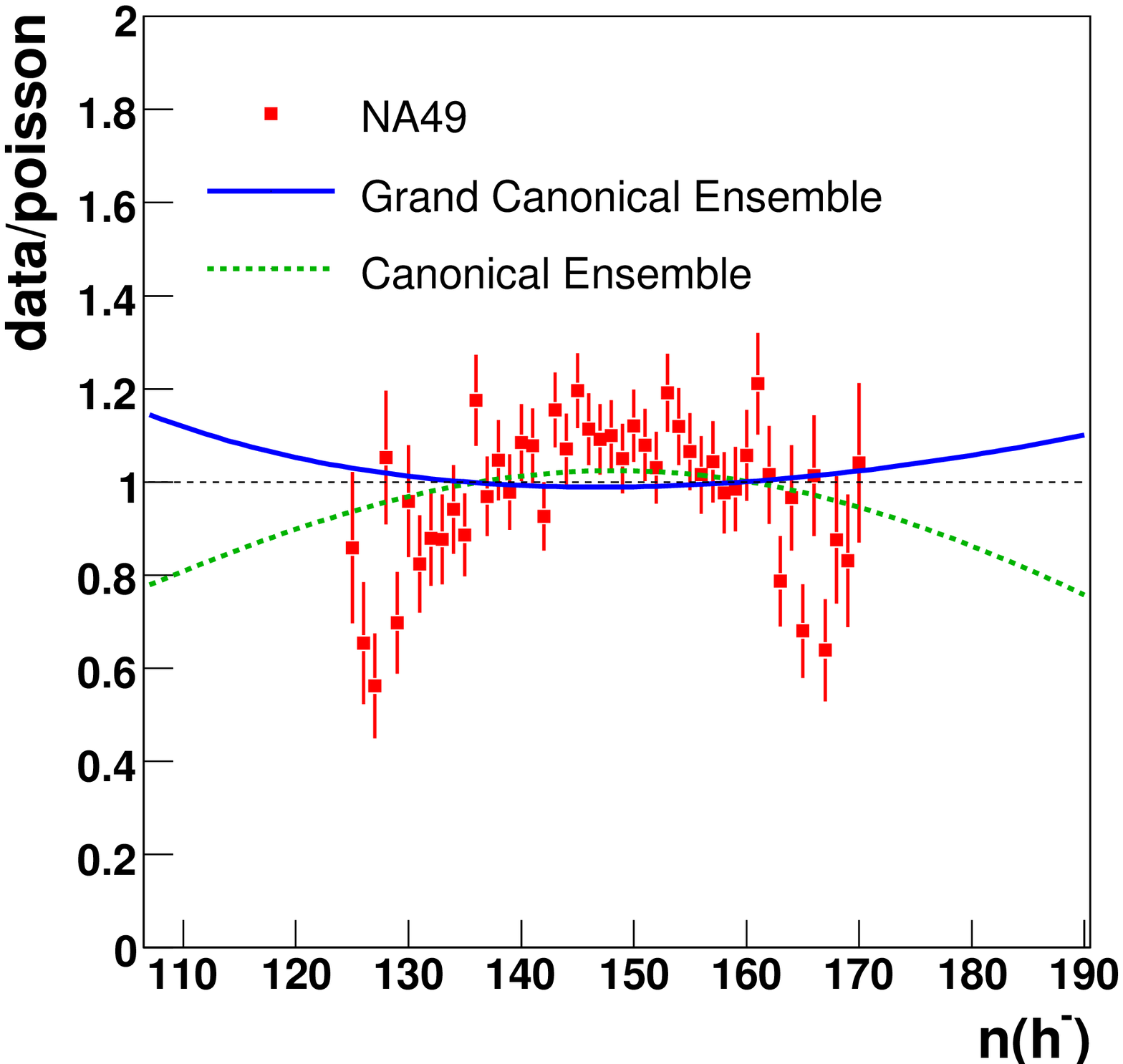}
\includegraphics[height=5.9cm]{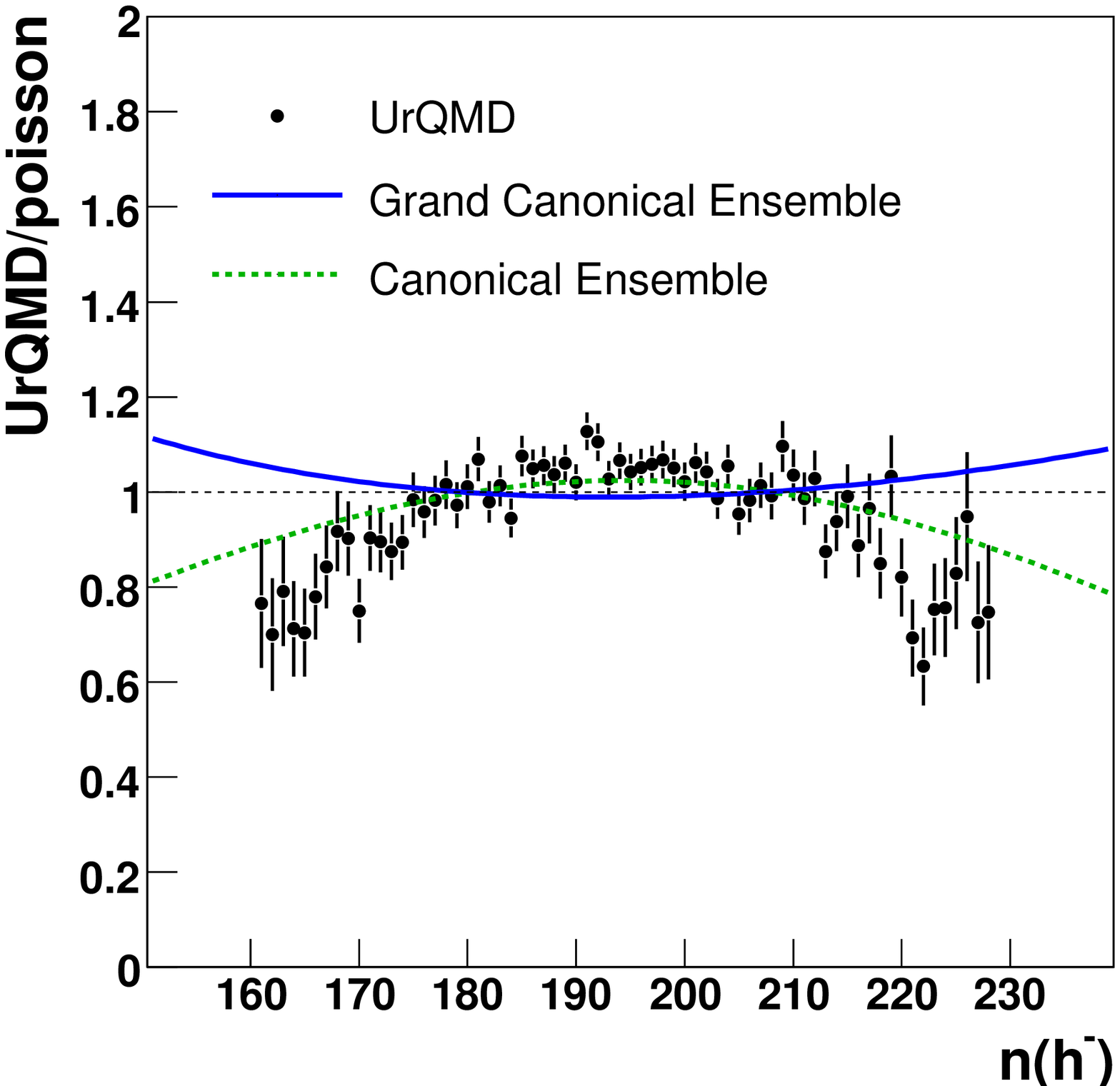}
\caption{\label{mult_rat}(Color online) Ratio of multiplicity distribution of experimental data (top) and UrQMD simulation (bottom) to a Poisson
distribution with the same mean value for negatively charged hadrons in Pb+Pb collisions at \eh in the forward acceptance. Only points with statistical errors smaller
than $20\%$ are shown.
Hadron gas model predictions in the grand-canonical and canonical ensemble with the same mean multiplicity and fraction of accepted
tracks are shown by lines.}
\end{figure}

The shape of the measured multiplicity distribution is compared to the hadron-resonance gas model prediction 
for negatively charged hadrons at \eh in the forward acceptance in Fig.~\ref{mult_rat} (top).
For this comparison the multiplicity distributions for the data and the model predictions are divided by Poisson distributions with the same mean multiplicities.
The hadron-resonance gas model predicts a Gaussian-like shaped multiplicity distribution in full phase space~\cite{Hauer:2007ju}. 
Since this model gives no prediction about the mean multiplicity, it is taken from data.
In order to calculate the multiplicity distribution in the limited experimental acceptance the distribution in the full phase space is 
folded with a Binomial distribution accepting the same fraction $p$ of tracks as the experimental acceptance:
\begin{equation}
B_N(n)=\frac{N!}{(N-n)! n!} p^n (1-p)^{N-n},
\end{equation}
where $N$ is the multiplicity in the full phase space and $n$ the multiplicity in the experimental acceptance.
The multiplicity distribution in the experimental acceptance is given by
\begin{equation}
P_{acc}(n)=\sum_N P_{4\pi}(N) B_N(n).
\end{equation}
Note that this procedure assumes that there are no correlations in momentum space.

The ratio for the grand-canonical ensemble has a concave shape, i.e. the multiplicity distribution is wider than a Poisson distribution. For the
canonical ensemble the shape is convex, showing that the distribution is narrower. 
The shape for the experimental data is more convex, demonstrating that 
the measured multiplicity distribution is even narrower
than the canonical one.

In the canonical and grand canonical ensembles of the hadron-resonance gas model 
no mechanisms are present which would introduce a strong
dependence of multiplicity fluctuations on  rapidity or transverse momentum,
which is observed in the data and in UrQMD (Figs.~\ref{ydep_hp}-\ref{ptdep}).
In a three-pion gas statistical model using the micro-canonical ensemble 
an increase of fluctuations near midrapidity and for
low $p_T$ was observed~\cite{Hauer:2007im} as an effect of energy and momentum conservation.

\subsection{String-hadronic models}\label{urqmd_s}

In this section
the experimental data on multiplicity fluctuations 
are compared to the outcome of 
string-hadronic model calculations, namely
of the Ultra-relativistic Quantum Molecular
Dynamics model (UrQMD v1.3)~\cite{Bleicher:1999xi,Bass:1998ca}
and the Hadron-String Dynamics model (HSD)~\cite{Ehehalt:1996uq}.

The UrQMD microscopic transport approach is based on the propagation of
constituent quarks and di-quarks accompanied by mesonic and baryonic
degrees of freedom. It simulates multiple interactions of
in-going and newly produced particles, the excitation
and fragmentation of colour strings and the formation and decay of
hadronic resonances.
Towards higher energies, the treatment of sub-hadronic degrees of freedom is
of major importance.
A phase transition to a quark-gluon state is
not incorporated explicitly into the model dynamics. 

\begin{figure}
\includegraphics[width=9cm]{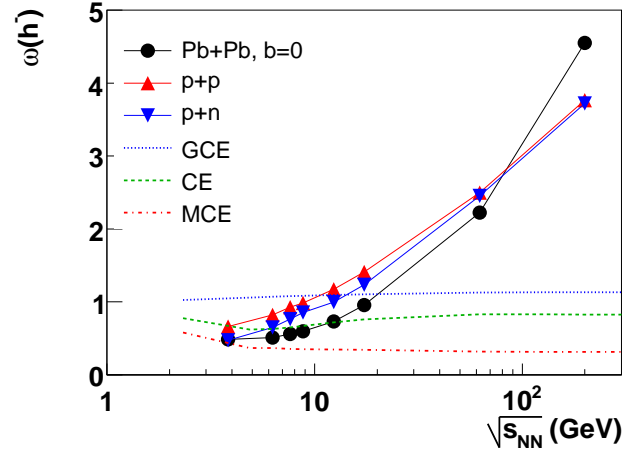}\\
\caption{\label{ed_w_urqmd}(Color online) UrQMD results of 
scaled variance $\omega$ of negatively charged hadrons in full phase space 
in inelastic p+p, p+n interactions and central \Pb 
collisions as a function of collision energy 
compared to hadron resonance gas model predictions~\cite{Begun:2006uu} 
for \Pb collisions. The plot is taken from Ref.~\cite{Lungwitz:2007uc}}
\end{figure}

The scaled variance $\omega$ of the multiplicity distribution of negatively charged hadrons for all inelastic 
p+p and p+n interactions as well as central ($b=0$) 
Pb+Pb collisions predicted by the UrQMD model~\cite{Lungwitz:2007uc} 
is shown in Fig.~\ref{ed_w_urqmd} in dependence of the collision energy.

The scaled variance of multiplicity fluctuations is similar 
in nucleon-nucleon interactions and central heavy ion collisions.
Thus with respect to the scaled variance of multiplicity distributions UrQMD
behaves like a superposition model.
The energy dependence of $\omega$ is different from the predictions of 
the hadron resonance gas model. $\omega$ in UrQMD
shows a strong increase with collision energy in accordance to the experimental p+p data, while the hadron resonance 
gas model has a much weaker energy dependence.

In order to compare the UrQMD model to the experimental data, 
both the acceptance and the centrality selection of 
the NA49 experiment have to be taken into account.
The predictions of the model, published in \cite{Lungwitz:2007uc}, 
are compared to the experimental data in Figs.~\ref{w_hp}-\ref{w_hpm}. 

Two different centrality selections (see section \ref{centsel})
are used in the model:
first, collisions with zero impact parameter (open circles), 
second the $1\%$ most central collisions selected in the same way as done in the experimental data 
using a simulation of the acceptance of the veto calorimeter (full dots). 

The UrQMD model with collisions selected by their energy in the 
veto calorimeter is mostly in agreement with data for all energies, acceptances and charges.
UrQMD simulation of events with zero impact parameter ($b=0$)
gives similar results in the forward rapidity region, whereas $\omega$ is 
smaller in the midrapidity and the full experimental regions, 
probably due to target participant fluctuations, which are still present for events selected by their forward going energy,
but not for collisions with
a zero impact parameter.

The deviation of the multiplicity distribution from a Poisson distribution 
is similar in the model and in the data 
(see Fig.~\ref{mult_rat}), but
the mean multiplicity is overpredicted in the UrQMD model for 
all rapidity intervals, charges and energies by about $20\%$. 
However, the scaled variance of the multiplicity distribution
is independent of mean multiplicity for superposition models. 
Since it was shown that UrQMD behaves like a superposition model for $\omega$, it is justified to compare
$\omega$ for data and UrQMD even though the mean multiplicities are different.
Within this framework one might speculate that the particle production sources in UrQMD are properly modeled but
the number of sources is overestimated in central Pb+Pb collisions.

In the experimental data an increase of fluctuations 
is observed when approaching midrapidity (Figs.~\ref{ydep_hp}-\ref{ydep_hpm}).
The UrQMD model reproduces this behavior 
when a similar centrality selection is used as in the data.

For the data an increase of $\omega$ is measured with decreasing 
transverse momentum at forward rapidity (Fig.~\ref{ptdep}).
In the UrQMD model a similar trend is observed, but $\omega$ 
is underpredicted at low transverse momenta.
This might be related to effects like Coulomb and Bose-Einstein correlations,
which are not implemented in the model.

The HSD transport approach, following a similar strategy as the UrQMD model,
yields similar results for $\omega$. The energy dependence 
for central ($b=0$) Pb+Pb collisions obtained by the HSD model are presented in 
\cite{Konchakovski:2007ss}. These predictions were compared to preliminary NA49 results on multiplicity
fluctuations in~\cite{Lungwitz:2006cx} and were found to agree in the forward acceptance.
Unfortunately HSD calculations for the larger acceptance used in this paper are not
available yet.

\subsection{Onset of Deconfinement}

In heavy ion collisions initial fluctuations in the stopped energy $E$ are expected to 
cause fluctuations in the entropy $S$ \cite{Gazdzicki:2003bb}.
The energy dependences of various hadron production properties, like the kaon to pion ratio, the inverse slope parameter of kaons 
and the pion multiplicity~\cite{Gazdzicki:2004ef,Alt:2007fe} show anomalies at low 
SPS energies which may be attributed to the onset of 
deconfinement~\cite{Gazdzicki:1998vd}. 
In \cite{Gazdzicki:2003bb} it is predicted that this
should lead to a non-monotonic behaviour of the ratio of fluctuations
of entropy to stopped energy 
\begin{equation}\label{r_e}
R_e=\frac{(\delta S)^2 / S^2}{(\delta E)^2 / E^2}.
\end{equation}
At intermediate SPS energies, where a mixed phase of hadron gas and QGP
is assumed, a "shark-fin" structure with a maximum near $80A$ GeV is predicted.
$R_e$ is approximately $0.6$ both in the hadron and quark gluon plasma phase, 
in the mixed phase it can reach values up to $0.8$.

In~\cite{Begun:2006uu} these relative fluctuations are related to multiplicity fluctuations
under the assumption of a proportionality of entropy to produced particle multiplicity, namely:
\begin{equation}\label{omega_de}
\omega_{\delta E} \approx \frac{(\delta E)^2}{E^2} \cdot \left<n\right> \cdot R_e.
\end{equation}
The fluctuations of thermalized energy are obtained by UrQMD and HSD simulations
and are found to be $\delta E/E<0.03$.

Using this result one can estimate the additional multiplicity fluctuations 
of negatively charged hadrons caused by the fluctuations of thermalized energy 
to be $\omega_{\delta E}(h^-) \approx 0.02$
for the pure hadron gas or quark gluon plasma phase.
In the mixed phase the expectation for $\omega_{\delta E}(h^-)$ amounts to $\approx 0.03$
at $80A$ GeV.
The predicted increase of $\omega$ by $0.01$ due
the mixed phase is smaller than the systematic 
error on the measurement of $\omega$. Therefore the data can neither support
nor disprove the existence of a mixed phase at SPS energies.

\subsection{First Order Phase Transition}

It is suggested in~\cite{Mishustin:2006ka} that droplets of hadronic matter should be 
formed in matter when the system crosses the first order phase transition line 
during cool-down. These droplets are expected to produce multiplicity 
fluctuations 10-100 times larger than the Poisson expectation in the full phase space.
No predictions of the increase of $\omega$ for the limited experimental acceptance are available, but naively it can be expected 
to be of the order of 1-10 (according to Eq.~\ref{wscale}).

In our acceptance an excess of multiplicity fluctuations with respect to the UrQMD baseline, 
which does not include an explicit phase transition,
of larger than $0.1$ can be excluded (see Fig.~\ref{w_mub}).

\subsection{Critical Point}

\begin{figure}
\includegraphics[height=6cm]{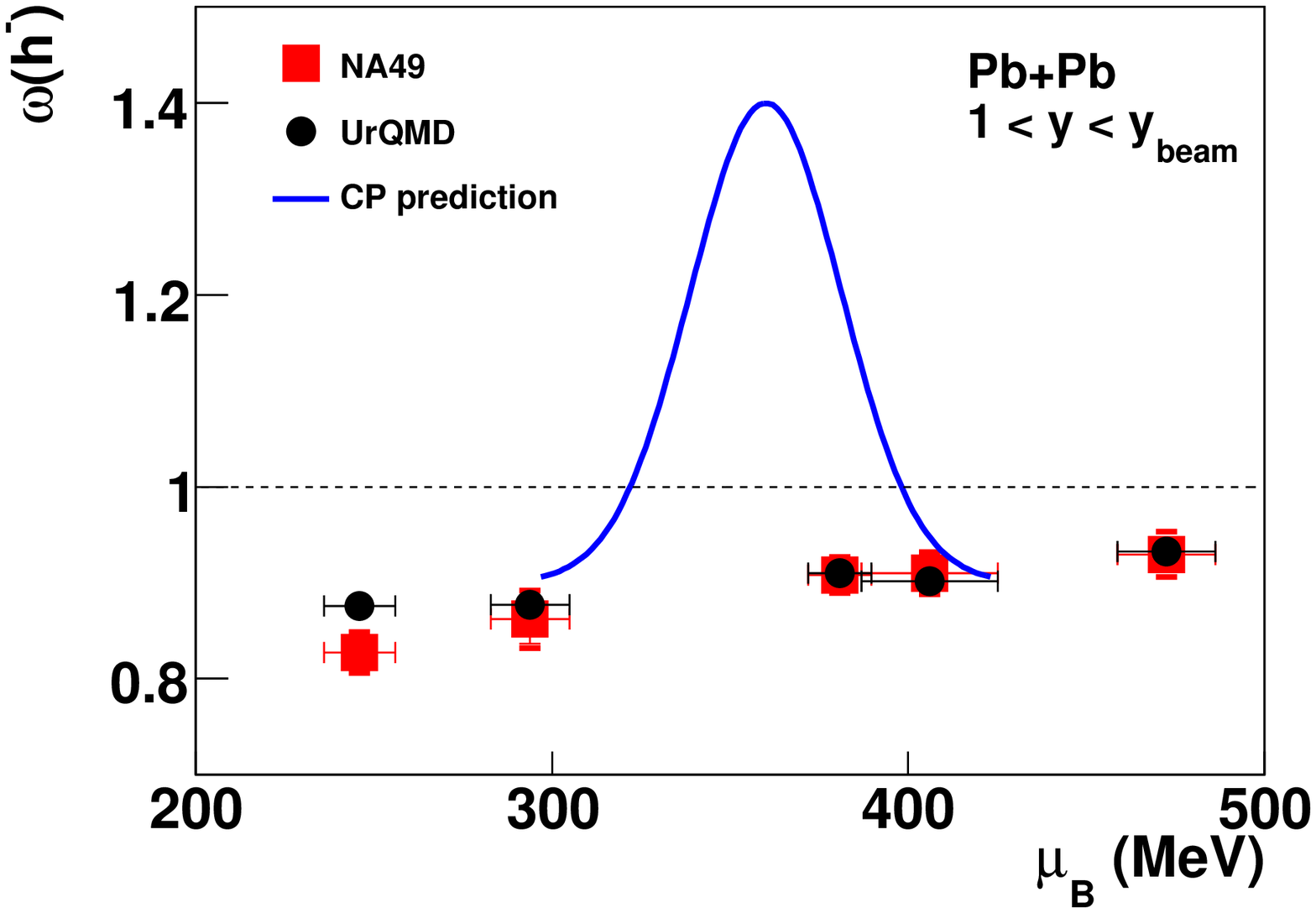}
\includegraphics[height=6cm]{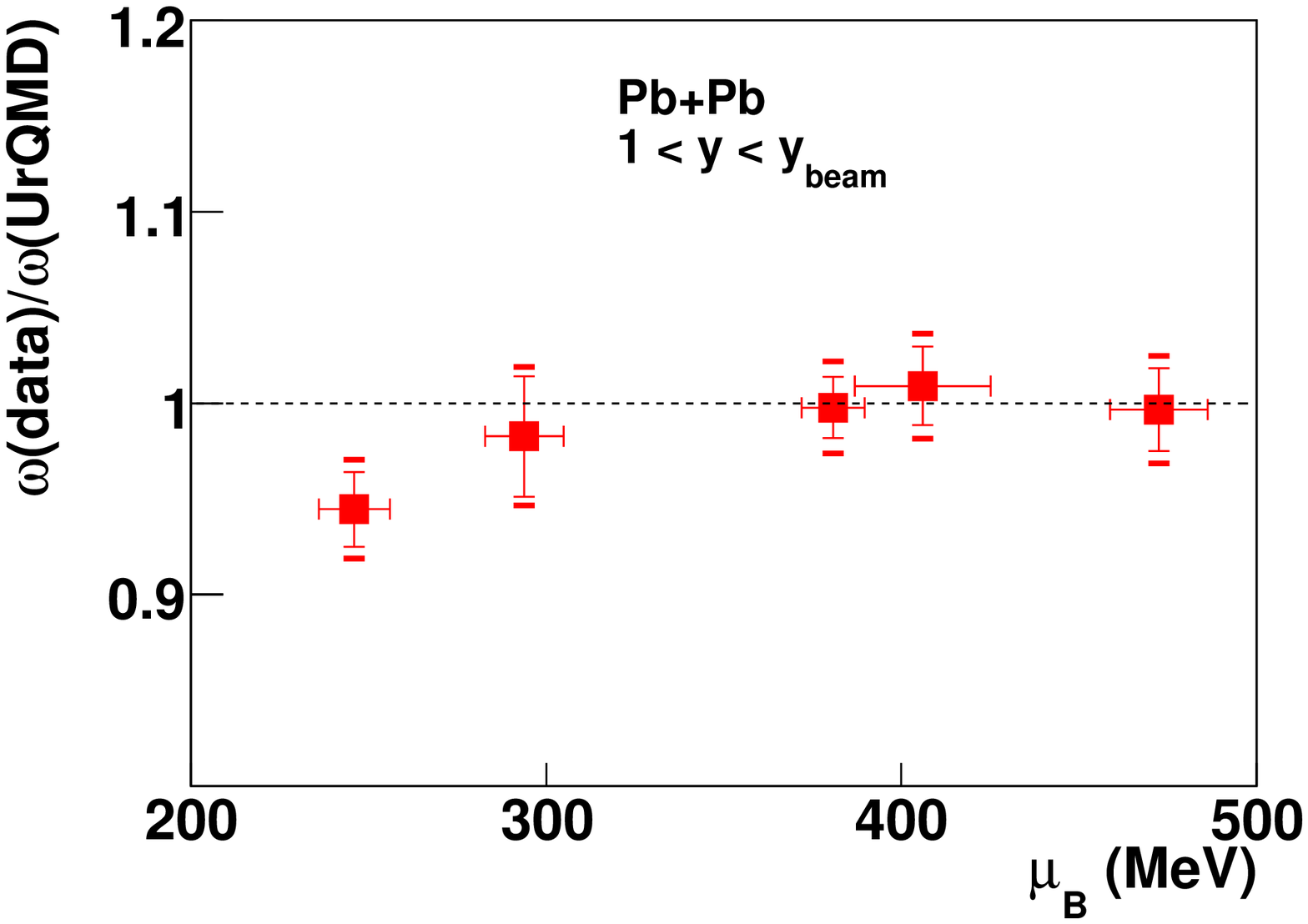}\\
\caption{\label{w_mub}(Color online) Top: Scaled variance $\omega$ of the multiplicity distribution
of negatively charged hadrons at forward rapidities
produced in central \Pb collisions as a function of the baryo-chemical potential $\mu_B$~\cite{Becattini:2005xt}.
{In addition a sketch of the expected increase of $\omega$ due to the critical point~\cite{Stephanov:1999zu,Stephanov:pc} is shown.}
The UrQMD results are given for a centrality selection similar to the experimental data. 
Bottom: Ratio of $\omega$ in data and UrQMD as a function of $\mu_B$.}
\end{figure}

It is expected that the hadron gas and quark gluon plasma regions
in the phase diagram of strongly interacting matter are separated by a first
order phase transition line at high baryo-chemical potentials 
and moderate temperatures. A crossover between both phases is predicted  
for high temperatures and low baryo-chemical potentials.
Then the first order phase transition line will end in a critical point.

If the freeze-out of matter happens near the critical point, 
large fluctuations, for instance in multiplicity and transverse momentum, are
expected.
In \cite{Stephanov:1999zu} it is estimated that the scaled variance of the distribution
of total multiplicity of single charged hadrons should increase by  
about $1$ near the critical point. However, this estimate has a large and
difficult to estimate systematic error. The limited acceptance should
reduce the critical point signal by a factor of about $2$. Consequnetly, the 
expected increase of the scaled variance in the vicinity of the
critical point is about $0.5$.

These critical fluctuations are expected to be located mainly at low transverse momenta~\cite{Stephanov:1999zu}.
The scaled variance as a function of the baryo-chemical potential is compared in Fig.~\ref{w_mub} to the UrQMD baseline.
As the increase of fluctuations due to the freeze-out in the vicinity of the critical point is expected to be restricted to a range in 
the baryo-chemical potential which is comparable to the difference in baryo-chemical potentials of the different collision energies~\cite{Hatta:2002sj},
the signature of the critical point is expected to increase $\omega$ at one collision energy only. 
A sketch of the expected increase of $\omega$ due to the critical point~\cite{Stephanov:1999zu} is shown in Fig.~\ref{w_mub}.
No significant increase of $\omega$ which may be attributed to the critical point is observed in the data. 
The scaled variance for low transverse momentum particles (see Fig.~\ref{w_lowpt}) does not show a significant non-monotonic structure or 
excess over the UrQMD baseline either.

\section{Summary}\label{c_summary}

The energy dependence of multiplicity fluctuations in central 
Pb+Pb collisions at $20A$, $30A$, $40A$, $80A$ and $158A$ GeV
was studied for positively, negatively and all charged hadrons. 
The total selected experimental acceptance ($0<y(\pi)<y_{beam}$) is divided 
into a midrapidity ($0<y(\pi)<1$) and a forward rapidity
region ($1<y(\pi)<y_{beam}$). 
At forward rapidity a suppression of fluctuations compared 
to a Poisson distribution is observed for positively and negatively charged hadrons.
At midrapidity and for all charged hadrons the fluctuations are higher.
Furthermore the rapidity dependence at all energies and the transverse momentum dependence 
at $158A$ GeV 
were studied. The scaled variance of the multiplicity distribution increases for decreasing rapidity and
transverse momentum.

The string-hadronic model UrQMD significantly overpredicts the mean multiplicities,
but approximately reproduces the scaled variance of the multiplicity 
distributions.

Multiplicity fluctuations predicted by
the grand-canonical and canonical formulations of the hadron resonance gas 
model~\cite{Begun:2006uu}  overpredict fluctuations in the forward acceptance.
The micro-canonical formulation predicts smaller fluctuations
and can qualitatively reproduce the increase of fluctuations for low rapidities and
transverse momenta. However no quantitative 
calculation is available yet for the limited experimental acceptance.

At RHIC and LHC energies the difference in $\omega$ for the string-hadronic and the hadron-gas 
models in the full phase space is much larger than for SPS energies and experimental
data should be able to distinguish between them rather easily. 

Narrower than Poissonian ($\omega < 1$) multiplicity fluctuations are measured in 
the forward kinematic region ($1<y(\pi)<y_{beam}$). They can be related to the reduced
fluctuations predicted for relativistic gases with imposed conservation laws. 
This general feature of relativistic gases may
be preserved also for some non-equilibrium systems as modeled by the string-hadronic approaches.

The predicted maximum in fluctuations due to a first order phase transition from hadron resonance 
gas to QGP~\cite{Gazdzicki:2003bb} is smaller than the 
experimental errors of the present measurements and can therefore neither be confirmed nor disproved.

No sign of increased fluctuations as expected for a freeze-out near the 
critical point of strongly interacting matter was observed. 
The future NA61 program~\cite{Gazdzicki:2006fy} will study both the energy 
and system size dependence of fluctuations with improved sensitivity 
in a systematic search for the critical point.

\begin{appendix}

\section{Derivations}

\subsection{Acceptance Dependence of $\omega$}\label{a_acc_dep}

Provided the particles are produced independently in momentum space and the form of the 
momentum distribution is independent of multiplicity, 
the scaled variance in a limited acceptance is related to the scaled variance 
in full phase-space ("4$\pi$") by an analytic formula.

Under these assumptions, having an experimental acceptance registering the fraction $p$ of the total number of 
tracks $N$ is equivalent to roll a dice for each particle in the full phase space and to accept it with
a probability of $p$.
Therefore the probability to measure a number of particles $n$ in a fixed acceptance follows a Binomial distribution:
\begin{equation}
B(n|N)=\frac{N!}{n! (N-n)!} p^n (1-p)^{N-n}.\label{binoE}
\end{equation}
For a number of particles varying in the full phase space according to $P_{4\pi}(N)$, 
the probability to measure a number of particles $n$ in the limited acceptance is:
\begin{equation}
P_A(n)=\sum_{N} B(n|N) P_{4\pi}(N).\label{A2E}
\end{equation}
From Eqs.~\ref{binoE},\ref{A2E} follow that the mean number of particles in the acceptance is:
\begin{equation}
<n>=p <N>,
\end{equation}
and the variance of the number $n$ of particles in the acceptance is given by:
\begin{equation}
\begin{split}
Var(n)&=<Var(n|N)>+Var(<n|N>)\\
&=<Var(n|N)>+Var(p N)\\
&=<N> p(1-p)+p^2 Var(N).
\end{split}
\end{equation}
Finally, the scaled variance in the limited acceptance $\omega_{acc}$ is related to the scaled variance in the
full phase space, $\omega_{4\pi}$, as:
\begin{equation}\label{wscale2}
\omega_{acc}=p \left( \omega_{4\pi} -1 \right) + 1.
\end{equation}

The acceptance dependence given by Eq.~\ref{wscale2} is not valid when effects like resonance decays, 
quantum statistics and energy- momentum conservation introduce correlations in momentum space.

\subsection{Participant Fluctuations}\label{a_partfluct}

In a superposition model the multiplicity $n$ is the sum of the number of particles 
produced by $k$ particle production sources:
\begin{equation}
n=\sum_{i=1}^{k}{n_{i}^{so}},
\end{equation}
where the summation index $i$ runs over the sources.
Assuming statistically identical sources the mean multiplicity is:
\begin{equation}
\left< n \right> = \left< k \right> \left< n^{so} \right>,
\end{equation}
and the variance reads:
\begin{equation}
Var(n) = \left< k \right> Var(n^{so}) + \left< n^{so} \right>^2 Var(k).
\end{equation}
Using these equations the scaled variance of $n$ can be expressed as:
\begin{equation}
\omega=\frac{\left< k \right> Var(n^{so})}{\left< k \right> \left< n^{so} \right>}
+ \frac{\left< n^{so} \right>^2 Var(k)}{\left< k \right> \left< n^{so} \right>}=
\omega^{so}+\left<n^{so}\right>\cdot \omega_k.
\end{equation}
For the case of a constant number of sources the scaled variance is independent of the number of sources.

\end{appendix}

\begin{acknowledgments}
Fruitful discussions with M. Bleicher, M. Hauer, E. Bratkovskaya, V. Konchakovski, 
V. Begun, M. Gorenstein and I. Mishustin are gratefully acknowledged.\\
This work was supported by the US Department of Energy
Grant DE-FG03-97ER41020/A000,
the Bundesministerium fur Bildung und Forschung, Germany (06F137), 
the Virtual Institute VI-146 of Helmholtz Gemeinschaft, Germany,
the Polish State Committee for Scientific Research (1 P03B 006 30, 1 P03B 097 29, 1 PO3B 121 29, 1 P03B 127 30),
the Hungarian Scientific Research Foundation (T032648, T032293, T043514),
the Hungarian National Science Foundation, OTKA, (F034707),
the Polish-German Foundation, the Korea Science \& Engineering Foundation (R01-2005-000-10334-0),
the Bulgarian National Science Fund (Ph-09/05) and the Croatian Ministry of Science, Education and Sport (Project 098-0982887-2878).
\end{acknowledgments}

\bibliography{biblio} 

\begin{thebibliography}{51}
\expandafter\ifx\csname natexlab\endcsname\relax\def\natexlab#1{#1}\fi
\expandafter\ifx\csname bibnamefont\endcsname\relax
  \def\bibnamefont#1{#1}\fi
\expandafter\ifx\csname bibfnamefont\endcsname\relax
  \def\bibfnamefont#1{#1}\fi
\expandafter\ifx\csname citenamefont\endcsname\relax
  \def\citenamefont#1{#1}\fi
\expandafter\ifx\csname url\endcsname\relax
  \def\url#1{\texttt{#1}}\fi
\expandafter\ifx\csname urlprefix\endcsname\relax\def\urlprefix{URL }\fi
\providecommand{\bibinfo}[2]{#2}
\providecommand{\eprint}[2][]{\url{#2}}

\bibitem[{\citenamefont{Collins and Perry}(1975)}]{Collins:1974ky}
\bibinfo{author}{\bibfnamefont{J.~C.} \bibnamefont{Collins}} \bibnamefont{and}
  \bibinfo{author}{\bibfnamefont{M.~J.} \bibnamefont{Perry}},
  \bibinfo{journal}{Phys. Rev. Lett.} \textbf{\bibinfo{volume}{34}},
  \bibinfo{pages}{1353} (\bibinfo{year}{1975}).

\bibitem[{\citenamefont{Shuryak}(1980)}]{Shuryak:1980tp}
\bibinfo{author}{\bibfnamefont{E.~V.} \bibnamefont{Shuryak}},
  \bibinfo{journal}{Phys. Rept.} \textbf{\bibinfo{volume}{61}},
  \bibinfo{pages}{71} (\bibinfo{year}{1980}).

\bibitem[{\citenamefont{Margetis et~al.}(1995)}]{Margetis:1994tt}
\bibinfo{author}{\bibfnamefont{S.}~\bibnamefont{Margetis}} \bibnamefont{et~al.}
  (\bibinfo{collaboration}{NA49}), \bibinfo{journal}{Phys. Rev. Lett.}
  \textbf{\bibinfo{volume}{75}}, \bibinfo{pages}{3814} (\bibinfo{year}{1995}).

\bibitem[{\citenamefont{Heinz and Jacob}(2000)}]{Heinz:2000bk}
\bibinfo{author}{\bibfnamefont{U.~W.} \bibnamefont{Heinz}} \bibnamefont{and}
  \bibinfo{author}{\bibfnamefont{M.}~\bibnamefont{Jacob}}
  (\bibinfo{year}{2000}), \eprint{nucl-th/0002042}.

\bibitem[{\citenamefont{Adams et~al.}(2005)}]{Adams:2005dq}
\bibinfo{author}{\bibfnamefont{J.}~\bibnamefont{Adams}} \bibnamefont{et~al.}
  (\bibinfo{collaboration}{STAR}), \bibinfo{journal}{Nucl. Phys.}
  \textbf{\bibinfo{volume}{A757}}, \bibinfo{pages}{102} (\bibinfo{year}{2005}),
  \eprint{nucl-ex/0501009}.

\bibitem[{\citenamefont{Adcox et~al.}(2005)}]{Adcox:2004mh}
\bibinfo{author}{\bibfnamefont{K.}~\bibnamefont{Adcox}} \bibnamefont{et~al.}
  (\bibinfo{collaboration}{PHENIX}), \bibinfo{journal}{Nucl. Phys.}
  \textbf{\bibinfo{volume}{A757}}, \bibinfo{pages}{184} (\bibinfo{year}{2005}),
  \eprint{nucl-ex/0410003}.

\bibitem[{\citenamefont{Back et~al.}(2005)}]{Back:2004je}
\bibinfo{author}{\bibfnamefont{B.~B.} \bibnamefont{Back}} \bibnamefont{et~al.},
  \bibinfo{journal}{Nucl. Phys.} \textbf{\bibinfo{volume}{A757}},
  \bibinfo{pages}{28} (\bibinfo{year}{2005}), \eprint{nucl-ex/0410022}.

\bibitem[{\citenamefont{Arsene et~al.}(2005)}]{Arsene:2004fa}
\bibinfo{author}{\bibfnamefont{I.}~\bibnamefont{Arsene}} \bibnamefont{et~al.}
  (\bibinfo{collaboration}{BRAHMS}), \bibinfo{journal}{Nucl. Phys.}
  \textbf{\bibinfo{volume}{A757}}, \bibinfo{pages}{1} (\bibinfo{year}{2005}),
  \eprint{nucl-ex/0410020}.

\bibitem[{\citenamefont{Alber et~al.}(2002)}]{Afanasiev:2002mx}
\bibinfo{author}{\bibfnamefont{T.}~\bibnamefont{Alber}} \bibnamefont{et~al.}
  (\bibinfo{collaboration}{NA49 Collaboration}), \bibinfo{journal}{Phys. Rev.}
  \textbf{\bibinfo{volume}{C66}}, \bibinfo{pages}{054902}
  (\bibinfo{year}{2002}), \eprint{nucl-ex/0205002}.

\bibitem[{\citenamefont{Alt et~al.}(2008)}]{Alt:2007fe}
\bibinfo{author}{\bibfnamefont{C.}~\bibnamefont{Alt}} \bibnamefont{et~al.}
  (\bibinfo{collaboration}{NA49}), \bibinfo{journal}{Phys. Rev.}
  \textbf{\bibinfo{volume}{C77}}, \bibinfo{pages}{024903}
  (\bibinfo{year}{2008}), \eprint{0710.0118}.

\bibitem[{\citenamefont{Gazdzicki and Gorenstein}(1999)}]{Gazdzicki:1998vd}
\bibinfo{author}{\bibfnamefont{M.}~\bibnamefont{Gazdzicki}} \bibnamefont{and}
  \bibinfo{author}{\bibfnamefont{M.~I.} \bibnamefont{Gorenstein}},
  \bibinfo{journal}{Acta Phys. Polon.} \textbf{\bibinfo{volume}{B30}},
  \bibinfo{pages}{2705} (\bibinfo{year}{1999}), \eprint{hep-ph/9803462}.

\bibitem[{\citenamefont{Gazdzicki
  et~al.}(2004{\natexlab{a}})\citenamefont{Gazdzicki, Gorenstein, and
  Mrowczynski}}]{Gazdzicki:2003bb}
\bibinfo{author}{\bibfnamefont{M.}~\bibnamefont{Gazdzicki}},
  \bibinfo{author}{\bibfnamefont{M.~I.} \bibnamefont{Gorenstein}},
  \bibnamefont{and}
  \bibinfo{author}{\bibfnamefont{S.}~\bibnamefont{Mrowczynski}},
  \bibinfo{journal}{Phys. Lett.} \textbf{\bibinfo{volume}{B585}},
  \bibinfo{pages}{115} (\bibinfo{year}{2004}{\natexlab{a}}),
  \eprint{hep-ph/0304052}.

\bibitem[{\citenamefont{Stephanov et~al.}(1999)\citenamefont{Stephanov,
  Rajagopal, and Shuryak}}]{Stephanov:1999zu}
\bibinfo{author}{\bibfnamefont{M.~A.} \bibnamefont{Stephanov}},
  \bibinfo{author}{\bibfnamefont{K.}~\bibnamefont{Rajagopal}},
  \bibnamefont{and} \bibinfo{author}{\bibfnamefont{E.~V.}
  \bibnamefont{Shuryak}}, \bibinfo{journal}{Phys. Rev.}
  \textbf{\bibinfo{volume}{D60}}, \bibinfo{pages}{114028}
  (\bibinfo{year}{1999}), \eprint{hep-ph/9903292}.

\bibitem[{\citenamefont{Begun et~al.}(2004)\citenamefont{Begun, Gazdzicki,
  Gorenstein, and Zozulya}}]{Begun:2004gs}
\bibinfo{author}{\bibfnamefont{V.~V.} \bibnamefont{Begun}},
  \bibinfo{author}{\bibfnamefont{M.}~\bibnamefont{Gazdzicki}},
  \bibinfo{author}{\bibfnamefont{M.~I.} \bibnamefont{Gorenstein}},
  \bibnamefont{and} \bibinfo{author}{\bibfnamefont{O.~S.}
  \bibnamefont{Zozulya}}, \bibinfo{journal}{Phys. Rev.}
  \textbf{\bibinfo{volume}{C70}}, \bibinfo{pages}{034901}
  (\bibinfo{year}{2004}), \eprint{nucl-th/0404056}.

\bibitem[{\citenamefont{Begun et~al.}(2005)\citenamefont{Begun, Gorenstein,
  Kostyuk, and Zozulya}}]{Begun:2004pk}
\bibinfo{author}{\bibfnamefont{V.~V.} \bibnamefont{Begun}},
  \bibinfo{author}{\bibfnamefont{M.~I.} \bibnamefont{Gorenstein}},
  \bibinfo{author}{\bibfnamefont{A.~P.} \bibnamefont{Kostyuk}},
  \bibnamefont{and} \bibinfo{author}{\bibfnamefont{O.~S.}
  \bibnamefont{Zozulya}}, \bibinfo{journal}{Phys. Rev.}
  \textbf{\bibinfo{volume}{C71}}, \bibinfo{pages}{054904}
  (\bibinfo{year}{2005}), \eprint{nucl-th/0410044}.

\bibitem[{\citenamefont{Begun et~al.}(2007)}]{Begun:2006uu}
\bibinfo{author}{\bibfnamefont{V.~V.} \bibnamefont{Begun}}
  \bibnamefont{et~al.}, \bibinfo{journal}{Phys. Rev.}
  \textbf{\bibinfo{volume}{C76}}, \bibinfo{pages}{024902}
  (\bibinfo{year}{2007}), \eprint{nucl-th/0611075}.

\bibitem[{\citenamefont{Lungwitz and Bleicher}(2007)}]{Lungwitz:2007uc}
\bibinfo{author}{\bibfnamefont{B.}~\bibnamefont{Lungwitz}} \bibnamefont{and}
  \bibinfo{author}{\bibfnamefont{M.}~\bibnamefont{Bleicher}},
  \bibinfo{journal}{Phys. Rev.} \textbf{\bibinfo{volume}{C76}},
  \bibinfo{pages}{044904} (\bibinfo{year}{2007}), \eprint{arXiv:0707.1788
  [nucl-th]}.

\bibitem[{\citenamefont{Konchakovski
  et~al.}(2007{\natexlab{a}})\citenamefont{Konchakovski, Lungwitz, Gorenstein,
  and Bratkovskaya}}]{Konchakovski:2007ah}
\bibinfo{author}{\bibfnamefont{V.~P.} \bibnamefont{Konchakovski}},
  \bibinfo{author}{\bibfnamefont{B.}~\bibnamefont{Lungwitz}},
  \bibinfo{author}{\bibfnamefont{M.~I.} \bibnamefont{Gorenstein}},
  \bibnamefont{and} \bibinfo{author}{\bibfnamefont{E.~L.}
  \bibnamefont{Bratkovskaya}} (\bibinfo{year}{2007}{\natexlab{a}}),
  \eprint{arXiv:0712.2044 [nucl-th]}.

\bibitem[{\citenamefont{Alt et~al.}(2007)}]{Alt:2006jr}
\bibinfo{author}{\bibfnamefont{C.}~\bibnamefont{Alt}} \bibnamefont{et~al.}
  (\bibinfo{collaboration}{NA49 Collaboration}), \bibinfo{journal}{Phys. Rev.}
  \textbf{\bibinfo{volume}{C75}}, \bibinfo{pages}{064904}
  (\bibinfo{year}{2007}), \eprint{nucl-ex/0612010}.

\bibitem[{\citenamefont{Aggarwal et~al.}(2002)}]{Aggarwal:2001aa}
\bibinfo{author}{\bibfnamefont{M.~M.} \bibnamefont{Aggarwal}}
  \bibnamefont{et~al.} (\bibinfo{collaboration}{WA98 Collaboration}),
  \bibinfo{journal}{Phys. Rev.} \textbf{\bibinfo{volume}{C65}},
  \bibinfo{pages}{054912} (\bibinfo{year}{2002}), \eprint{nucl-ex/0108029}.

\bibitem[{\citenamefont{Mitchell}(2005)}]{Mitchell:2005at}
\bibinfo{author}{\bibfnamefont{J.~T.} \bibnamefont{Mitchell}}
  (\bibinfo{collaboration}{PHENIX Collaboration}) (\bibinfo{year}{2005}),
  \eprint{nucl-ex/0510076}.

\bibitem[{\citenamefont{Homma}(2006)}]{Homma:2007qh}
\bibinfo{author}{\bibfnamefont{K.}~\bibnamefont{Homma}}
  (\bibinfo{collaboration}{PHENIX Collaboration}), \bibinfo{journal}{PoS}
  \textbf{\bibinfo{volume}{CPOD2006}}, \bibinfo{pages}{007}
  (\bibinfo{year}{2006}), \eprint{nucl-ex/0703046}.

\bibitem[{\citenamefont{Anticic et~al.}(2004)}]{Anticic:2003fd}
\bibinfo{author}{\bibfnamefont{T.}~\bibnamefont{Anticic}} \bibnamefont{et~al.}
  (\bibinfo{collaboration}{NA49 Collaboration}), \bibinfo{journal}{Phys. Rev.}
  \textbf{\bibinfo{volume}{C70}}, \bibinfo{pages}{034902}
  (\bibinfo{year}{2004}), \eprint{hep-ex/0311009}.

\bibitem[{\citenamefont{Mrowczynski et~al.}(2004)\citenamefont{Mrowczynski,
  Rybczynski, and Wlodarczyk}}]{Mrowczynski:2004cg}
\bibinfo{author}{\bibfnamefont{S.}~\bibnamefont{Mrowczynski}},
  \bibinfo{author}{\bibfnamefont{M.}~\bibnamefont{Rybczynski}},
  \bibnamefont{and}
  \bibinfo{author}{\bibfnamefont{Z.}~\bibnamefont{Wlodarczyk}},
  \bibinfo{journal}{Phys. Rev.} \textbf{\bibinfo{volume}{C70}},
  \bibinfo{pages}{054906} (\bibinfo{year}{2004}), \eprint{nucl-th/0407012}.

\bibitem[{\citenamefont{Cunqueiro et~al.}(2005)\citenamefont{Cunqueiro,
  Ferreiro, del Moral, and Pajares}}]{Cunqueiro:2005hx}
\bibinfo{author}{\bibfnamefont{L.}~\bibnamefont{Cunqueiro}},
  \bibinfo{author}{\bibfnamefont{E.~G.} \bibnamefont{Ferreiro}},
  \bibinfo{author}{\bibfnamefont{F.}~\bibnamefont{del Moral}},
  \bibnamefont{and} \bibinfo{author}{\bibfnamefont{C.}~\bibnamefont{Pajares}},
  \bibinfo{journal}{Phys. Rev.} \textbf{\bibinfo{volume}{C72}},
  \bibinfo{pages}{024907} (\bibinfo{year}{2005}), \eprint{hep-ph/0505197}.

\bibitem[{\citenamefont{Grebieszkow et~al.}(2007)}]{Grebieszkow:2007xz}
\bibinfo{author}{\bibfnamefont{K.}~\bibnamefont{Grebieszkow}}
  \bibnamefont{et~al.} (\bibinfo{year}{2007}), \eprint{arXiv:0707.4608
  [nucl-ex]}.

\bibitem[{\citenamefont{Lungwitz et~al.}(2007)}]{Lungwitz:2007cy}
\bibinfo{author}{\bibfnamefont{B.}~\bibnamefont{Lungwitz}} \bibnamefont{et~al.}
  (\bibinfo{collaboration}{NA49 Collaboration}) (\bibinfo{year}{2007}),
  \eprint{arXiv:0709.1646 [nucl-ex]}.

\bibitem[{\citenamefont{Bass et~al.}(1998)}]{Bass:1998ca}
\bibinfo{author}{\bibfnamefont{S.~A.} \bibnamefont{Bass}} \bibnamefont{et~al.},
  \bibinfo{journal}{Prog. Part. Nucl. Phys.} \textbf{\bibinfo{volume}{41}},
  \bibinfo{pages}{255} (\bibinfo{year}{1998}), \eprint{nucl-th/9803035}.

\bibitem[{\citenamefont{Mishustin}(2006)}]{Mishustin:2006ka}
\bibinfo{author}{\bibfnamefont{I.~N.} \bibnamefont{Mishustin}},
  \bibinfo{journal}{Eur. Phys. J.} \textbf{\bibinfo{volume}{A30}},
  \bibinfo{pages}{311} (\bibinfo{year}{2006}), \eprint{hep-ph/0609196}.

\bibitem[{\citenamefont{Konchakovski
  et~al.}(2007{\natexlab{b}})\citenamefont{Konchakovski, Gorenstein, and
  Bratkovskaya}}]{Konchakovski:2007ss}
\bibinfo{author}{\bibfnamefont{V.~P.} \bibnamefont{Konchakovski}},
  \bibinfo{author}{\bibfnamefont{M.~I.} \bibnamefont{Gorenstein}},
  \bibnamefont{and} \bibinfo{author}{\bibfnamefont{E.~L.}
  \bibnamefont{Bratkovskaya}}, \bibinfo{journal}{Phys. Lett.}
  \textbf{\bibinfo{volume}{B651}}, \bibinfo{pages}{114}
  (\bibinfo{year}{2007}{\natexlab{b}}), \eprint{nucl-th/0703052}.

\bibitem[{\citenamefont{Heiselberg}(2001)}]{Heiselberg:2000fk}
\bibinfo{author}{\bibfnamefont{H.}~\bibnamefont{Heiselberg}},
  \bibinfo{journal}{Phys. Rept.} \textbf{\bibinfo{volume}{351}},
  \bibinfo{pages}{161} (\bibinfo{year}{2001}), \eprint{nucl-th/0003046}.

\bibitem[{\citenamefont{Hauer}(2007)}]{Hauer:2007im}
\bibinfo{author}{\bibfnamefont{M.}~\bibnamefont{Hauer}} (\bibinfo{year}{2007}),
  \eprint{arXiv:0710.3938 [nucl-th]}.

\bibitem[{\citenamefont{Afanasiev et~al.}(1999)}]{Afanasev:1999iu}
\bibinfo{author}{\bibfnamefont{S.}~\bibnamefont{Afanasiev}}
  \bibnamefont{et~al.} (\bibinfo{collaboration}{NA49 Collaboration}),
  \bibinfo{journal}{Nucl. Instrum. Meth.} \textbf{\bibinfo{volume}{A430}},
  \bibinfo{pages}{210} (\bibinfo{year}{1999}).

\bibitem[{\citenamefont{Appelshauser et~al.}(1998)}]{Appelshauser:1998tt}
\bibinfo{author}{\bibfnamefont{H.}~\bibnamefont{Appelshauser}}
  \bibnamefont{et~al.} (\bibinfo{collaboration}{NA49 Collaboration}),
  \bibinfo{journal}{Eur. Phys. J.} \textbf{\bibinfo{volume}{A2}},
  \bibinfo{pages}{383} (\bibinfo{year}{1998}).

\bibitem[{\citenamefont{De~Marzo et~al.}(1983)}]{DeMarzo:1983gd}
\bibinfo{author}{\bibfnamefont{C.}~\bibnamefont{De~Marzo}}
  \bibnamefont{et~al.}, \bibinfo{journal}{Nucl. Instrum. Meth.}
  \textbf{\bibinfo{volume}{217}}, \bibinfo{pages}{405} (\bibinfo{year}{1983}).

\bibitem[{\citenamefont{Lungwitz}(2007{\natexlab{a}})}]{accTables}
\bibinfo{author}{\bibfnamefont{B.}~\bibnamefont{Lungwitz}}
  (\bibinfo{year}{2007}{\natexlab{a}}),
  \urlprefix\url{https://edms.cern.ch/document/885236/1}.

\bibitem[{\citenamefont{Dementev and Sobolevsky}(1997)}]{Dementev:1997ca}
\bibinfo{author}{\bibfnamefont{A.~V.} \bibnamefont{Dementev}} \bibnamefont{and}
  \bibinfo{author}{\bibfnamefont{N.~M.} \bibnamefont{Sobolevsky}}
  (\bibinfo{year}{1997}), \bibinfo{note}{prepared for 3rd Workshop on
  Simulating Accelerator Radiation Environments (SARE3), Tsukuba, Japan, 7-9
  May 1997}.

\bibitem[{\citenamefont{Konchakovski et~al.}(2006)}]{Konchakovski:2005hq}
\bibinfo{author}{\bibfnamefont{V.~P.} \bibnamefont{Konchakovski}}
  \bibnamefont{et~al.}, \bibinfo{journal}{Phys. Rev.}
  \textbf{\bibinfo{volume}{C73}}, \bibinfo{pages}{034902}
  (\bibinfo{year}{2006}), \eprint{nucl-th/0511083}.

\bibitem[{\citenamefont{Gazdzicki and Gorenstein}(2006)}]{Gazdzicki:2005rr}
\bibinfo{author}{\bibfnamefont{M.}~\bibnamefont{Gazdzicki}} \bibnamefont{and}
  \bibinfo{author}{\bibfnamefont{M.~I.} \bibnamefont{Gorenstein}},
  \bibinfo{journal}{Phys. Lett.} \textbf{\bibinfo{volume}{B640}},
  \bibinfo{pages}{155} (\bibinfo{year}{2006}), \eprint{hep-ph/0511058}.

\bibitem[{\citenamefont{Rybczynski and Wlodarczyk}(2005)}]{Rybczynski:2004zi}
\bibinfo{author}{\bibfnamefont{M.}~\bibnamefont{Rybczynski}} \bibnamefont{and}
  \bibinfo{author}{\bibfnamefont{Z.}~\bibnamefont{Wlodarczyk}},
  \bibinfo{journal}{J. Phys. Conf. Ser.} \textbf{\bibinfo{volume}{5}},
  \bibinfo{pages}{238} (\bibinfo{year}{2005}), \eprint{nucl-th/0408023}.

\bibitem[{\citenamefont{Werner}(1993)}]{Werner:1993uh}
\bibinfo{author}{\bibfnamefont{K.}~\bibnamefont{Werner}},
  \bibinfo{journal}{Phys. Rept.} \textbf{\bibinfo{volume}{232}},
  \bibinfo{pages}{87} (\bibinfo{year}{1993}).

\bibitem[{\citenamefont{Lungwitz et~al.}(2006)}]{Lungwitz:2006cx}
\bibinfo{author}{\bibfnamefont{B.}~\bibnamefont{Lungwitz}} \bibnamefont{et~al.}
  (\bibinfo{collaboration}{NA49 Collaboration}), \bibinfo{journal}{PoS}
  \textbf{\bibinfo{volume}{CFRNC2006}}, \bibinfo{pages}{024}
  (\bibinfo{year}{2006}), \eprint{nucl-ex/0610046}.

\bibitem[{\citenamefont{Lungwitz}(2007{\natexlab{b}})}]{Lungwitz:2006cy}
\bibinfo{author}{\bibfnamefont{B.}~\bibnamefont{Lungwitz}},
  \bibinfo{journal}{AIP Conf. Proc.} \textbf{\bibinfo{volume}{892}},
  \bibinfo{pages}{400} (\bibinfo{year}{2007}{\natexlab{b}}),
  \eprint{nucl-ex/0610047}.

\bibitem[{\citenamefont{Hauer et~al.}(2007)\citenamefont{Hauer, Begun, and
  Gorenstein}}]{Hauer:2007ju}
\bibinfo{author}{\bibfnamefont{M.}~\bibnamefont{Hauer}},
  \bibinfo{author}{\bibfnamefont{V.~V.} \bibnamefont{Begun}}, \bibnamefont{and}
  \bibinfo{author}{\bibfnamefont{M.~I.} \bibnamefont{Gorenstein}}
  (\bibinfo{year}{2007}), \eprint{arXiv:0706.3290 [nucl-th]}.

\bibitem[{\citenamefont{Bleicher et~al.}(1999)}]{Bleicher:1999xi}
\bibinfo{author}{\bibfnamefont{M.}~\bibnamefont{Bleicher}}
  \bibnamefont{et~al.}, \bibinfo{journal}{J. Phys.}
  \textbf{\bibinfo{volume}{G25}}, \bibinfo{pages}{1859} (\bibinfo{year}{1999}),
  \eprint{hep-ph/9909407}.

\bibitem[{\citenamefont{Ehehalt and Cassing}(1996)}]{Ehehalt:1996uq}
\bibinfo{author}{\bibfnamefont{W.}~\bibnamefont{Ehehalt}} \bibnamefont{and}
  \bibinfo{author}{\bibfnamefont{W.}~\bibnamefont{Cassing}},
  \bibinfo{journal}{Nucl. Phys.} \textbf{\bibinfo{volume}{A602}},
  \bibinfo{pages}{449} (\bibinfo{year}{1996}).

\bibitem[{\citenamefont{Gazdzicki
  et~al.}(2004{\natexlab{b}})}]{Gazdzicki:2004ef}
\bibinfo{author}{\bibfnamefont{M.}~\bibnamefont{Gazdzicki}}
  \bibnamefont{et~al.} (\bibinfo{collaboration}{NA49 Collaboration}),
  \bibinfo{journal}{J. Phys.} \textbf{\bibinfo{volume}{G30}},
  \bibinfo{pages}{S701} (\bibinfo{year}{2004}{\natexlab{b}}),
  \eprint{nucl-ex/0403023}.

\bibitem[{\citenamefont{Becattini et~al.}(2006)\citenamefont{Becattini,
  Manninen, and Gazdzicki}}]{Becattini:2005xt}
\bibinfo{author}{\bibfnamefont{F.}~\bibnamefont{Becattini}},
  \bibinfo{author}{\bibfnamefont{J.}~\bibnamefont{Manninen}}, \bibnamefont{and}
  \bibinfo{author}{\bibfnamefont{M.}~\bibnamefont{Gazdzicki}},
  \bibinfo{journal}{Phys. Rev.} \textbf{\bibinfo{volume}{C73}},
  \bibinfo{pages}{044905} (\bibinfo{year}{2006}), \eprint{hep-ph/0511092}.

\bibitem[{\citenamefont{Stephanov}(2008)}]{Stephanov:pc}
\bibinfo{author}{\bibfnamefont{M.~A.} \bibnamefont{Stephanov}},
  \bibinfo{journal}{private communication}  (\bibinfo{year}{2008}).

\bibitem[{\citenamefont{Hatta and Ikeda}(2003)}]{Hatta:2002sj}
\bibinfo{author}{\bibfnamefont{Y.}~\bibnamefont{Hatta}} \bibnamefont{and}
  \bibinfo{author}{\bibfnamefont{T.}~\bibnamefont{Ikeda}},
  \bibinfo{journal}{Phys. Rev.} \textbf{\bibinfo{volume}{D67}},
  \bibinfo{pages}{014028} (\bibinfo{year}{2003}), \eprint{hep-ph/0210284}.

\bibitem[{\citenamefont{Gazdzicki et~al.}(2006)}]{Gazdzicki:2006fy}
\bibinfo{author}{\bibfnamefont{M.}~\bibnamefont{Gazdzicki}}
  \bibnamefont{et~al.} (\bibinfo{collaboration}{NA49-future Collaboration}),
  \bibinfo{journal}{PoS} \textbf{\bibinfo{volume}{CPOD2006}},
  \bibinfo{pages}{016} (\bibinfo{year}{2006}), \eprint{nucl-ex/0612007}.

\end{thebibliography}

\clearpage

\begin{table}[h]
\begin{flushleft}
\begin{tabular}{|c||c|c|c|}
\hline
energy&\multicolumn{3}{|c|}{$\omega(h^+)$}\\
\hline
(GeV)&	$0<y(\pi)<y_{beam}$&	$0<y(\pi)<1$&	$1<y(\pi)<y_{beam}$\\
\hline
\hline
$20A$&	$0.88 \pm 0.02 \pm 0.02$&	      $0.99 \pm 0.02 \pm 0.02$&   $0.86 \pm 0.02 \pm 0.02$\\
\hline
$30A$&	$0.85 \pm 0.01 \pm 0.02$&	      $0.96 \pm 0.02 \pm 0.02$&   $0.84 \pm 0.01 \pm 0.02$\\
\hline
$40A$&	$0.89 \pm 0.01 \pm 0.02$&	      $1.01 \pm 0.01 \pm 0.02$&   $0.87 \pm 0.01 \pm 0.02$\\
\hline
$80A$&	$0.93 \pm 0.03 \pm 0.02$&	      $1.04 \pm 0.03 \pm 0.03$&   $0.89 \pm 0.03 \pm 0.02$\\
\hline
$158A$&	$0.89 \pm 0.02 \pm 0.02$&	      $1.00 \pm 0.02 \pm 0.02$&   $0.84 \pm 0.02 \pm 0.02$\\
\hline
\hline
energy&\multicolumn{3}{|c|}{$\omega(h^-)$}\\ 
\hline
(GeV)&	$0<y(\pi)<y_{beam}$&	$0<y(\pi)<1$&	$1<y(\pi)<y_{beam}$\\
\hline
\hline
$20A$&	$0.94 \pm 0.02 \pm 0.02$&	      $1.01 \pm 0.02 \pm 0.02$&   $0.93 \pm 0.02 \pm 0.02$\\
\hline
$30A$&	$0.91 \pm 0.01 \pm 0.02$&	      $1.01 \pm 0.02 \pm 0.02$&   $0.91 \pm 0.01 \pm 0.02$\\
\hline
$40A$&	$0.92 \pm 0.01 \pm 0.02$&	      $1.02 \pm 0.01 \pm 0.02$&   $0.91 \pm 0.01 \pm 0.02$\\
\hline
$80A$&	$0.88 \pm 0.03 \pm 0.02$&	      $1.05 \pm 0.03 \pm 0.02$&   $0.86 \pm 0.03 \pm 0.02$\\
\hline
$158A$&	$0.90 \pm 0.02 \pm 0.02$&	      $1.05 \pm 0.02 \pm 0.02$&   $0.83 \pm 0.02 \pm 0.01$\\
\hline
\hline
energy&\multicolumn{3}{|c|}{$\omega(h^{\pm})$}\\
\hline
(GeV)&	$0<y(\pi)<y_{beam}$&	$0<y(\pi)<1$&	$1<y(\pi)<y_{beam}$\\
\hline
\hline
$20A$&	$1.01 \pm 0.02 \pm 0.04$&	      $1.10 \pm 0.02 \pm 0.04$&   $0.94 \pm 0.02 \pm 0.04$\\
\hline
$30A$&	$1.01 \pm 0.02 \pm 0.04$&	      $1.07 \pm 0.02 \pm 0.04$&   $0.94 \pm 0.01 \pm 0.04$\\
\hline
$40A$&	$1.10 \pm 0.01 \pm 0.04$&	      $1.15 \pm 0.01 \pm 0.04$&   $1.01 \pm 0.01 \pm 0.04$\\
\hline
$80A$&	$1.21 \pm 0.04 \pm 0.05$&	      $1.22 \pm 0.04 \pm 0.05$&   $1.07 \pm 0.03 \pm 0.04$\\
\hline
$158A$&	$1.24 \pm 0.03 \pm 0.05$&	      $1.20 \pm 0.02 \pm 0.05$&   $1.09 \pm 0.02 \pm 0.04$\\
\hline
\end{tabular}
\caption{\label{w_num}Scaled variance of the multiplicity distribution of positively (top), 
negatively and all (bottom) charged hadrons as a function of energy. The first error is the statistical
and the second error the systematical one.}
\end{flushleft}
\end{table}

\pagebreak

\begin{figure}[h!]
\includegraphics[width=8.2cm]{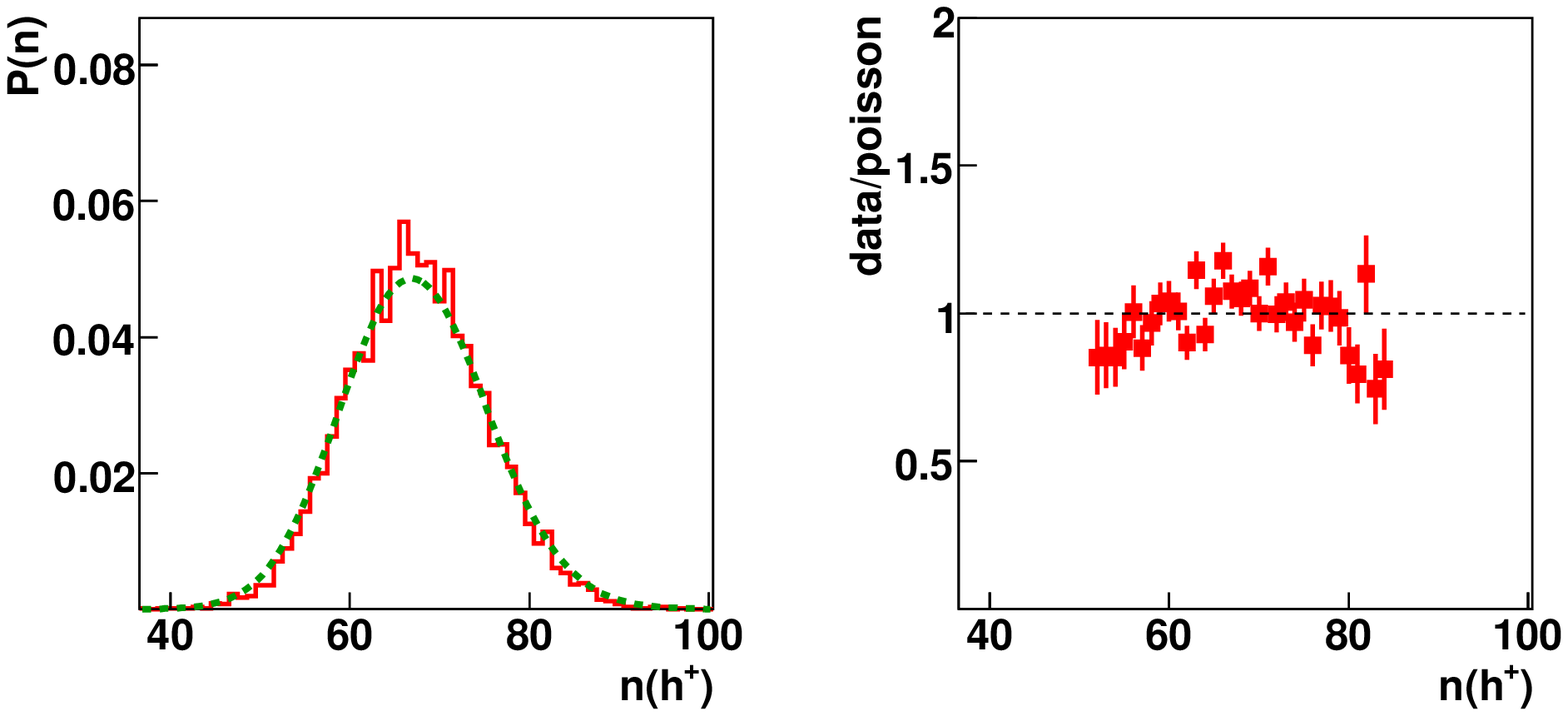}
\includegraphics[width=8.2cm]{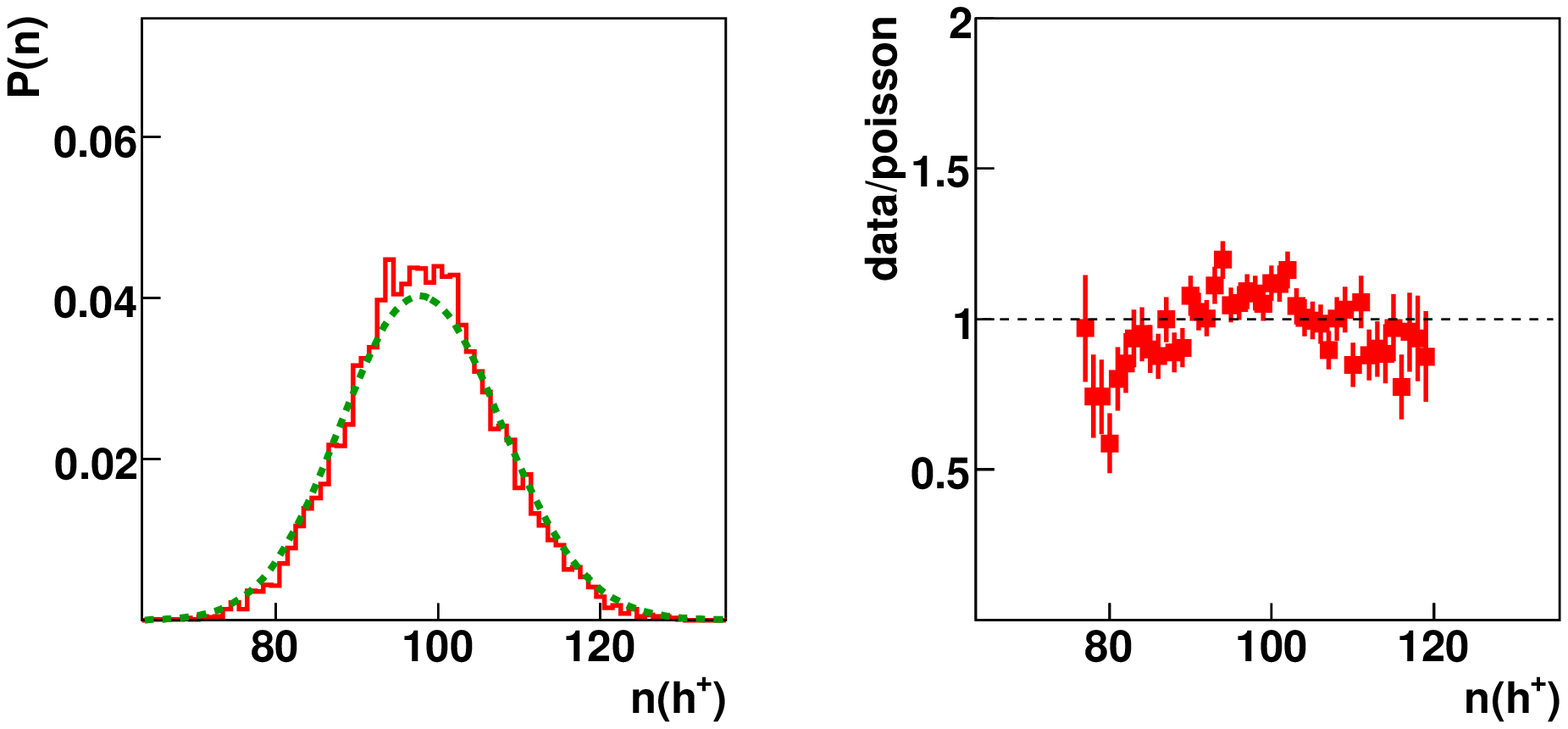}
\includegraphics[width=8.2cm]{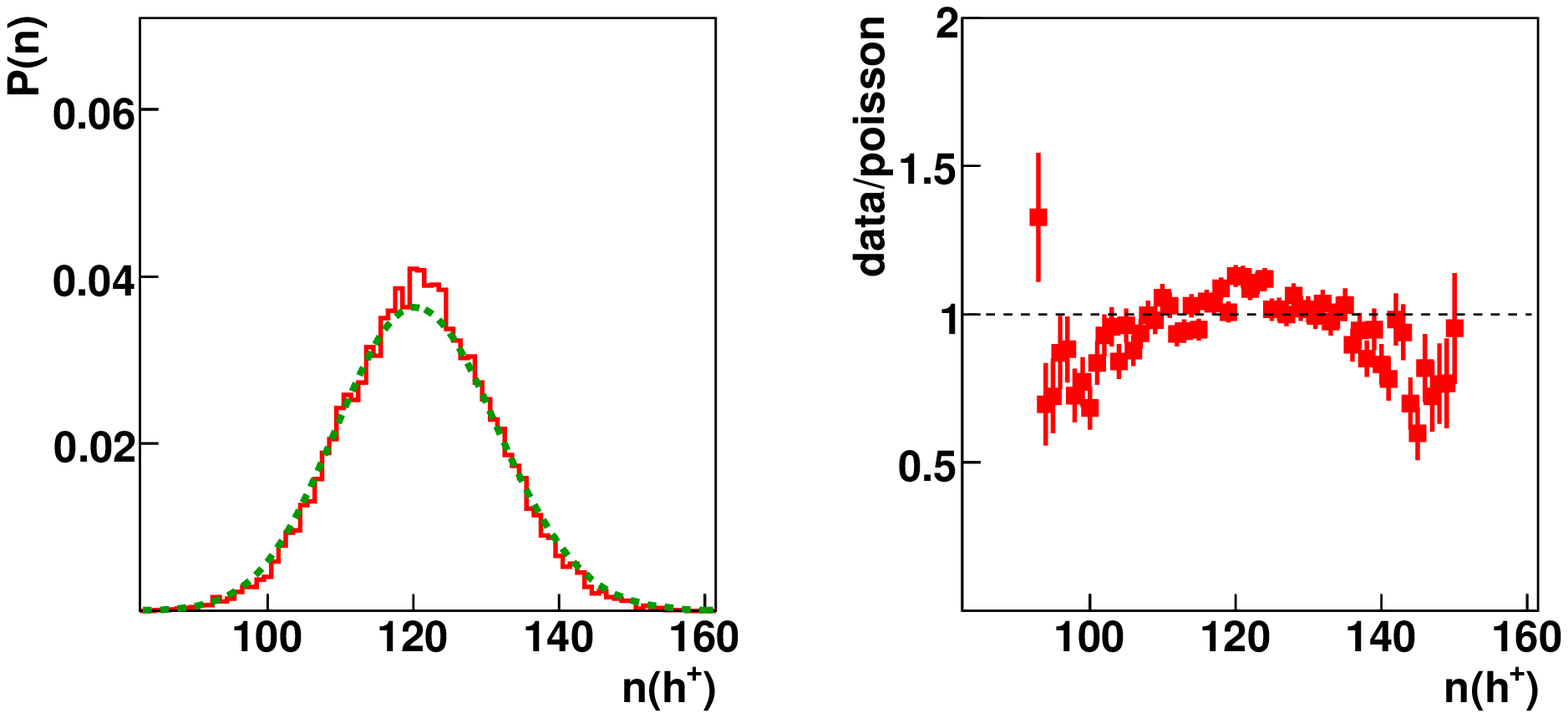}
\includegraphics[width=8.2cm]{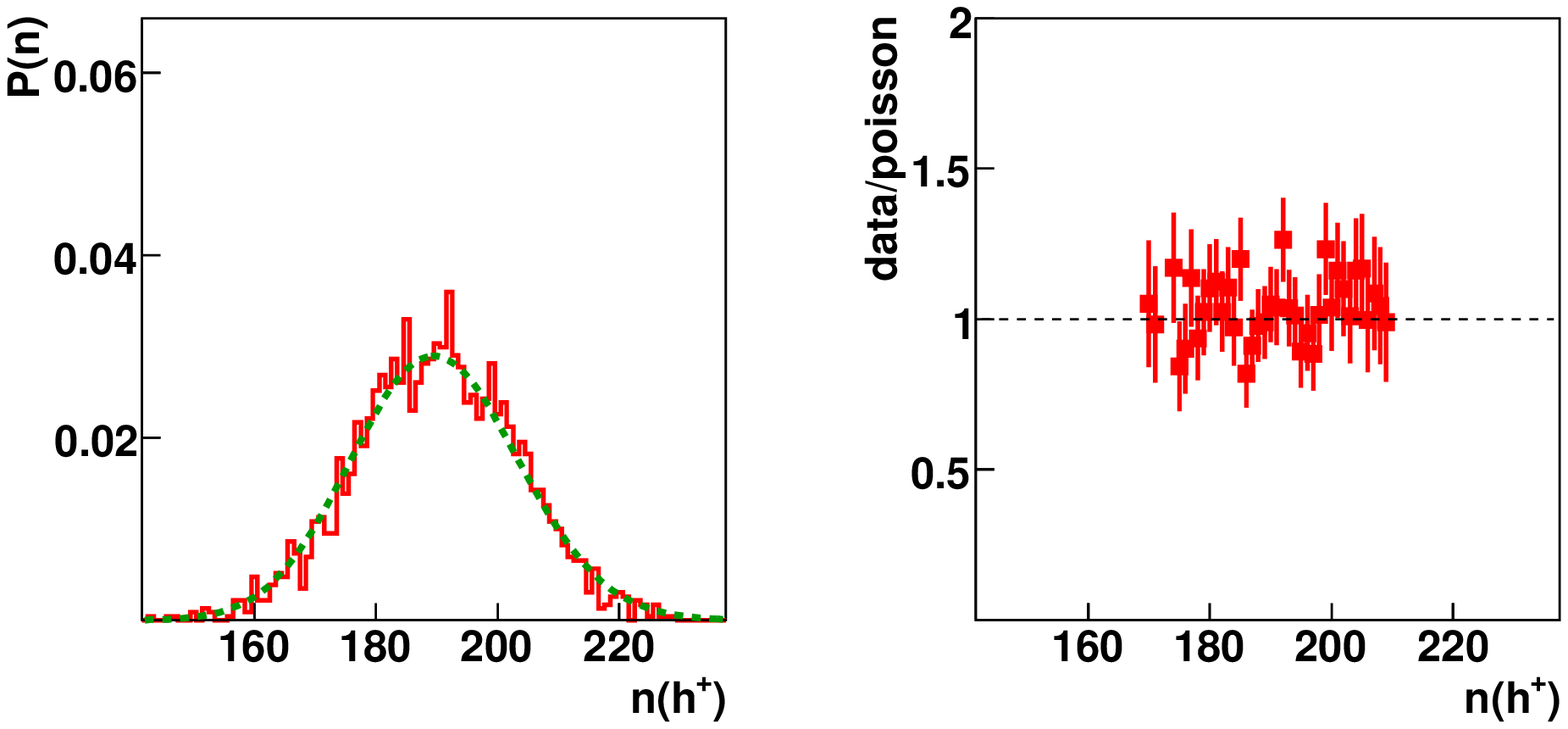}
\includegraphics[width=8.2cm]{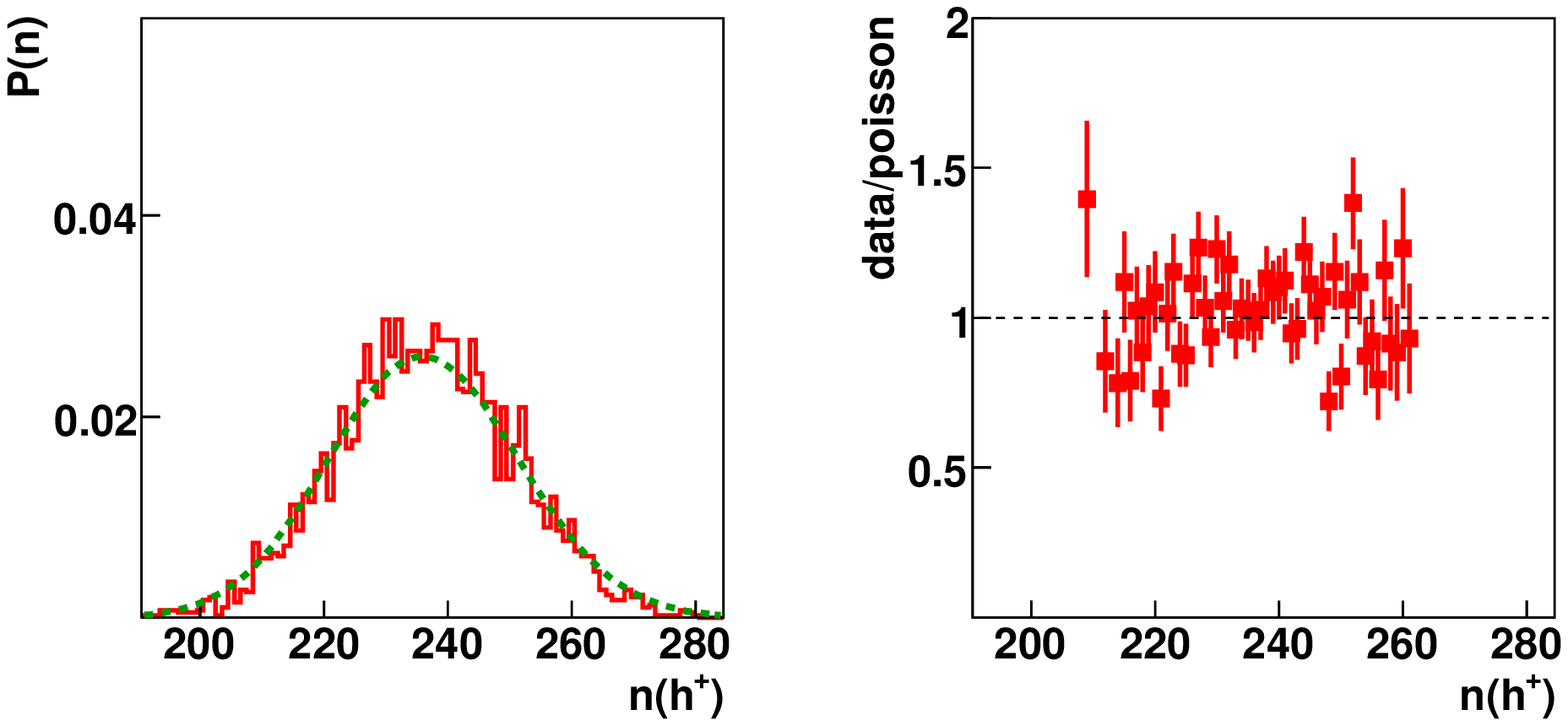}
\caption{\label{mult_dist_facc_hp}(Color online) Left: multiplicity distributions 
of positively charged  hadrons in full experimental acceptance in the 1\% most
central Pb+Pb collisions from $20A$ (top) to  $158A$ GeV (bottom).
The dashed lines indicate Poisson distributions with the same mean multiplicity as in data.
Right: the ratio of the measured multiplicity distribution to
the corresponding Poisson one.}
\end{figure}

\pagebreak

\begin{figure}[h!]
\includegraphics[width=8.2cm]{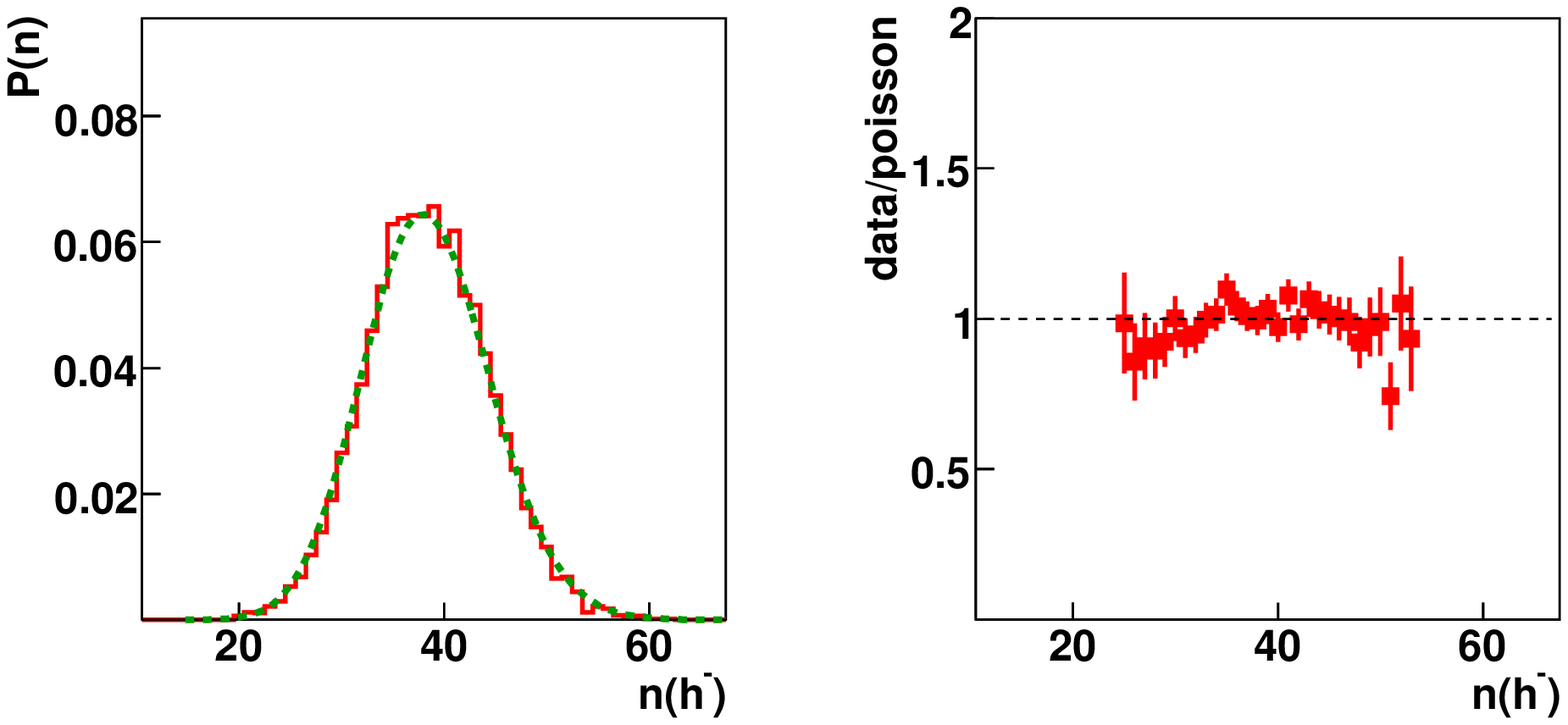}
\includegraphics[width=8.2cm]{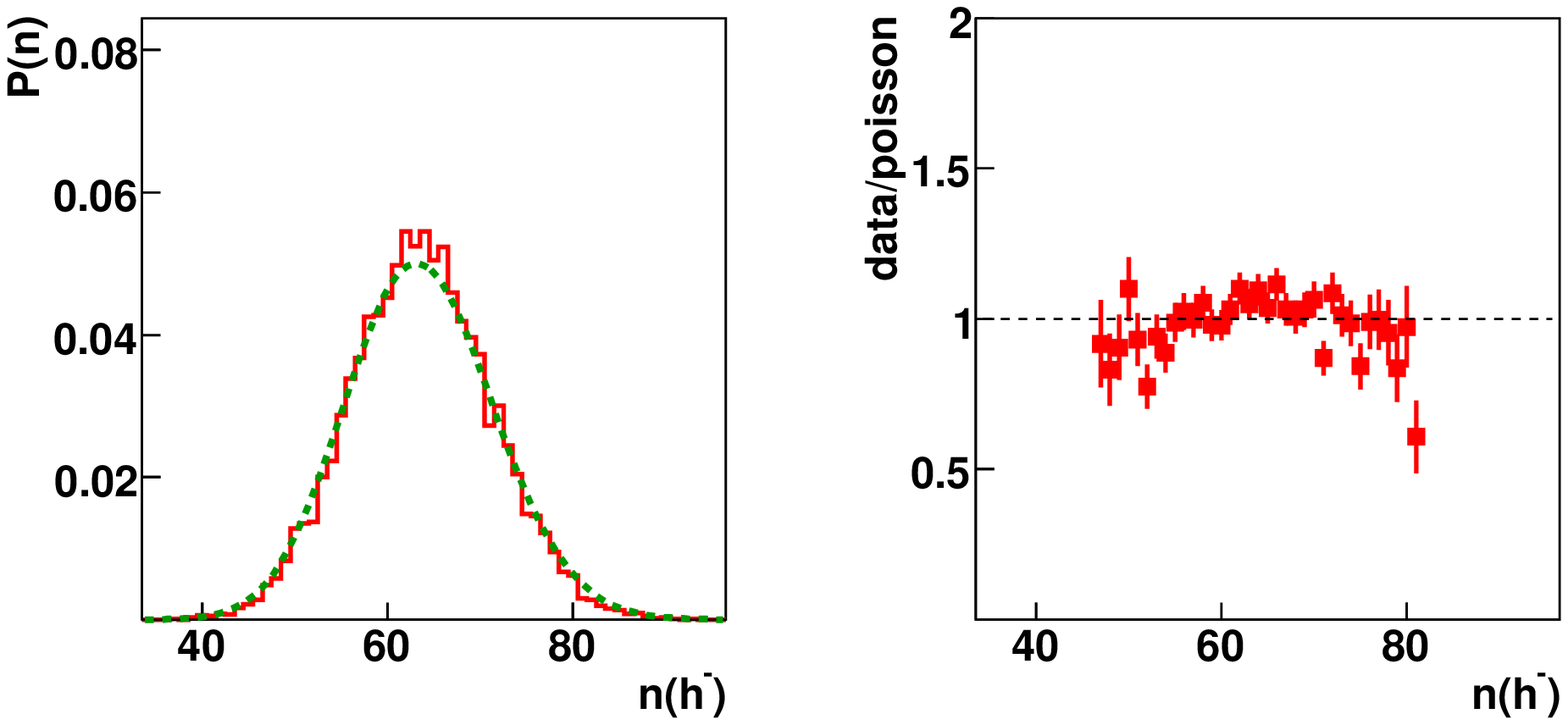}
\includegraphics[width=8.2cm]{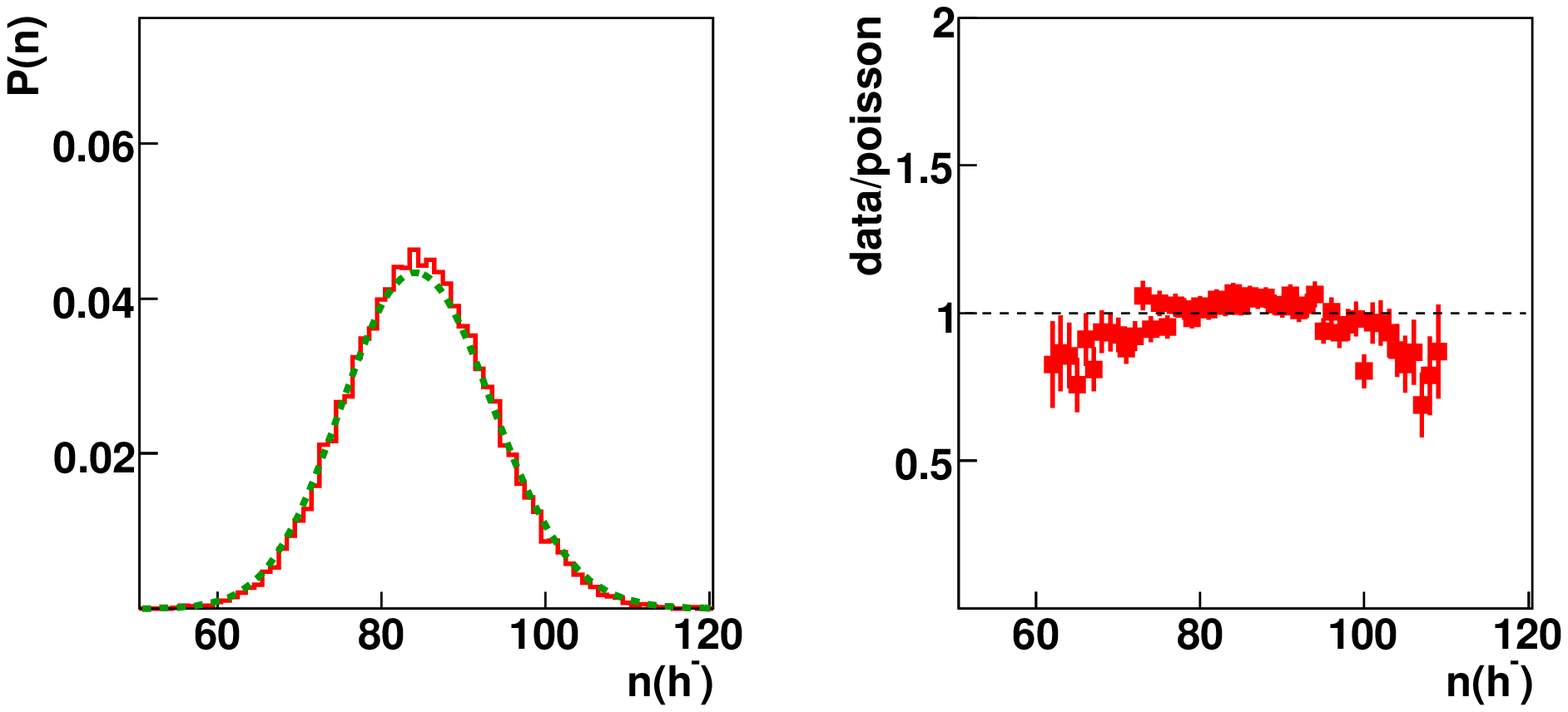}
\includegraphics[width=8.2cm]{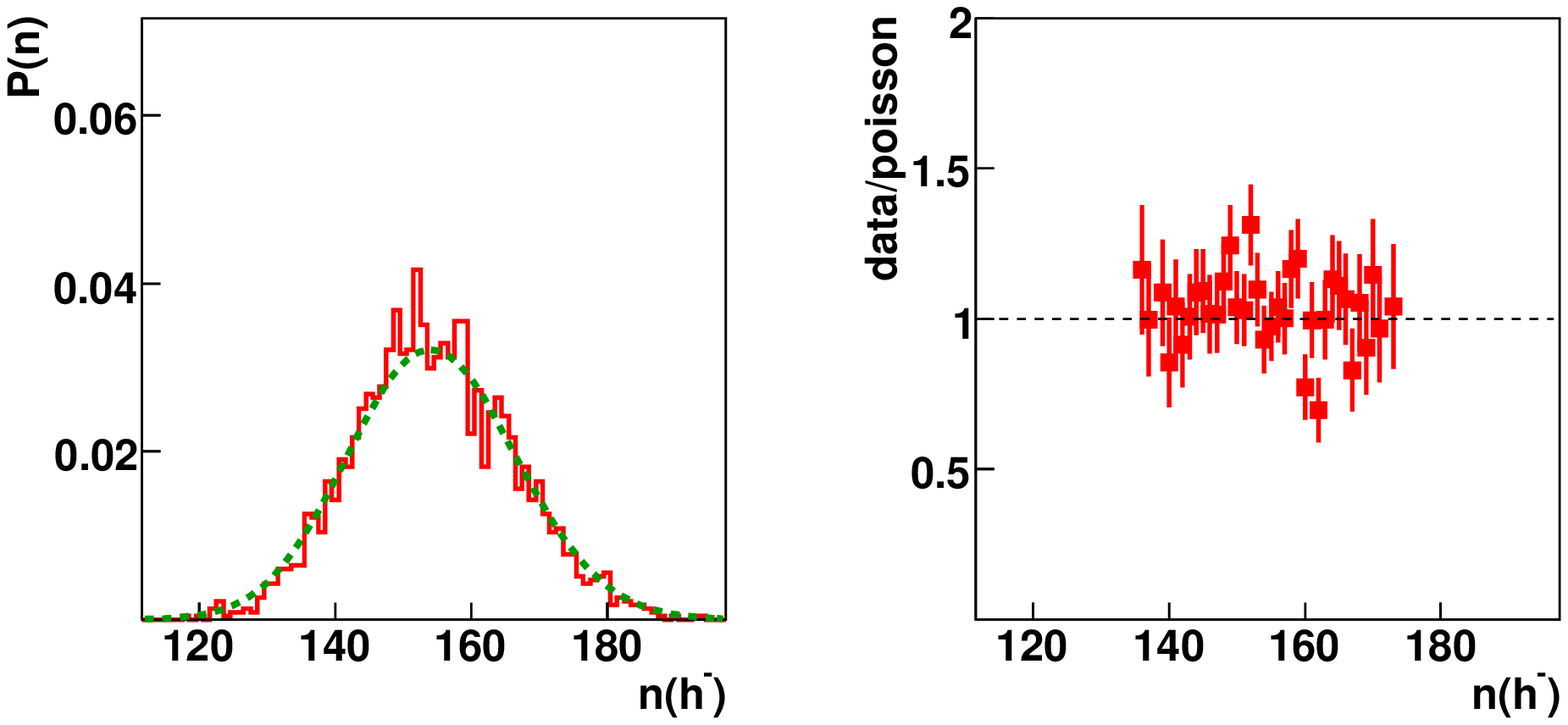}
\includegraphics[width=8.2cm]{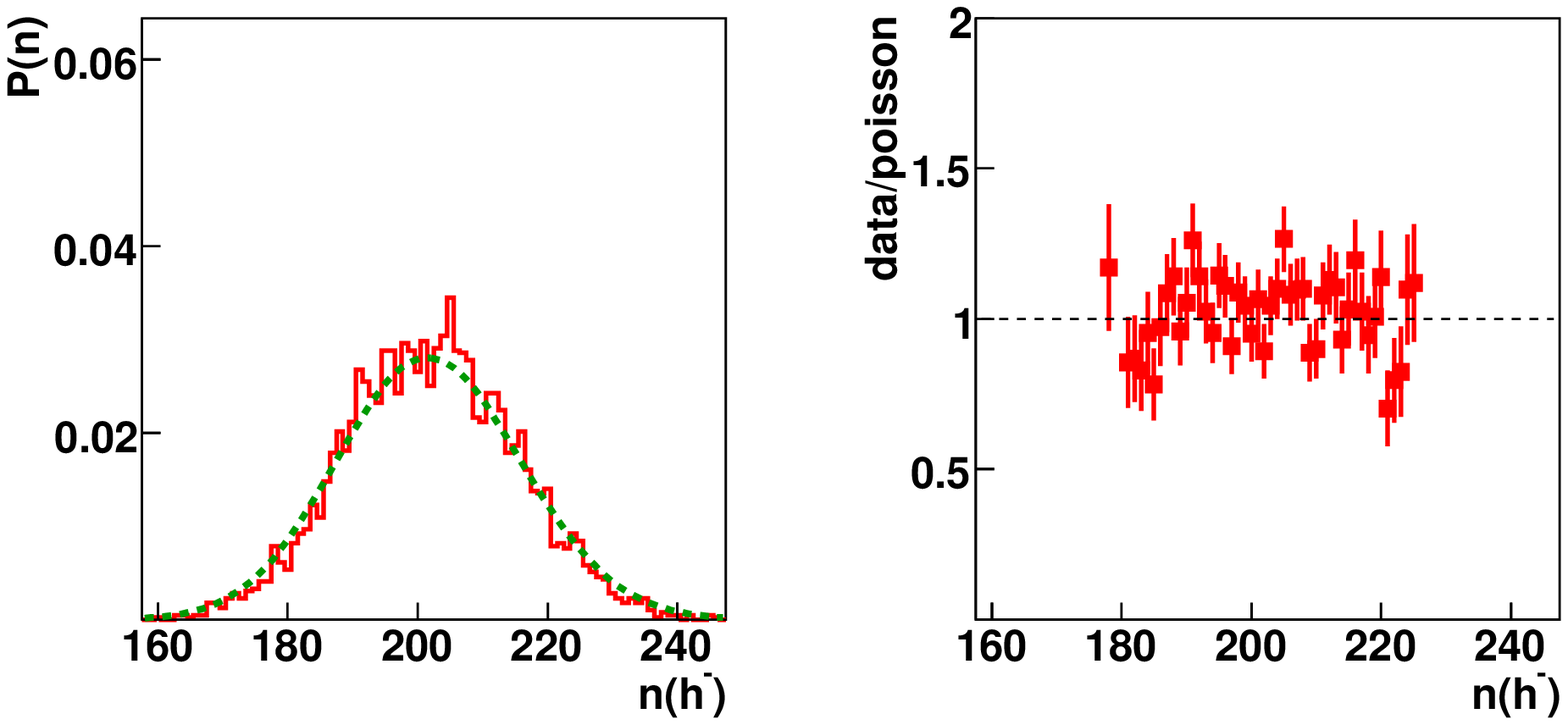}
\caption{\label{mult_dist_facc}(Color online) Left: multiplicity distributions 
of negatively charged  hadrons in full experimental acceptance in in the 1\% most
central Pb+Pb collisions from $20A$ (top) to  $158A$ GeV (bottom).
The dashed lines indicate Poisson distributions with the same mean multiplicity as in data.
Right: the ratio of the measured multiplicity distribution to
the corresponding Poisson one.}
\end{figure}

\pagebreak

\begin{figure}[h!]
\includegraphics[width=8.2cm]{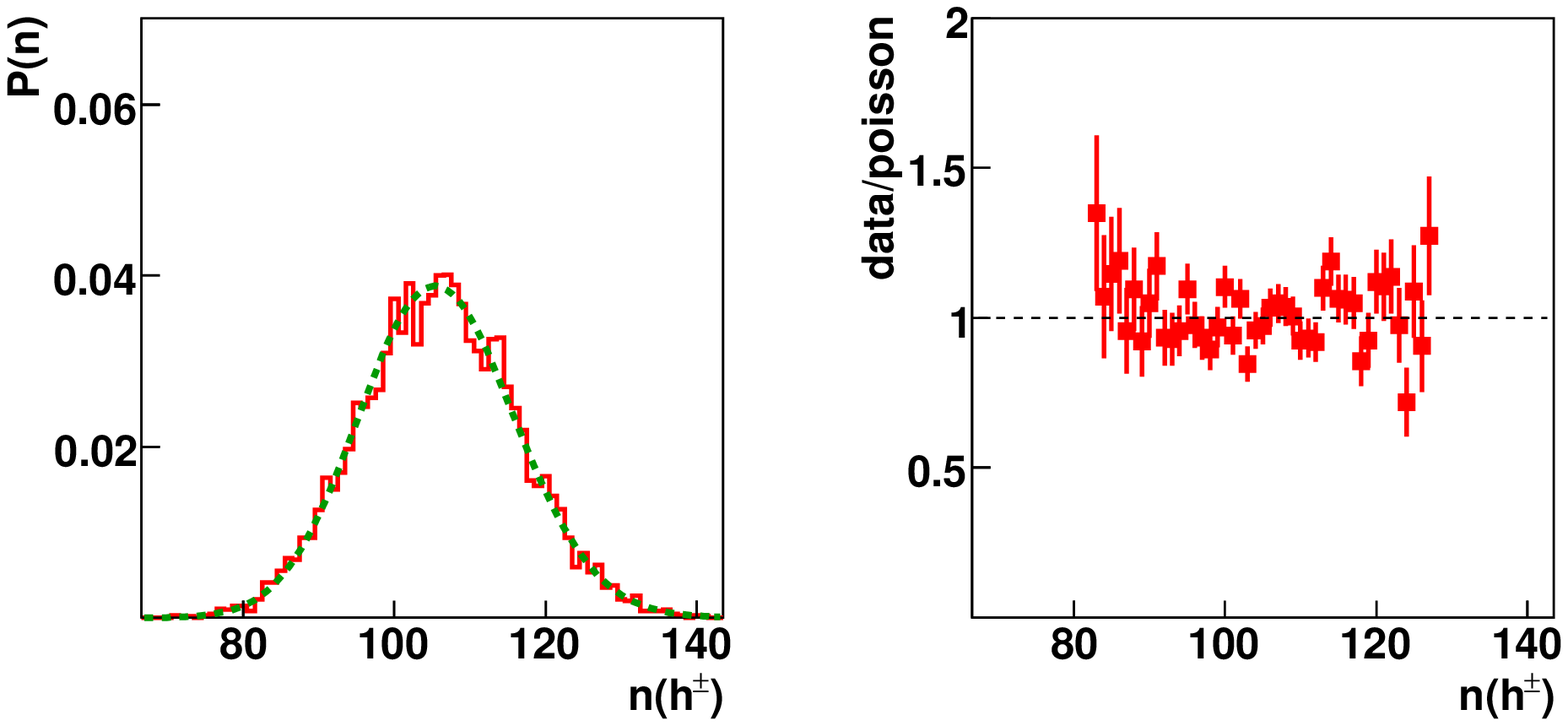}
\includegraphics[width=8.2cm]{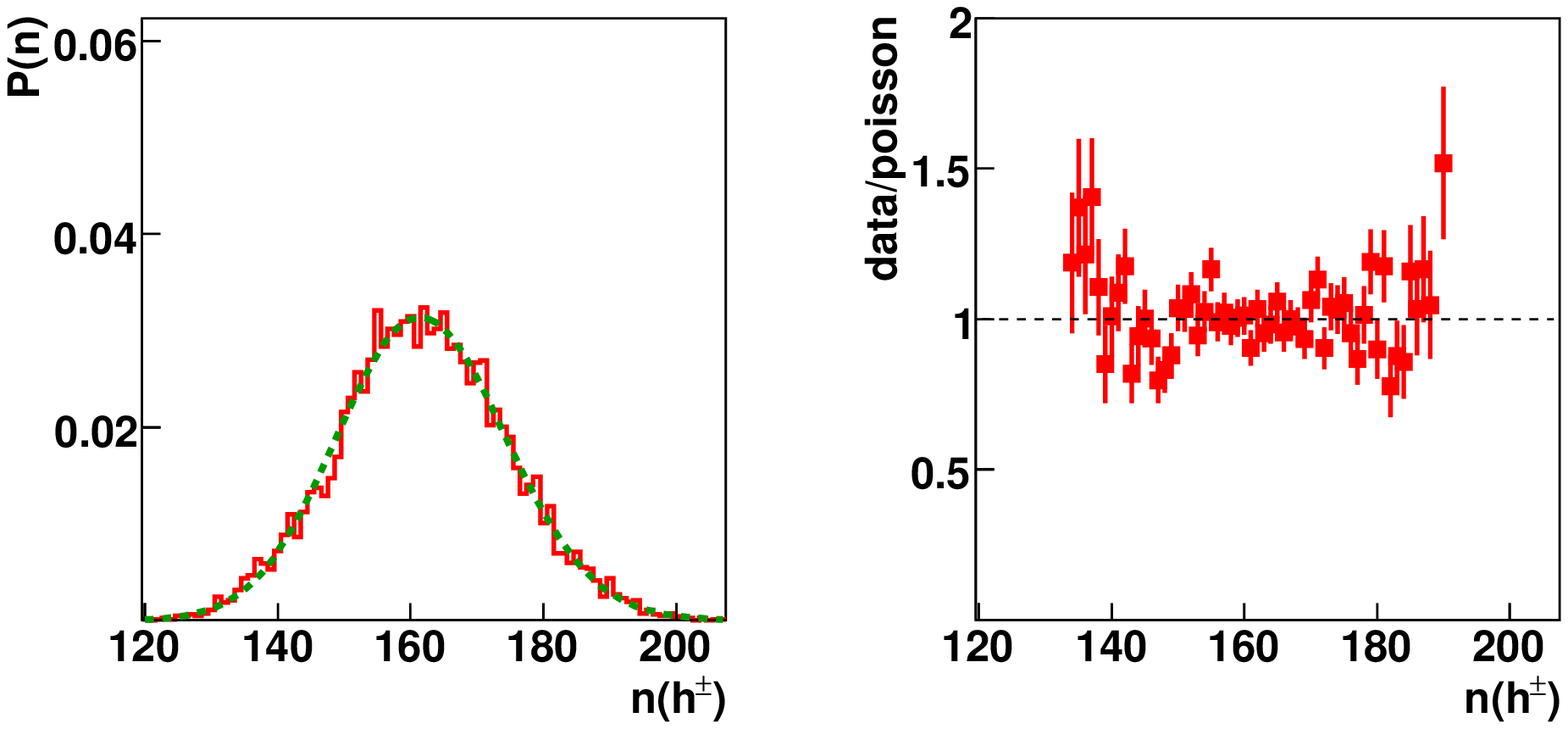}
\includegraphics[width=8.2cm]{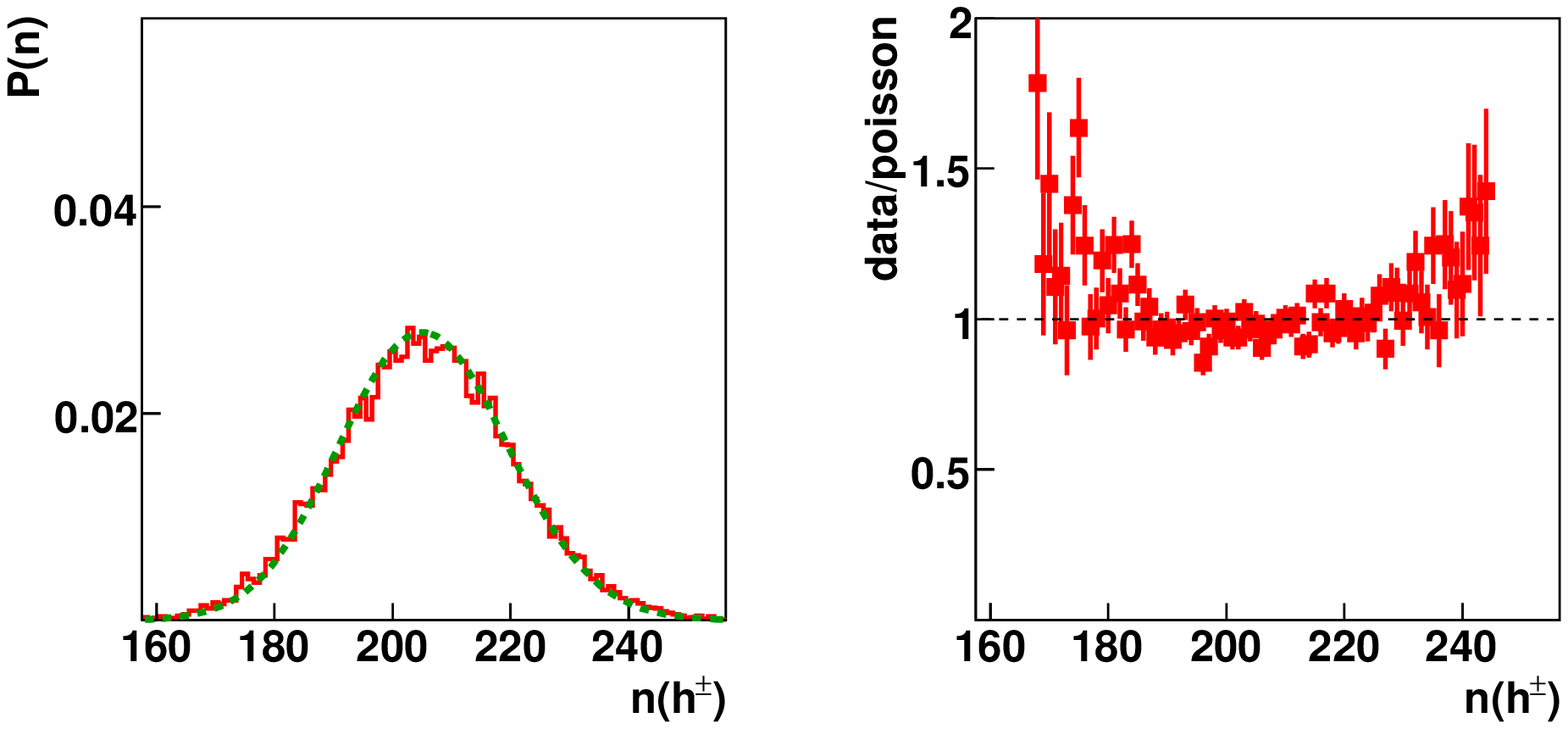}
\includegraphics[width=8.2cm]{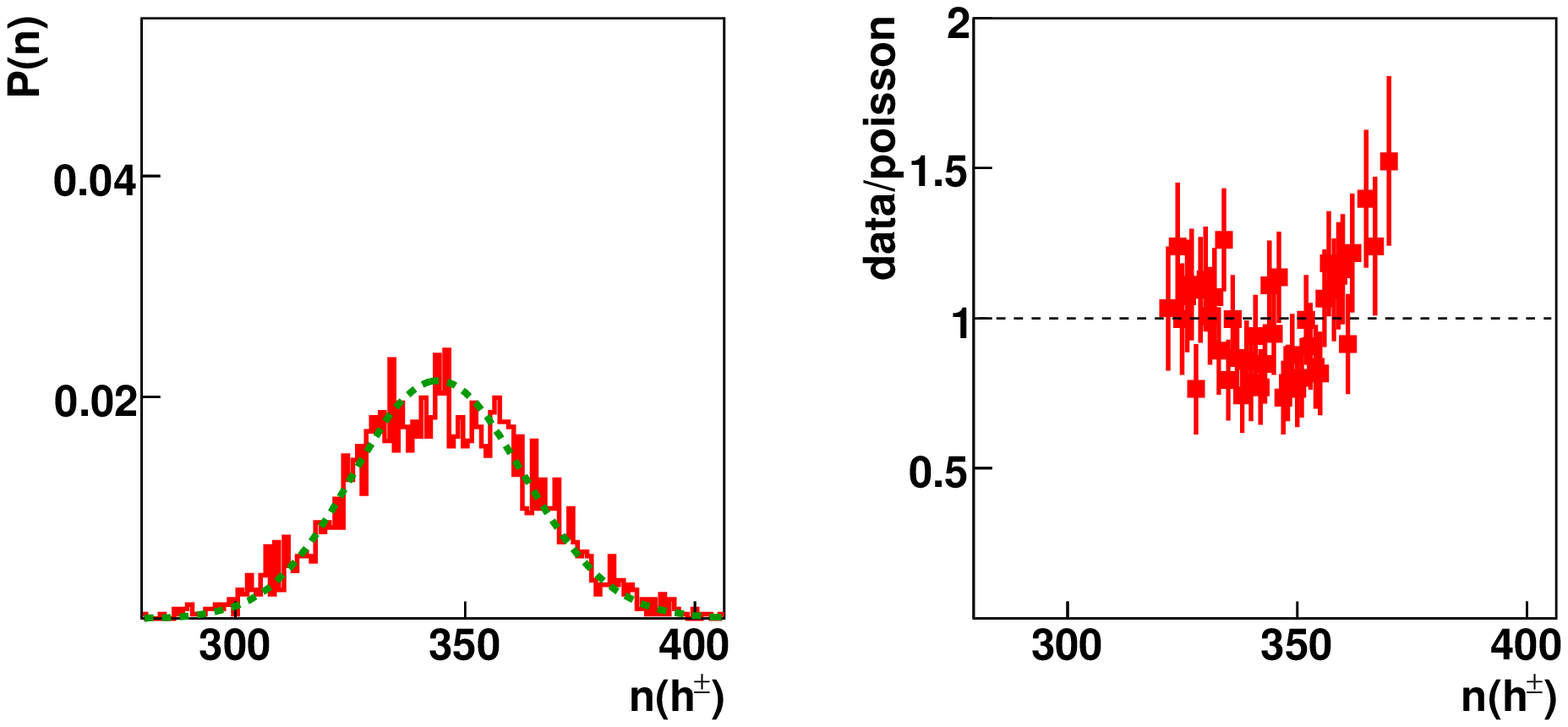}
\includegraphics[width=8.2cm]{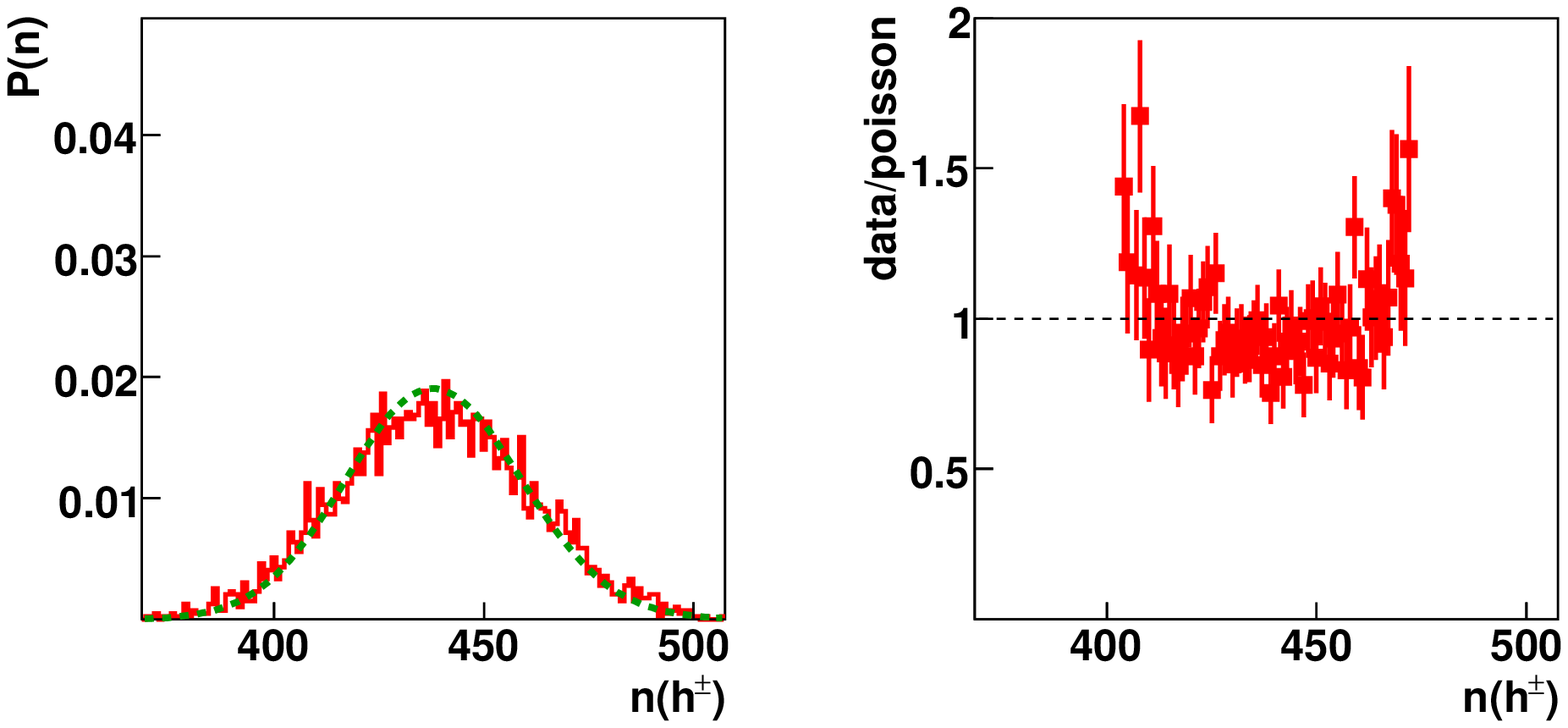}
\caption{\label{mult_dist_facc_hpm}(Color online) Left: multiplicity distributions 
of all charged  hadrons in full experimental acceptance in the 1\% most
central Pb+Pb collisions from $20A$ (top) to  $158A$ GeV (bottom).
The dashed lines indicate Poisson distributions with the same mean multiplicity as in data.
Right: the ratio of the measured multiplicity distribution to
the corresponding Poisson one.}
\end{figure}

\pagebreak

\begin{figure}[h!]
\includegraphics[width=8.2cm]{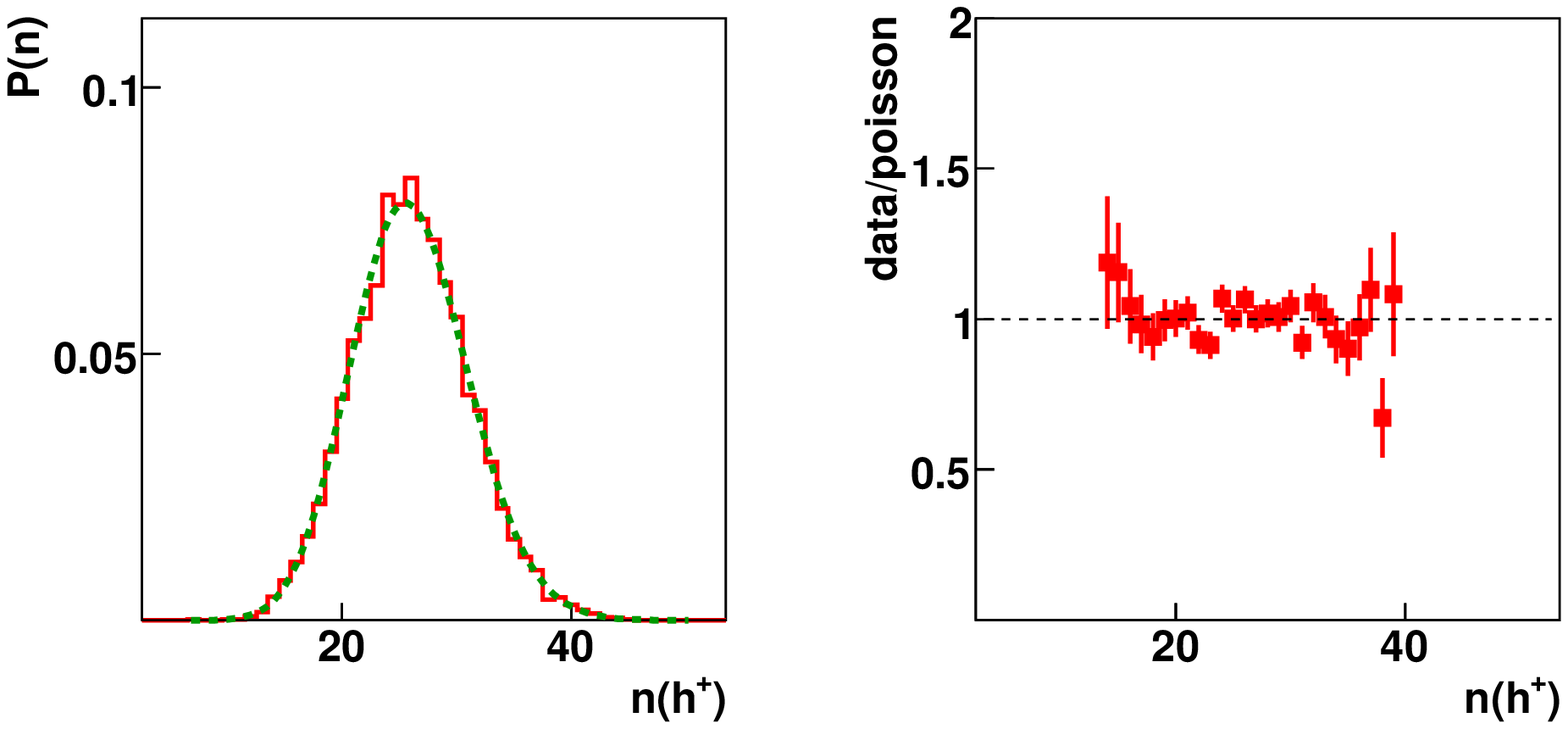}
\includegraphics[width=8.2cm]{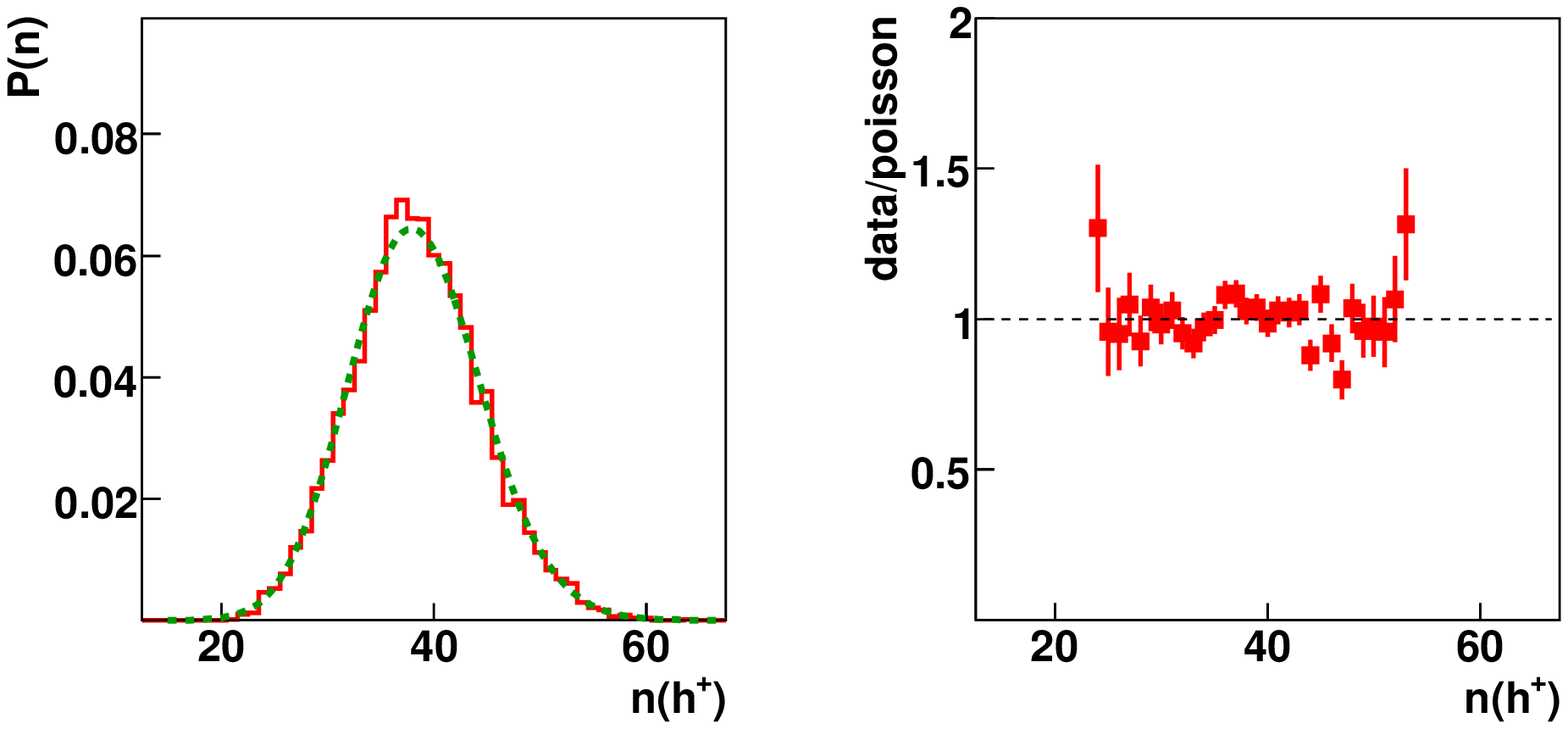}
\includegraphics[width=8.2cm]{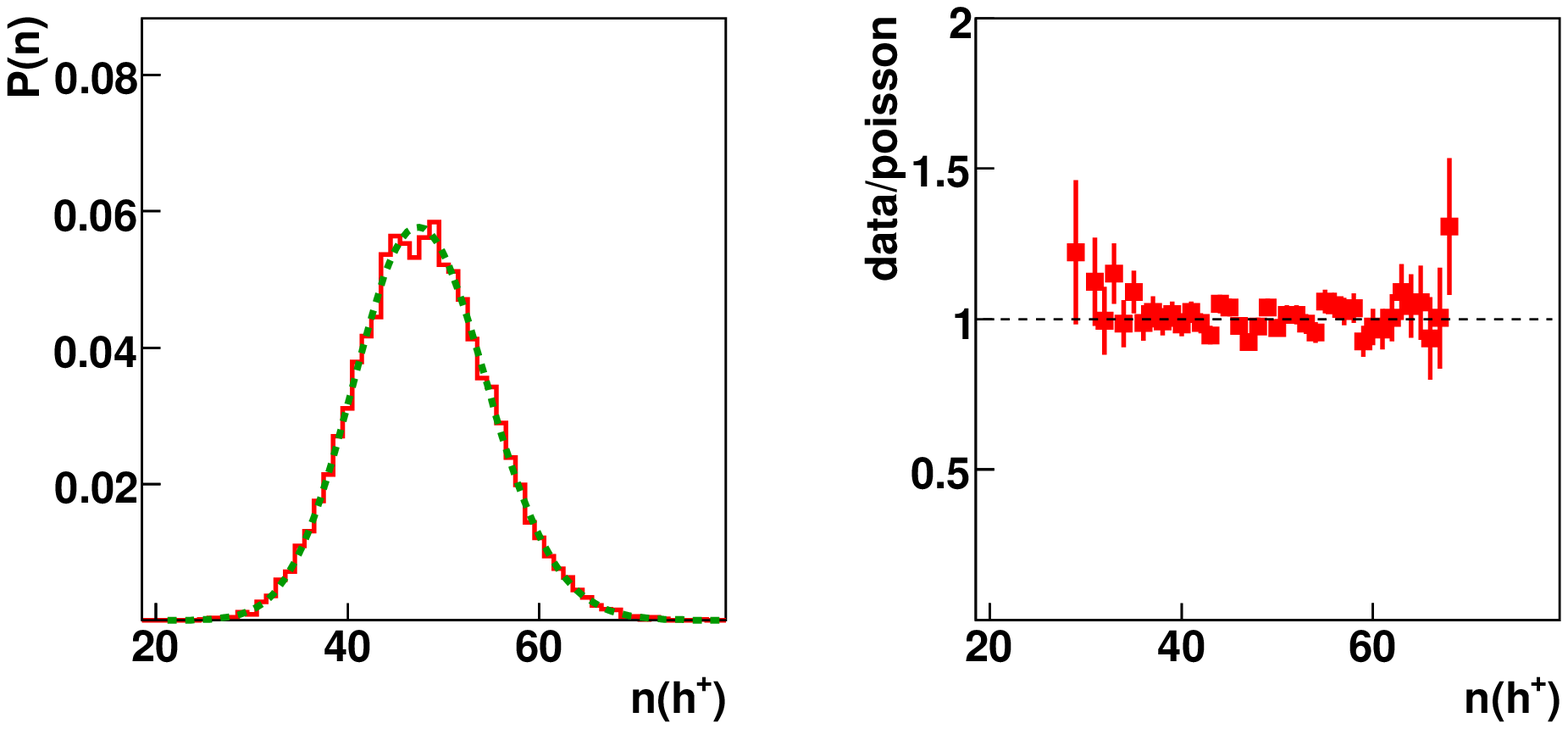}
\includegraphics[width=8.2cm]{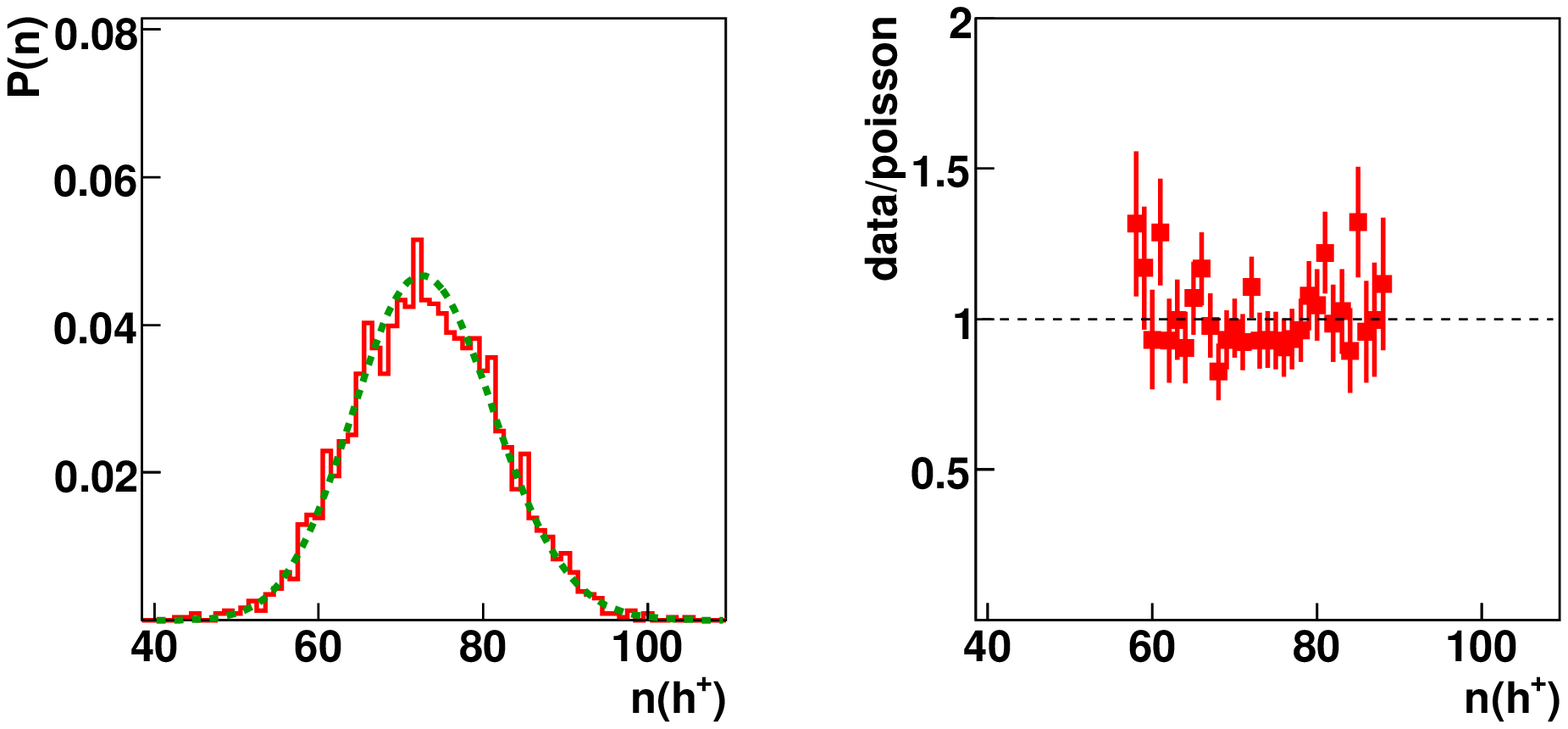}
\includegraphics[width=8.2cm]{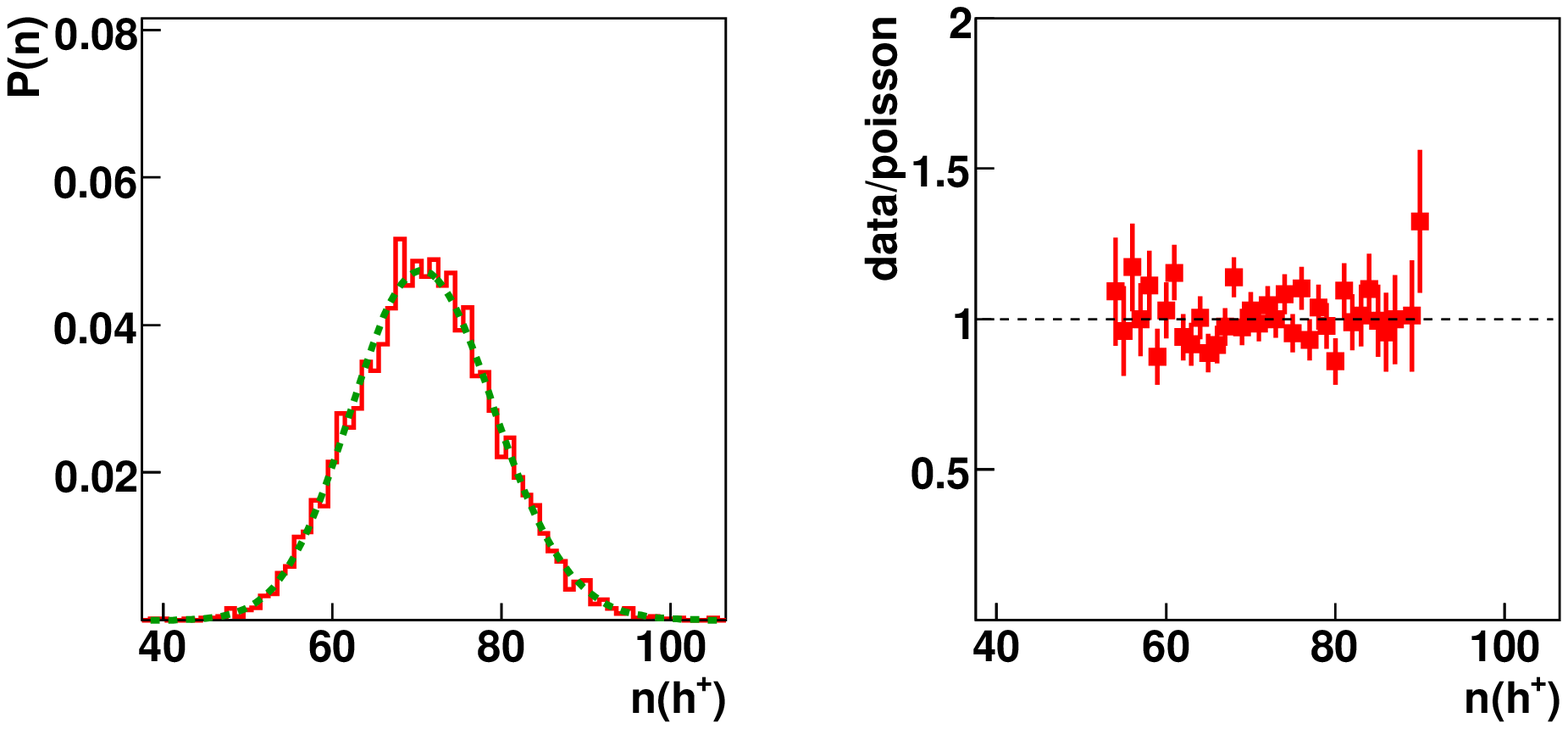}
\caption{\label{mult_dist_midr_hp}(Color online) Left: multiplicity distributions 
of positively charged  hadrons in midrapidity acceptance in the 1\% most
central Pb+Pb collisions from $20A$ (top) to  $158A$ GeV (bottom).
The dashed lines indicate Poisson distributions with the same mean multiplicity as in data.
Right: the ratio of the measured multiplicity distribution to
the corresponding Poisson one.}
\end{figure}

\pagebreak

\begin{figure}[h!]
\includegraphics[width=8.2cm]{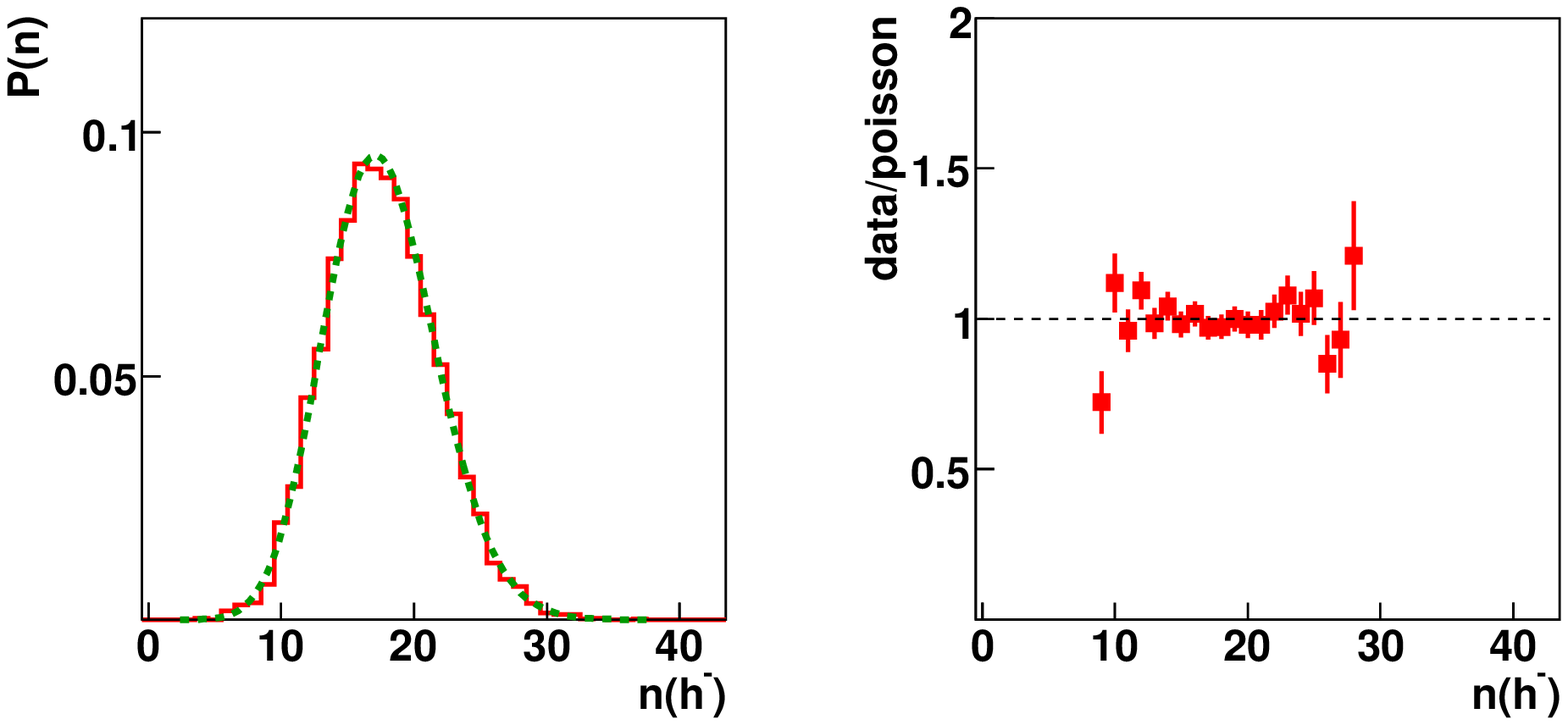}
\includegraphics[width=8.2cm]{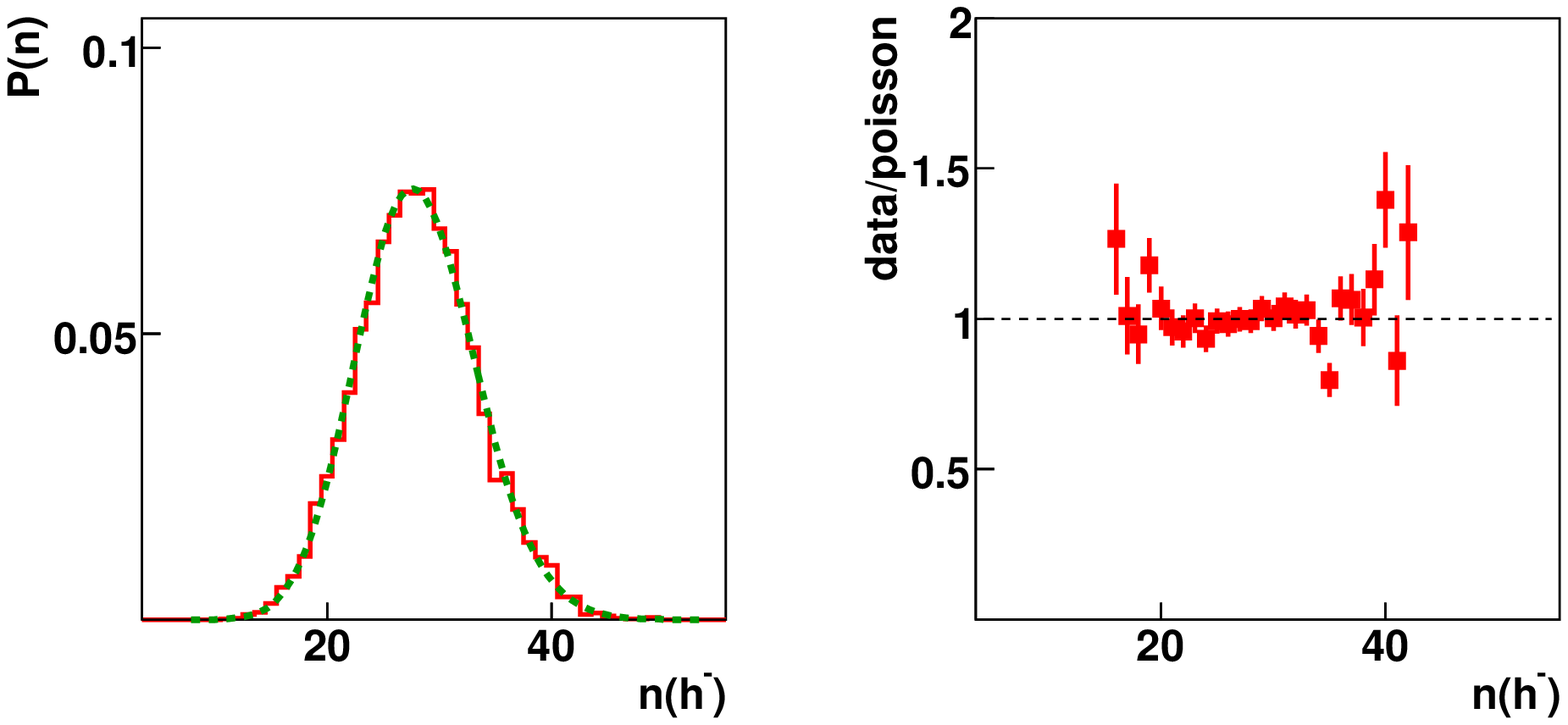}
\includegraphics[width=8.2cm]{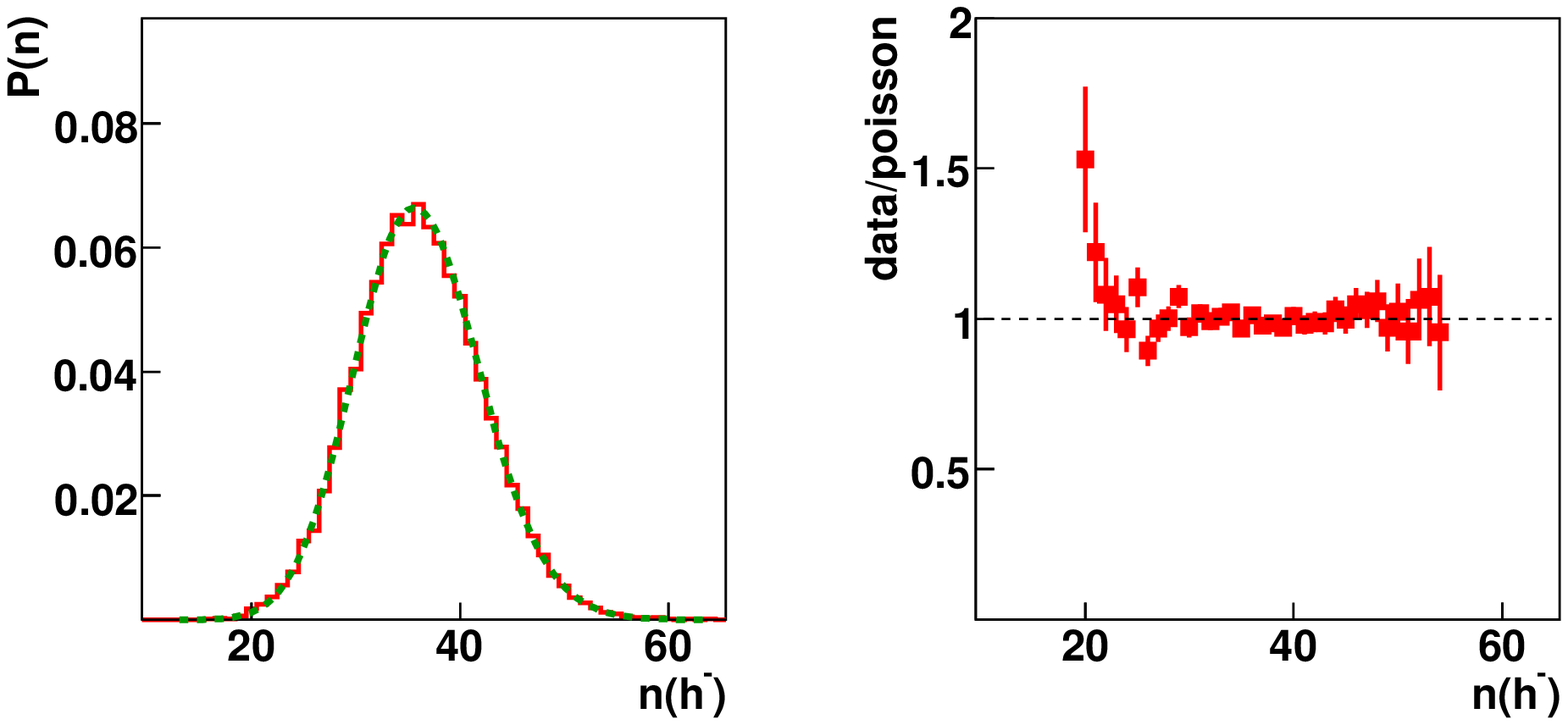}
\includegraphics[width=8.2cm]{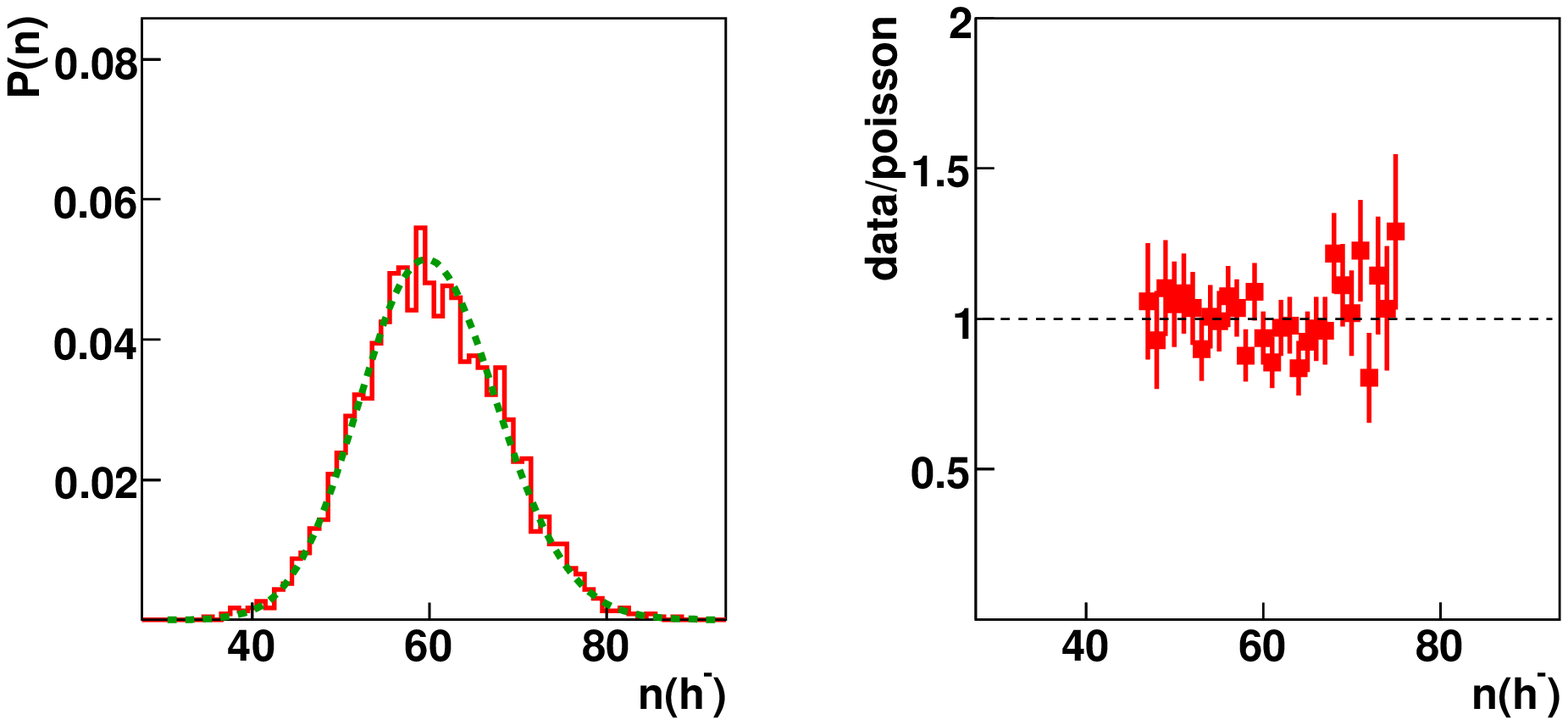}
\includegraphics[width=8.2cm]{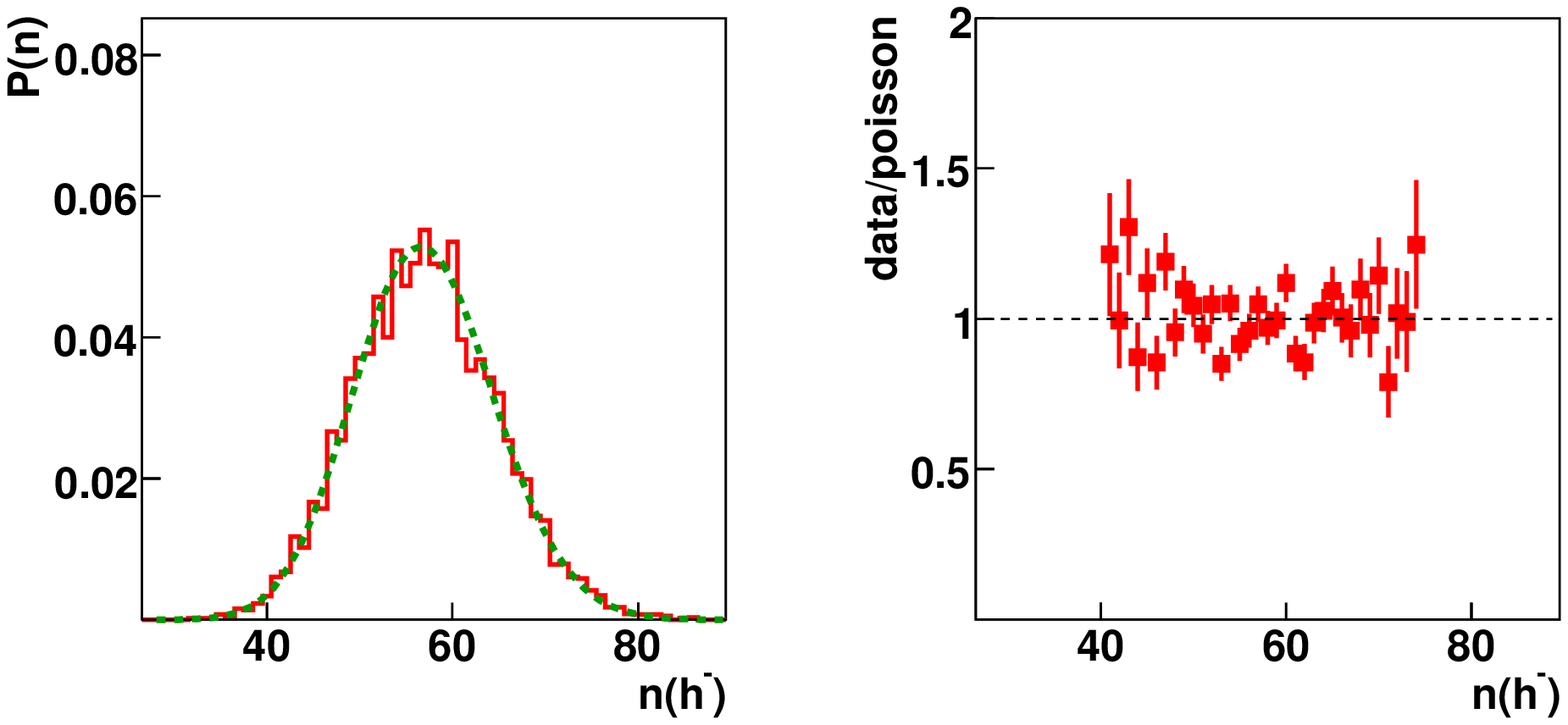}
\caption{\label{mult_dist_midr}(Color online) Left: multiplicity distributions 
of negatively charged  hadrons in midrapidity acceptance in the 1\% most
central Pb+Pb collisions from $20A$ (top) to  $158A$ GeV (bottom).
The dashed lines indicate Poisson distributions with the same mean multiplicity as in data.
Right: the ratio of the measured multiplicity distribution to
the corresponding Poisson one.}
\end{figure}

\pagebreak

\begin{figure}[h!]
\includegraphics[width=8.2cm]{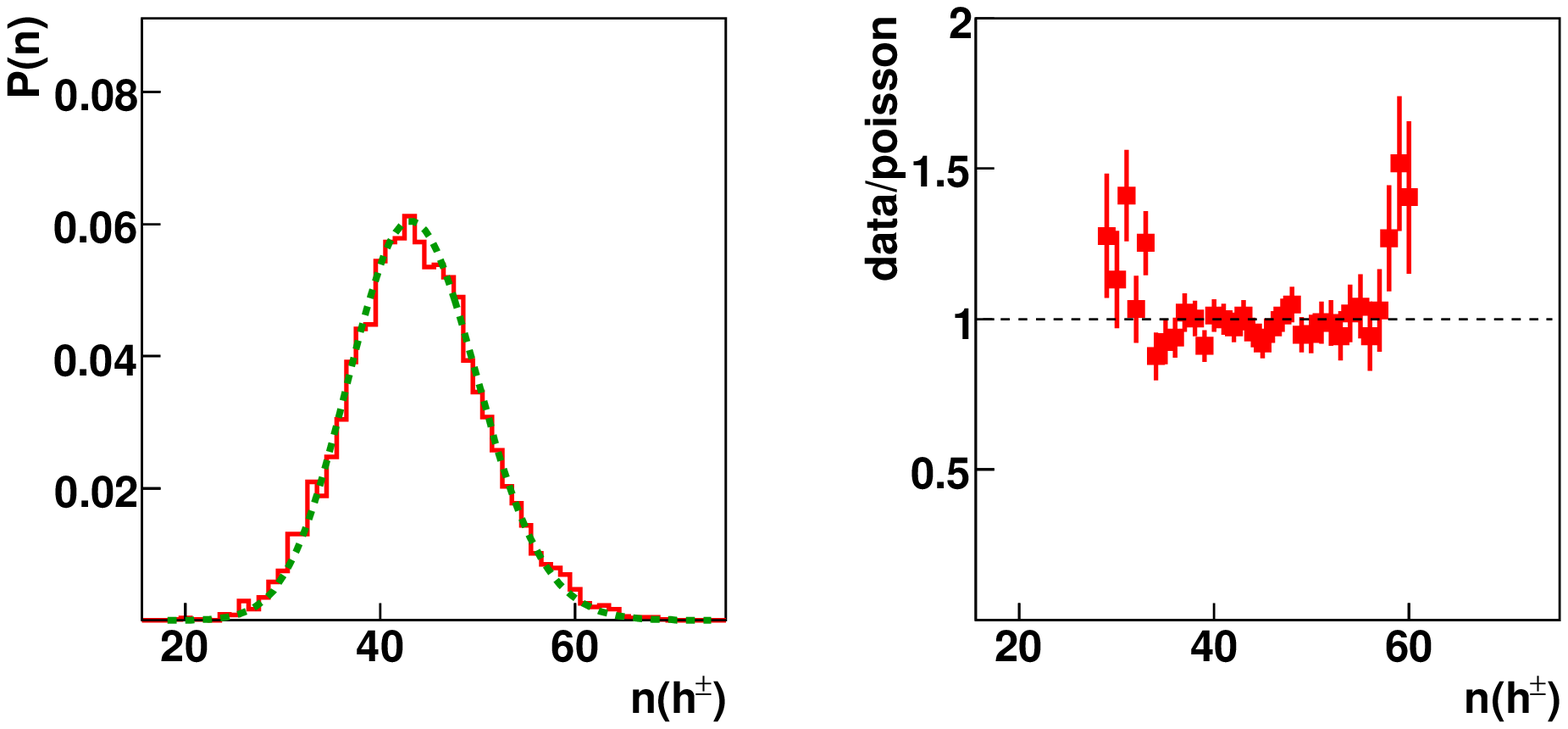}
\includegraphics[width=8.2cm]{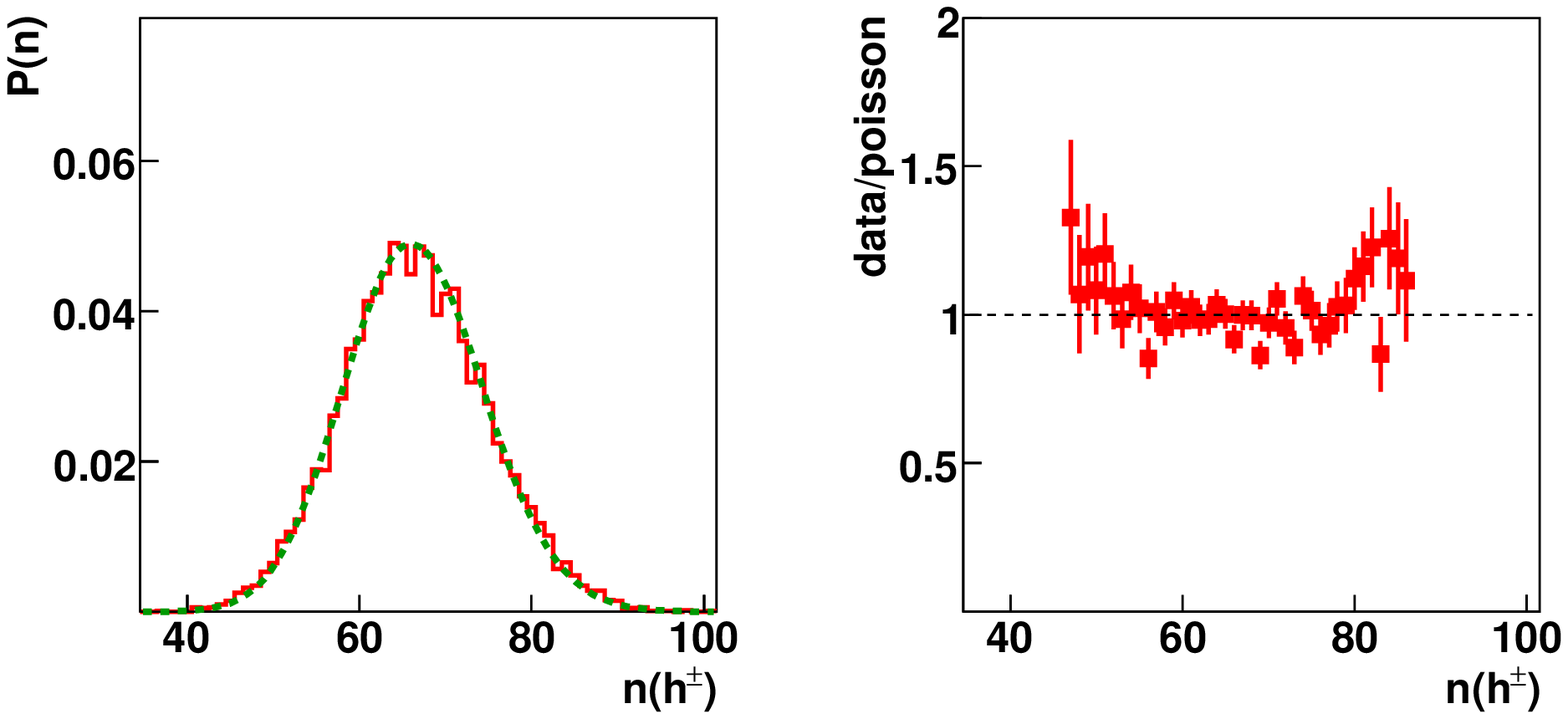}
\includegraphics[width=8.2cm]{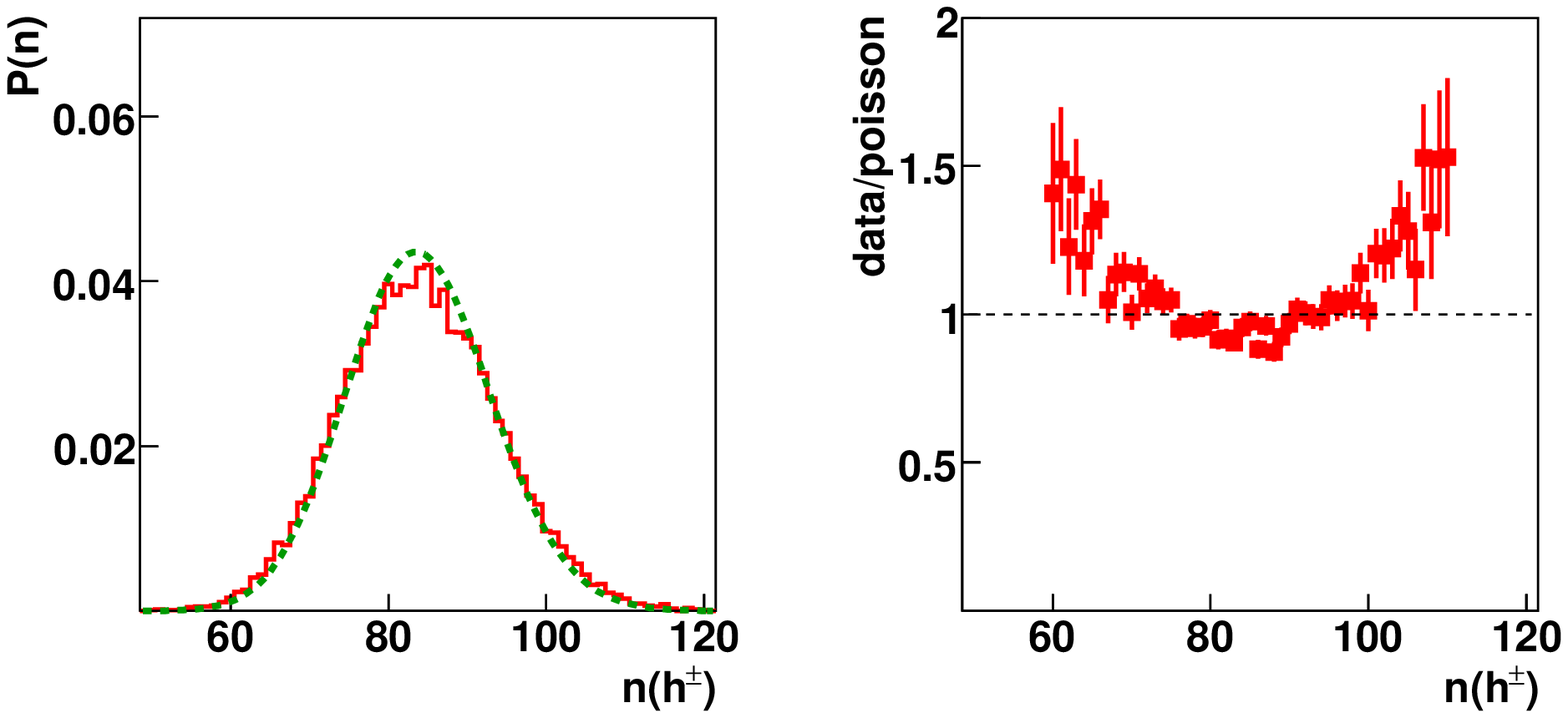}
\includegraphics[width=8.2cm]{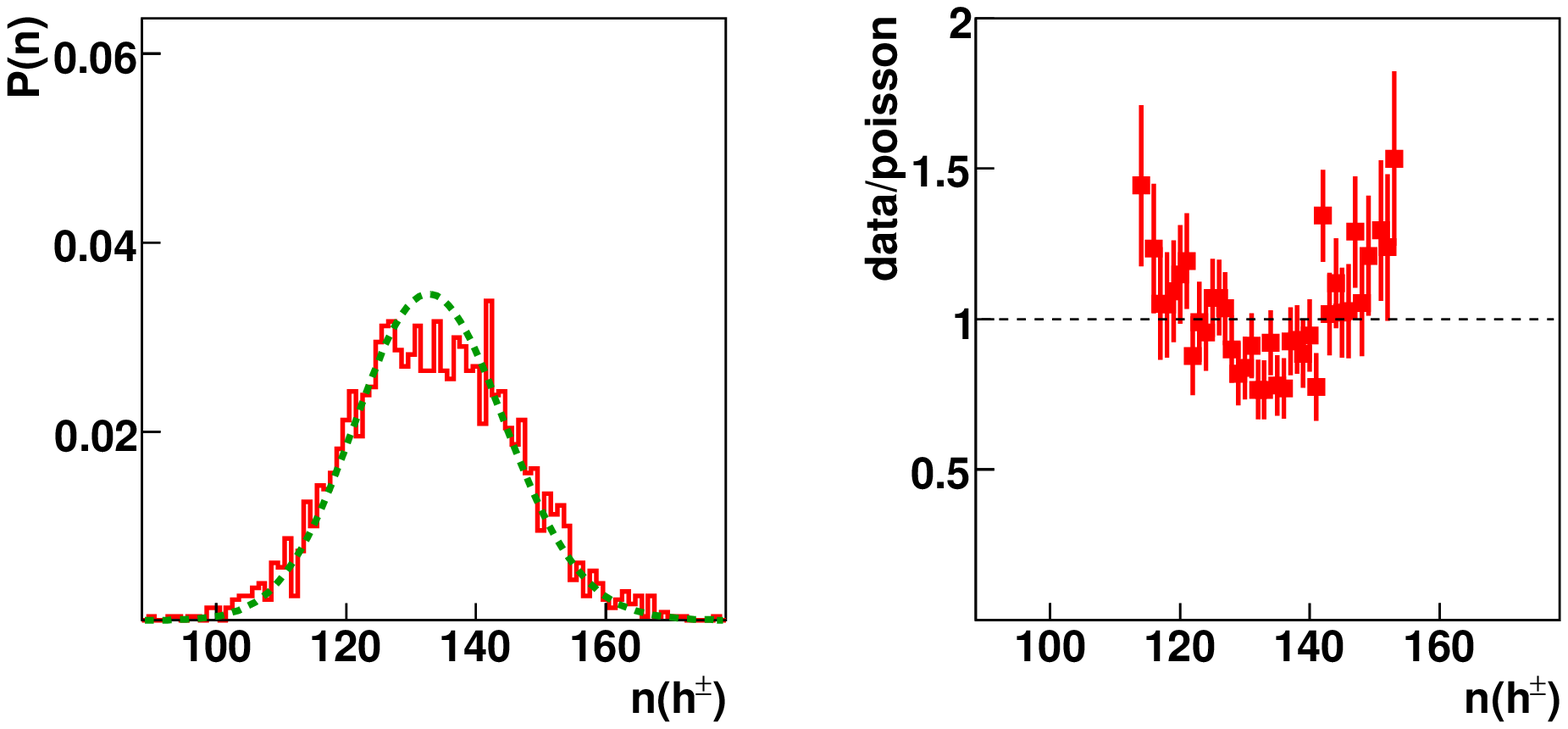}
\includegraphics[width=8.2cm]{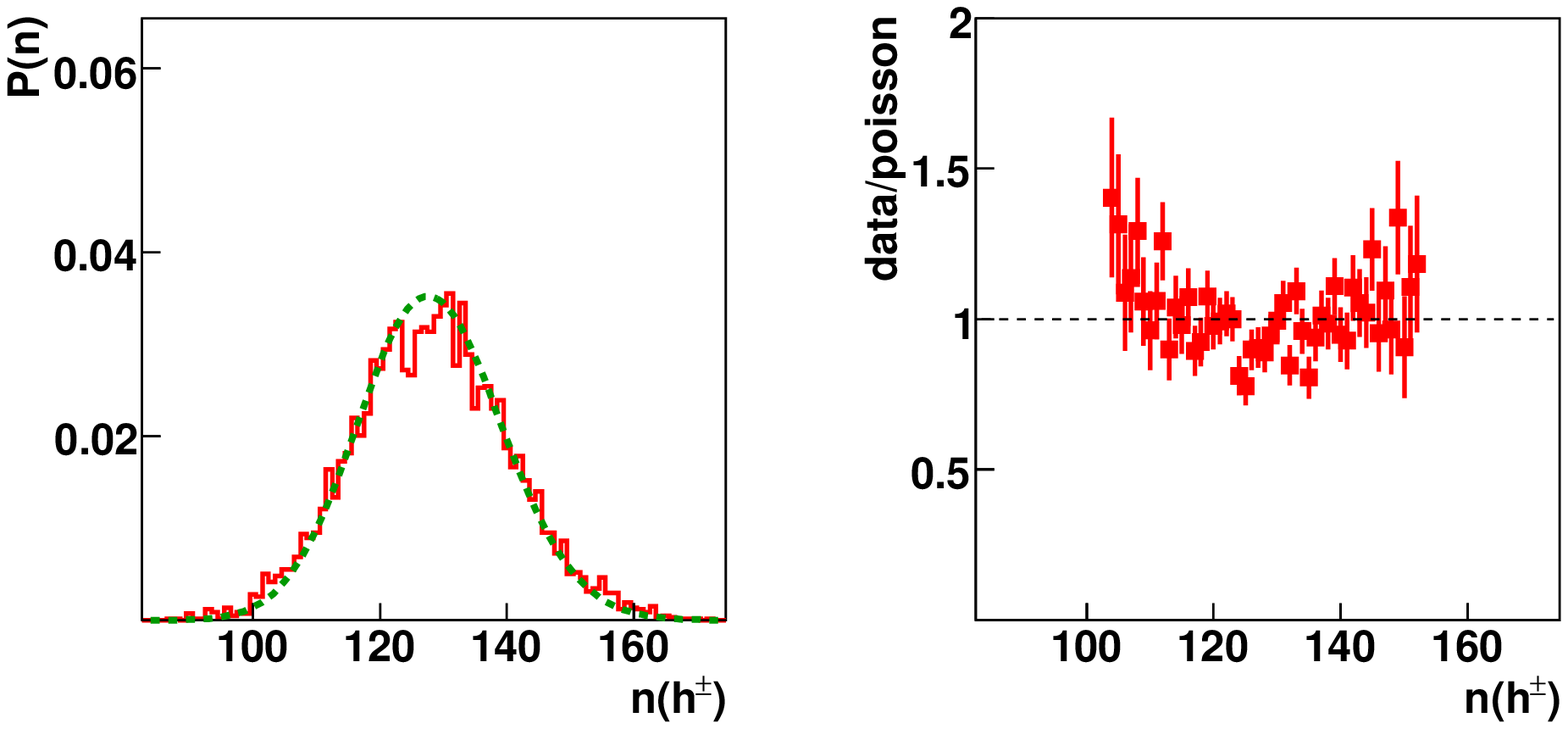}
\caption{\label{mult_dist_midr_hpm}(Color online) Left: multiplicity distributions 
of all charged  hadrons in midrapidity acceptance in the 1\% most
central Pb+Pb collisions from $20A$ (top) to  $158A$ GeV (bottom).
The dashed lines indicate Poisson distributions with the same mean multiplicity as in data.
Right: the ratio of the measured multiplicity distribution to
the corresponding Poisson one.}
\end{figure}

\pagebreak

\begin{figure}[h!]
\includegraphics[width=8.2cm]{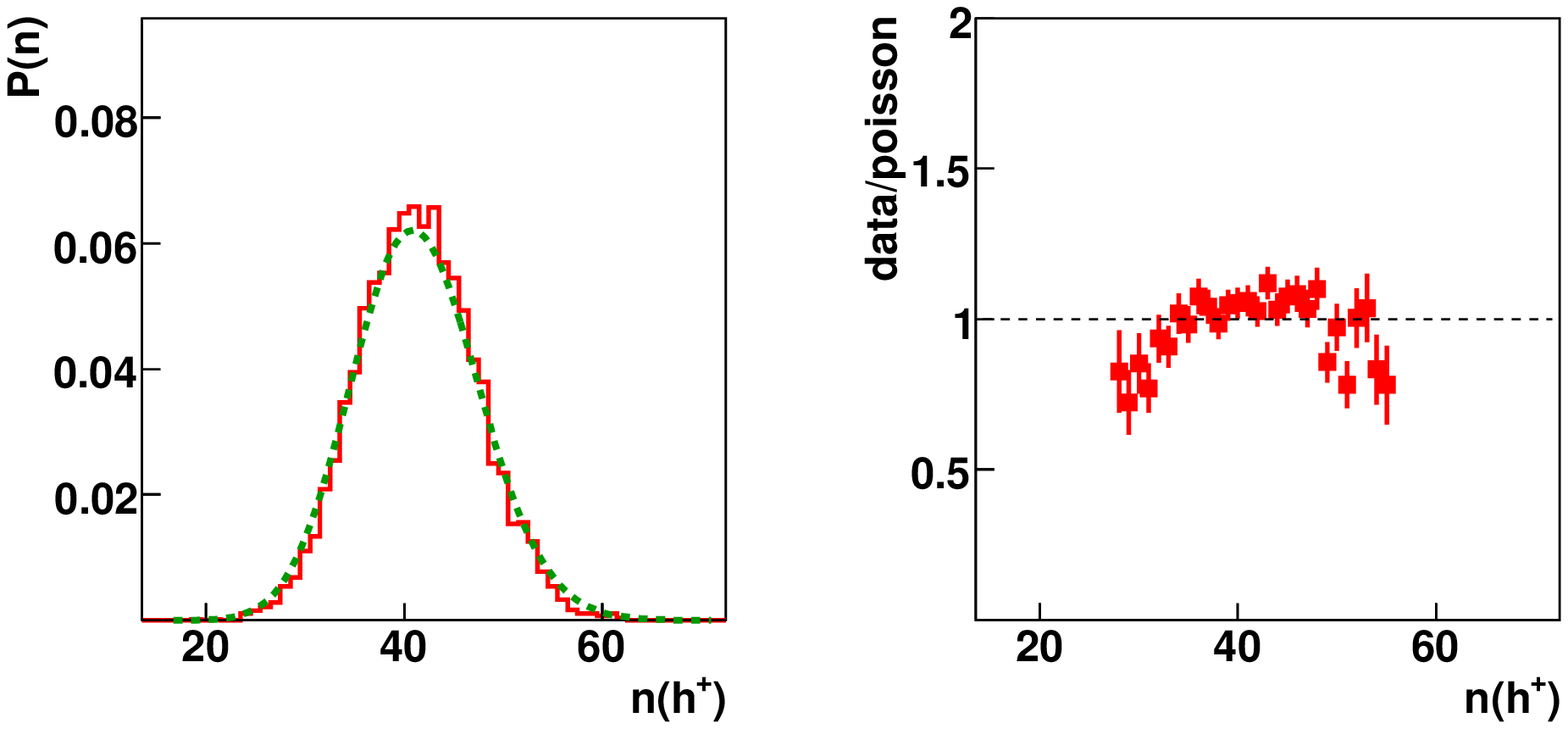}
\includegraphics[width=8.2cm]{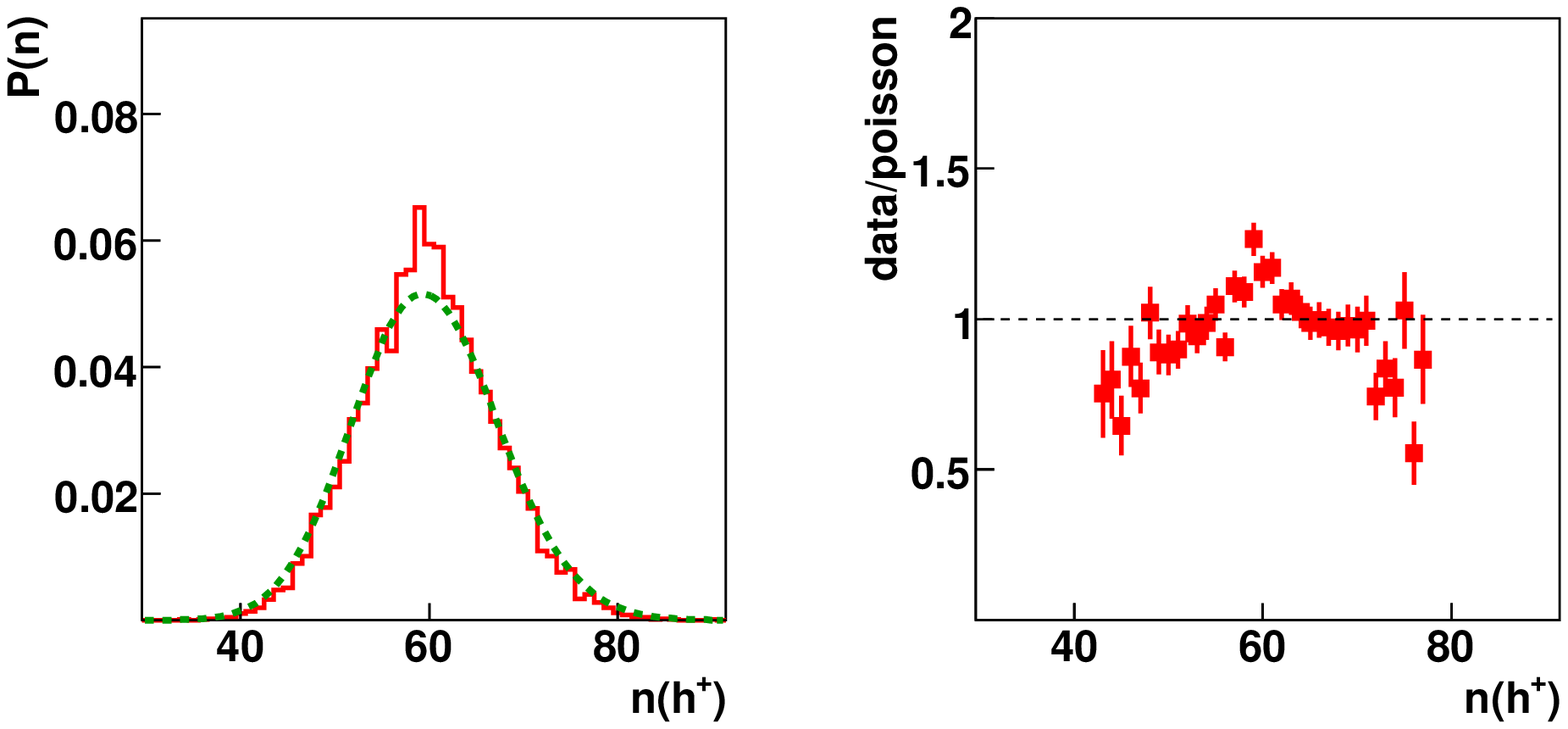}
\includegraphics[width=8.2cm]{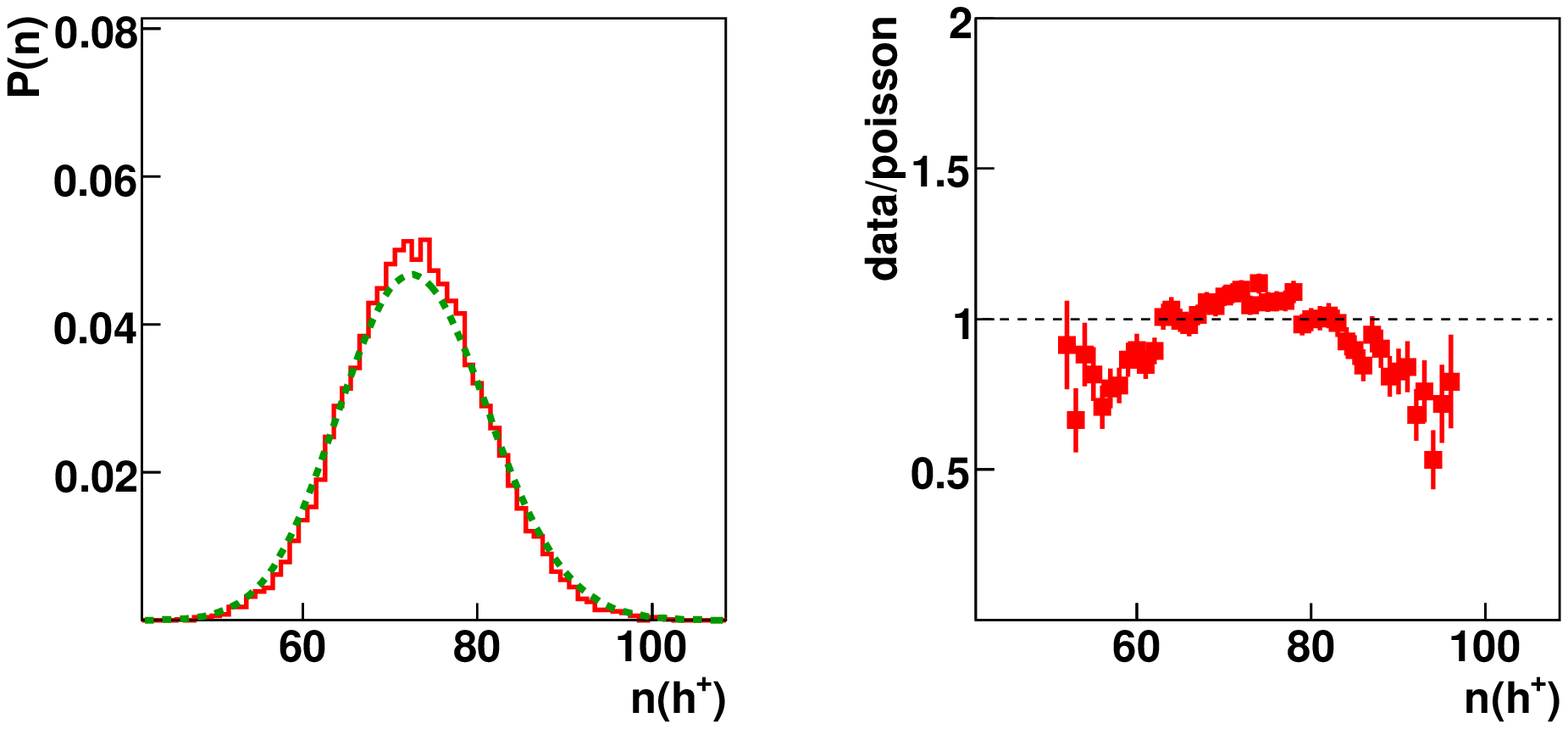}
\includegraphics[width=8.2cm]{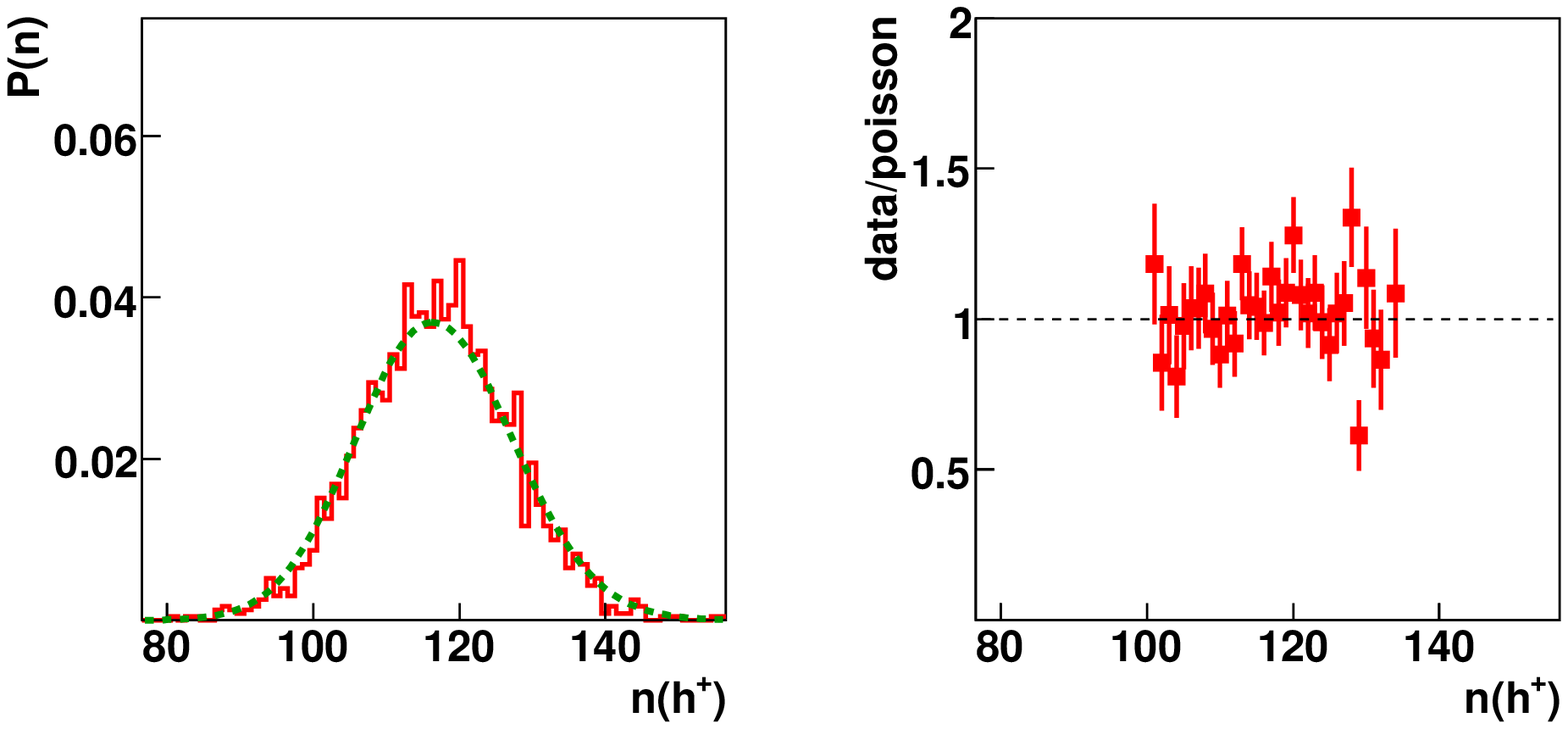}
\includegraphics[width=8.2cm]{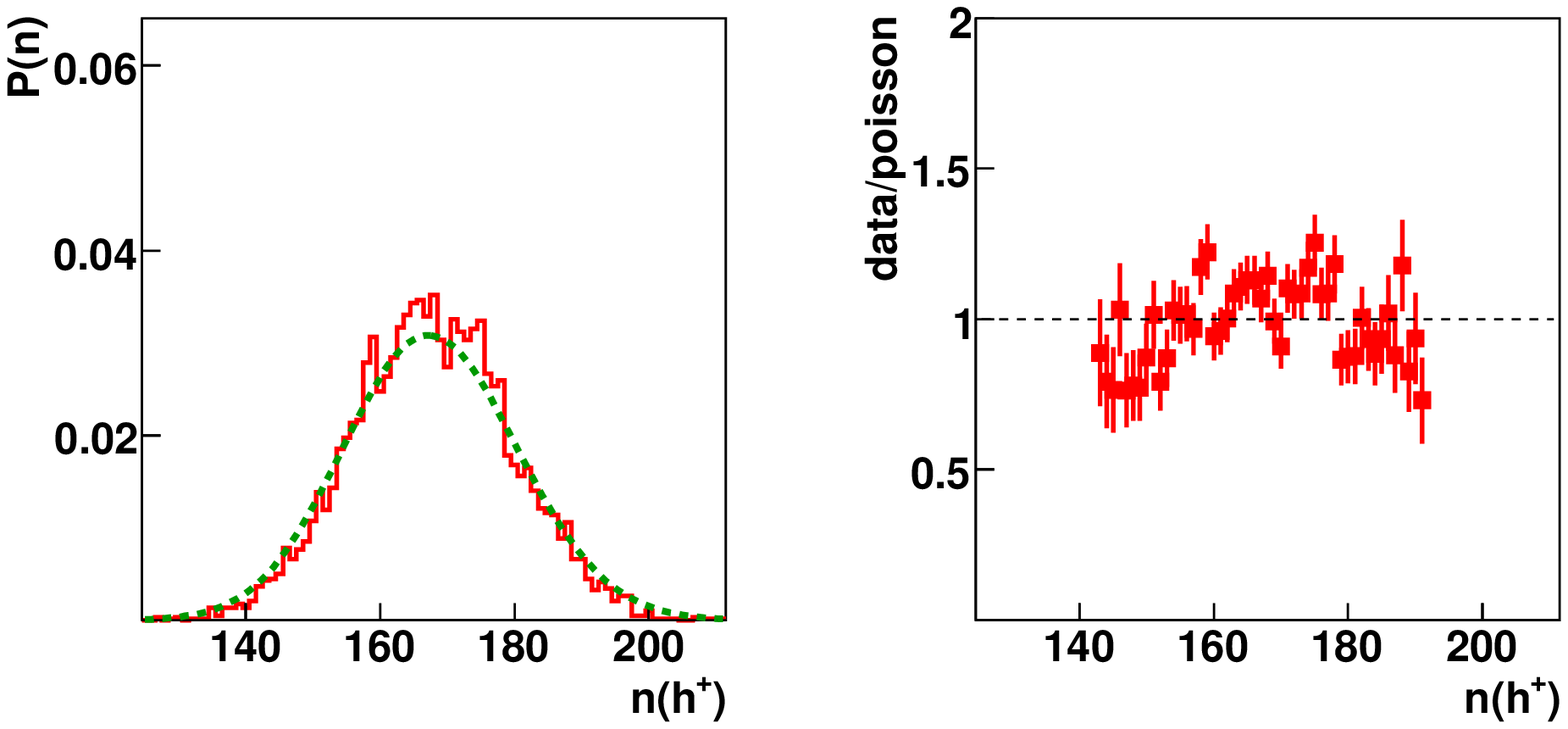}
\caption{\label{mult_dist_hp}(Color online) Left: multiplicity distributions 
of positively charged  hadrons in forward acceptance in the 1\% most
central Pb+Pb collisions from $20A$ (top) to  $158A$ GeV (bottom).
The dashed lines indicate Poisson distributions with the same mean multiplicity as in data.
Right: the ratio of the measured multiplicity distribution to
the corresponding Poisson one.}
\end{figure}

\pagebreak

\begin{figure}[h!]
\includegraphics[width=8.2cm]{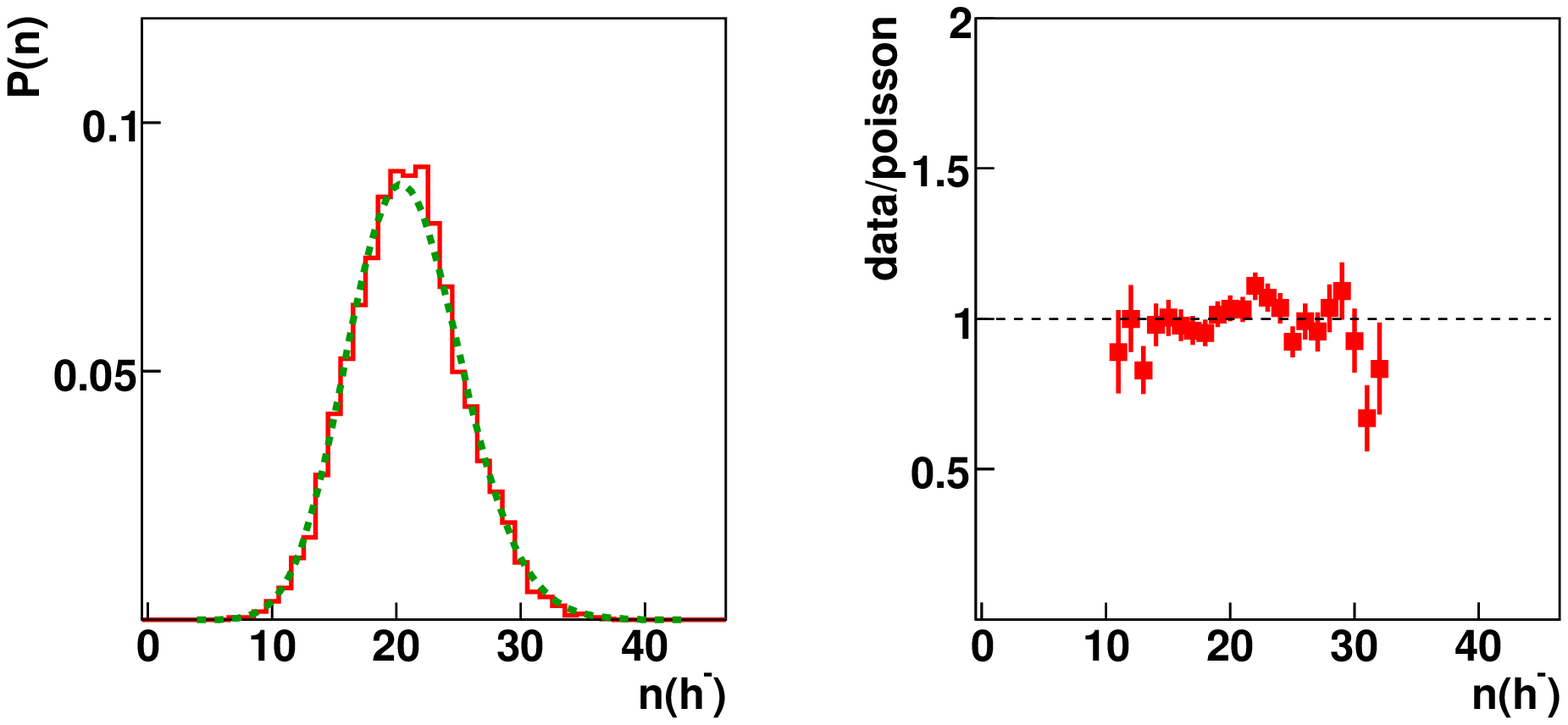}
\includegraphics[width=8.2cm]{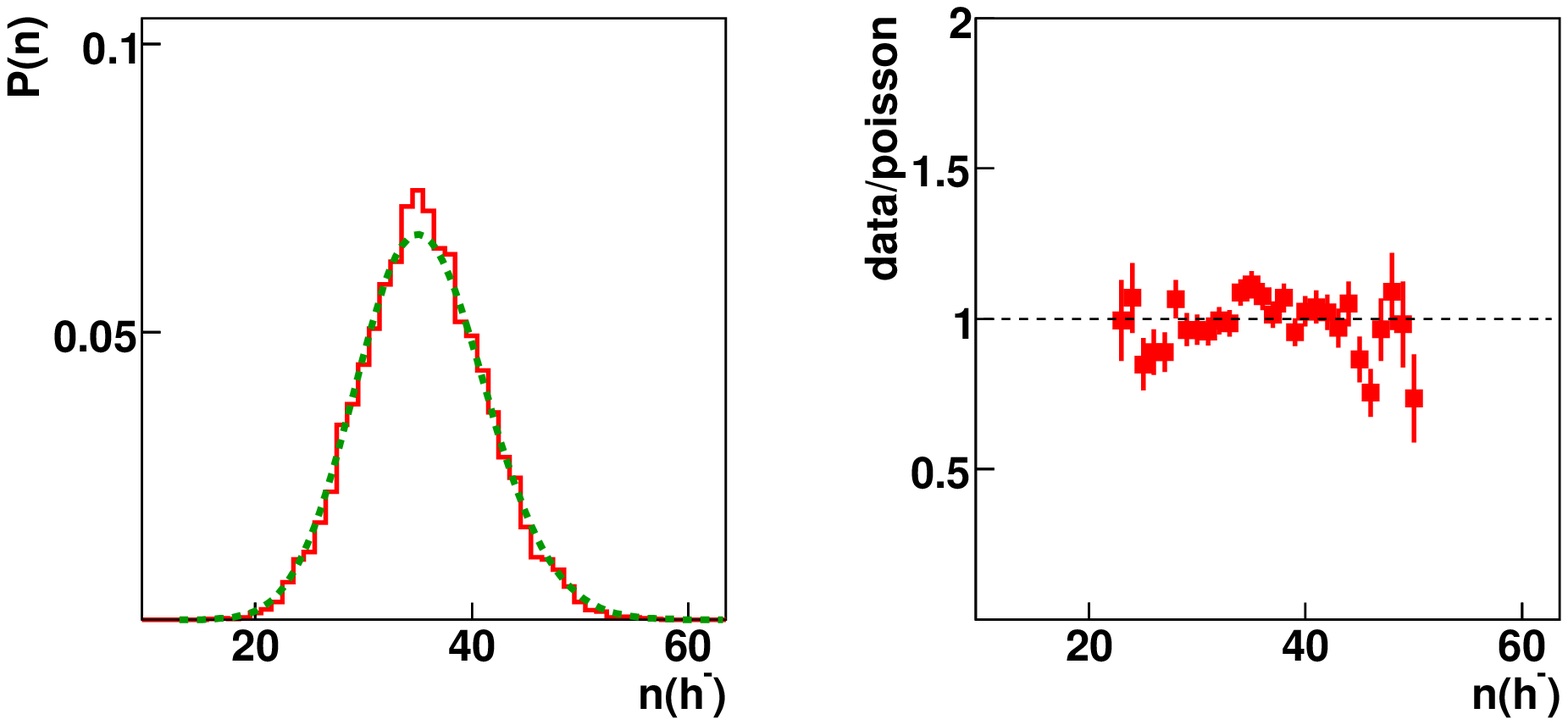}
\includegraphics[width=8.2cm]{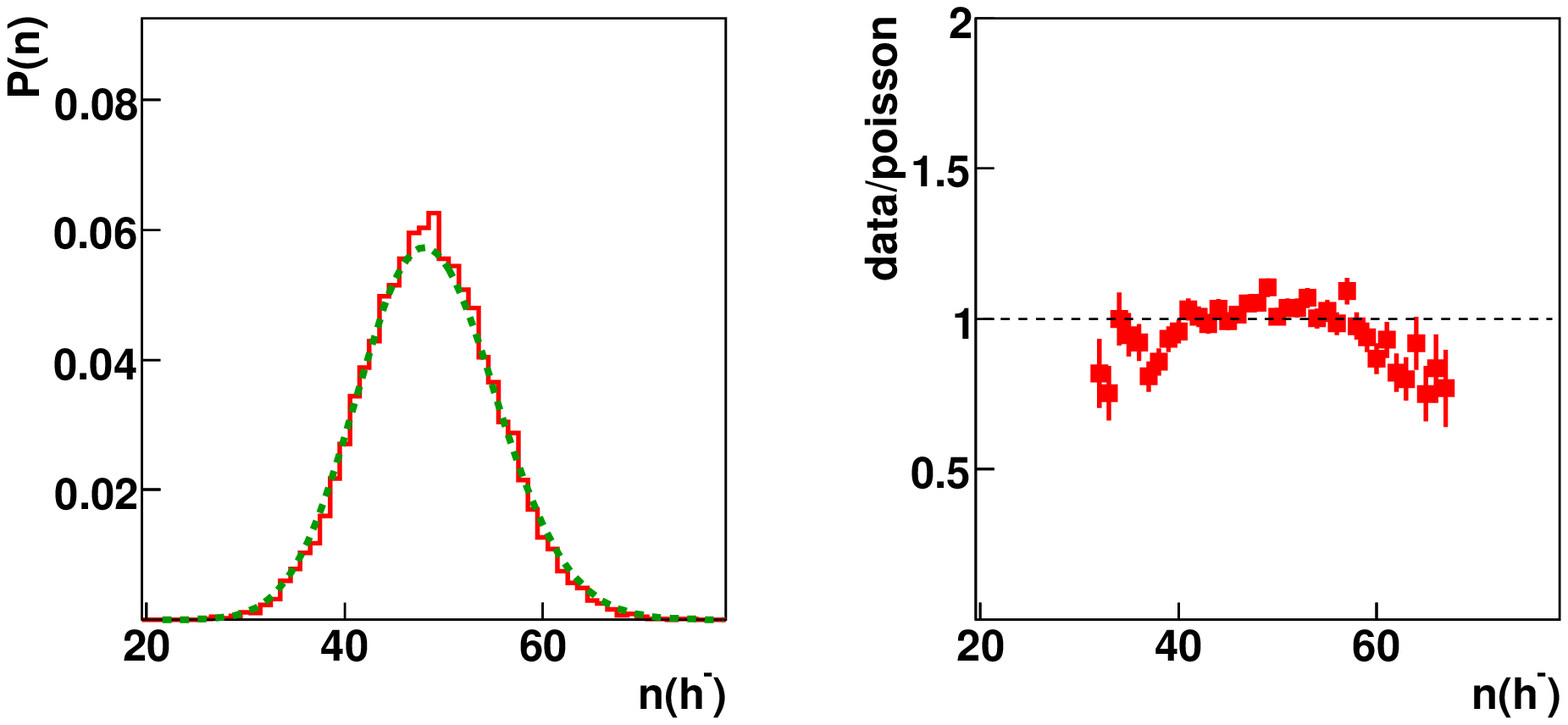}
\includegraphics[width=8.2cm]{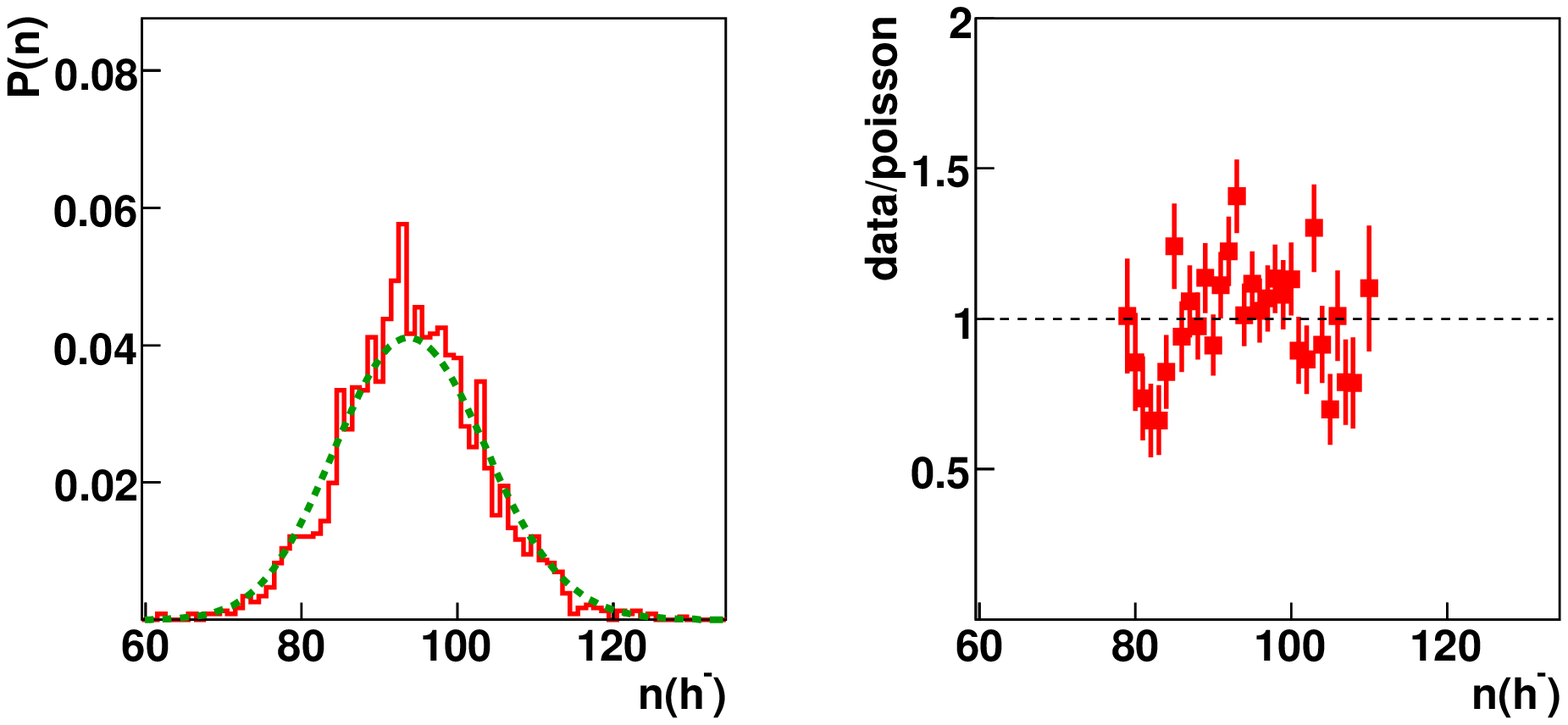}
\includegraphics[width=8.2cm]{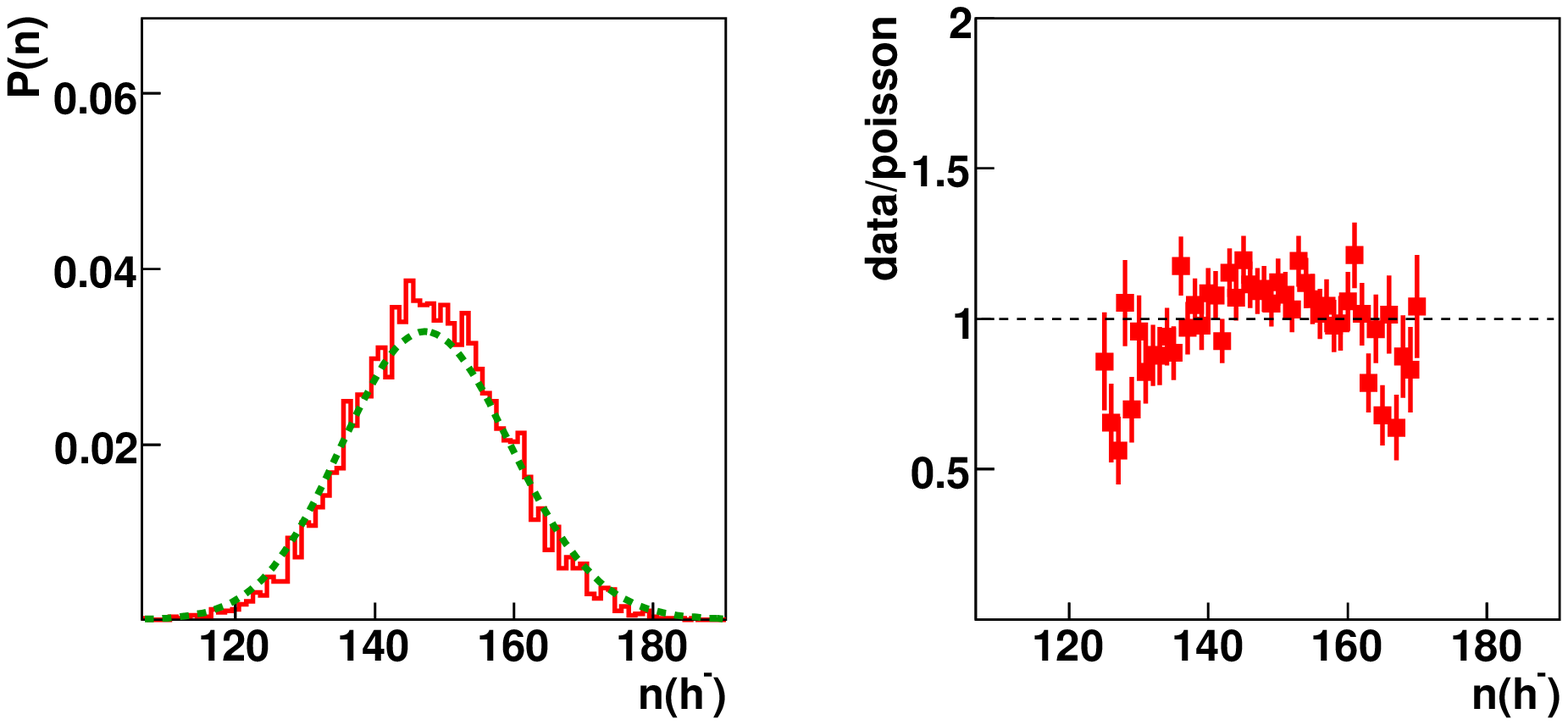}
\caption{\label{mult_dist}(Color online) Left: multiplicity distributions 
of negatively charged  hadrons in forward acceptance in the 1\% most
central Pb+Pb collisions from $20A$ (top) to  $158A$ GeV (bottom).
The dashed lines indicate Poisson distributions with the same mean multiplicity as in data.
Right: the ratio of the measured multiplicity distribution to
the corresponding Poisson one.}
\end{figure}

\pagebreak

\begin{figure}[h!]
\includegraphics[width=8.2cm]{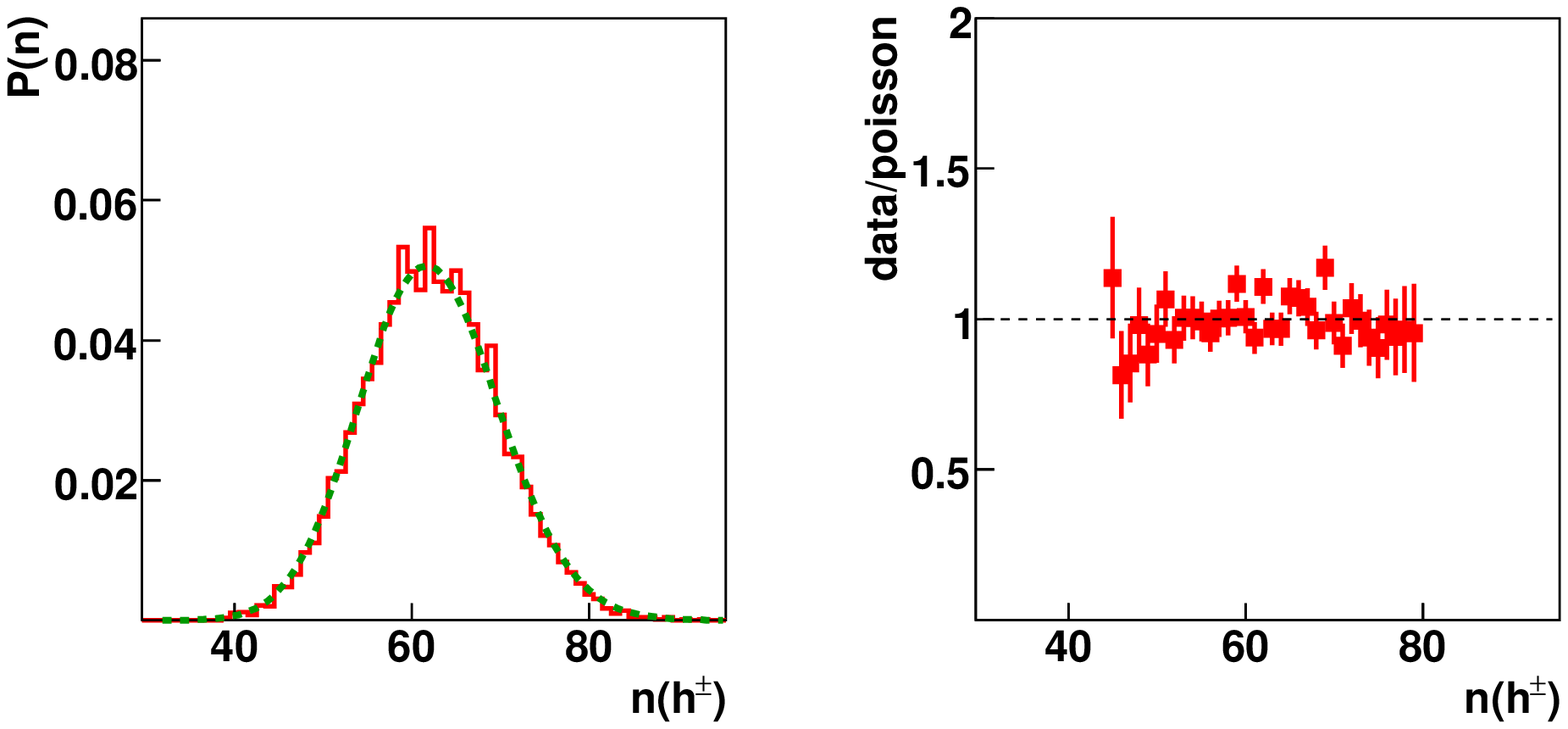}
\includegraphics[width=8.2cm]{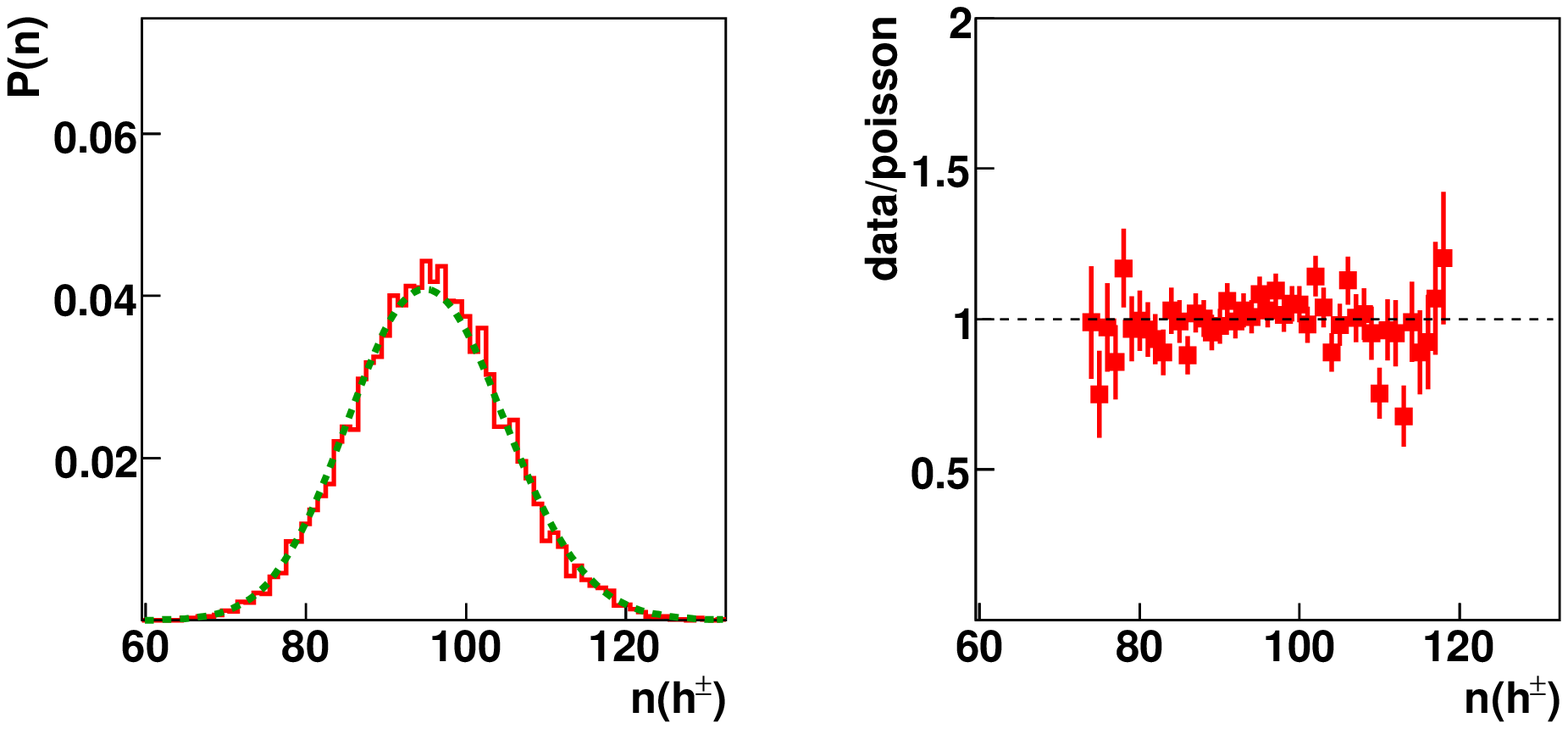}
\includegraphics[width=8.2cm]{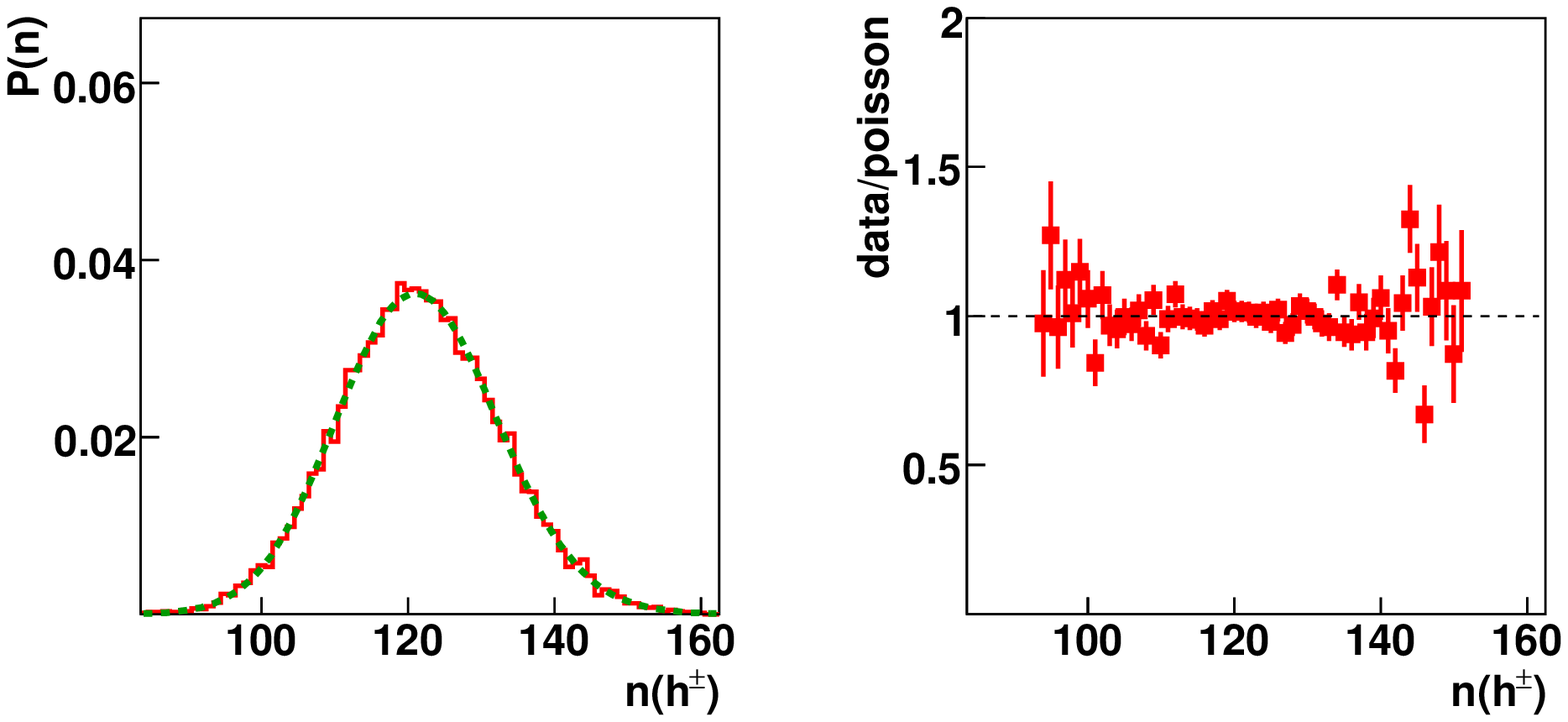}
\includegraphics[width=8.2cm]{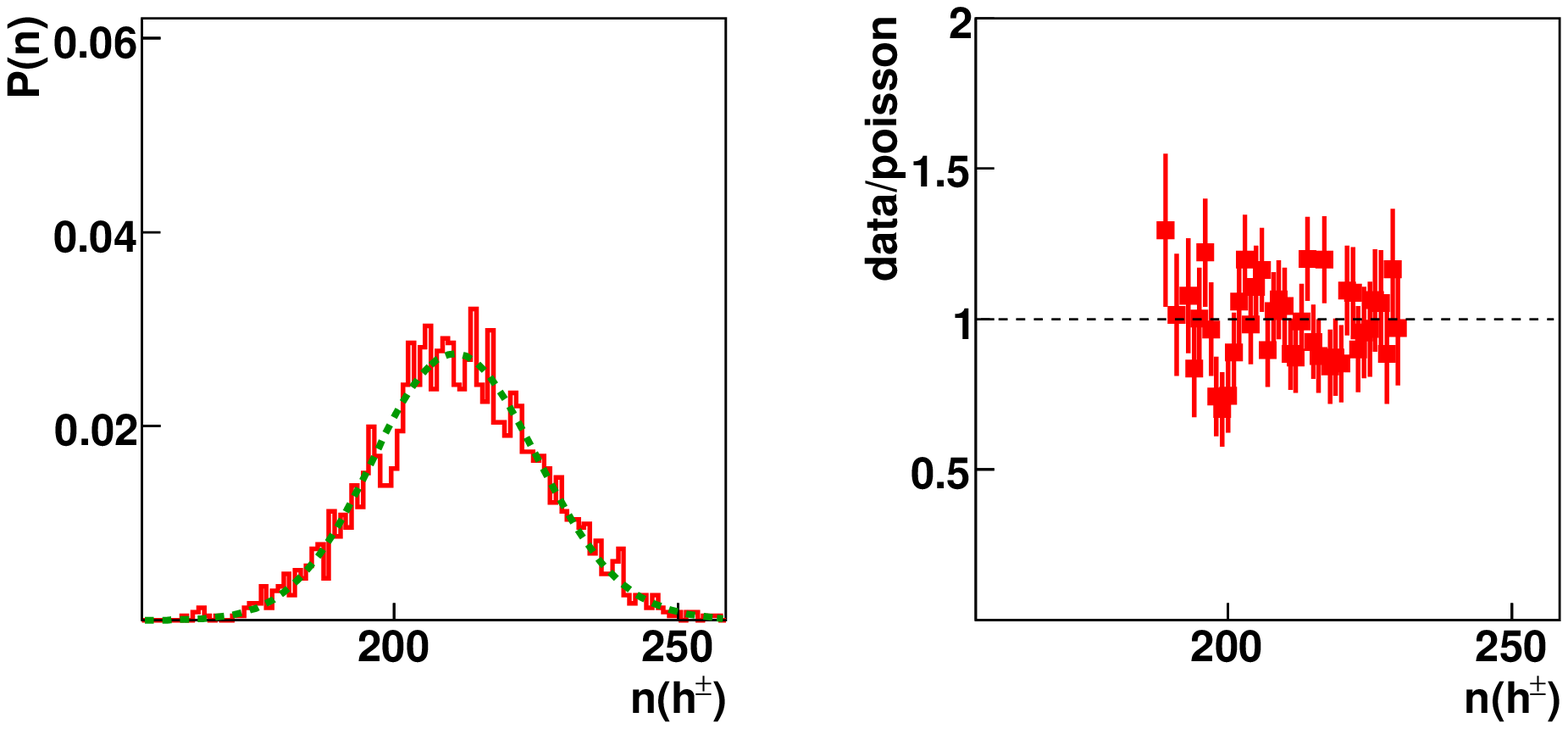}
\includegraphics[width=8.2cm]{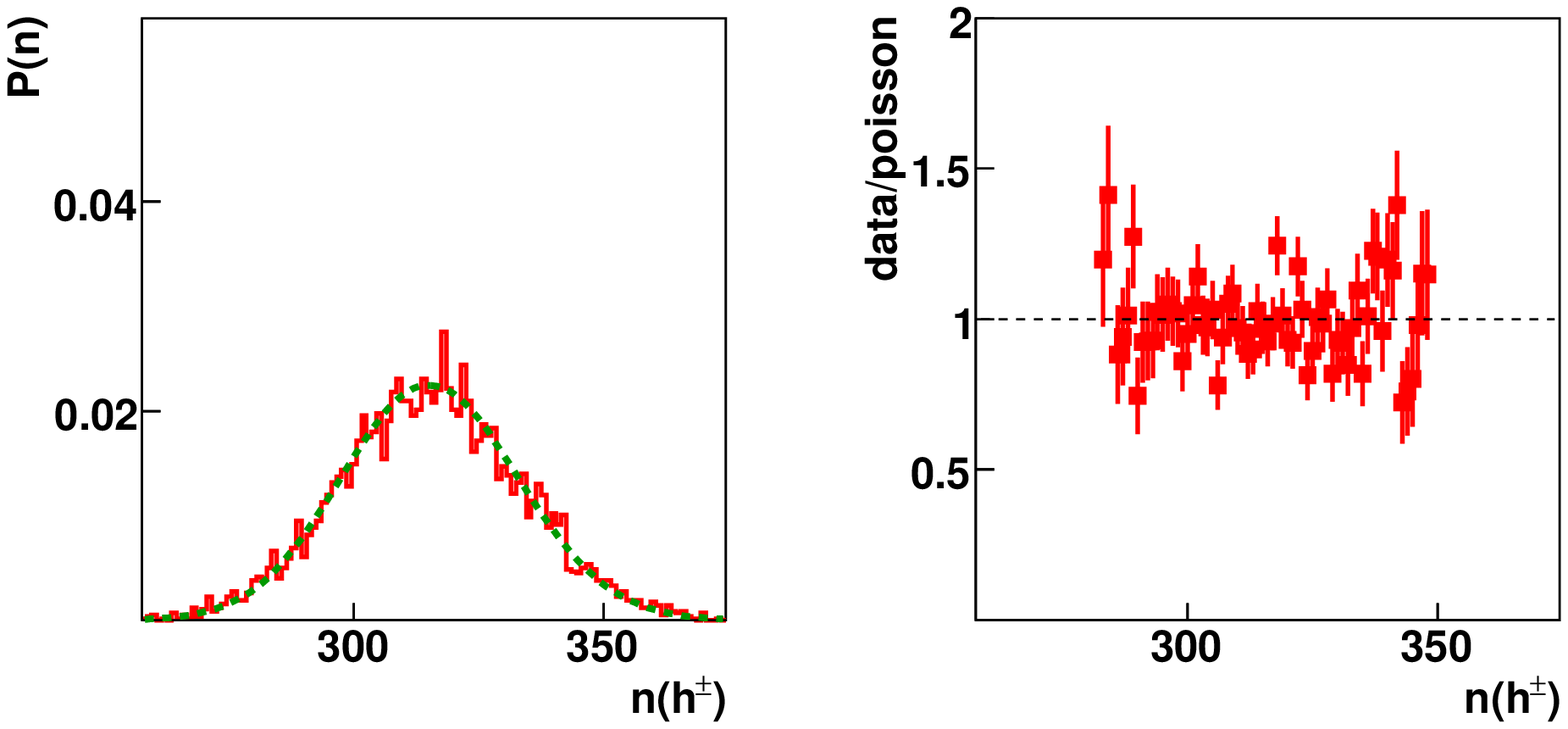}
\caption{\label{mult_dist_hpm}(Color online) Left: multiplicity distributions 
of all charged  hadrons in forward acceptance in the 1\% most
central Pb+Pb collisions from $20A$ (top) to  $158A$ GeV (bottom).
The dashed lines indicate Poisson distributions with the same mean multiplicity as in data.
Right: the ratio of the measured multiplicity distribution to
the corresponding Poisson one.}
\end{figure}

\pagebreak

\end{document}